\documentclass[12pt]{article}
\usepackage{amsmath, amssymb}
\usepackage{graphicx}
\usepackage{float}
\usepackage{empheq}
\usepackage[caption=false,labelformat=simple]{subfig}
\DeclareGraphicsExtensions{.pdf,.png,.jpg}

\numberwithin{equation}{section}

\newcommand{\bea}{\begin{eqnarray}}
\newcommand{\eea}{\end{eqnarray}}

\newcommand{\ret}{\nonumber \\}
\newcommand{\nn}{\nonumber}

\newcommand{\refb}[1]{(\ref{#1})}

\newcommand{\cA}{ {\cal{A}} }

\newcommand{\bra}[1]{\langle#1|}
\newcommand{\ket}[1]{|#1\rangle}

\newcommand{\cor}[1]{\langle#1 \rangle}

\newcommand{\tr}{\mbox{tr}}

\newcommand{\lsu}{su(2)}
\newcommand{\qsu}{U_q(su(2))}

\newcommand{\wigthreej}[6]{\left(\begin{array}{ccc} #1&#2&#3\\#4&#5&#6\end{array}\right)}
\newcommand{\wsj}[6]{\left| \begin{array}{ccc} #1&#2&#3\\#4&#5&#6\end{array}\right|}
\newcommand{\wsjj}[3]{\left| \begin{array}{ccc} #1&#2&#3\\j&j&j\end{array}\right|}
\newcommand{\wsjq}[6]{\left| \begin{array}{ccc} #1&#2&#3\\#4&#5&#6\end{array} \right|_q}
\newcommand{\wsjjq}[3]{\left| \begin{array}{ccc} #1&#2&#3\\j&j&j\end{array} \right|_q}

\newcommand{\sG}{S[{\cal G}]}
\newcommand{\sGl}{S[{\cal G}, \{l_i\}]}

\newcommand{\mR}{\mathbb{R}}

\newcommand{\mG}{{\cal M}[{\cal G}]}

\newcommand{\m}[1]{{\cal M}\left[  \raisebox{-12pt}{\includegraphics[height=32pt]{#1}} \right]}
\newcommand{\mtwo}[1]{{\cal M}\left[  \raisebox{-28pt}{\includegraphics[height=64pt]{#1}} \right]}

\newcommand{\rG}{{\cal R [G]}}

\newcommand{\ra}[1]{{\cal R}\left[  \raisebox{-12pt}{\includegraphics[height=32pt]{#1}} \right]}
\newcommand{\rb}[1]{{\cal R}\left[  \raisebox{-28pt}{\includegraphics[height=64pt]{#1}} \right]}

\newcommand{\G}{{\cal G}}
\newcommand{\R}{{\cal R}}

\newcommand{\boxeq}[1]{\begin{empheq}[box={\fboxsep=10pt\fbox}]{align} #1 \end{empheq}}

\begin{document}

\rightline{QMUL-PH-13-03}

\vspace*{2cm} 

{\LARGE{  
\centerline{   \bf Holographic Hierarchy in the Gaussian Matrix } 
 \centerline {\bf  Model via the Fuzzy Sphere  } 
}}  

\vskip.5cm 

\thispagestyle{empty} \centerline{
    {\large \bf David Garner
${}^{a,} $\footnote{ {\tt d.p.r.garner@qmul.ac.uk}}}
   {\large \bf and Sanjaye Ramgoolam
               ${}^{a,}$\footnote{ {\tt s.ramgoolam@qmul.ac.uk}}   }
                                                       }

\vspace{.4cm}
\centerline{{\it ${}^a$ Centre for Research in String Theory,}}
\centerline{ {\it School of Physics and Astronomy},}
\centerline{{ \it Queen Mary University of London},} 
\centerline{{\it    Mile End Road, London E1 4NS, UK}}

\vspace{1.4truecm}

\thispagestyle{empty}

\centerline{\bf ABSTRACT}
\vspace*{1cm} 

The Gaussian Hermitian matrix model  was recently proposed 
to have a dual string description with worldsheets mapping to a sphere target space. 
The correlators were  written as  sums over holomorphic (Belyi) maps 
from worldsheets to the two-dimensional sphere, branched over three points. 
We express the matrix model correlators by using the  fuzzy sphere construction of matrix algebras, which 
can be interpreted as a string field theory description of  the Belyi strings. 
This  gives the correlators in terms of 
 trivalent ribbon  graphs that represent the couplings of  irreducible representations of $\lsu$, 
which can be evaluated in terms of $3j$ and $6j$ symbols. The Gaussian model perturbed by 
a cubic potential is then recognised as  a generating function for Ponzano-Regge partition functions 
for 3-manifolds having the worldsheet as boundary, and equipped  with boundary data determined 
by the ribbon graphs. This can be viewed as a holographic extension of the Belyi string worldsheets to membrane 
worldvolumes, forming part of a  holographic hierarchy linking,
 via the large $N$ expansion,  the  zero-dimensional  QFT of the 
  Matrix model to 2D strings and 3D membranes. 

\vskip.4cm

\setcounter{page}{0}
\setcounter{tocdepth}{2}

\newpage 

\tableofcontents

\section{Introduction}

The correlators of products of traces in the Gaussian Hermitian matrix model can be expressed in terms of 
triples of permutations that multiply to the identity. This fact, known in the early nineties
\cite{BaIt,Looijenga}, was recently revisited in \cite{dMR2010}. 
It was used to propose that the Hermitian matrix model has a dual string theory with the 2-sphere  
$S^2 $ as the target space. The matrix model can be viewed as a zero-dimensional quantum field theory, 
with  the Hermitian matrix field living on a point. This duality relies on the connection between permutations
and branched coverings of Riemann surfaces, which leads naturally to an $S^2$ target space. 
Holomorphic maps branched at three points on the sphere are called Belyi maps. They are 
determined by graphs embedded on the covering  Riemann surface, also known as ribbon graphs which are related to the
double-line diagrams of large N expansions.  In the mathematics literature on Belyi maps, these graphs are also called 
Grothendieck's dessins d'enfants \cite{joubp, schneps}.  The idea that the worldsheet string theory is the standard A-model topological string with sphere target has been 
developed in \cite{Gop,GopPius} for genus zero (planar) worldsheets. Refined counting of these graphs was developed in \cite{RRW}.  

In this paper, we develop another approach for arguing in favour of 
$S^2$ as a target space in the dual string theory of the Hermitian matrix model. 
It is known that the algebra of $N\times N $ matrices, for any positive integer $N$, 
can be viewed as the algebra generated by matrices $J_i$, representing 
$\lsu$ in the $N=2j+1$ dimensional irreducible representation of spin  $j$. 
This is the fuzzy sphere construction \cite{madore}, in which the Casimir equation 
$ J_1^2 + J_2^2 + J_3^2 = j ( j+1)$ is viewed as a matrix version of 
the equation defining the sphere through its embedding in Euclidean 3-space $\mR^3$.
In this construction, with $N$ taken to approach $\infty$, one can recover 
standard  field theory actions on the sphere, with an appropriate choice of matrix action. 
The matrix action of interest to us, which is the Gaussian action perturbed by 
other traces weighted by small couplings, can be viewed as a simple topological version of  scalar 
field theory on the sphere. Since quantum field theory on the target space of strings 
is precisely what string field theory attempts to construct, we may view the fuzzy sphere construction 
as providing the string field theory for the string theory of Belyi maps. The fuzzy sphere is distinguished among fuzzy 
geometry constructions in that it uses the matrix algebra for any positive integer $N$, in contrast to fuzzy projective spaces
and other fuzzy co-adjoint orbits which use $N$ equal to sequences of dimensions of representations of higher rank 
groups (see for example \cite{fuzzyrev}).  

The fuzzy sphere construction uses fuzzy spherical harmonics which give an $\lsu$ covariant basis for the 
matrix algebra. The matrix model 
correlators can be expressed  in terms of Wigner $3j$ and $6j$ symbols arising in the product of 
fuzzy spherical harmonics \cite{scalinglimits}. One of our main results is to show that the correlators 
can in fact be expressed exclusively in terms of sums of $6j$ symbols. This result will 
be no surprise to readers familiar with spin networks, as our method is in fact 
an adaptation of arguments from the spin network literature. The analogous result in the context
of spin networks was proved in \cite{moussouris}, and discussed further in \cite{dowdall, hcruiz}.

It is convenient, for a compact statement of our next result, 
to restrict attention to correlators of cubic traces, or equivalently 
to the Gaussian action perturbed by $\tr X^3$. We will also restrict 
attention, in the first instance, to the leading large $N$ limit, where only spherical worldsheets contribute. 
The matrix model computation of these correlators is a sum over planar ribbon graphs, 
which are graphs embedded on the sphere. Each ribbon graph evaluates to 
a power of $N$. From the fuzzy sphere connection, each ribbon graph can be expressed 
as a sum of $6j$ symbols with a structure related to the ribbon graph. 

Our second main result is that the sums of $6j$ symbols, 
which compute the matrix model correlators, can be viewed as partition functions 
of the Ponzano-Regge model for a 3-manifold with a boundary. 
The 3-manifold is topologically a ball $B^3$ and the boundary is 
the 2-sphere containing the embedded ribbon graph. 
The Ponzano-Regge state sum model
is a model of Euclidean gravity in three dimensions which 
is known to be related to Chern-Simons theory with $ISO(3)$ gauge group. It is the $q\rightarrow 1 $ limit of 
the Turaev-Viro model, which generates invariants of 3-manifolds using sums over representations
of the quantum deformation of $\lsu$. The computation of 
the Ponzano-Regge invariant, with the ribbon graph boundary data, chooses 
a cell decomposition of the 3-manifold in terms of tetrahedra which determine the 
$6j$ symbols being summed. Our construction builds this tetrahedral cell complex
(which we call the {\it Belyi 3-complex}) 
by extending to the 3D bulk a triangulation of the boundary sphere $S^2$ which is 
well-studied in the context of the Belyi map literature. 

Our third main result is that a membrane extension of a Belyi map can also be found for non-planar ribbon graphs.
A non-planar ribbon graph can be embedded without intersection on a higher genus closed oriented
surface, and a triangulation of this surface can be extended to a tetrahedral decomposition of a handlebody
in three dimensions. We can therefore relate all ribbon graphs generated by the Hermitian matrix model to 
partition functions of the Ponzano-Regge model.

The Gaussian Hermitian matrix model, which can be viewed as a zero dimensional quantum 
field theory, has a two-dimensional dual string theory with 2D worldsheets and 2D target. 
The 2D string theory worldsheet can be lifted to a triangulated ball or handlebody in 3D.
It is therefore appropriate to interpret the 
3D space as a holographic lift of the 2D string worldsheet to a 3D membrane worldvolume. 
We thus have a heirarchy of holographies, linking 
\vskip.3cm 
\centerline{ 
\fbox{ 0D Matrix model }   $\leftrightarrow$  \fbox{ 2D string } $\leftrightarrow$ \fbox{ 3D membrane } 
} 
\vskip.3cm 
This lifting shares similarities with the construction presented in \cite{dm07}, since the 
hologram is also a string worldsheet. It is noteworthy that in  the context of M-theory there are also conjectured hierachies of holographies, which can be viewed by analogy 
as a precedent for the above hierachy. 
 Eleven dimensional M-theory on a 4-torus has a dual which is 5-dimensional $(0,2)$ theory \cite{BFSS,seiberg97,Sen97}. This in turn has a dual in terms of a large $N$ matrix quantum mechanics \cite{abs97}.  
Other lower-dimensional formulations of $(0,2)$ theory in 5D and 4D are also reviewed in \cite{Lambert:2012qy}.
It would be fascinating to embed the holographic hierarchy of the Gaussian Matrix model in a precise manner in M-theory.

The paper is organised as follows. In the review Section \ref{sec:hmm} we recall the
connection between Hermitian matrix model correlators and permutation triples. 
We explain how this leads to the Belyi map interpretation. Finally we review 
triangulations associated to Belyi maps, which will play an important role subsequently.  
In particular, for the case where the ribbon graph has trivalent vertices, we distinguish 
two such triangulations, which we call inner and outer in anticipation of their roles in the three dimensional picture. 
The {\it inner Belyi triangulation}  is the dual of the ribbon graph. The {\it outer Belyi triangulation}  contains the ribbon graph 
itself, in addition to extra vertices and  edges added according to specified rules. 

Section \ref{sec:fuzzysphere} reviews the relevant facts about fuzzy spheres 
and the connection to quantum field theory on the 2-sphere.
In Section \ref{sec:HMMandFS} we explain the calculation of correlators in the Gaussian matrix model 
in terms of the fuzzy sphere. This leads to sums involving $3j$ and $6j$ symbols.
We show that the $3j$s can be summed to give expressions in terms
of $6j$ symbols only. The $6j$s are the basic building blocks of the Ponzano-Regge model.

In Section \ref{sec:HMMPR} we introduce the Ponzano-Regge model 
and its $q$-deformed version, the Turaev-Viro model, which serves as a regulator. We then explain a prescription for 
constructing a complex, which is a triangulation of the ball $B^3$, for each planar ribbon graph.
The complex is built by associating a constituent 3-complex to each 
vertex of the ribbon graph and gluing these constituent complexes together.  
Since the gluing is determined by the data of the Belyi map, we call the complete complex a {\it Belyi 3-complex}.
The boundary of the complex is the outer Belyi triangulation of $S^2$ associated to the Belyi map, which in particular includes
a copy of the ribbon graph itself. The interior of the complex contains the inner Belyi triangulation, which is the dual of the ribbon graph. 
We prove, using $6j$ identities, 
that the Ponzano-Regge partition function of the Belyi 3-complex thus constructed 
gives the same answer as the contribution to the Hermitian matrix integral from the specified ribbon graph.  
In Section \ref{sec:PRNP} we extend our construction of 3-complexes to non-planar ribbon graphs,
and prove that the contribution of any ribbon graph matches the Ponzano-Regge partition function
of the complex constructed from the ribbon graph.
Section~\ref{sec:discussions} discusses avenues for further research. 

\section{Review: The Hermitian matrix model and Belyi maps }\label{sec:hmm}

We start by reviewing the Gaussian Hermitian matrix model, following \cite{dMR2010} and \cite{ginsparg:1993}. The Hermitian matrix model  can be thought of as a quantum field theory in zero space-time dimensions, where the observables 
are correlators of traces of the Hermitian matrix $X$, invariant under $ X \rightarrow U X U^{\dagger}$ for unitary matrices $U$. 
 It captures the non-trivial combinatoric structure of higher dimensional 
theories with gauge symmetry, e.g. the  gauged Hermitian matrix quantum mechanics discussed in \cite{itzmcg}.  
It is also closely related to the combinatorics of the half-BPS sector of $N=4$ super-Yang Mills theory \cite{gg,ber}. 
We review the description of correlators in terms of sums over conjugacy classes of
permutation groups, and exhibit some equivalent diagrammatic methods of calculating 
these correlators. We then discuss the string dual of this theory via the counting of Belyi maps, 
and introduce dessins d'enfants as an important tool in visualising this duality.

\subsection{Generating functionals}
 We first consider the free Gaussian integral over the $N\times N$ Hermitian matrices
\bea Z[0] = \int DX e^{-\frac{1}{2}\tr X^2}, \eea
where $X^\dagger = X$, and where the integral is performed over all the real degrees of freedom of the Hermitian matrices,
\bea\label{Hmeasure} 
 DX = \prod_{k=1}^{N}dX_{\ k}^k\prod_{1\leq i < j \leq N}d(\mathrm{Re} X^i_{\ j})d(\mathrm{Im} X^i_{\ j}). \eea
As the functional integration is performed over a finite number of variables weighted by an exponentially decaying factor, it is well-defined and convergent even after insertion of polynomials in $X$. It is also invariant under the adjoint action of U($N$), as the action $X\to UXU^\dagger$ preserves the trace of any product of the matrices.

We can follow the standard procedure for generating functionals of field theories and introduce a source term $J$, which is also a Hermitian matrix,
\bea Z[J] = \int DX e^{-\frac{1}{2}\tr X^2 + \tr(JX)} \eea
which leads to the propagator
\bea
\cor{X^i_{\ j}X^k_{\ l}} &:=& \frac{ \int DX e^{-\frac{1}{2}\tr X^2}X^i_{\ j}X^k_{\ l} }{ \int DX e^{-\frac{1}{2}\tr X^2} } \ret
&=& \delta^i_l\delta^j_k.
\eea
Correlators with more matrix insertions can be calculated using Wick's theorem. For example, 
\bea \cor{X^{i_1}_{\ j_1}X^{i_2}_{\ j_2}X^{i_3}_{\ j_3}X^{i_4}_{\ j_4}} = \delta^{i_1}_{j_2} \delta^{i_2}_{j_1} \delta^{i_3}_{j_4} \delta^{i_4}_{j_3} + \delta^{i_1}_{j_3}\delta^{i_3}_{j_1}\delta^{i_2}_{j_4}\delta^{i_4}_{j_2} + \delta^{i_1}_{j_4}\delta^{i_4}_{j_1}\delta^{i_2}_{j_3}\delta^{i_3}_{j_2}. \eea
The general form of Wick's theorem here is
\bea \cor{X^{i_1}_{\ j_1}X^{i_2}_{\ j_2}\ldots X^{i_{2n}}_{\ j_{2n}}} = \sum_{\tau\in [2^n]}\delta^{i_1}_{j_{\tau(1)}}\delta^{i_2}_{j_{\tau(2)}}\ldots\delta^{i_{2n}}_{j_{\tau(2n)}},\eea
where the sum is performed over all the permutations in the permutation group on $2n$ elements that are products of $n$ disjoint 2-cycles.

\subsection{Combinatoric and diagrammatic methods}

There is a useful method of computing the correlators by thinking of the matrices as linear operators with a basis on an $N$-dimensional vector space, and representing the operators and contractions diagrammatically \cite{dMR2010}. 

\begin{figure}[h]
\centering
\includegraphics[width=0.5\textwidth]{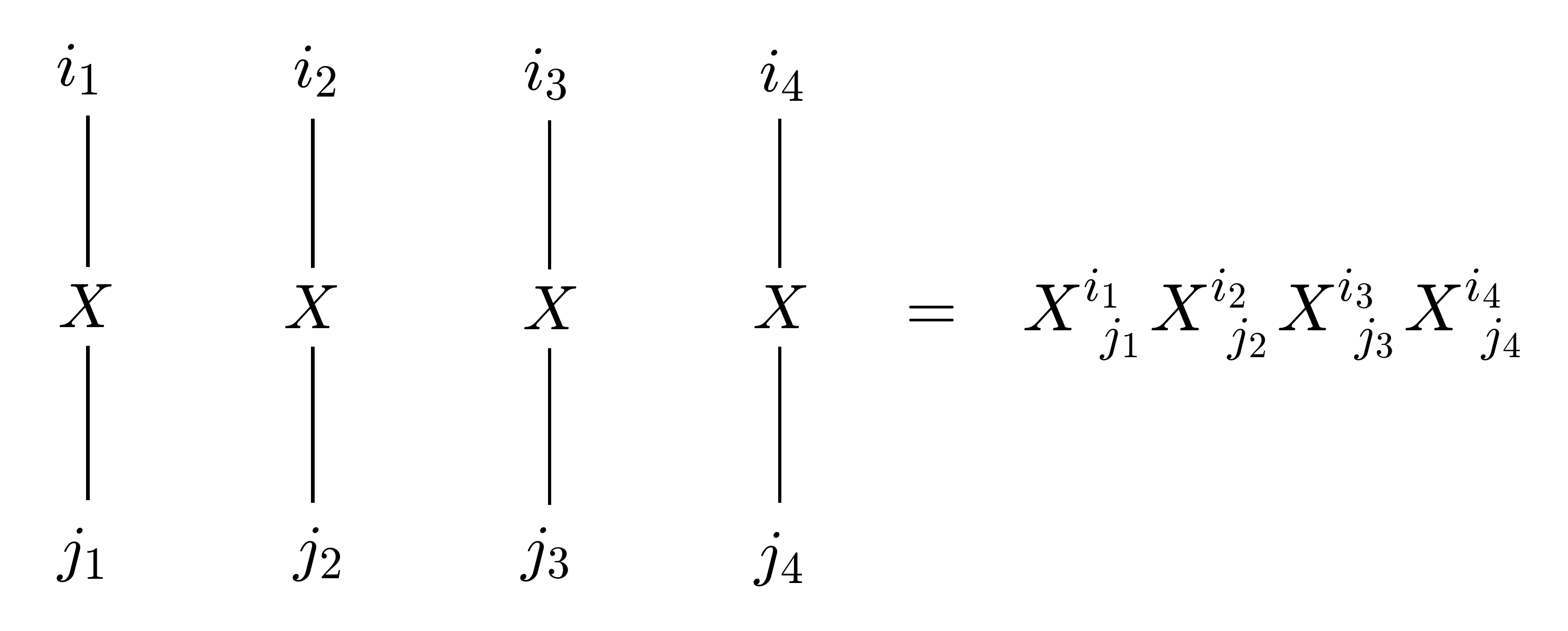}
\caption{$X^{\otimes4}$ diagram.}\label{fig:hmm1}
\end{figure}

Consider an $N$-dimensional space $V$ with an orthonormal basis $\ket{e_i}$. The linear operator associated to the matrix $X^i_{\ j}$ is
\bea X\ket{e_i} = X^j_{\ i}\ket{e_j}. \eea
By extending to the multilinear operator acting on $V^{\otimes n}$, we have
\bea (X \otimes X \otimes \ldots X)\ket{e_{i_1}\otimes e_{i_2} \otimes \ldots e_{i_n}} = X^{j_1}_{\ i_1}X^{j_2}_{\ i_2}\ldots X^{j_n}_{\ i_n}\ket{e_{j_1}\otimes e_{j_2} \otimes \ldots e_{j_n}}, \eea
and by considering the dual vectors, we can write
\bea \bra{e^{i_1}\otimes e^{i_2}\otimes \ldots e^{i_n}}(X \otimes X \otimes \ldots X) \ket{e_{j_1}\otimes e_{j_2} \otimes \ldots e_{j_n}} = X_{\ j_1}^{i_1}X_{\ j_2}^{i_2}\ldots X_{\ j_n}^{i_n}. \eea
We can introduce a diagrammatic notation for such products of matrices by drawing lines, representing the operators, that join labelled points together, representing the vectors and dual vectors, as in Figure \ref{fig:hmm1}.
We can also denote the contractions as straight lines joining different vectors and dual vectors. 

\begin{figure}[H]
\centering
\includegraphics[width=0.5\textwidth]{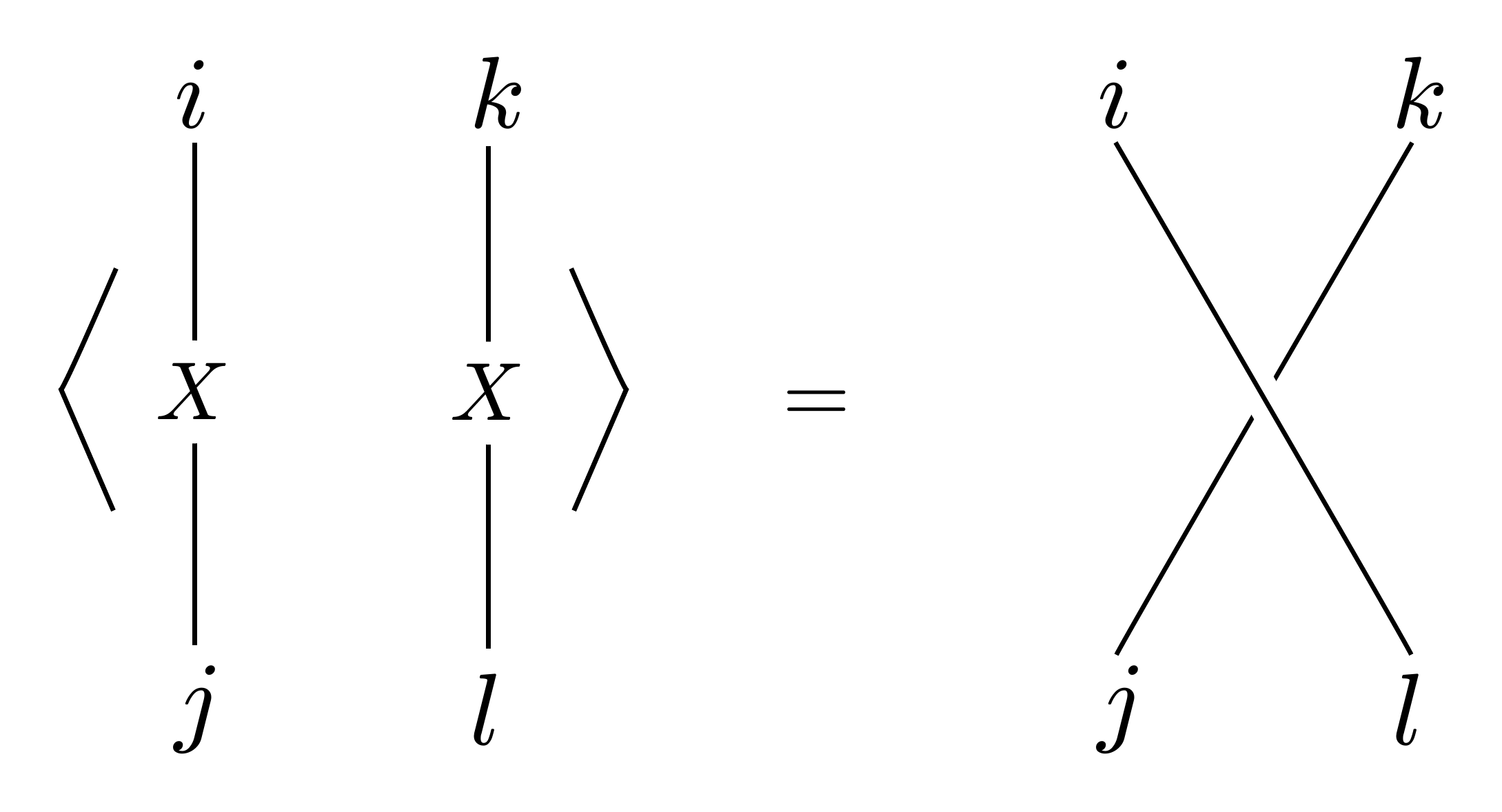}
\caption{Diagram of a propagator and contraction.}\label{fig:hmm2}
\end{figure}

\begin{figure}[H]
\centering
\includegraphics[width=1\textwidth]{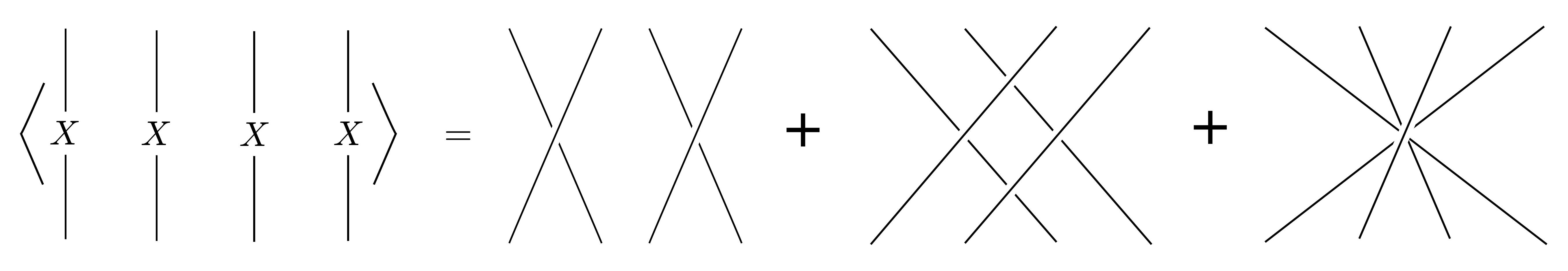}
\caption{The general $X^{\otimes4}$ correlator, with the indices suppressed.}\label{fig:hmm3}
\end{figure}

\begin{figure}[h]
\centering
\includegraphics[height=0.23\textwidth]{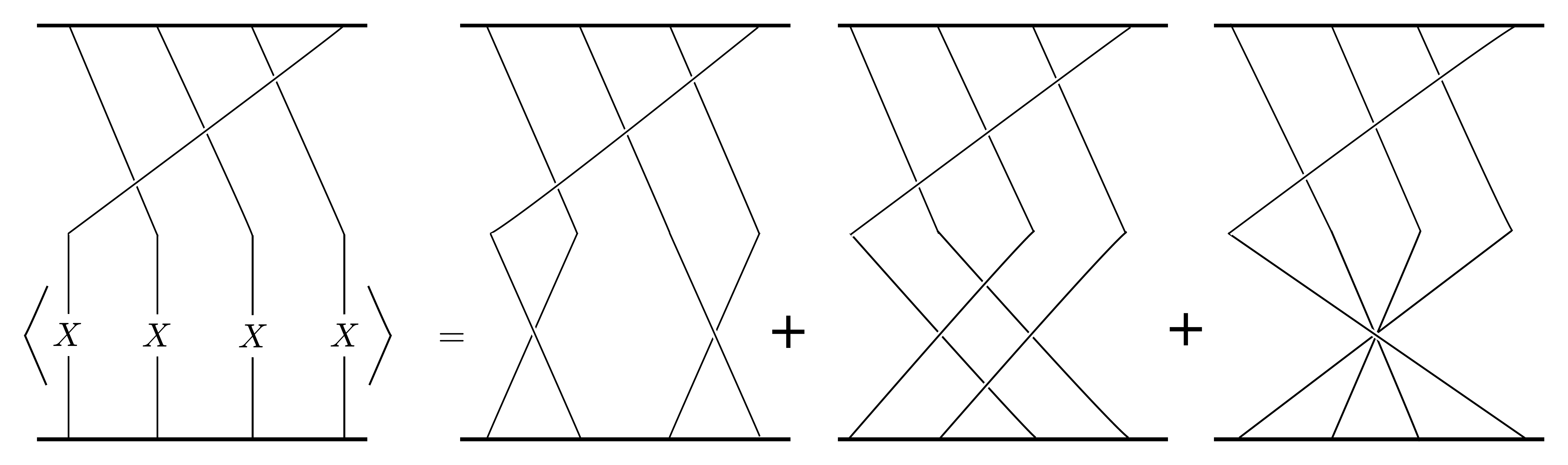}
\caption{The $\cor{\tr X^4}$ correlator.}\label{fig:hmm4}
\end{figure}

The natural gauge invariant operators in this theory are products of traces, as these are invariant under the adjoint action of U($N$) on $X$. Hence, we contract the free labelled indices to form correlators of products of traces. For example, 
\bea \cor{\tr X^4} = \delta_{j_1}^{i_2}\delta_{j_2}^{i_3}\delta_{j_3}^{i_4}\delta_{j_4}^{i_1}\cor{X^{i_1}_{\ j_1}X^{i_2}_{\ j_2}X^{i_3}_{\ j_3}X^{i_4}_{\ j_4}}, \quad \cor{\tr X^2\tr X^2} = \delta_{j_1}^{i_2}\delta_{j_2}^{i_1}\delta_{j_3}^{i_4}\delta_{j_4}^{i_3}\cor{X^{i_1}_{\ j_1}X^{i_2}_{\ j_2}X^{i_3}_{\ j_3}X^{i_4}_{\ j_4}}. \quad \eea
The calculation of the correlators can be more clearly performed diagrammatically by adding the contractions determining the product of traces to the top of the diagram and identifying the upper and lower sides of the diagram. 
In such diagrams, each loop represents a contraction of the form $\delta_i^i = N$. For example, the $\tr X^4$  operator corresponds to the cyclic contraction of four indices, and forms the top half of each of the four diagrams in Figure \ref{fig:hmm4}. Wick's theorem generates a sum of three diagrams, and as there are three loops in the first and third diagrams, and one loop in the second diagram, the correlator $\cor{\tr(X^4)}$ evaluates to $2N^3+N$.

The data contained in the different Wick contractions and products of traces can be expressed in terms of permutations. We define the action of a permutation $\sigma \in S_n$ on $V^{\otimes n}$ as
\bea \sigma \ket{e_{j_1}\otimes e_{j_2} \otimes \ldots e_{j_n}} = \ket{e_{j_{\sigma(1)}}\otimes e_{j_{\sigma(2)}} \otimes \ldots e_{j_{\sigma(n)}}}. \eea
For example,
\bea
\tr(X^4) &=& \bra{e^{i_1}\otimes e^{i_2}\otimes e^{i_3} \otimes e^{i_4}}(X\otimes X\otimes X\otimes X)\ket{e_{i_2}\otimes e_{i_3} \otimes e_{i_4} \otimes e_{i_1}} \ret
&=& \tr_{V^{\otimes4}}((X\otimes X\otimes X\otimes X)(1234) ) \ret
&=& \tr_{V^{\otimes4}}((1234)(X\otimes X\otimes X\otimes X) )
\eea 
We can therefore write a general multi-trace correlator as a sum over Wick contractions,
\bea
\cor{\tr_{2n}( \mathbf{X}\sigma)} &=& \sum_{\tau\in[2^n]}\tr_{2n}(\tau\sigma) \ret
&=& \sum_{\tau\in[2^n]} N^{C_{\tau\sigma}},
\eea
where we have used the abbreviations $\mathbf{X}=X^{\otimes 2n}$ and $\tr_{2n}=\tr_{V^{\otimes 2n}}$, and where $C_{\gamma}$ is the number of disjoint cycles in the permutation $\gamma$, e.g. $C_{(12)(3)(4)} = 3$ for the permutation $(12)(3)(4)\in S_4$.

We define the delta function $\delta(\sigma)$ on a permutation group by setting $\delta(\sigma) = 1$ if $\sigma$ is the identity
permutation  and $\delta(\sigma)=0$ otherwise. Using $C_{\gamma} = C_{\gamma^{-1}}$, this allows us to express the multi-trace correlator as 
\bea \cor{\tr_{2n}( \mathbf{X}\sigma)} = \sum_{\gamma \in S_{2n}}\sum_{\tau\in[2^n]} \delta(\sigma\tau\gamma)N^{C_{\gamma}}. \eea
We note that the correlator is invariant under the action of conjugacy on $\sigma$, i.e. $\sigma \to \tau \sigma \tau^{-1}$ for $\tau  \in S_{2n}$. Hence, we could replace $\sigma$ in the delta function with any $\sigma'$  in the conjugacy class of $\sigma$ (denoted $[\sigma]$) and perform the sum over the conjugacy class weighted by its size $|[\sigma]|$,
\bea\label{eq:triplesum}
\cor{\tr_{2n}( \mathbf{X}\sigma)} = \frac{1}{|[\sigma]|}\sum_{\sigma'\in[\sigma]} \sum_{\gamma \in S_{2n}}\sum_{\tau\in[2^n]} \delta(\sigma'\tau\gamma)N^{C_{\gamma}}.
\eea 

From the above, we conclude that the observables of this theory have a purely group theoretic  description, as sums over triples of permutations that multiply to the identity. The combinatoric data can be described diagrammatically in several ways. 
The traditional physics way is to use double line diagrams \cite{tHooft} and the closely related  ribbon graphs and 
Grothendieck's dessins d'enfants \cite{Grothendieck}.

In the double line description, for a contribution to a correlator $\tr_{2n}(\tau\sigma)$, $\tau \in [2^n]$, $\sigma\in S_{2n}$, we associate with each disjoint $k$-cycle in $\sigma$ a vertex of order $k$, with each connecting half-edge labelled by the numbers in the cycle. We then connect these vertices together with edges corresponding to the disjoint 2-cycles in $\tau$. The closed loops formed by the double line graphs now correspond to the permutation $\gamma$ such that $\sigma\tau\gamma = 1$, and hence the evaluation of a ribbon graph is $N$ to the power of the number of closed loops in the double line graph.

\begin{figure}[H]
\centering
\includegraphics[height=0.23\textwidth]{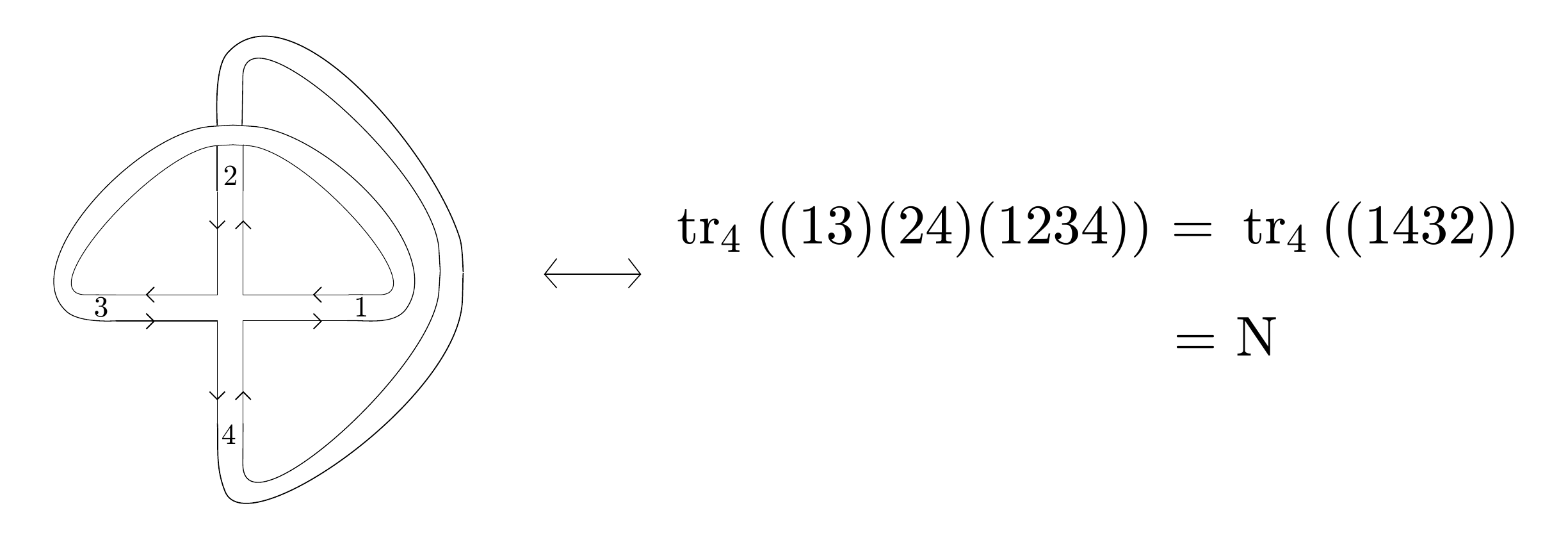}
\caption{The graph associated with the non-planar contraction of $\tr(X^4)$.}\label{fig:hmm7}
\end{figure}

The double lines can be shrunk to single lines, and there is no loss of information if 
we keep track of a local cyclic orientation at each vertex. This cyclic orientation can be viewed as 
being derived from an embedding of the graph on a Riemann surface, of the smallest genus 
that will allow the graph to be embedded without intersections.  This gives the ribbon graph description consisting 
of vertices and edges, along with cyclic order at the vertices.  
This leads directly to the description as a  dessin d'enfant when we subdivide the edges of the ribbon 
graph by introducing a new type of vertex in the middle of each edge. 
A dessin d'enfant is a bipartite graph, 
i.e a graph with two types of vertices distinguished as black and white with edges only linking black to white, that has 
 cyclic order at the vertices.  By labelling the edges, we can associate a permutation $\sigma $  to the black vertices 
 and a permutation $ \tau $   to the white vertices. 
 In the case at hand, the first permutation determines the trace structure and  the second permutation is a member of the conjugacy class $[2^n]$. Hence the white vertices are always bivalent and the graph is called a clean dessin d'enfant. 

\begin{figure}[h]
\centering
\includegraphics[height=0.23\textwidth]{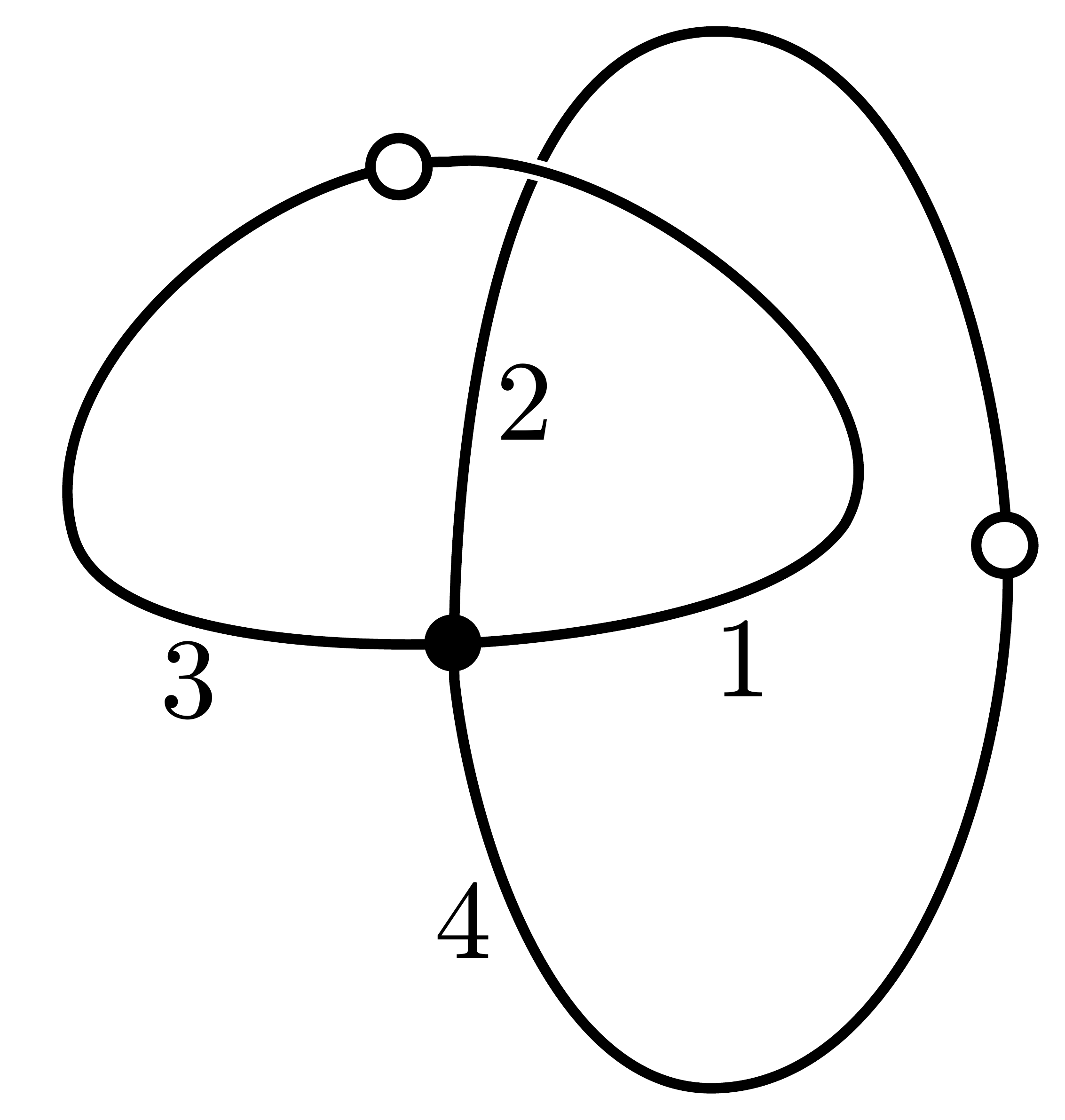}
\caption{The dessin d'enfant equivalent to the above ribbon graph.}\label{fig:hmm8}
\end{figure}

If the connected components of a dessin are drawn without intersection on a collection of surfaces of possibly non-zero genus, then the graph partitions the surfaces into distinct faces. We can then add to the dessin a third type of vertex, one for each face, that describes the permutation $\gamma$ such that $\gamma\sigma\tau = 1$. Each face corresponds to a cycle of $\gamma$ determined by the ordering of the half-edges on its boundary.

Dessins can also be used to describe Belyi maps, which are holomorphic maps from Riemann surfaces onto a sphere branched at three
points. Hence, the counting of triples of permutation is equivalent to the counting of Belyi maps. In the following section, we review this construction and its interpretation as a string theory.

\subsection{Belyi maps}\label{sec:belyi}

The Gaussian Hermitian matrix model has a dual string theory in a similar manner to the AdS/CFT correspondence. This correspondence is exact, as the combinatoric data of the matrix model correlators can be encoded exactly in the branching of holomorphic maps from worldsheets to a target space.

Consider a surjective holomorphic map $f$ from a Riemann surface $\Sigma$, consisting of a collection of connected components of genus $g_i$, to the complex projective line, or Riemann sphere, $\mathbb{P}^1$. For a generic point on the sphere, there are $2n$ preimages on $\Sigma$, where $2n$ is the degree of the map, but for a finite set of points on the sphere there are fewer inverse images. These points on the sphere are the {\bf branch points} of the map, and their preimages on $\Sigma$ are the {\bf ramification points} of the map. 
Now consider a base point on the sphere away from the branch points, and draw closed paths from the marked point around each of the branch points. We label the preimages of the punctured sphere by the natural numbers $\{1, 2, \ldots, 2n\}$. 
The holomorphic map can be characterised by a permutation in $S_{2n}$ by  for each branch point on the target space. 
The permutation is constructed by following the inverse images of the path from the base point 
around  the branch point. 

We now specialise to the case where there are three branch points on the sphere. In this case, the holomorphic map $f$ is called a {\bf Belyi map}, and the pair $(\Sigma, f)$ is called a {\bf Belyi pair}. 
If we form a simple loop by combining the loops around each of the three branch points into a single loop, then this loop is contractible on the punctured sphere. The preimage of this loop must therefore be a collection of $2n$ disjoint loops, and can be interpreted as the identity permutation acting on the set of $2n$ elements. Therefore, we can characterise the branching of a holomorphic map from a Riemann surface to a sphere by a triple of permutations in $S_{2n}$ that multiply to the identity permutation, 
\bea\label{eq:triple}
\sigma_1\sigma_2\sigma_3=1.
\eea

This equation still holds after a conjugacy transformation on each of the three elements $\sigma_i\to\gamma\sigma_i\gamma^{-1}$, which reflects the arbitrariness of our choice of labelling of the punctured spheres.
The cycle structure of these permutations is equivalent to the branching profiles about the branch points, and hence the Riemann-Hurwitz formula for the covering of a sphere can be written as the sum over the Euler characters of the connected components of $\Sigma$,
\bea \sum_i (2-2g_i) = 4n - (2n - C_{\sigma_1}) - (2n - C_{\sigma_2}) - (2n - C_{\sigma_3}) \eea
For the case when one permutation is a product of $n$ 2-cycles, the Riemann-Hurwitz formula is
\bea \sum_i (2-2g_i) = C_\gamma + C_\sigma - n. \eea

We can thus interpret the sums over triples of permutations from the previous section as sums over holomorphic maps from a worldsheet to a target space. By using a different normalisation of the correlators, we can write \refb{eq:triplesum} as 
\bea \frac{|[\sigma]|}{2n!}N^{C_\sigma-n}\cor{\tr_{2n}( \mathbf{X}\sigma)} = \frac{1}{2n!}\sum_{\sigma'\in[\sigma]} \sum_{\gamma \in S_{2n}}\sum_{\tau\in[2^n]} \delta(\sigma'\tau\gamma)N^{C_\gamma+C_\sigma-n} \eea
\bea = \sum_{ f ( [ \sigma ], [2^n] ) : \Sigma \to {\mathbb{P}}^1} \frac{1}{|Aut (f)|}\prod_i N^{2-2g_i}. \eea
This notation denotes a sum over maps with branching profiles given by $[\sigma]$ and $[2^n]$, weighted by $N$ raised to the power of the Euler characters $2-2g_i$ of the connected components of the Riemann surface, and where $|Aut(f)|$ is the order of the group of maps $\phi$ from the Riemann surface to itself that satisfy $f\circ\phi = f$.

The structure of the maps from a Riemann surface to the sphere can be visualised by using the notion of dessins d'enfants introduced in the previous section. We introduced a dessin as representing the data of a triple of permutations that multiply to the identity, but it also has a natural interpretation in terms of Belyi maps. Without loss of generality, we can set the branch points on the target sphere to be at $\{0, 1, \infty\}$. We associate the permutation $\sigma$ in the above expressions with the branching profile of 0, and the permutation $\tau\in [2^n]$ with the branching at 1. If the interval $[0,1]$ on the real line is drawn on the sphere, then the preimage of this interval on the Riemann surface produces the dessin associated to the triple of permutations, where the black vertices are the preimages of 0, and the white vertices the preimages of 1.

\begin{figure}[h]
\centering
\includegraphics[height=0.23\textwidth]{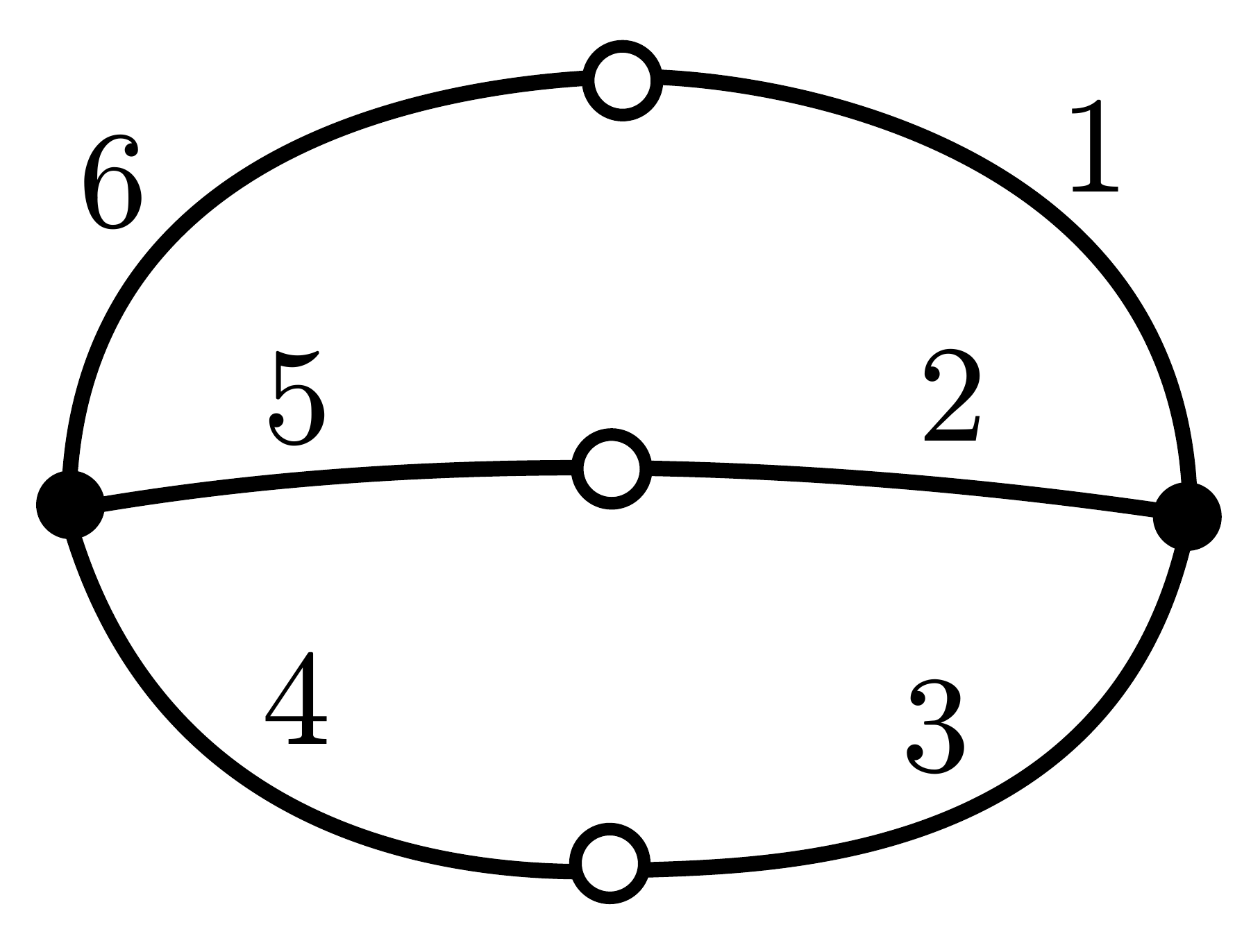}
\caption{The dessin d'enfant for the permutations $\sigma = (123)(456)$, $\tau = (16)(25)(34)$ in $S_{6}$.}\label{fig:belyi3}
\end{figure}

We can also clarify this mapping and visualise the third permutation by considering the preimage of the branch point at infinity. Draw the intervals $[1, \infty]$ on the positive real axis and $[\infty, 0]$ on the negative real axis on the sphere. Denote the preimages of $\infty$ by a cross, and colour the preimages of the interval $[0,1]$ in blue, $[1,\infty]$ in black, and $[\infty, 0]$ in red. The real axis partitions the Riemann sphere into two triangles, and the preimages of these triangles forms a triangulation of the Riemann surface $\Sigma$.

The triangles labelled by a plus are preimages of the same triangle on the Riemann sphere, and the triangles labelled by a minus are preimages of the other triangle on the Riemann sphere. The numbered labelling of the triangle comes from the labels assigned to the preimage of $[0,1]$. We can read off the permutation $\gamma = (\sigma\tau)^{-1}$ from a dessin by writing down the anticlockwise cyclic ordering of the triangles labelled with a minus around each preimage of $\infty$.

\begin{figure}[h]
\centering
\includegraphics[height=0.23\textwidth]{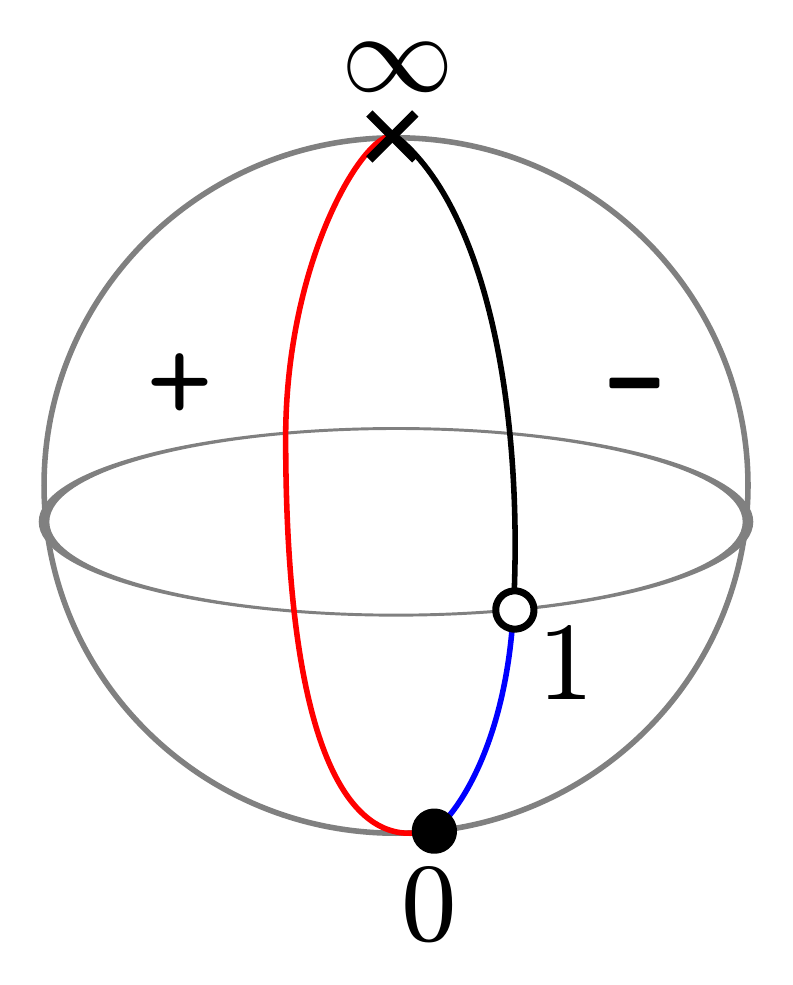}
\caption{The image of any Belyi triangulation is a Riemann sphere triangulated by two triangles.}\label{fig:riemann}
\end{figure}

\begin{figure}[h]
\centering
\includegraphics[height=0.23\textwidth]{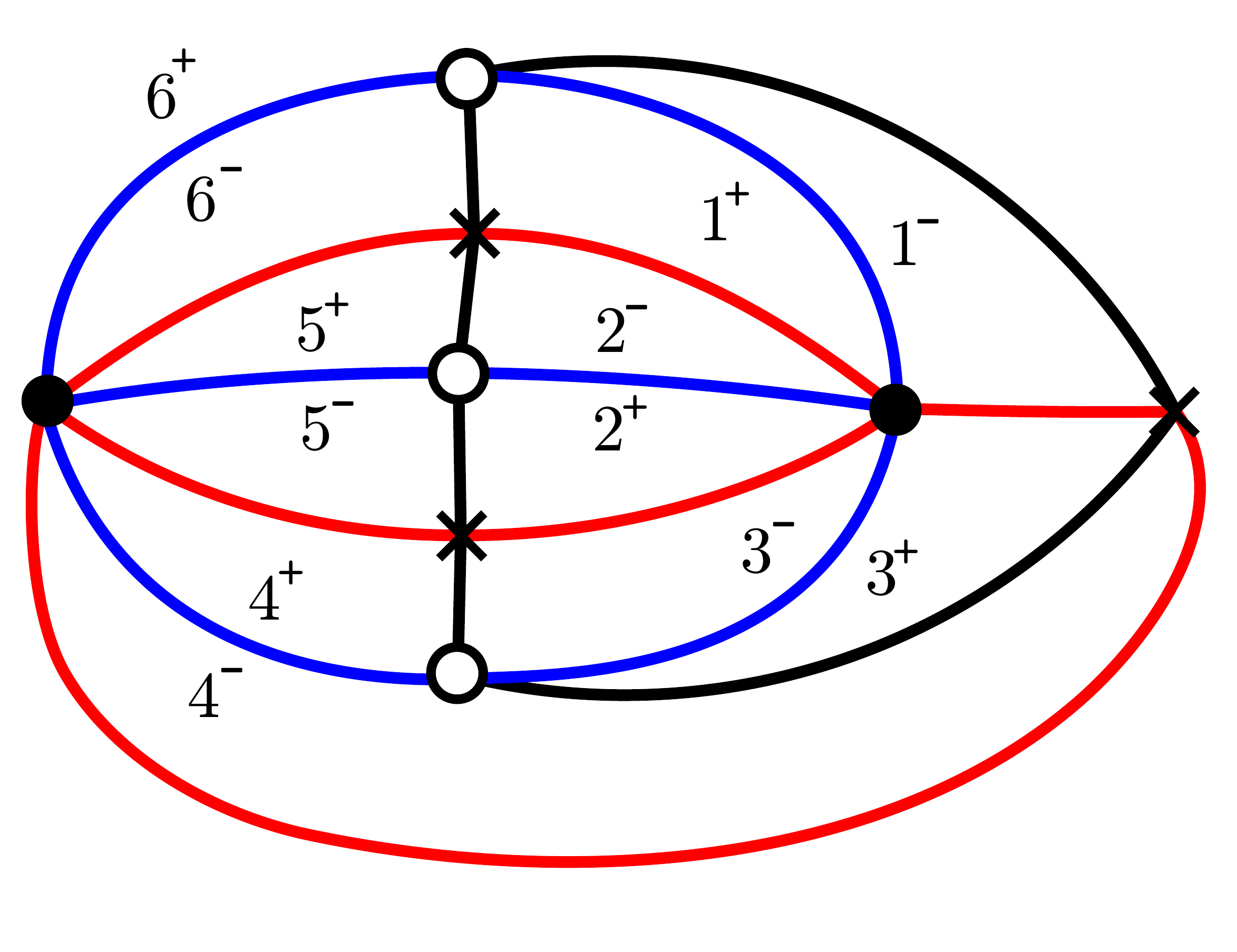}
\caption{The Belyi triangulation of a sphere generated by the permutations $\sigma = (123)(456)$,  $\tau = (16)(25)(34)$, and $\gamma = (14)(26)(35)$.}\label{fig:belyi4}
\end{figure}

\begin{figure}[h]
\centering
\includegraphics[height=0.23\textwidth]{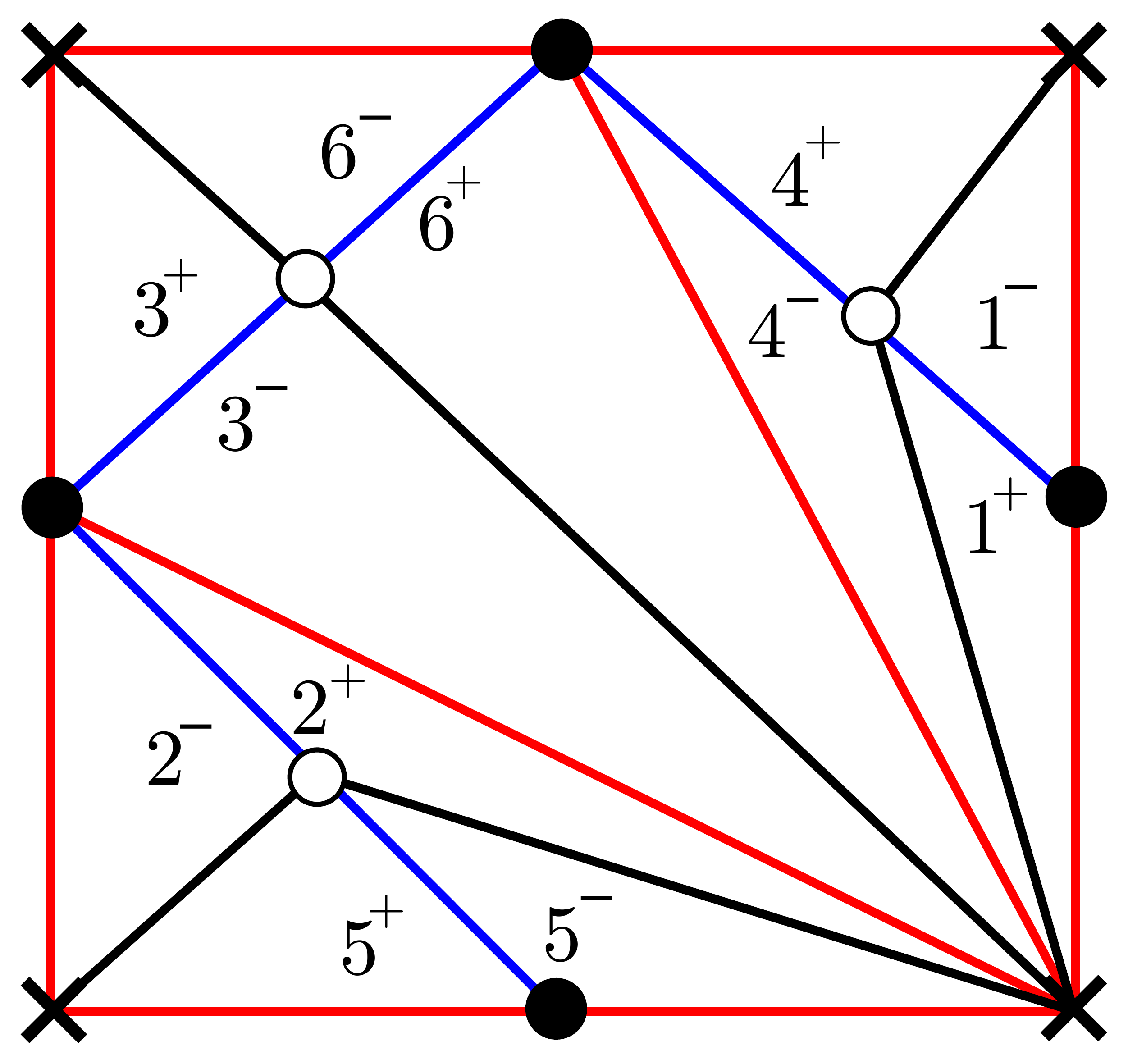}
\caption{The Belyi triangulation of a torus generated by the permutations $\sigma = (123)(456)$,  $\tau = (14)(25)(36)$, and $\gamma = (162435)$. The opposite edges bounding the rectangle are identified.}\label{fig:belyitorus}
\end{figure}

Examples of this generated triangulation are given in Figures \ref{fig:belyi4} and \ref{fig:belyitorus} for triples of permutations $(\sigma,\tau,\gamma)$  where $\sigma$ is a product of 3-cycles. The permutation $\gamma$ that satisfies $\sigma\tau\gamma = 1$ can be read off by observing the ordering of the minus-labelled triangles around the preimages of $\infty$, denoted by crosses. Note that in Figure \ref{fig:belyitorus}, the opposite edges on the boundary are identified, so there is only one preimage of $\infty$.  

We conclude this section by discussing a partitioning of the Belyi triangulation into two separate triangulations that will prove useful later.
In the main body of this paper, we shall only consider connected Ribbon graphs with trivalent vertices, which are equivalent to dessins d'enfants
specified by one permutation $\sigma$ that is a product of 3-cycles and another permutation $\tau$ that is a product of 2-cycles.

\begin{figure}[h]
\centering
\subfloat[]{\label{fig:belyi5a}\includegraphics[width=0.25\textwidth]{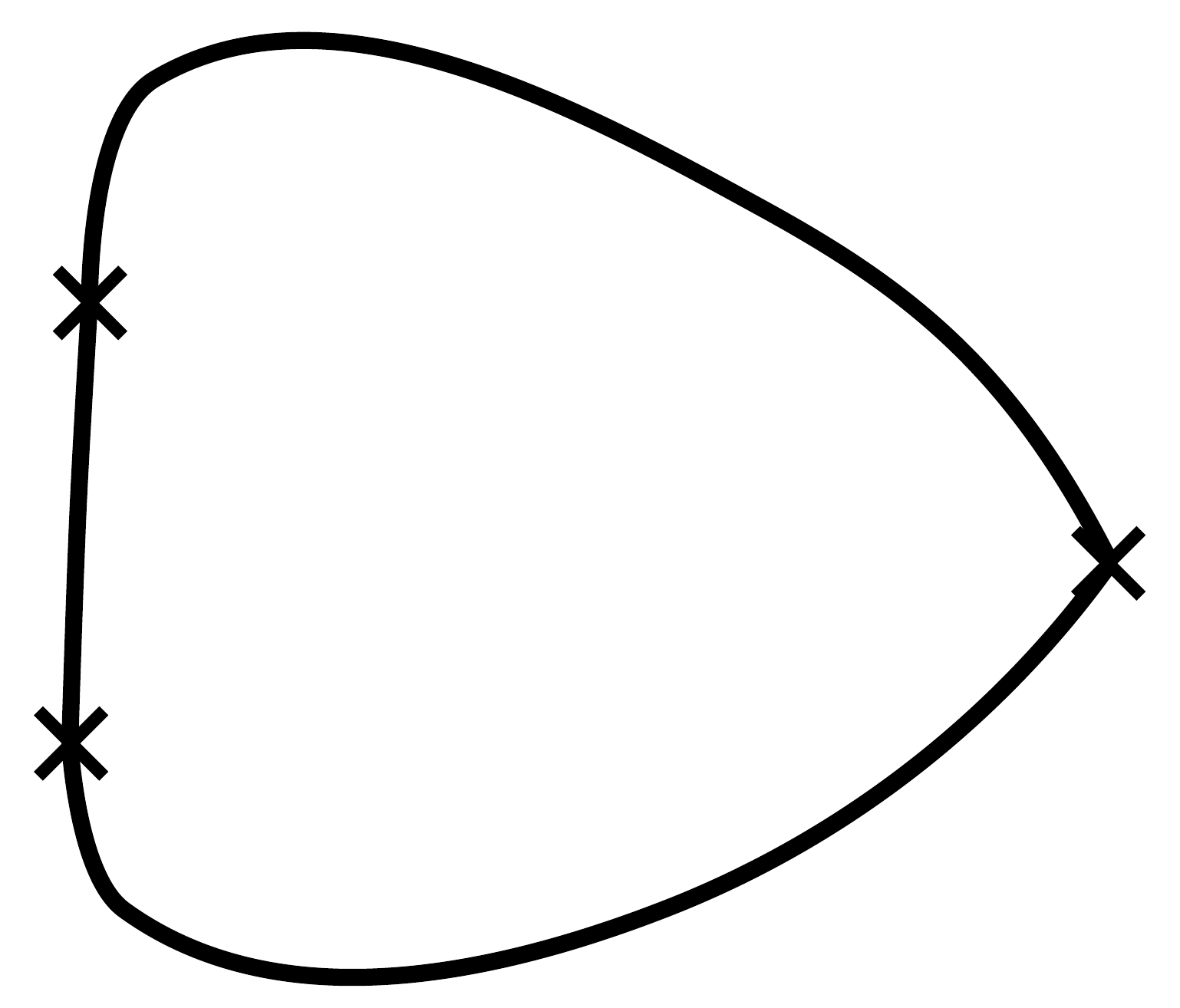}}\hspace{1cm}
\subfloat[]{\label{fig:belyi5b}\includegraphics[height=0.25\textwidth]{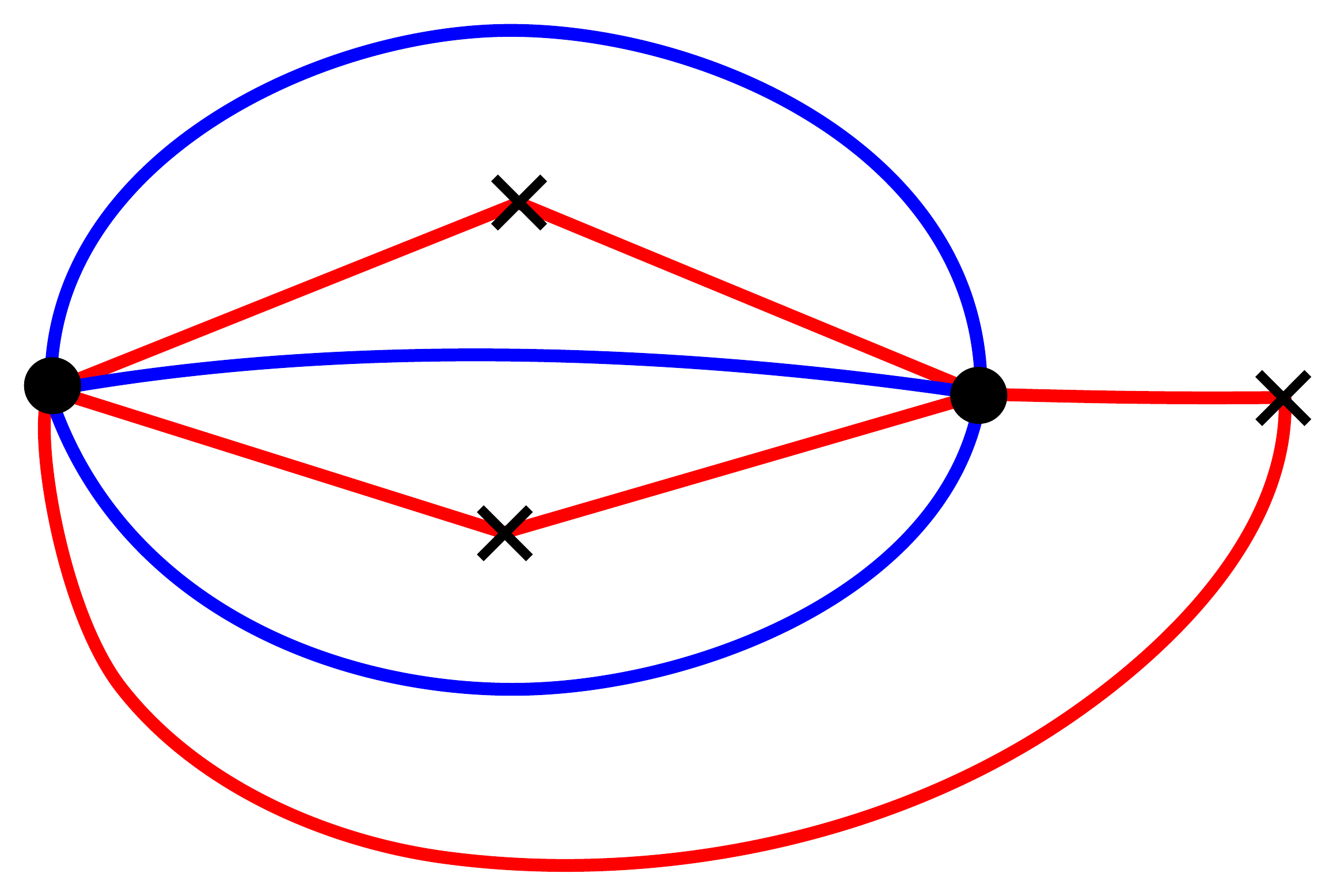}}
\caption{The inner and outer triangulations generated from the Belyi triangulation in~Figure~\ref{fig:belyi4}.}
\label{fig:belyi5}
\end{figure}

\begin{figure}[h]
\centering
\subfloat[]{\label{fig:belyi6a}\includegraphics[width=0.25\textwidth]{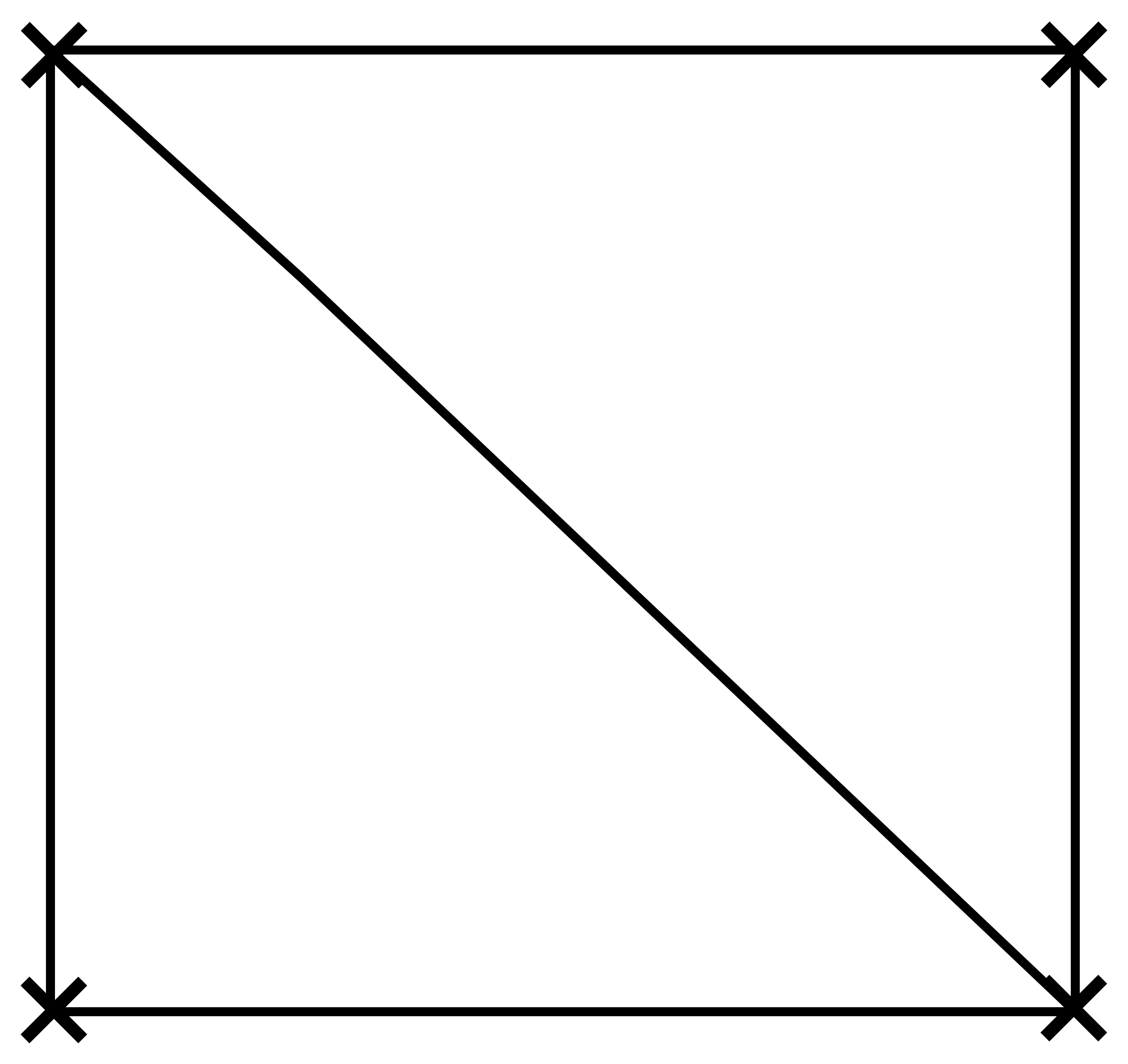}}\hspace{1cm}
\subfloat[]{\label{fig:belyi6b}\includegraphics[height=0.25\textwidth]{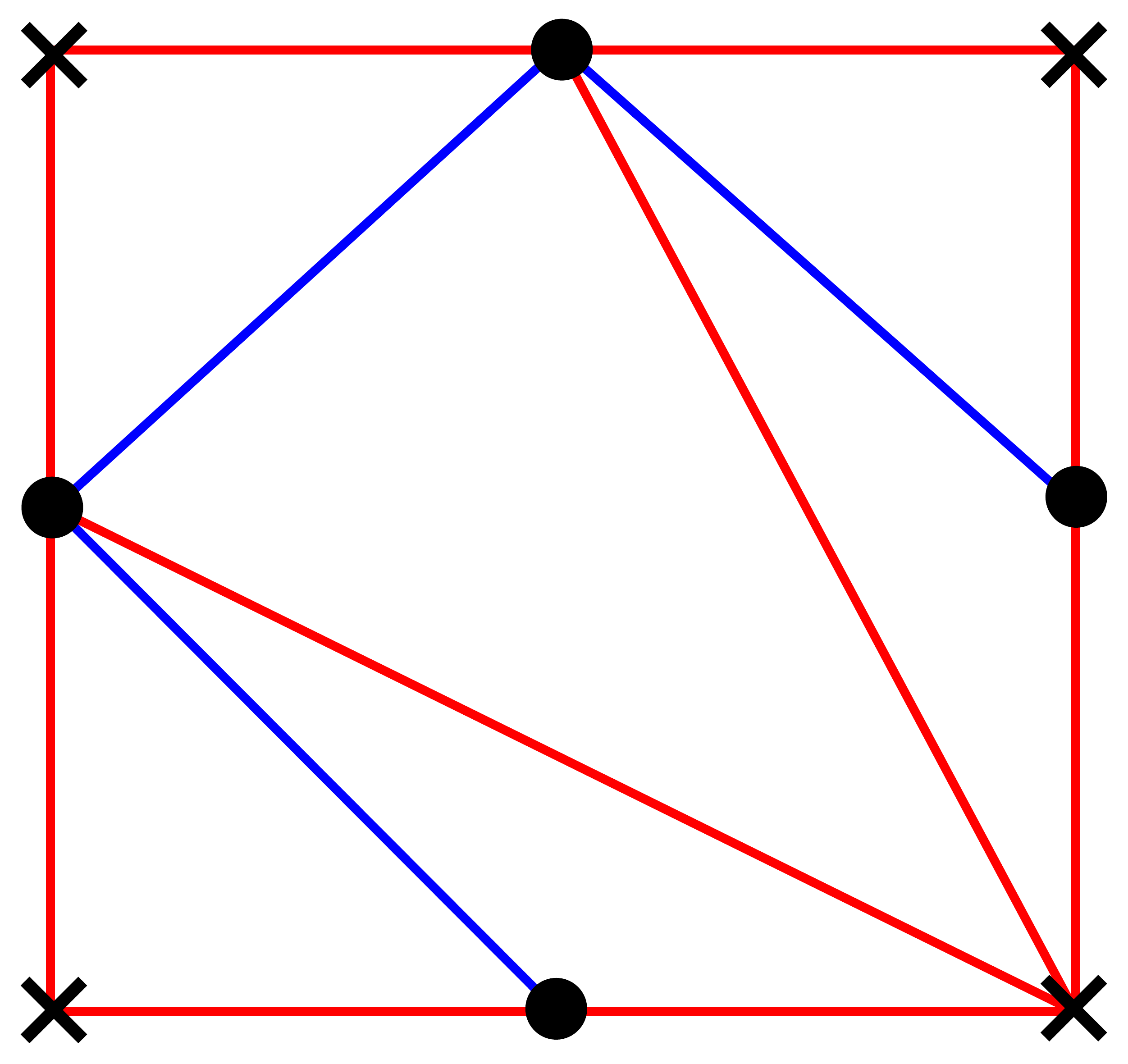}}
\caption{The inner and outer triangulations generated from the Belyi triangulation in~Figure \ref{fig:belyitorus}. Opposite edges of the boundary rectangle are identified.}
\label{fig:belyi6}
\end{figure}

Consider the white vertices of a Belyi triangulation, associated to a product of 2-cycles $\tau$. Each white vertex connects to a pair of black edges that connect to cross vertices associated to $\gamma$, and each white vertex also connects to a pair of blue edges connected to black vertices associated to $\sigma$. Hence, we can remove the white vertices from the diagram by combining their connecting pairs of edges of the same colour to generate a new graph with intersecting blue and black edges. The new blue edges and black vertices trace out the ribbon graph version of the dessin.

Next, consider only the (new) black edges and their boundary cross vertices. These edges and vertices partition the surface of genus $g$ into disjoint contractible faces, one for each black vertex in the Belyi triangulation. Since each black vertex is trivalent, and its three connecting edges intersect each bordering black edge, we conclude that the black edges partition the surface into triangles, and hence also form a triangulation of the surface of genus $g$.\footnote{Note that if, after removing the white vertices, the graph contains a blue edge connecting to the same black vertex at both ends, then the triangulation generated from the black edges will contain faces that resemble cut discs. These faces are triangles with two of the edges identified.} We call the triangulation of the surface of genus $g$ generated from the black edges the {\bf inner Belyi triangulation}. This triangulation is the 2D dual of the ribbon graph.

Finally, consider again the full Belyi triangulation, but with the white vertices and black edges removed. This is a graph containing red and blue edges, bounded by black and cross vertices. Since the removal of white vertices and black edges from the original Belyi triangulation is essentially combining pairs of adjacent triangles into new triangles, this process generates a new triangulation of the genus $g$ surface. We call this triangulation the {\bf outer Belyi triangulation}. We have given two examples of the inner and outer triangulations in Figures \ref{fig:belyi5} and \ref{fig:belyi6}. 

In general, any Belyi map generates a full Belyi triangulation of a Riemann surface, but the preimages of $[1, \infty]$ and of $[\infty, 0]\cup[0,1]$ only form the inner and outer Belyi triangulations respectively if the branching data at 0 and 1 is specified by a product of 3-cycles and a product of 2-cycles. However, this class of Belyi maps will be the ones generated by the correlators of the Hermitian matrix model on the fuzzy sphere, and the inner and outer Belyi triangulations will eventually lead to an interpretation of this model in three dimensions.

\section{The fuzzy sphere}\label{sec:fuzzysphere} 

In this section we review the relevant necessary facts in the construction of the fuzzy sphere, focusing on its description in terms of the fuzzy spherical harmonics, taking definitions and results from \cite{madore, quantph, scalinglimits}.

The fuzzy sphere is a family of noncommutative deformations of the algebra of functions on the sphere $S^2$ \cite{madore}. The deformation replaces the commuting coordinate functions of the sphere with noncommuting operators acting on a vector space. The deformation is performed in a manner that preserves the $SO(3)$ symmetry group of the sphere, but loses the notion of a base manifold with a continuum of points.

\subsection{Fuzzy spherical harmonics}\label{sec:fsh}

The generators of the fuzzy sphere can be defined as operator versions of the coordinate functions of a 2-sphere embedded in three-dimensional Euclidean space. 
These generators are the $N\times N$ matrix generators of the Lie algebra $\lsu$ in an $N$-dimension representation, with a specific choice of normalisation, defined by the commutation relations
\bea
[x_i, x_j] = i\lambda_N\epsilon_{ijk}x_k. \label{eq:sphnc}
\eea
Here, $\lambda_N$ is a noncommutativity parameter defined by
\bea \lambda_N := R\left[\frac{1}{4}\left(N^2-1\right)\right]^{-\frac{1}{2}}  = R[j(j+1)]^{-\frac{1}{2}}. \eea
In particular, this choice of normalisation of the Lie algebra generators ensures that the quadratic Casimir $x_ix_i$ is equal to $R^2$, and that the coefficient in the commutator tends to zero as the dimension of the representation $N$ becomes large. 
We have also introduced the half-integer noncommutativity parameter $j$, related to the dimension of the representation $N$ by
\bea N \equiv 2j+1, \quad  j\in \frac{\mathbb{Z}}{2}. \eea
We will use these two noncommutativity parameters interchangeably throughout.

The three operators $x_i$ generate the complex matrix algebra of the fuzzy sphere ${\cal A}_N$. Every element $f\in{\cal A}_N$ has a unique expansion
\bea
f = \sum_{l=0}^{N-1} \frac{1}{l!}f_{i_1\ldots i_l}x^{i_1}\ldots x^{i_l}, 
\label{eq:sumnc}
\eea 
where the quadratic constraint $x_ix_i=R^2$ allows the coefficients $f_{i_1\ldots i_l}$ to be taken to be traceless and symmetric. This algebra consists of $N\times N$ matrices and has dimension $N^2$, so it is equivalent to the algebra of complex $N\times N$ matrices.

An alternative basis of the algebra ${\cal A}_N$ is given by the fuzzy spherical harmonics $Y_{lm}$, which are deformations of the classical spherical harmonics. The classical spherical harmonics are eigenfunctions of the angular momentum operators $J^2$ and $J_3$ and are labelled by their eigenvalues, so we expect their fuzzy sphere counterparts to satisfy similar relations. Unlike the classical spherical harmonics, there are only finitely many linearly independent eigenfunctions of $J^2$ in the fuzzy sphere algebra, and so there are are only finitely many linearly independent fuzzy spherical harmonics. 

Denoting the orthonormal basis vectors in a $(2j+1)$ dimensional representation of $\lsu$ by $\ket{j\mu}$, where $j$ is half-integer, we define the fuzzy spherical harmonics by
\bea Y_{lm}(j) =  \sqrt{2l+1}\sum_{\mu \nu} C_{j\nu\  lm}^{j\mu}\ket{j\mu}\bra{j\nu}, \label{eq:nsphdef} \eea
where $C_{j\nu\  lm}^{j\mu}$ are real Clebsch-Gordan coefficients, corresponding to the coupling of a pair of irreducible representations of $\lsu$ \cite{quantph}. By using the orthonormality of the $\ket{j\mu}$ and some standard relations involving sums of Clebsch-Gordan coefficients given in the appendix, it can be shown that these operators are eigenfunctions of the angular momentum operators $J^2$ and $J_3$ under the adjoint action, that is, they obey 
\begin{eqnarray}
\left[J_i, \left[J_i, Y_{lm}\right]\right] =& \mathrm{ad}_{J_i}(\mathrm{ad}_{J_i}(Y_{lm})) &= l(l+1)Y_{lm}, \label{eq:adjjj} \ret
\left[ J_{3}, Y_{lm}\right] =& \mathrm{ad}_{J_3}(Y_{lm}) &= mY_{lm}. \label{eq:adjj3}
\end{eqnarray}
We define the inner product on the algebra using the matrix trace, 
\bea \label{eq:inner} \cor{f, g} := \frac{1}{N}\tr( f^\dagger g). \eea
Using the Clebsch-Gordan identities \refb{eq:orthoga} and \refb{eq:rose} from the Appendix,
we can see that 
\bea \cor{Y_{lm}, Y_{l'm'}} =\delta_{l,l'}\delta_{m,m'},\eea
hence the fuzzy spherical harmonics are an orthonormal basis of the algebra with respect to this
inner product. From the definition \refb{eq:nsphdef}, it can be seen that the Hermitian conjugate of a fuzzy
spherical harmonic is
\bea \label{eq:herm} Y_{lm}^\dagger  = (-)^mY_{l-m}, \eea
so we can state that the trace of a product of fuzzy spherical harmonics is
\bea \label{eq:trace} \tr{(Y_{l_1m_1}Y_{l_2m_2})} = N(-)^{m_1}\delta_{l_1,l_2}\delta_{m_1,-m_2}. \eea
This relation will be more frequently used than the inner product \refb{eq:inner} in the following.

By calculating the product of a pair of fuzzy spherical harmonics, we can write 
\bea Y_{l_1m_1}Y_{l_2m_2} = \sum_{l_3 m_3} \cA_{l_1 m_1 l_2 m_2}^{l_3 m_3} Y_{l_3m_3} \eea
\bea \cA_{l_1 m_1 l_2 m_2}^{l_3 m_3} = \sqrt{N}\sum_{l_3,\ m_3}(-)^{3j+m_3}\sqrt{(2l_1+1)(2l_2+1)(2l_3+1)}\wigthreej{l_1}{l_2}{l_3}{m_1}{m_2}{-m_3}\wsjj{l_1}{l_2}{l_3}\qquad
\eea
where the symbols in braces are the Wigner $3j$ and $6j$ symbols that describe the coupling of irreducible representations of $\lsu$ \cite{scalinglimits}. More information and background on these symbols is given in Appendix \ref{sec:awig}.

This construction has shown that the basis of fuzzy spherical harmonics decomposes into a sum of irreducible matrix representations of $\lsu$ as
\bea \mathbf{1}\oplus\mathbf{3}\oplus\mathbf{5}\oplus\ldots\oplus\mathbf{(2N+1)}. \eea
The asymptotic formula for the 6$j$ symbol 
\bea\label{eq:asymsixj} \lim_{j\to\infty}(-)^{3j}\sqrt{2j}\wsjj{l_1}{l_2}{l_3} = \wigthreej{l_1}{l_2}{l_3}{0}{0}{0}, \eea
shows that the coefficient $\cA_{l_1 m_1 l_2 m_2}^{l_3 m_3}$ in the fuzzy algebra reproduces the coefficient for the classical spherical harmonics algebra in the large $N$ limit.

\subsection{Quantum field theory on the fuzzy sphere }

Quantum field theories can be constructed on the fuzzy sphere by analogy with those on the commutative sphere, see for example 
 \cite{scalinglimits, vaidya, ncgauge}. 
 We can construct the standard complex scalar quantum field theory on the fuzzy sphere by expanding a general member of ${\cal A}_N$ as
\bea X = \sum_{l=0}^{N-1}\sum_{m=-l}^{l} a_{lm}Y_{lm}. \eea
As the fuzzy spherical harmonics satisfy \refb{eq:herm}, we can set the matrix $X$ to be Hermitian by demanding that the complex conjugate of $a_{lm}$ satisfies
\bea a_{lm}^* = (-)^ma_{l-m}. \eea
The partition function is 
\bea Z = \int D\Phi e^{-S[\Phi]}, \eea
where we integrate over the $N^2$ real degrees of freedom of the fuzzy sphere with the measure
\bea\label{FSmeasure}
 D\Phi = \prod_{l=0}^{N} \left[ da_{l0}\prod_{m=1}^l da_{lm}da_{lm}^* \right]. \eea
The more commonly considered action for scalar field theory on the fuzzy sphere is
\bea S[X] = \frac{1}{N}\tr\left( \frac{1}{2}X[J_i, [J_i, X]] + \frac{1}{2}\mu^2X^2 + V(X) \right).  \eea
This expression includes a Laplacian, a mass term, and a general potential term, and 
results in the propagator 
\bea \cor{a_{lm}^*a_{l'm'}} := \frac{\int DX e^{-S}a_{lm}^*a_{l'm'}}{\int DX e^{-S}} = \frac{\delta_{l,l'}\delta_{m,m'}}{l(l+1) + \mu^2}. \eea

\section{The Gaussian Hermitian matrix model as a fuzzy sphere}\label{sec:HMMandFS} 

The dynamics of scalar field theories on the fuzzy sphere with Laplacians and other terms have been considered in several papers. 
Here, however, we confine ourselves to considering a topological theory on the quantum sphere where the Laplacian vanishes. As we could consider the terms in a perturbative expansion of the potential to be operator insertions in the correlators, we also set the potential to zero, leaving only a mass term. We therefore set $\mu^2=N^2$ to arrive at the generating functional
\bea Z = \int DX e^{-\frac{N}{2}\tr X^2}. \eea
This partition function is a Gaussian Hermitian matrix model like the one discussed in Section \ref{sec:hmm}, but with the $N^2$ real degrees of freedom rewritten in the $\lsu$ covariant form $a_{lm}$ instead of the $U(N)$ covariant form $X^i_{\ j}$. In the next section, we will prove that the matrix models are equivalent. Note that we have chosen a different factor in front of the action for this generating function. This will result in different powers of $N$ appearing in the evaluations of the diagrams than those of the previous section, but this is just a different choice of normalisation.

By using \refb{eq:trace}, we see that 
\bea
\tr{(X^2)} &=& \sum_{l, l'=0}^{N-1}\sum_{m=-l}^l\sum_{m'=-l'}^{l'} a_{lm}a_{l'm'}\tr{(Y_{lm}Y_{l'm'})} \ret
&=& N\sum_{l=0}^{N-1}\sum_{m=-l}^l a_{lm}^*a_{lm},
\eea
and hence calculate the propagator 
\bea
\cor{a_{lm}a_{l'm'}} &=& \frac{\int DX e^{-\frac{N}{2}\tr X^2}a_{lm}a_{l'm'}}{\int DX e^{-\frac{N}{2}\tr X^2}} \ret
&=& \frac{(-)^m \delta_{l,l'}\delta_{m,-m'}}{N^2}.
\eea
We can again retain the $U(N)$ invariance of the original Hermitian matrix model by again considering only correlators of products of traces. Using the fuzzy algebra
\bea Y_{l_1m_1}Y_{l_2m_2} = \sum_{l_3 m_3} \cA_{l_1 m_1 l_2 m_2}^{l_3 m_3}Y_{l_3m_3} \eea
and the orthonormality of the spherical harmonics, along with Wick's theorem for the variables $a_{lm}$, 
\bea \cor{a_{l_1m_1}\ldots a_{l_{2n}m_{2n}}} =
\sum_{\tau\in [2^n]} \prod_{\substack{\mathrm{disjoint\ cycles}\\ (ij)\ \mathrm{in\ } \tau}}
\cor{a_{l_im_i}a_{l_jm_j}} \eea
we can calculate any correlator by writing out explicitly the factors of  $\cA_{l_i m_i l_j m_j}^{l_k m_k}$ and performing the sums over all labels $l$ and $m$. Alternatively, we can perform the calculations diagrammatically as in the original Hermitian matrix model. 

\subsection{Equivalence of matrix integration and fuzzy sphere path integral}\label{sec:equiv}

We introduce a new notation to more clearly exhibit the cyclic structure of the fuzzy spherical harmonics, and to simplify the calculations.
In Section \ref{sec:fsh}, a derivation of the fuzzy sphere algebra is given that results in the relation
\bea Y_{l_1m_1}{Y}_{l_2 m_2} = \sum_{l_3 m_3} A_{l_1m_1 l_2m_2}^{l_3m_3}{Y}_{l_3m_3}, \eea 
where
\bea A_{l_1m_1l_2m_2}^{l_3m_3} = \sqrt{N}(-)^{3j+m_3}\sqrt{(2l_1+1)(2l_2+1)(2l_3+1)}\wigthreej{l_1}{l_2}{l_3}{m_1}{m_2}{-m_3} \wsjj{l_1}{l_2}{l_3}. \quad \eea

We abbreviate this relation to $Y_1Y_2=A_{12}^{\ \ 3}Y_3$, where $i=1,2,3$ represents a pair of indices $l_i, m_i$, and the repeated upper and lower indices are summed over. The above expression for $A_{12}^{\ \ 3}$ can be put into a cyclically symmetric form by defining the lowering and raising operators
\bea\eta_{12}= N(-)^{m_1}\delta_{l_1,l_2} \delta_{m_1,-m_2}, \quad \eta^{12}= N^{-1}(-)^{m_1}\delta_{l_1,l_2} \delta_{m_1,-m_2}, \quad \eta_{12}\eta^{23} = \delta_1^3 \equiv \delta_{l_1,l_3}\delta_{m_1,m_3} \eea
We thus have \bea A_{123} = A_{12}^{\ \ 4}\eta_{34} = N^{\frac{3}{2}}(-)^{3j}\sqrt{(2l_1+1)(2l_2+1)(2l_3+1)}\wigthreej{l_1}{l_2}{l_3}{m_1}{m_2}{m_3} \wsjj{l_1}{l_2}{l_3}. \quad \eea
This now has manifest cyclic symmetry, as it is the trace of three fuzzy spherical harmonics. The expressions for the traces in the expansion of correlators can be simplified using this notation. For example, the expression for the trace of four spherical harmonics can be written
\bea \tr(Y_{l_1m_1}Y_{l_2m_2}Y_{l_3m_3}Y_{l_4m_4}) = A_{125}A_{34}^{\ \ 5},\eea
where the repeated $5$ represents the sum over the representation labels $l_5$ and representation states $m_5$. In addition, a propagator can be written as
\bea \cor{a_{l_1m_1}a_{l_2m_2}} = N^{-1}\eta^{12}, \eea
and so a general correlator of the model will be a sum over Wick contractions weighted by factors of $N$, $\eta^{ij}$ and $A_{ijk}$.

We have written the Gaussian Hermitian matrix model variables in terms of fuzzy spherical harmonics, but 
have not derived the form of the Hermitian matrix model measure \refb{Hmeasure} in terms of the new 
variables. To show that the Jacobian for the change of variables is trivial, we will show that  
arbitrary correlators computed in  the fuzzy sphere picture,  using the standard fuzzy sphere measure \refb{FSmeasure}, 
give the same answer as the standard matrix model computation.

First we recall from the definition \refb{eq:nsphdef} that the fuzzy spherical harmonics act on the $N$-dimensional $\lsu$  irrep ${\ket{j\mu}}$, $\mu=-j,-j+1,\ldots, +j$. We abbreviate these vectors to $\ket{\mu}$ in the following.
We next note that a general product of traces can be expressed by a permutation $\sigma$ by writing
\bea 
\tr_{2n}(\mathbf{X}\sigma) &=& \sum_{\mu_1,\ldots,\mu_{2n}}\bra{\mu_1\mu_2\ldots\mu_{2n}} X\otimes\ldots\otimes X\sigma\ket{\mu_1\mu_{2}\ldots\mu_{2n}} \ret
&:=& \sum_{\mu_1,\ldots,\mu_{2n}}\bra{\mu_1\mu_2\ldots\mu_{2n}} X\otimes\ldots\otimes X\ket{\mu_{\sigma(1)}\mu_{\sigma(2)}\ldots\mu_{\sigma(2n)}} \ret
\eea
where the permutation $\sigma\in S_{2n}$ acts on the tensor product of vectors in the same way as in the Hermitian matrix model. 

Next, we consider a general contribution to the correlator $\cor{\tr_{2n}(\mathbf{X}\sigma)} = \cor{\tr_{2n}(\sigma\mathbf{X})}$. Wick's theorem states that
\begin{eqnarray} \cor{a_{l_1m_1}\ldots a_{l_{2n}m_{2n}}} &=& \sum_{\tau\in[2^n]}\frac{1}{N^{2n}}(-)^{m_{i_1}+\ldots m_{i_n}}\delta_{l_{i_1},l_{j_1}}\delta_{m_{i_1},-m_{j_1}}\ldots \delta_{l_{i_n},l_{j_n}}\delta_{m_{i_n},-m_{j_n}} 
\label{eq:wick}\end{eqnarray}
where $\tau = (i_1j_1)\ldots(i_nj_n)$ is summed over all products of disjoint 2-cycles, i.e. the $i_k, j_k$ are all distinct integers from 1 to $2n$.

Next, we consider the action of contraction upon a tensor product of spherical harmonics. Using the explicit expression \refb{eq:nsphdef} for the fuzzy spherical harmonics in terms of the $\ket{\mu}$ basis, we write
\bea
\eta^{ij}Y_i\otimes Y_j &=& \sum_{\substack{l_i m_i\ l_j m_j \\ \mu_1\mu_2\ \nu_1\nu_2}}\eta^{ij}\sqrt{(2l_i+1)(2l_j+1)}C_{j\mu_2\ l_im_i}^{j\mu_1}C_{j\nu_2\ l_jm_j}^{j\nu_1}\ket{\mu_1}\bra{\mu_2}\otimes\ket{\nu_1}\bra{\nu_2} \ret
&=& \sum_{\substack{l_i m_i \\ \mu_1\mu_2\ \nu_1\nu_2}} \frac{(2l_i+1)}{(2j+1)}(-)^{m_i}C_{j\mu_2\ l_im_i}^{j\mu_1}C_{j\nu_2\ l_i-m_i}^{j\nu_1}\ket{\mu_1}\bra{\mu_2}\otimes\ket{\nu_1}\bra{\nu_2}.
\eea
We can simplify this expression using the properties of the Clebsch-Gordan coefficients \refb{eq:orthogb} and \refb{eq:rose} from the appendix. Hence we calculate
\bea
\eta^{ij}Y_i\otimes Y_j &=& \sum_{\substack{l_i m_i \\ \mu_1\mu_2\ \nu_1\nu_2}}(-)^{2j-\mu_2-\nu_2+m_i}C_{j\mu_2\ j-\mu_1}^{l_i-m_i} C_{j\nu_2\ j-\nu_1}^{l_im_i} \ket{\mu_1}\bra{\mu_2}\otimes\ket{\nu_1}\bra{\nu_2} \ret
&=& \sum_{\substack{l_i m_i \\ \mu_1\mu_2\ \nu_1\nu_2}}(-)^{2(j-\mu_2)+\mu_1-\nu_2} C_{j\mu_1\ j-\mu_2}^{l_im_i} C_{j\nu_2\ j-\nu_1}^{l_im_i} \ket{\mu_1}\bra{\mu_2}\otimes\ket{\nu_1}\bra{\nu_2} \ret
&=& \sum_{\mu_1\mu_2} \ket{\mu_1}\bra{\mu_2}\otimes\ket{\mu_2}\bra{\mu_1},
\eea
where we have also used the facts that 
\bea C_{j\mu_2\ j-\mu_1}^{l_i-m_i}(-)^{m_i} = C_{j\mu_2\ j-\mu_1}^{l_i-m_i}(-)^{\mu_1 - \mu_2} \eea
and that $(j-\mu_2)$ is always an integer.
We therefore see that the contraction of a pair of indices acts like a transposition on the basis vectors. With the complete
tensor product of spherical harmonics, a contraction of a pair of indices gives
\bea \eta^{ij}Y_1 \otimes \ldots Y_i\otimes \ldots \otimes Y_j \otimes \ldots \otimes Y_{2n} = \sum_{\nu_i\nu_j}Y_1 \otimes \ldots \otimes \ket{\nu_i}\bra{\nu_j} \otimes \ldots \otimes \ket{\nu_j}\bra{\nu_i} \otimes \ldots \otimes Y_{2n}. \quad \eea

We now consider a general contraction $\tau = (i_1j_1)\ldots(i_nj_n)$, where the numbers $i_k$, $j_k$ represent distinct integers between 1 and $2n$. Using the above relation for all $1\leq i_k, j_k\leq 2n$ with the implicit contraction of the $\eta^{ij}$ with the $Y_i$, we deduce that
\bea \eta^{i_1j_1}\ldots\eta^{i_nj_n}\tr_{2n}(Y_1\otimes\ldots\otimes Y_{2n} \sigma) = \sum_{\mu_1,\ldots, \mu_{2n}}\bra{\mu_1\ldots\mu_{2n}} \eta^{i_1j_1}\ldots\eta^{i_nj_n} Y_1\otimes\ldots\otimes Y_{2n} \ket{\mu_{\sigma(1)}\ldots\mu_{\sigma(2n)}} \nn \eea \bea
&=&  \sum_{\substack{\mu_1\ldots \mu_{2n} \\ \nu_1\ldots\nu_{2n}}} \bra{\mu_1\ldots\mu_{2n}} \left(\ket{\nu_1}\bra{\nu_{\tau(1)}}\otimes\ldots\otimes\ket{\nu_{2n}}\bra{\nu_{\tau(2n)}} \right) \ket{\mu_{\sigma(1)}\ldots\mu_{\sigma(2n)}} \ret
&=& \sum_{\mu_1\ldots\mu_{2n}}\bra{\mu_1\ldots\mu_{2n}}\tau \ket{\mu_{\sigma(1)}\ldots\mu_{\sigma(2n)}} \ret
&=& \sum_{\mu_1\ldots\mu_{2n}}\bra{\mu_1\ldots\mu_{2n}}\tau\sigma \ket{\mu_1\ldots\mu_{2n}} \ret
&=&  \tr_{2n}(\tau\sigma).
\eea
This is the same result as for the general correlator in the original Hermitian matrix model. We thus see that changing pictures is merely a change of basis, and that changing the variables of integration results in a trivial Jacobian. In particular, the ribbon graph and dessins d'enfants methods of representing correlator contributions diagrammatically is still valid, where the edges now represent pairs of spin labels ($l$, $m$).
We can therefore conclude that if $\sigma$ and $\tau$ represent a Wick contraction of a correlator, and the associated ribbon graph (or dessin d'enfant) has $F$ faces and $E$ edges, then the evaluation of the contribution to the correlator is
\bea N^{-E}\eta^{i_1j_1}\ldots\eta^{i_Ej_E}\tr_{2E}(Y_1\otimes\ldots\otimes Y_{2E} \sigma) = N^{F-E}. \eea

A natural normalisation to use for traces of permutations is to weight a correlator described by the permutation $\sigma\in S_{2E}$ by $N^{C_\sigma}$. For the ribbon graph associated to such a correlator, $C_\sigma$ is the number of vertices $V$. Therefore, using the formula for the Euler character of a closed two-dimensional surface, we can write the {\bf ribbon graph evaluation} ${\cal R}$ of a connected ribbon graph $\G$ in terms of its genus $g$,
\bea \rG := N^{V-E}\eta^{i_1j_1}\ldots\eta^{i_Ej_E}\tr_{2E}(Y_1\otimes\ldots\otimes Y_{2E} \sigma) = N^{2-2g}. \label{eq:rg} \eea
Denoting by $g_i$ the genus of each connected component of a ribbon graph $\G$, we can expand a general correlator in terms of Ribbon graphs with

\bea N^V\cor{\tr(\mathbf{X}\sigma)} &=& \sum_{\substack{\mathrm{Wick}\\ \mathrm{contractions}}} \rG = \sum_{\substack{\mathrm{Wick}\\ \mathrm{contractions}}} \prod_{i} N^{2-2g_i}. \eea
In this sum, each Wick contraction determines a ribbon graph $\G$. The evaluation of the graph $\G$, denoted by $\rG$, is $\prod_i N^{2-2g_i}$. In the following, we will specialise to connected ribbon graphs with genus $g$, which evaluate to $N^{2-2g}$. 
The focus of this paper is on developing a three-dimensional interpretation of $\rG$ using the Ponzano-Regge model. 
 
\subsection{Trivalent ribbon graphs}

In this section we demonstrate that the contributions to correlators in the fuzzy sphere matrix model are naturally understood using trivalent connected ribbon graphs.
We have shown that there is a correspondence between contributions to correlators in the fuzzy sphere and contributions to correlators in the matrix model, as they are both represented by ribbon graphs expressing the combinatorial data. For the fuzzy sphere ribbon graphs, each vertex represents a trace of fuzzy spherical harmonics, and each edge represents a sum over spin labels weighted by a factor of $N^{-1}\eta^{ij}$. The traces of general products of fuzzy spherical harmonics are built up by the contractions of factors of $A_{ijk}$ with each other using the raising operators $\eta^{ij}$, where the factor $A_{ijk}$ is a trace of three spherical harmonics. Thus we see that any contribution to a correlator can be expressed in terms of products of traces of triples of spherical harmonics. This suggests that the ribbon graph of a general correlator has an equivalent ribbon graph where all the vertices are trivalent. We show that this indeed the case. 

\begin{figure}[H]
\centering
\includegraphics[width=0.7\textwidth]{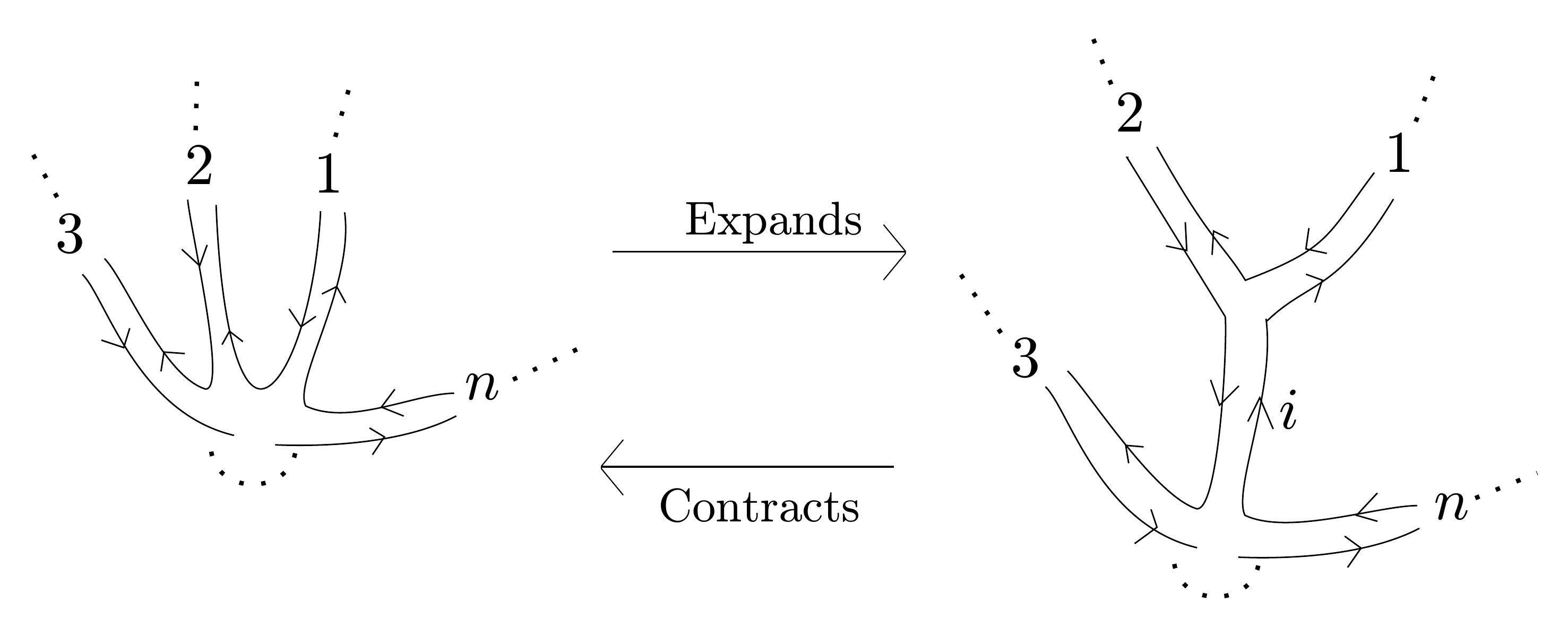}
\caption{Contraction and expansion of a ribbon graph}\label{fig:contexp}
\end{figure}

A general property of ribbon graphs is that a collection of edges around a vertex can be `pulled off' in a manner preserving the cyclic ordering at each vertex to generate a new graph with an extra edge and vertex, but preserving the number of faces and the genus of the graph. Hence the correlator contributions associated to a ribbon graph before and after `expansion' are equivalent. For example, in the expansion shown in Figure \ref{fig:contexp}, a vertex with $n$ outgoing half-edges has been expanded into two vertices with 3 and $(n-1)$ outgoing half-edges respectively, connected by a new edge. 

In particular, this expansion can be performed repeatedly until all vertices in the graph are trivalent. In such a ribbon graph, all vertices are associated with the factors $A_{ijk}$, and all edges with the factors $\eta^{ij}$. We can therefore restrict our attention to correlators of the form $N^V\cor{(\tr X^3)^{V}}$, which are generated by the expression

\bea Z = \int DX e^{-\frac{N}{2}\tr X^2 - N\tr X^3 } = \sum_V \frac{N^V}{V!}\cor{(\tr X^3)^{V}} =
\sum_{\substack{\mathrm{Wick}\\ \mathrm{contractions}}} N^{2-2g}. \eea

To conclude this section, we consider the example of a ribbon graph in the expansion of $N\cor{\tr X^6}$.

\begin{figure}[H]
\centering
\subfloat[]{\label{fig:egnp}\includegraphics[width=0.3\textwidth]{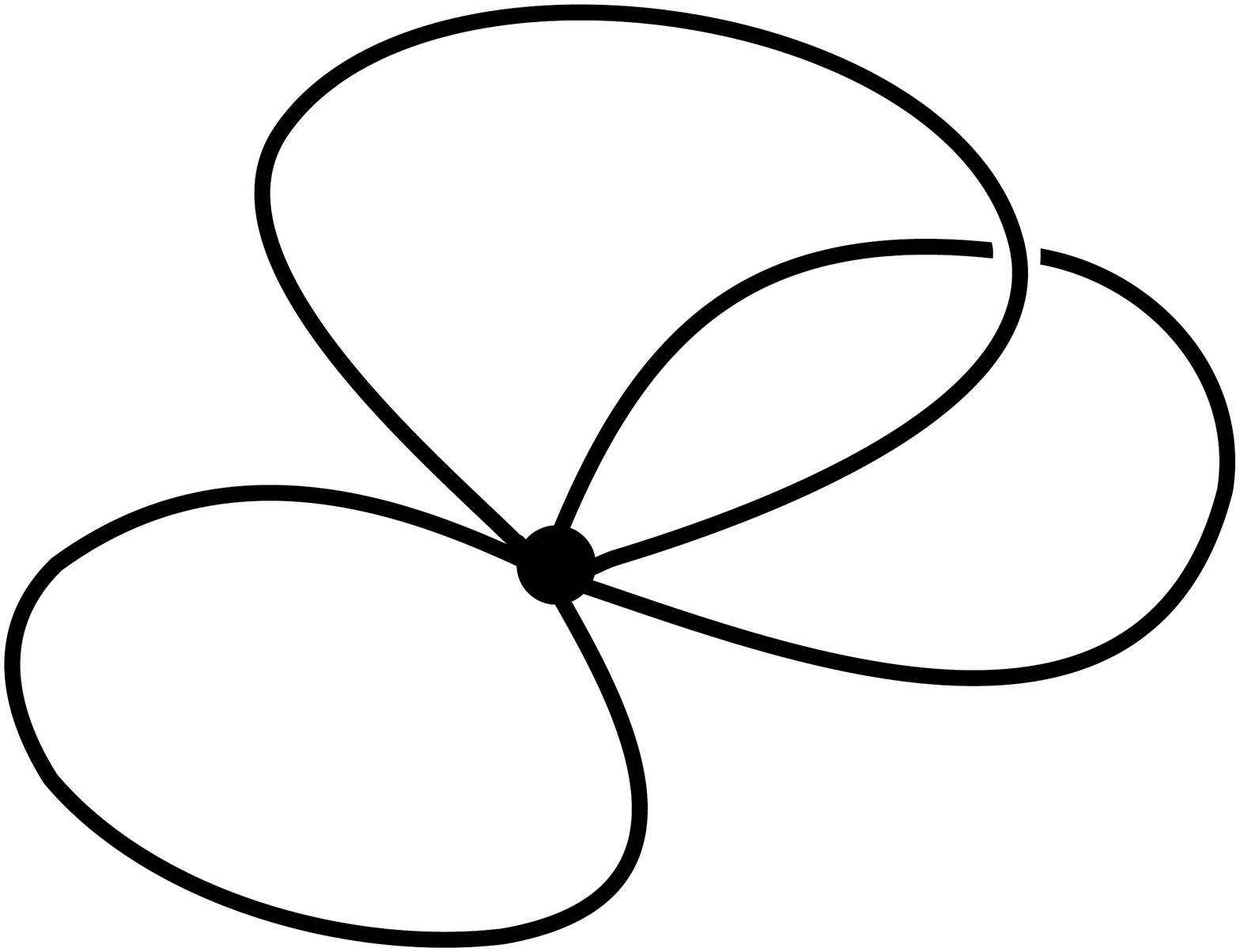}}
\subfloat[]{\label{fig:egnpe}\includegraphics[height=0.3\textwidth]{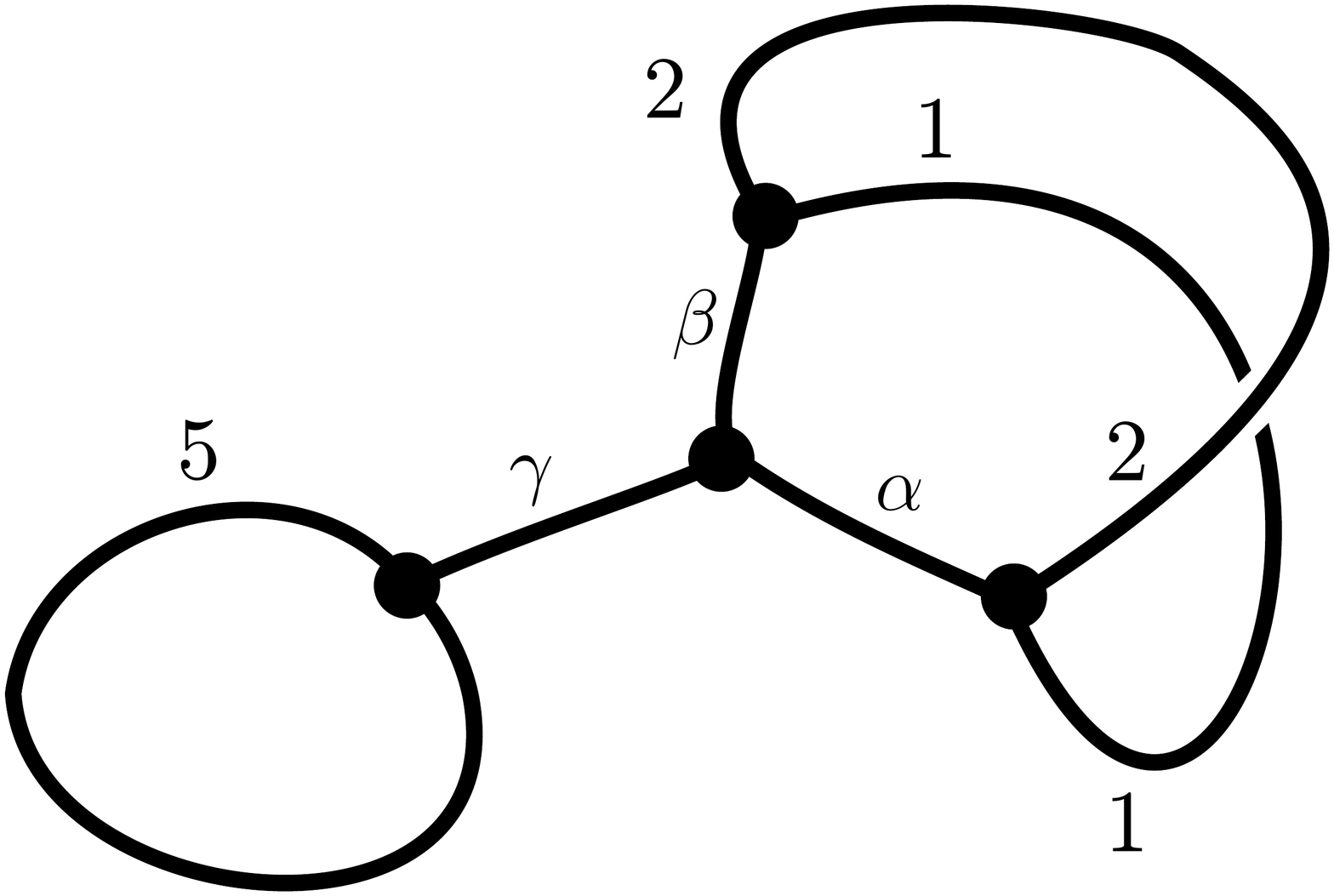}}
\caption{A non-planar $\tr(X^6)$ graph and an expansion of the same graph.}
\label{fig:triexpansion}
\end{figure}

The ribbon graph in Figure \refb{fig:egnp} represents the contribution to the correlator \bea N^{-2}\eta^{13}\eta^{24}\eta^{56}\tr (Y_1Y_2Y_3Y_4Y_5Y_6) = N^0 = 1. \eea
The products of the spherical harmonics in the trace can be written in terms of $A_{ijk}$ symbols by using the product rule of the algebra
\bea Y_1 Y_2 = A_{12}^{\ \ 3}Y_3, \eea
and the choice of which products to take is equivalent to the choice of expansion of the ribbon graph. For example, the expansion shown in Figure \refb{fig:egnpe} corresponds to

\bea
N^{-2}\eta^{13}\eta^{24}\eta^{56}\tr (Y_1Y_2Y_3Y_4Y_5Y_6) &=& N^{-2}\tr((Y_1Y_2)(Y^1Y^2)(Y_5Y^5)) \ret
&=& N^{-2}A_{12}^{\ \ \alpha}A^{12\beta}A_{5}^{\ 5\gamma}\tr(Y_\alpha Y_\beta Y_\gamma) \ret
&=& N^{-2} A_{12}^{\ \ \alpha}A^{12\beta}A_{5}^{\ 5\gamma}A_{\alpha\beta\gamma}.
\eea

We can develop more understanding of the fuzzy sphere interpretation of the matrix model by considering the structure of the sums over these factors $A_{ijk}$ - in particular, their constituent Wigner $3j$ and $6j$ symbols. 

\subsection{Separating the Wigner $3j$s and $6j$s}

The sums corresponding to the trivalent graphs are performed over different $\lsu$ representations labelled by $l_i$ and by the states in the representations $m_i$. It is possible to separate the sums out into an $m_i$-dependent part and a $l_i$-dependent part, and to perform the sums over the $m_i$ first to arrive at an expression that has no dependence on the states within the $\lsu$ representations, but only on the representation labels $l_i$. Although we know that the final evaluation of a ribbon graph sum will always be $N^{2-2g}$, this decomposition of the sum is still a useful approach to take because it results in a link between the Hermitian matrix model and theories involving spin networks and the Ponzano-Regge model.

A general correlator contribution on the fuzzy sphere can be expressed entirely as a sum with weights $NA_{123}$ and $N^{-1}\eta^{12}$, where each number $i$ represents a pair of angular momentum variables $l_i$, $m_i$. The contractions between these symbols can be encoded by permutations, or by cyclically ordered trivalent graphs, with a factor $A_{123}$ assigned to each vertex and a propagator $\eta^{12}$ to each line. This sum could also be written out fully in terms of Wigner $3j$ and $6j$ symbols with phases and representation dimension weights as follows: recall the definitions

 \bea A_{123} = N^{\frac{3}{2}}(-)^{3j}\sqrt{(2l_1+1)(2l_2+1)(2l_3+1)}\wigthreej{l_1}{l_2}{l_3}{m_1}{m_2}{m_3} \wsjj{l_1}{l_2}{l_3}, \eea
\bea \eta^{12}= N^{-1}(-)^{m_1}\delta_{l_1,l_2} \delta_{m_1,-m_2}. \eea

Now, since we are considering only trivalent graphs with no exterior edges, we know that the number of edges $E$ and number of vertices $V$ satisfies $2E=3V$, which means that the total factor of $N$ at the front of the expression is $N^{V+\frac{3}{2}V-2E} = N^{V-E}$. Hence, we can associate a factor of
\bea N(-)^{3j}\wigthreej{l_1}{l_2}{l_3}{m_1}{m_2}{m_3}\wsjj{l_1}{l_2}{l_3} \eea
to each vertex, a factor
\bea \sqrt{\frac{2l_1+1}{N}}(-)^{m_1}\delta_{l_1,l_2}\delta_{m_1,-m_2} \eea
to each half-edge, and perform the sum over all $l_i$, $m_i$.
The $2E$ sets of spin labels ($l_i$, $m_i$) correspond to the different half-edges of the trivalent graph, so summing exactly half of these labels can immediately reduce the sum to $E$ sets of spin labels, introducing minus signs in the $3j$ symbols. This can be represented diagrammatically by assigning orientations to the edges. Hence, for a general ribbon graph $\G$, we arrive at the expression for a correlator contribution,
\bea \rG :=  (-)^{2jE}N^{V-E} \sum_{l_i m_i} \prod_{\mathrm{edges}}(-)^{m_i}(2l_i+1)\prod_{\mathrm{vertices}}\wsjj{l_i}{l_j}{l_k}\wigthreej{l_i}{l_j}{l_k}{\pm m_i}{\pm m_j}{\pm m_k},\quad  \eea
where $m_i$ appears with a positive sign in the Wigner $3j$ if the edge is directed towards the vertex. 

We can partition this sum into two parts by considering just the $m_i$-dependent terms, which are the $3j$s and phase factors $(-)^m$, and performing the sums over the labels $m_i$. This expression depends purely on the structure of the graph $\G$ and the spin labels $l_i$ assigned to the edges, and is essentially the evaluation of a spin network discussed in \cite{major:1999md}. Hence, we call this part of the sum the {\bf spin network state sum}, or alternatively the $3j$ sum, associated to a graph with $l_i$-labelled edges.
The $m_i$-dependent part of the sum for a graph ${\cal G}$ is
\bea \label{eq:sgee}
S[{\cal G},\{ l_i\}] := \sum_{m_i} \prod_{\mathrm{edges}}(-)^{m_i}\prod_{\mathrm{vertices}}\wigthreej{l_i}{l_j}{l_k}{\pm m_i}{\pm m_j}{\pm m_k}.
\eea 
This sum is invariant under the interchange $m_i\to-m_i$, so the orientations assigned to edges in a ribbon graph are arbitrary, and only relevant when constructing and evaluating the expression \refb{eq:sgee}.
We can write the total trivalent ribbon graph evaluation,
\boxeq{ \label{eq:totalsum}
\rG = (-)^{2jE}N^{V-E} \sum_{l_i}S[{\cal G}, \{l_i\}]\prod_{\mathrm{edges}}(2l_i+1)\prod_{\mathrm{vertices}}\wsjj{l_i}{l_j}{l_k}. } 
 The ribbon graph contribution to a correlator,  $\rG$, is here 
 expressed in  a factorised form containing the spin network state sum.
The evaluation of the spin network state sum $\sG$ cannot usually be performed by inspection, but can be deduced in all cases by employing an algorithm of $3j$ identities in a systematic manner. These $3j$ identities correspond to operations on the trivalent graphs which we call \textbf{trivalent graph moves}, or $3j$ moves, and will also prove to be useful in understanding the 3D interpretation of the graphs in the next section. 

Before presenting the trivalent graph moves, we first discuss two special graphs that are relevant to the algorithic evaluation of $\sG$.
The simplest trivalent graph with no external edges or self-connecting vertices is the `theta' graph, denoted $\Theta$,

\begin{eqnarray}
S\left[  \raisebox{-35pt}{\includegraphics[width=0.25\textwidth]{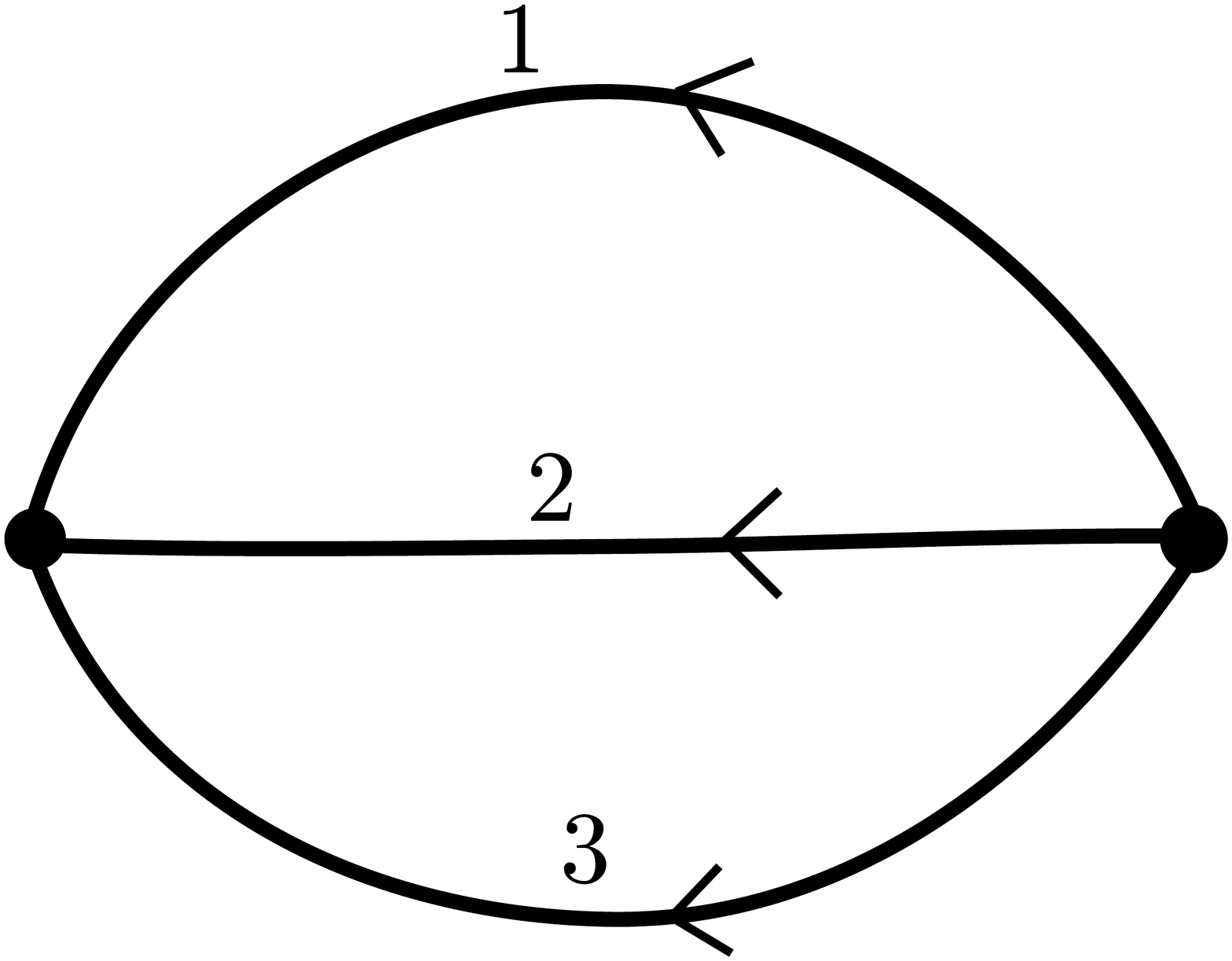}} \right] &=& \sum_{m_i}(-)^{m_1+m_2+m_3}\wigthreej{l_1}{l_3}{l_2}{m_1}{m_3}{m_2}\wigthreej{l_1}{l_2}{l_3}{-m_1}{-m_2}{-m_3} \nonumber \ret
&=& \sum_{m_1,m_2,m_3}\wigthreej{l_1}{l_2}{l_3}{m_1}{m_2}{m_3}\wigthreej{l_1}{l_2}{l_3}{m_1}{m_2}{m_3} \nonumber \ret & & \nonumber \ret
&=& 1.\label{eq:theta}
\end{eqnarray}
Here we have used only the symmetries and orthogonality properties of the $3j$s to evaluate $S[\Theta]$. 
We have also implicitly assumed that the labels $(l_1, l_2, l_3)$ satisfy a \emph{triangle constraint} for the $3j$s to be non-vanishing (see Appendix \ref{sec:awig}), which is enforced by the factor $\wsjj{l_1}{l_2}{l_3}$ in (\ref{eq:totalsum}). 

The second graph is the tetrahedral network, whose spin network state sum evaluates to a $6j$ symbol purely by definition.
\begin{multline}
S\left[  \raisebox{-45pt}{\includegraphics[width=0.25\textwidth]{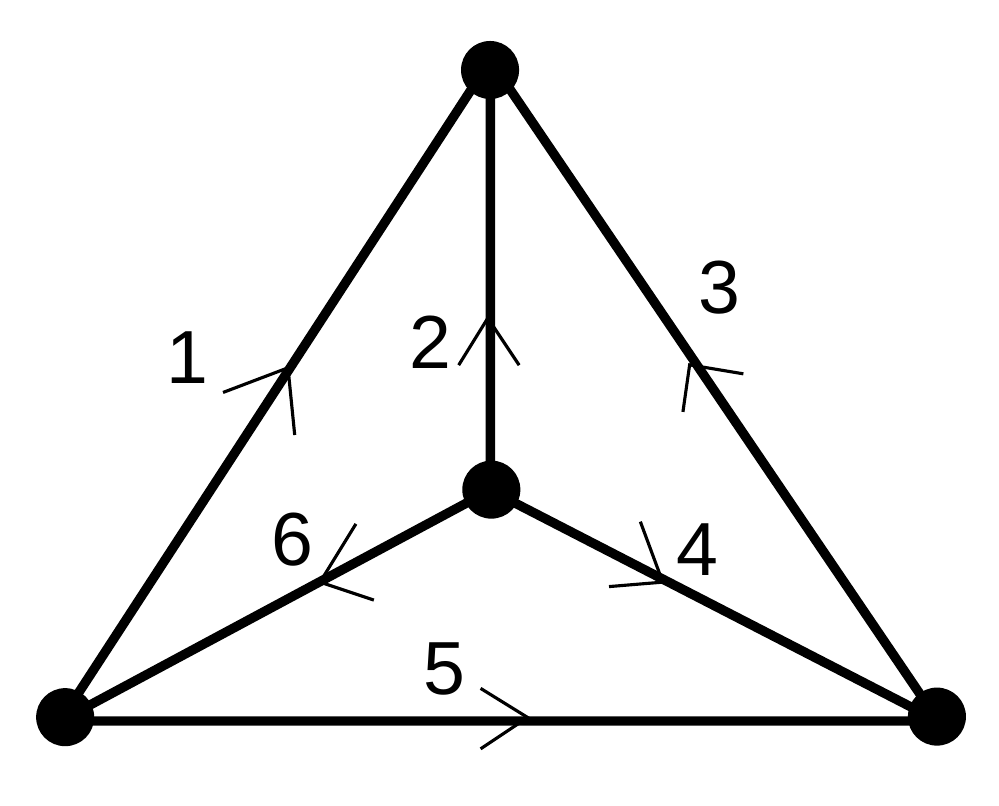}} \right] = \sum_{m_i}(-)^{m_1+m_2+m_3+m_4+m_5+m_6}
\wigthreej{l_1}{l_2}{l_3}{m_1}{m_2}{m_3} \times  \\ \\
\times \wigthreej{l_1}{l_5}{l_6}{-m_1}{-m_5}{m_6} \wigthreej{l_3}{l_4}{l_5}{-m_3}{m_4}{m_5}\wigthreej{l_2}{l_6}{l_4}{-m_2}{-m_6}{-m_4} \quad :=  \wsj{l_1}{l_2}{l_3}{l_4}{l_5}{l_6}. \quad \label{eq:sixj} \end{multline}

\subsection{The trivalent graph moves}\label{sec:moves}

To present the moves on trivalent graphs associated to $3j$ identities more clearly, we extend the definition of $S$ to include graphs with external edges.
By convention, we do not assign a weight of $(-)^{m_i}$ to the external edges, and do not sum over their labels, but will reintroduce these required weights and sums when we connect all the external edges together to create a complete graph.

\begin{enumerate}
\item \textbf{The orthogonality relation} between two $3j$s can be expressed as

\bea
S\left[  \raisebox{-25pt}{\includegraphics[width=0.25\textwidth]{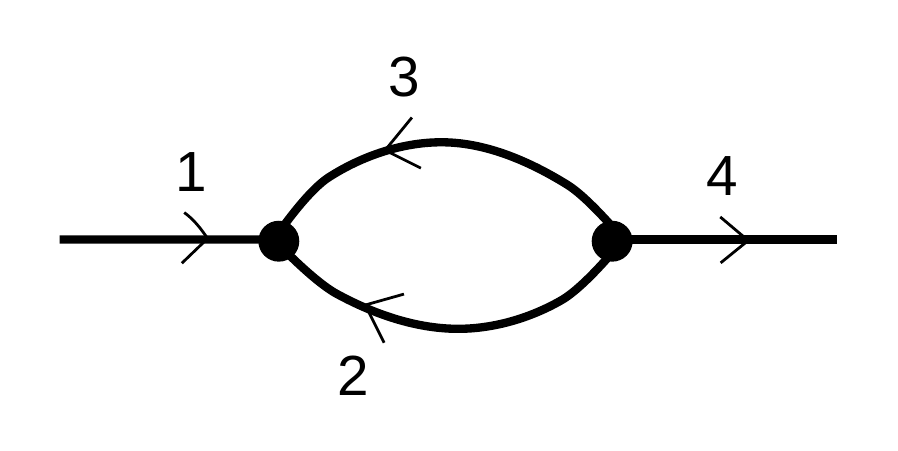}} \right] &=& \sum_{m_2,m_3}(-)^{m_2+m_3}\wigthreej{l_1}{l_2}{l_3}{m_1}{m_2}{m_3}\wigthreej{l_2}{l_4}{l_3}{-m_2}{-m_4}{-m_3} \ret
&=& \frac{(-)^{-m_1}}{(2l_1+1)}\delta_{l_1,l_4}\delta_{m_1,m_4}
\eea
By writing the factor $(2l_1+1)$ diagrammatically as a loop, and denoting a delta function (on $l$ and $m$ labels) by a straight line, we can write this expression as
\begin{eqnarray} S\left[  \raisebox{-35pt}{\includegraphics[width=0.35\textwidth]{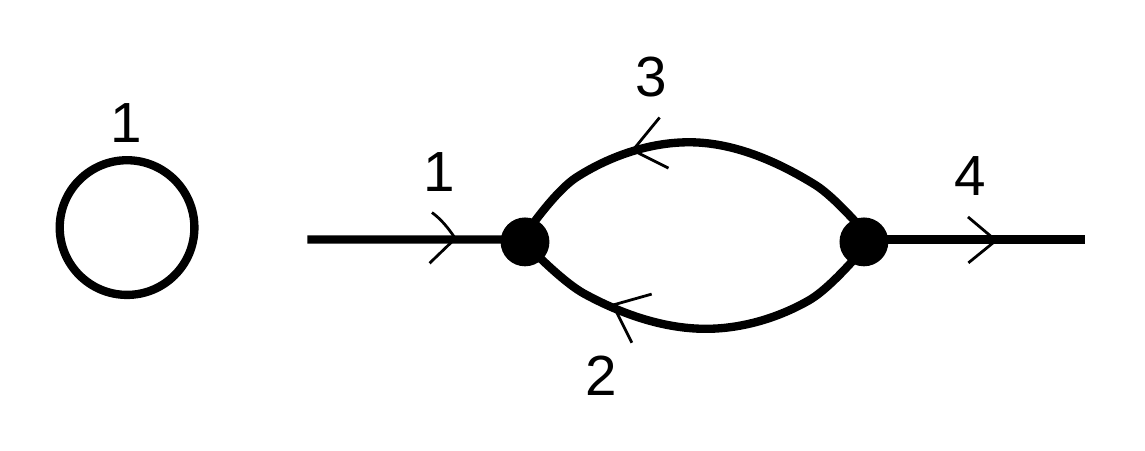}} \right] = S\left[  \raisebox{-25pt}{\includegraphics[width=0.25\textwidth]{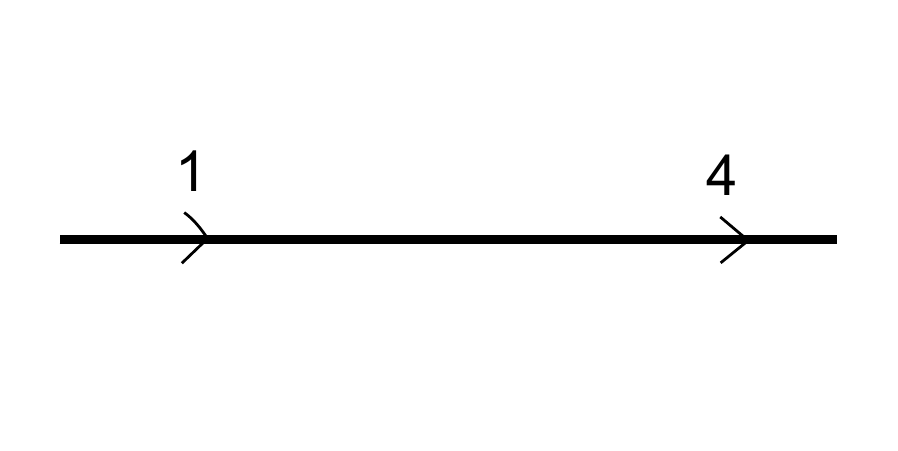}}\right] \label{eq:orthog}
\end{eqnarray}
We note that $S[{\cal G H}] = S[{\cal G}]S[{\cal H}]$ for a pair of disconnected graphs ${\cal G}, {\cal H}$. The $(-)^{-m}$ factor that appears in this identity can be interpreted as ensuring that each line in the reduced graph has just a single associated factor of $(-)^m$. In addition, by setting $l_1=l_4$ and introducing the required factor of $(-)^{m_1}$ we can recover \refb{eq:theta}. We also note that this identity cannot reduce down the tetrahedral network from \refb{eq:sixj} to a simpler evaluation, since there are no two edges in the graph connected to the same two vertices.

\item The $3j$ symbols also satisfy an identity corresponding to \textbf{the `2-2' move}

\begin{multline}
\sum_{m_3}(-)^{m_3}\wigthreej{l_1}{l_2}{l_3}{m_1}{m_2}{-m_3}\wigthreej{l_4}{l_5}{l_3}{m_4}{m_5}{m_3} = \\
= \sum_{l_6,m_6}(-)^{m_6}(2l_6+1) \wigthreej{l_5}{l_1}{l_6}{m_5}{m_1}{m_6}\wigthreej{l_2}{l_4}{l_6}{m_2}{m_4}{-m_6}\wsj{l_1}{l_2}{l_3}{l_4}{l_5}{l_6}.
\end{multline}
This can be expressed diagrammatically as
\begin{eqnarray} 
\hspace{-10mm} S\left[  \raisebox{-35pt}{\includegraphics[width=0.25\textwidth]{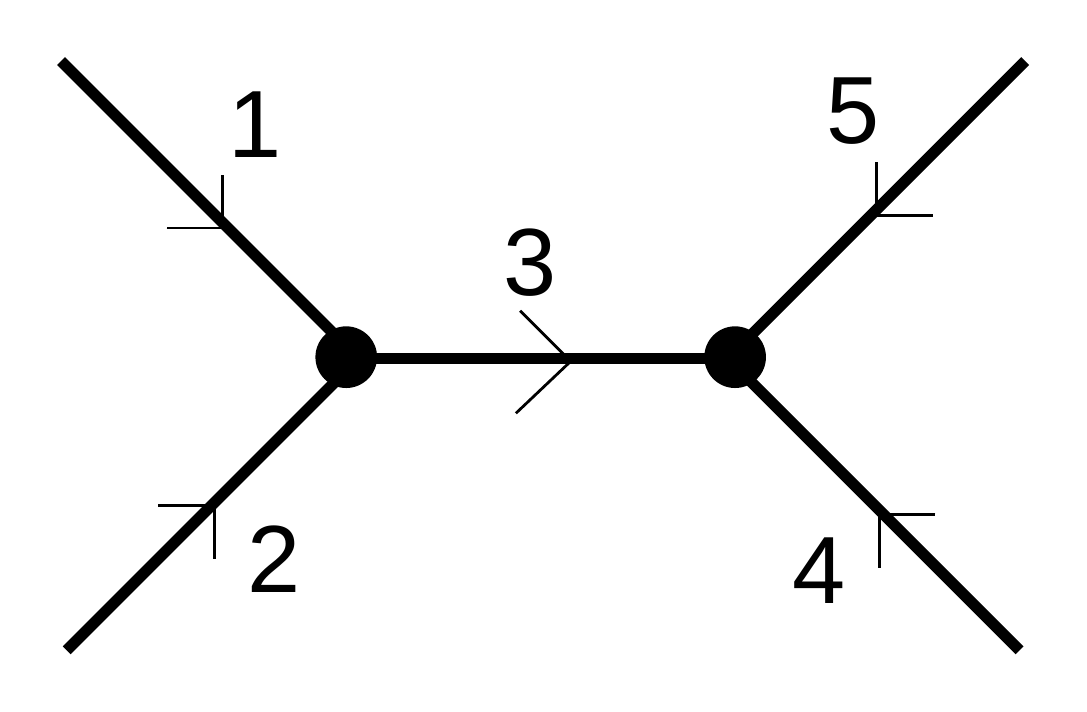}} \right] = \sum_{l_6}S\left[  \raisebox{-55pt}{\includegraphics[width=0.55\textwidth]{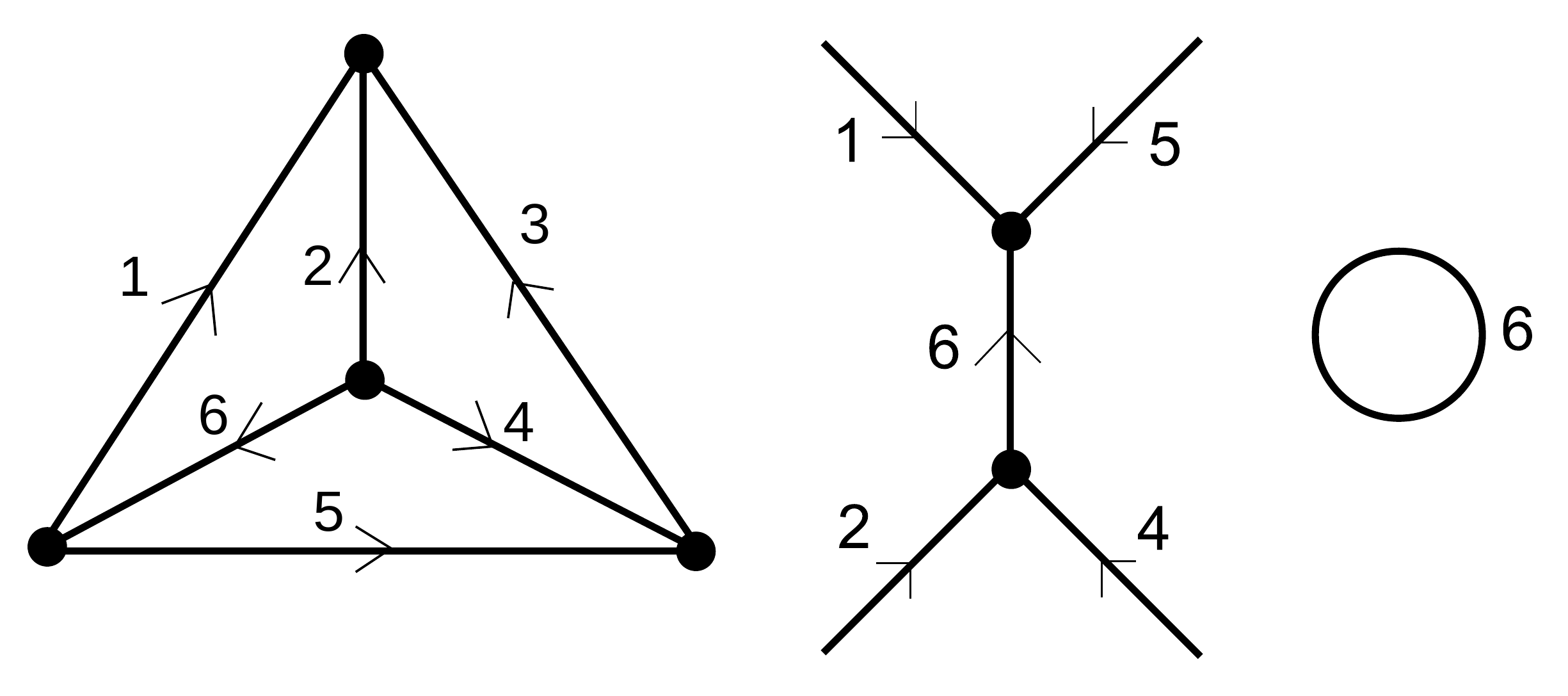}}\right] \label{eq:four}\qquad
\end{eqnarray}

\item The third $3j$ identity is associated to \textbf{the `3-1' move}, which reduces three $3j$ symbols to a single $3j$ and a $6j$,

\begin{multline} \sum_{m_4,m_5,m_6}(-)^{m_4+m_5+m_6} \wigthreej{l_5}{l_1}{l_6}{m_5}{m_1}{-m_6} \wigthreej{l_4}{l_3}{l_5}{m_4}{m_3}{-m_5} \wigthreej{l_6}{l_2}{l_4}{m_6}{m_2}{-m_4} = \\ 
= \wigthreej{l_1}{l_2}{l_3}{m_1}{m_2}{m_3}\wsj{l_1}{l_2}{l_3}{l_4}{l_5}{l_6} \quad \end{multline}
\bea 
S\left[  \raisebox{-0.08\textheight}{\includegraphics[height=0.16\textheight]{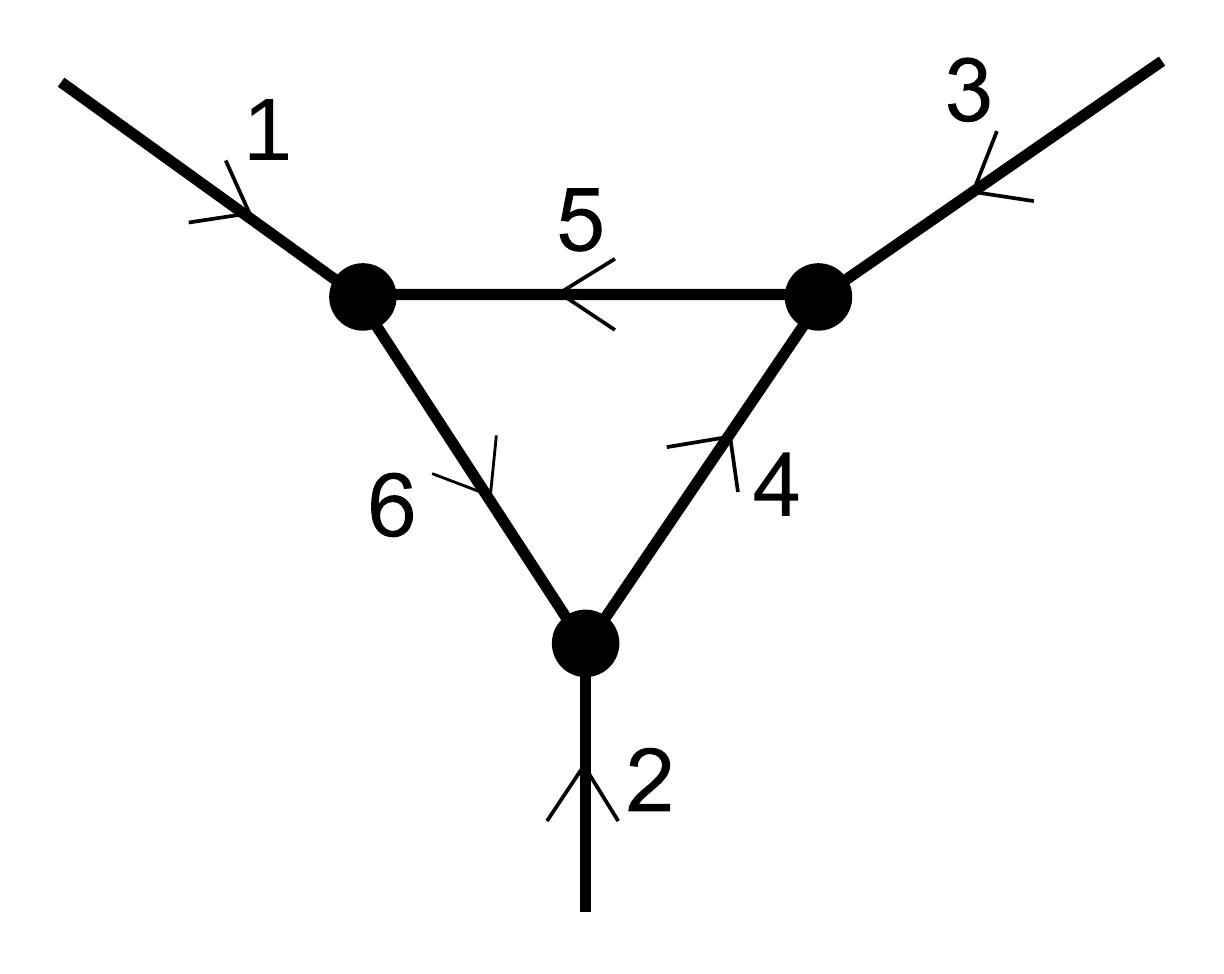}} \right] = S\left[ \raisebox{-0.08\textheight}{\includegraphics[height=0.16\textheight]{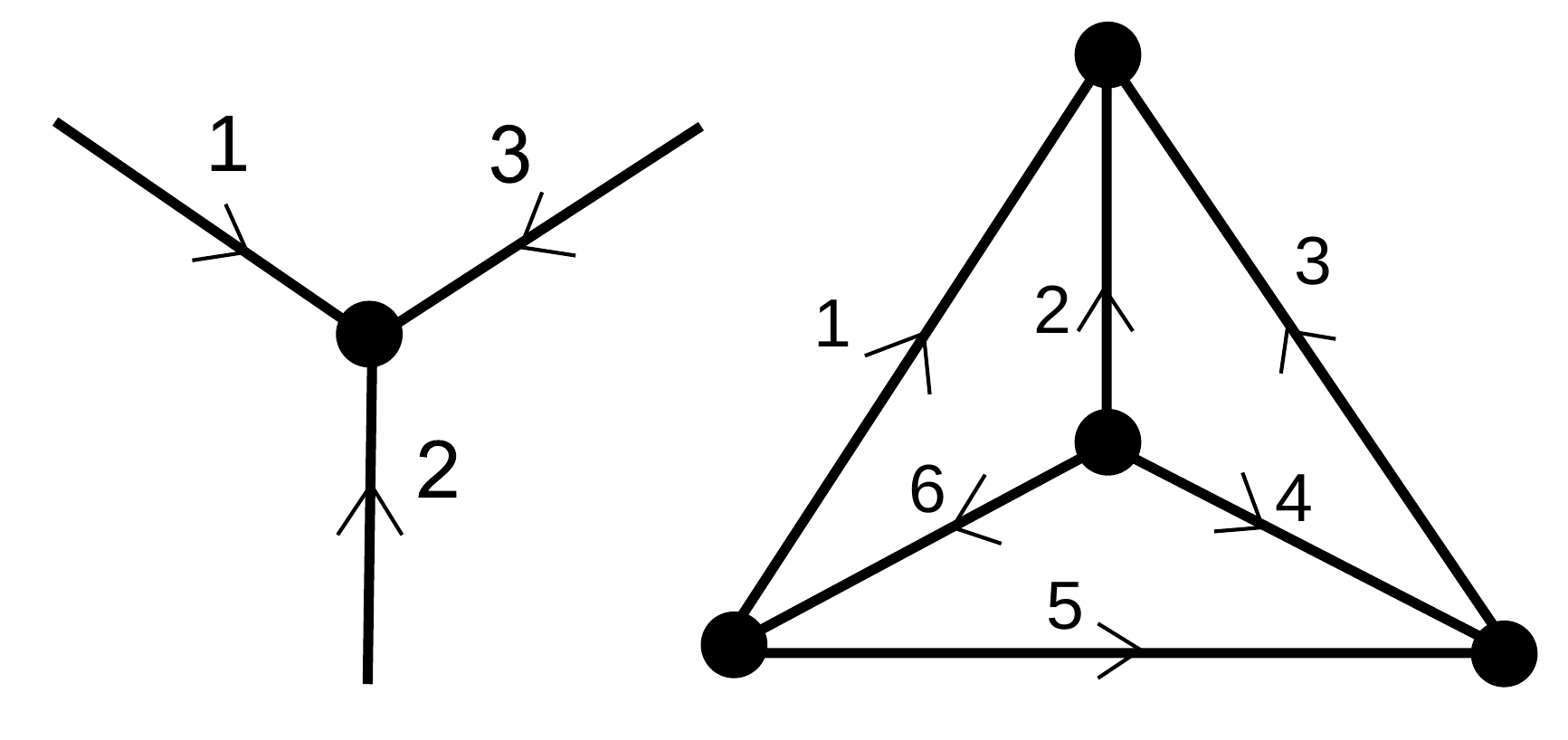}}\right] \label{eq:loop}
\eea
The inverse of this move is called the `1-3' move.

We note that the 3-1, 2-2 and orthogonality moves are not independent, as the 3-1 move could be deduced from the application of the 2-2 move and orthogonality. Alternatively, the orthogonality relation could be deduced from the 2-2 and 3-1 moves, provided the graph considered has more than two vertices. However, it is useful to include all these moves in the set for later applications.

\item The final necessary $3j$ move is \textbf{the `parity' move}, which is the permutation of a pair of edges at a vertex. This move is necessary to reduce down non-planar graphs to planar graphs, and is not needed for the reduction of planar graphs.

\bea  \wigthreej{l_1}{l_2}{l_3}{m_1}{m_2}{m_3} =   (-)^{l_1+l_2+l_3}\wigthreej{l_2}{l_1}{l_3}{m_2}{m_1}{m_3}. \eea  
\bea 
S\left[  \raisebox{-0.08\textheight}{\includegraphics[height=0.16\textheight]{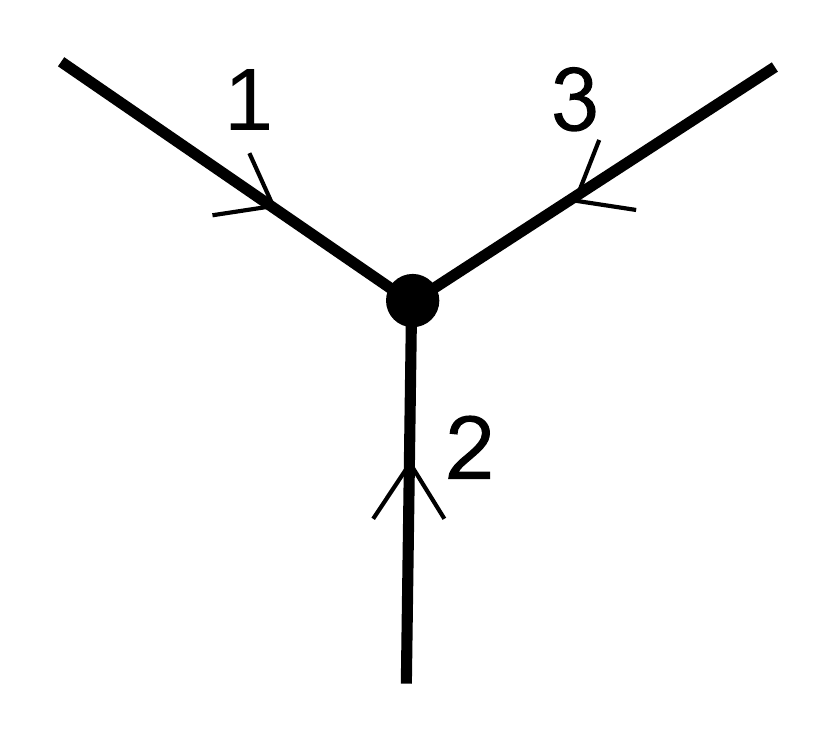}} \right] = (-)^{l_1+l_2+l_3}\quad S\left[ \raisebox{-0.08\textheight}{\includegraphics[height=0.16\textheight]{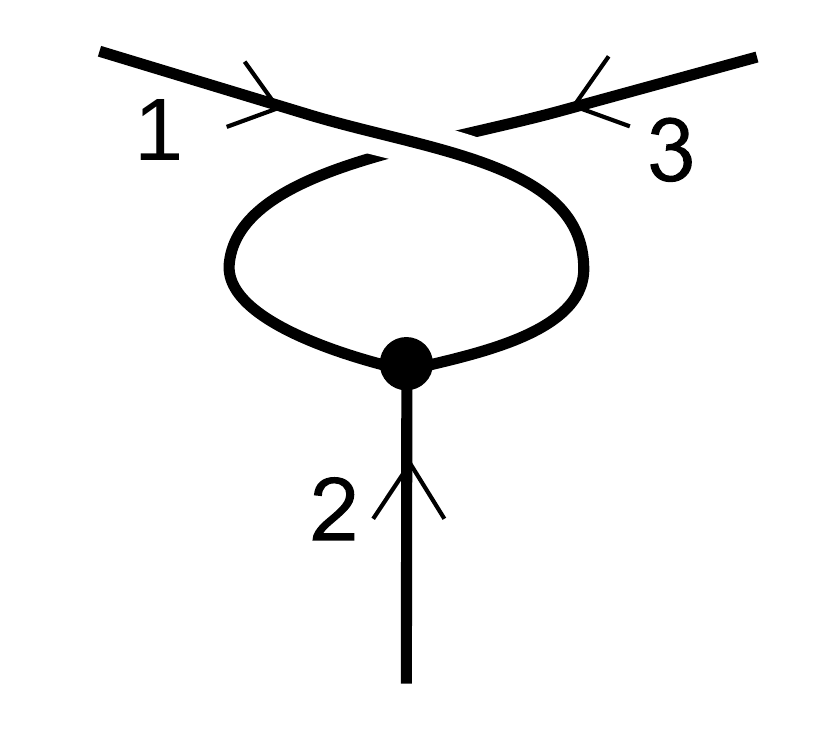}}\right] \label{eq:np}
\eea
\end{enumerate}

\begin{figure}[h]
\centering 
\includegraphics[width=0.6\textwidth]{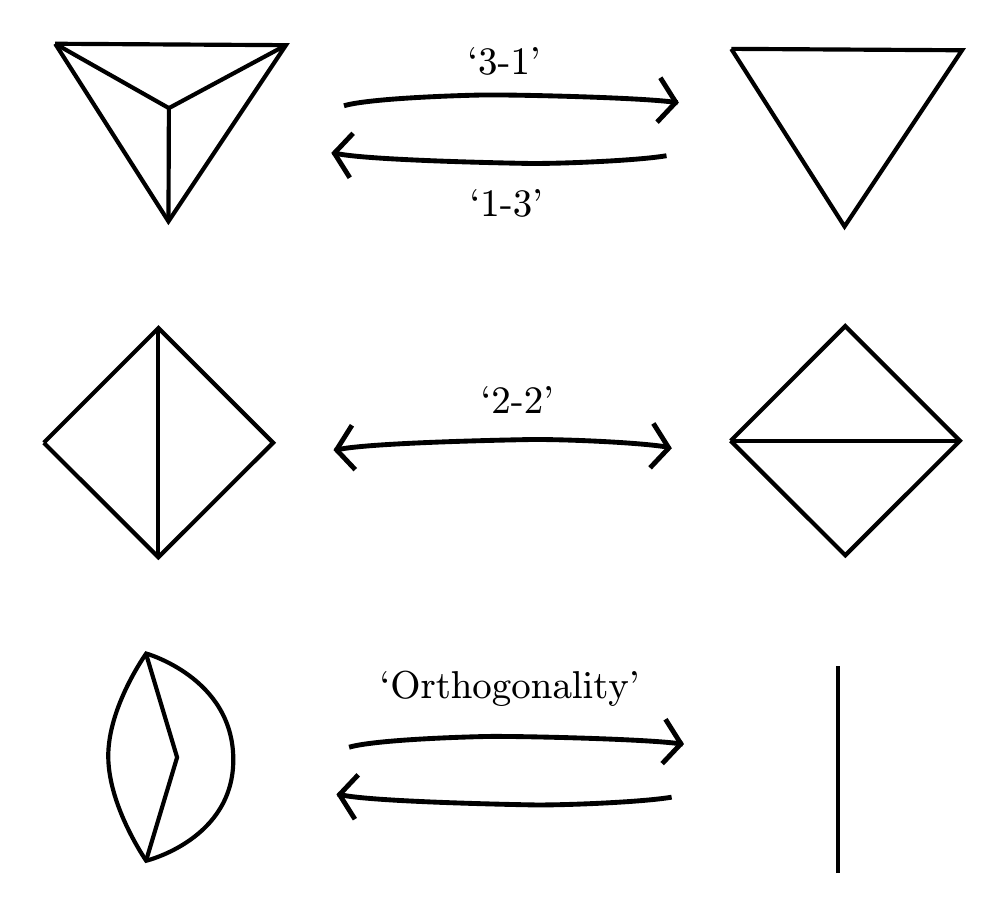}
\caption{The 3-1, 2-2, and orthogonality Alexander moves are dual to the genus-preserving trivalent graph moves.}\label{fig:alexander}
\end{figure}

In Section \ref{sec:PRNP}, it will also be useful to apply the 2D duals of the first three moves, which are called the {\bf Alexander moves}, or 2D Pachner moves. The 2D dual of a trivalent graph is a triangulation of a surface, and as the first three moves do not alter the genus of a graph, the Alexander moves relate triangulations of the same surface. It was proved in \cite{alexander} that any two
triangulations of a surface of the same genus can be related by a finite series of the Alexander moves. These moves are listed in
Figure \ref{fig:alexander}.

\subsection{Algorithmic evaluation of $\sG$}\label{sec:algsg}

The action of these moves on a labelled trivalent graph will generate in the spin network state sum evaluation $\sG$ a string of factors of $(2l+1)$, $(-)^{l_1+l_2+l_3}$, and sums over new labels, as well as the $6j$ symbols, which are the evaluations of tetrahedral networks. The tetrahedral networks are irreducible under the trivalent graph moves, as any application of the moves on them will generate the same $6j$ or an expression which evaluates to the same $6j$. 
Therefore, the evaluation of a general graph by reduction must be a sum of a product of these factors. We show algorithmically that all trivalent graphs can be reduced down in this manner.

A trivalent ribbon graph partitions a surface into vertices, edges, and faces. Each vertex is incident to either one, two, or three faces, and each edge is incident to either one or two faces. We say that a face is a {\bf polygon} if it is homeomorphic to a disc when considered with its bounding edges and vertices. A necessary and sufficient condition for a face to be a polygon is for each edge bounding the face to be incident to two distinct faces.

The first step in the algorithm is to isolate a polygon of the ribbon graph. Not all ribbon graphs possess a polygonal face, but it is always possible to generate such a face from any ribbon graph by applying a single parity move. To see this, follow the boundary of a face around 
until a vertex is visited twice, and apply the parity move at this vertex. This will always produce a polygon from a non-polygonal face.
Since a planar graph always has a polygonal face, a planar graph can be reduced without applying the parity move. Also, a non-planar graph will eventually reduce down to graph with no polygonal faces, so it is always necessary to apply a parity move at least once to evaluate a non-planar graph.
Once a polygon has been isolated, we can use a combination of the remaining trivalent moves to remove the polygon from the graph and reduce the number of vertices by two. 

\begin{figure}[h]
\centering 
\includegraphics[width=0.6\textwidth]{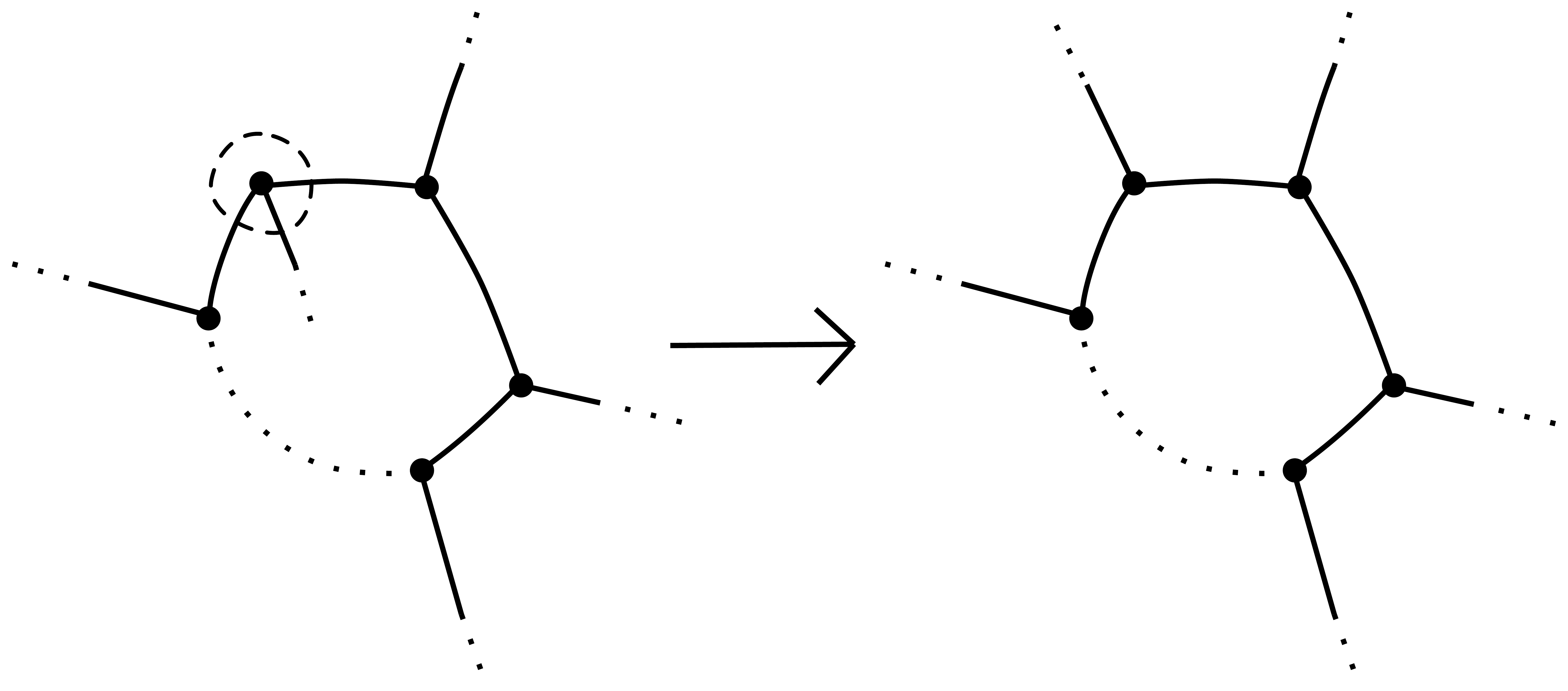}
\caption{The parity move produces a polygon from a face of a non-planar ribbon graph.}\label{fig:poly2}
\end{figure}

If the polygon is bounded by a single edge, then it is a `tadpole', and can be removed using the 2-2 move and orthogonality relation. Applying the 2-2 move on the edge that connects the vertex to a different vertex, as in \refb{eq:tadpole1},

\begin{eqnarray} S\left[  \raisebox{-40pt}{\includegraphics[width=0.20\textwidth]{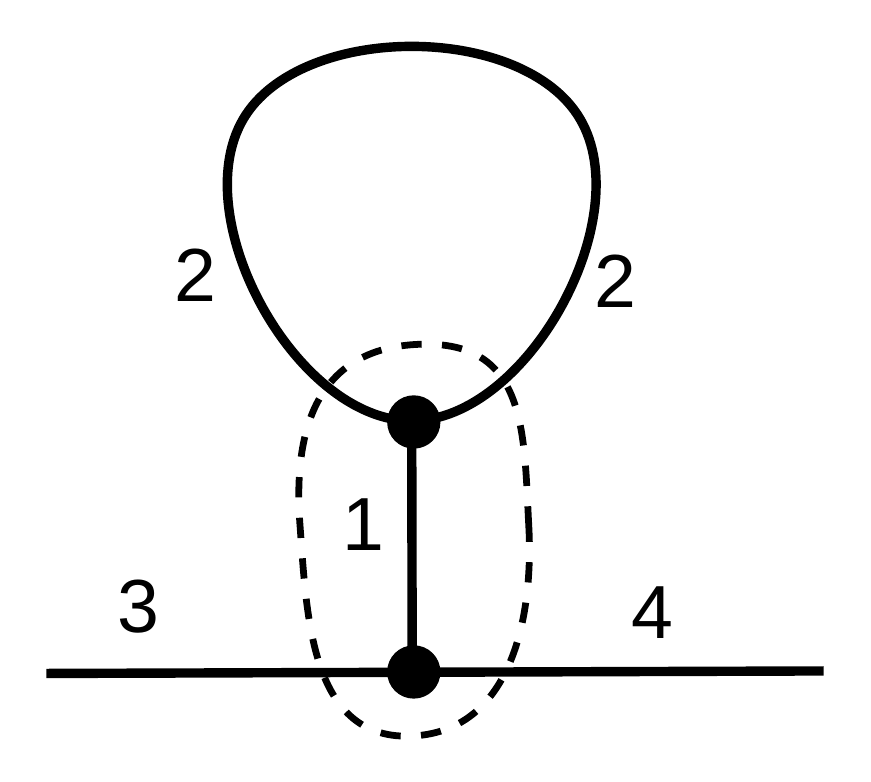}} \right] = \sum_{l_5}S\left[  \raisebox{-30pt}{\includegraphics[width=0.50\textwidth]{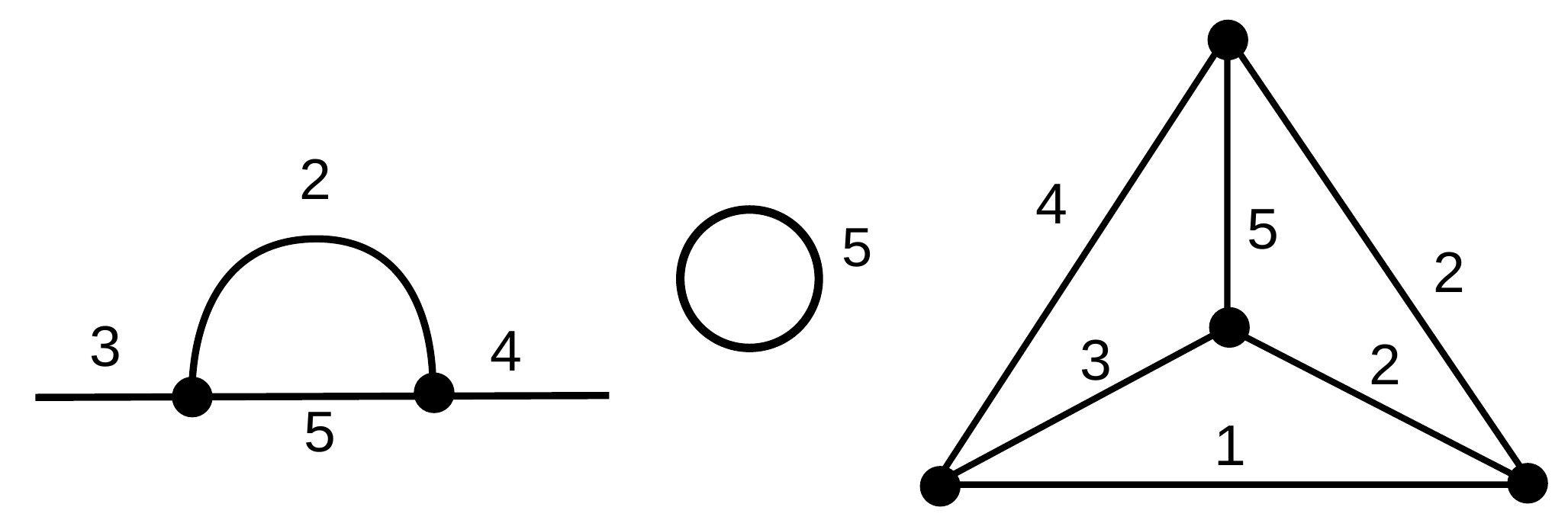}}\right] \label{eq:tadpole1}
\end{eqnarray}
Now, using the orthogonality move, and the $6j$ identity
\bea \sum_{l_5}\left(\frac{2l_5+1}{2l_3+1}\right)\wsj{l_1}{l_2}{l_2}{l_5}{l_3}{l_3} = (-)^{l_2+l_3}\delta_{l_1,0}\sqrt{\frac{2l_2+1}{2l_3+1}}, \eea
we can evaluate the sum directly and deduce that
\bea S\left[  \raisebox{-30pt}{\includegraphics[width=0.20\textwidth]{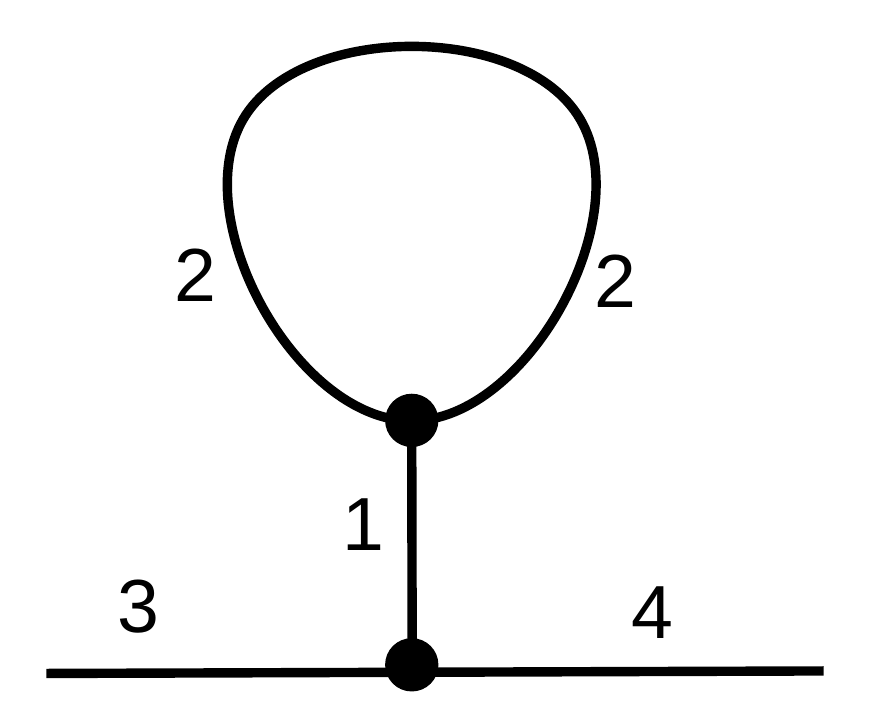}} \right] = \delta_{l_1,0}(-)^{l_2+l_3}\sqrt{\frac{2l_2+1}{2l_3+1}}\quad S\left[  \raisebox{-20pt}{\includegraphics[width=0.20\textwidth]{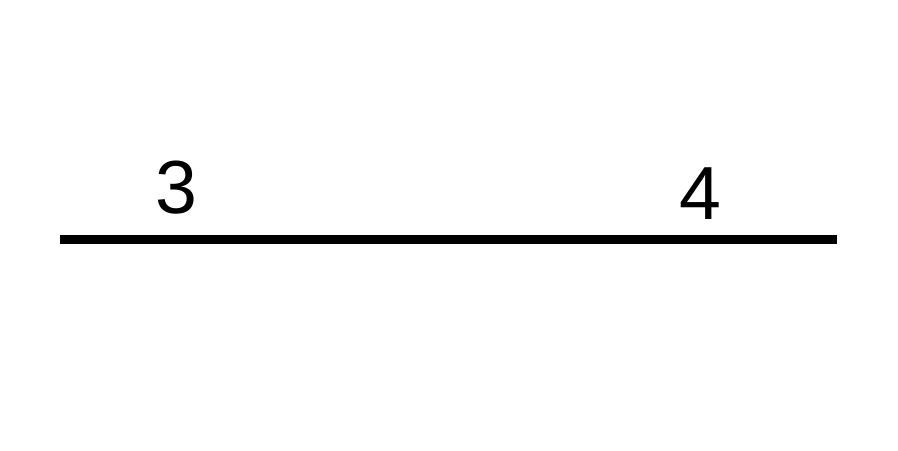}}\right].
\eea

If the polygon is bounded by two edges, then it can be removed using the orthogonality relation \refb{eq:orthog}. 
If the polygon has three edges, then it can be reduced to a vertex using the 3-1 move, as in \refb{eq:loop}.
Otherwise, the polygon has four or more edges, and applying the 2-2 move on adjacent vertices of the face will reduce 
the number of edges (and vertices) bounding the face by one. Performing this move repeatedly will eventually result in a polygon
with three edges, which can be reduced to a vertex using the 3-1 move.

\begin{figure}[H]
\centering
\includegraphics[width=0.6\textwidth]{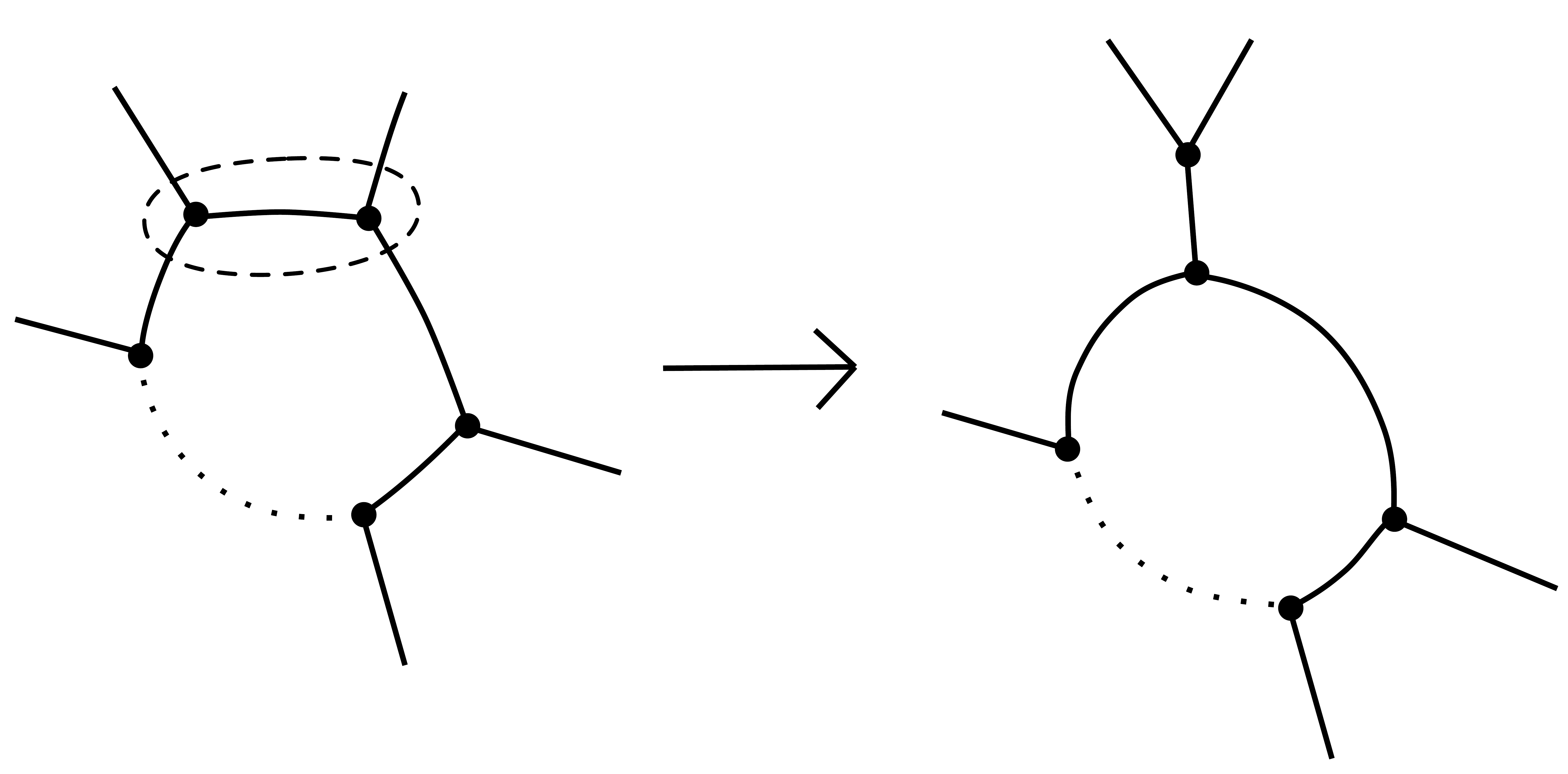}
\caption{The 2-2 move on a polygon reduces the number of edges bounding the polygon by one.}\label{fig:poly1}
\end{figure}

The generation and reduction of a polygon will always reduce the number of vertices of the graph by two. Therefore, this procedure will eventually reduce the graph down to a trivial loop, which evaluates to a dimension factor $(2l+1)$. The string of factors and sums that are produced in performing these moves gives the final evaluation of the spin network state sum $\sG$, which is manifestly independent of the labels $m_i$.
In summary, the algorithm is:

\begin{enumerate}
\item Choose a face that is homeomorphic to a disc. If no face is homeomorphic to a disc, then apply the parity move to construct such a face.
\item Apply the 3-1 and 2-2 moves and the orthogonality relation to remove the face.
\item Repeat these steps until the graph is reduced to a single loop.
\end{enumerate}

\subsection{Examples of $3j$ and $6j$ sums}\label{sec:sixjex}

In this section we have expanded the Gaussian Hermitian matrix model on the fuzzy sphere and found that the correlators are described
by sums over trivalent graphs. Each trivalent graph corresponds to a sum over spin labels $(l_i, m_i)$ weighted by $3j$ and $6j$ symbols, and the sum over $m_i$s weighted by $3j$s can be evaluated algorithmically.

We conclude this section by presenting some explicit $3j$ and $6j$ sums associated to some trivalent graphs. In these cases, we can perform the sum over all the labels using $6j$ identities to confirm the result \refb{eq:rg}. The simplest graph to consider is the theta ribbon graph given in Equation \refb{eq:theta}. For this graph, the $3j$ sum is trivial, and so the total $6j$ sum is

\bea \label{eq:thetaribbon} \hspace{-10mm} \R \left[  \raisebox{-35pt}{\includegraphics[width=0.25\textwidth]{theta}} \right] = (-)^{2j}N^{-1}\sum_{l_1l_2l_3}(2l_1+1)(2l_2+1)(2l_3+1)\wsjj{l_1}{l_2}{l_3}\wsjj{l_1}{l_2}{l_3}. \qquad \eea
Orthogonality of the $6j$s, and the contraint on the range of summation $0\leq l_i \leq 2j$, give the expected final answer,
\bea \rG = \frac{1}{(2j+1)^2}\sum_{l_1l_2}(2l_1+1)(2l_2+1) = N^2. \eea

Another simple planar graph to consider is the tetrahedral network given in \refb{eq:sixj}. For this graph, we can state that
\begin{multline} \label{eq:tetribbon} \R \left[  \raisebox{-45pt}{\includegraphics[width=0.25\textwidth]{6jno}} \right] = N^{-2}\sum_{\substack{l_1l_2l_3\\ l_4l_5l_6}}(2l_1+1)(2l_2+1)(2l_3+1)(2l_4+1)(2l_5+1)(2l_6+1)\times \\ 
\times \wsj{l_1}{l_2}{l_3}{l_4}{l_5}{l_6}\wsjj{l_1}{l_2}{l_3}\wsjj{l_2}{l_4}{l_6}\wsjj{l_1}{l_5}{l_6}\wsjj{l_3}{l_4}{l_5}. \end{multline}
This $6j$ sum can be evaluated explicitly by using the Biedenharn-Elliot identity \refb{eq:bhid2}, given in the appendix, to elimate a spin label and a $6j$. The remaining sums can be performed by using the orthogonality relation.

Next, we consider the following non-planar graph,
\begin{multline} \R \left[  \raisebox{-30pt}{\includegraphics[width=0.25\textwidth]{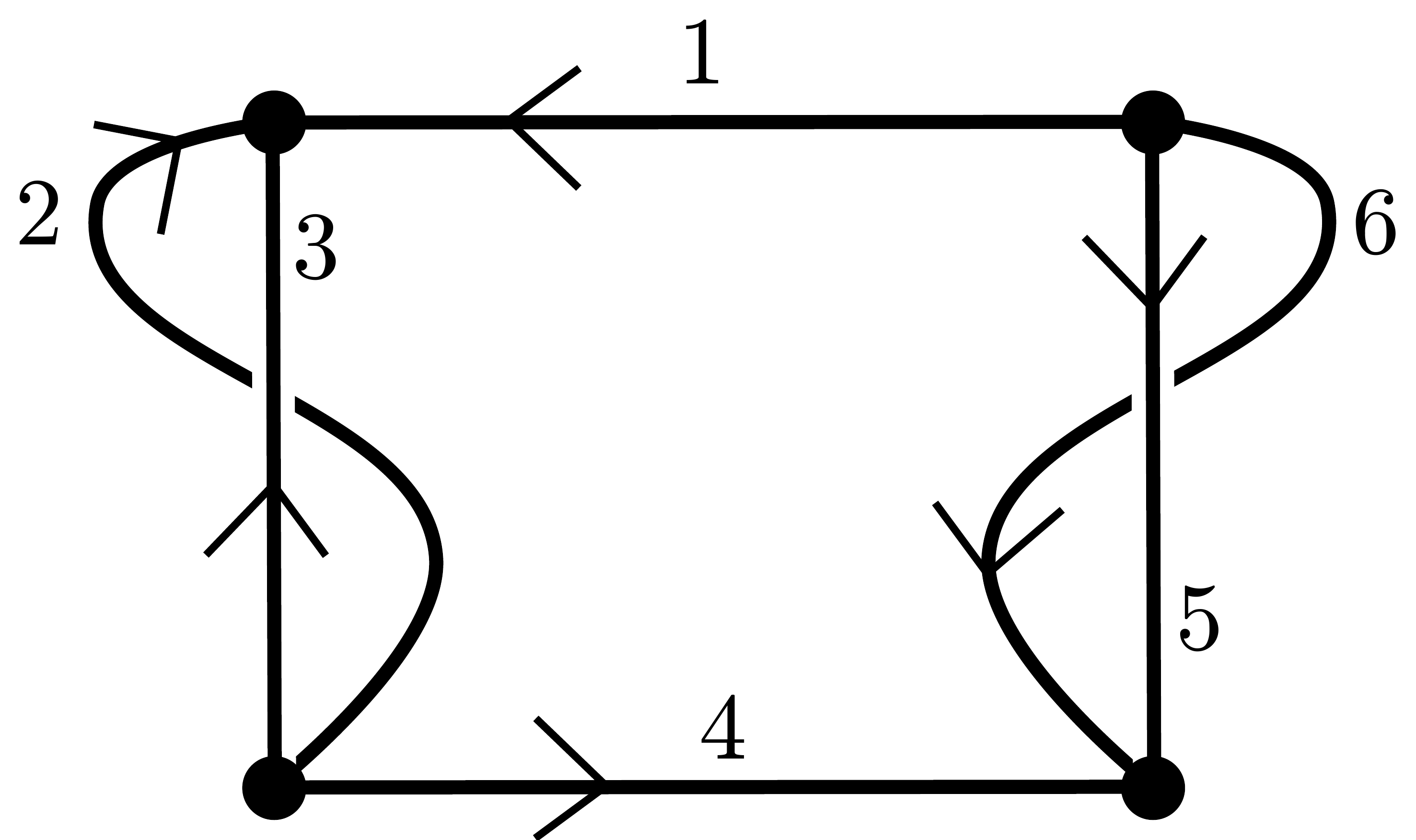}} \right] = N^{-2}\sum_{\substack{l_1l_2l_3\\ l_4l_5l_6}}(2l_1+1)(2l_2+1)(2l_3+1)(2l_4+1)(2l_5+1)(2l_6+1)\times \\
\times \sGl \wsjj{l_1}{l_2}{l_3}\wsjj{l_2}{l_3}{l_4}\wsjj{l_4}{l_5}{l_6}\wsjj{l_1}{l_5}{l_6}, \end{multline}

\begin{multline} \sGl = \sum_{m_1\ldots m_6} (-)^{m_1+m_2+m_3+m_4+m_5+m_6}\wigthreej{l_1}{l_2}{l_3}{m_1}{m_2}{m_3} \wigthreej{l_2}{l_3}{l_4}{-m_2}{-m_3}{-m_4} \times \\
\times \wigthreej{l_4}{l_5}{l_6}{m_4}{m_5}{m_6}  \wigthreej{l_1}{l_5}{l_6}{-m_1}{-m_5}{-m_6}. \end{multline}
We apply the algorithm to find $\sG$. This ribbon graph has no polygonal faces, so we apply a parity move on a vertex to deduce that

\bea \sG = (-)^{l_4+l_5+l_6}\quad S\left[  \raisebox{-35pt}{\includegraphics[width=0.25\textwidth]{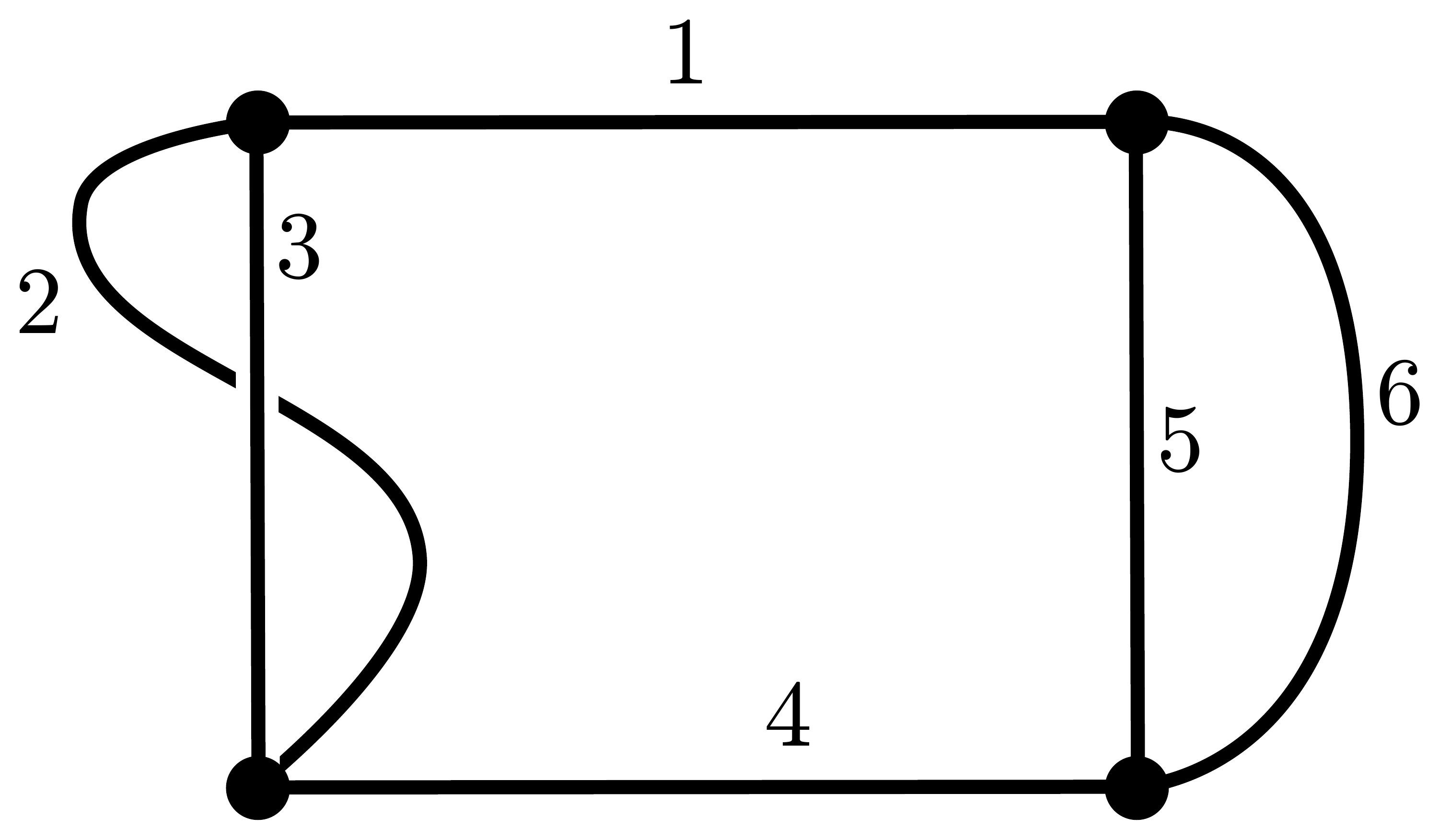}} \right] \eea
We can now apply the orthogonality relation on this bubble to deduce that
\bea \sG =\frac{ (-)^{l_4+l_5+l_6}}{(2l_4+1)}\delta_{l_1,l_4} \quad S\left[  \raisebox{-35pt}{\includegraphics[width=0.25\textwidth]{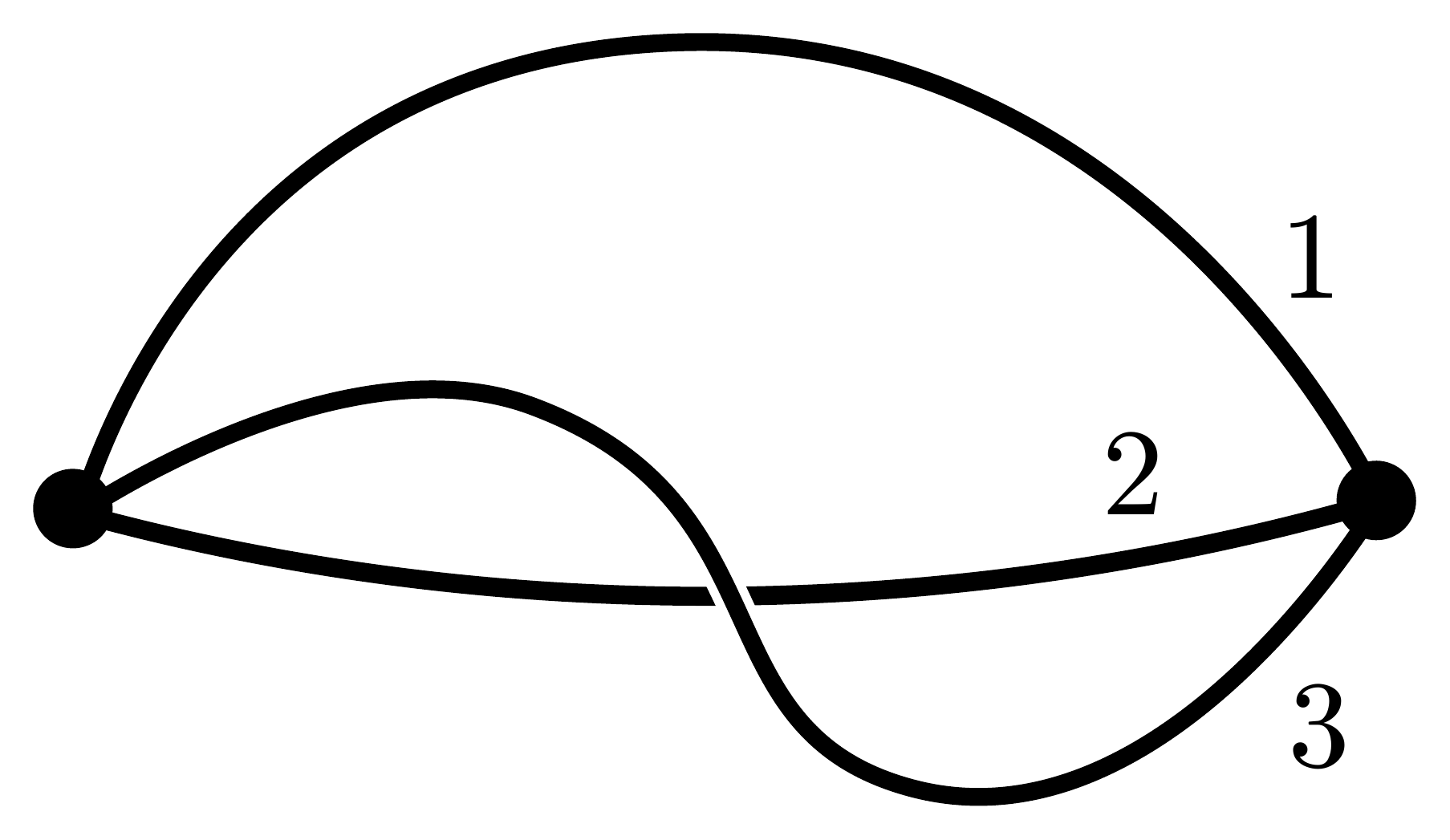}} \right] \eea
A second parity move on the graph will reduce this trivalent network to a theta network with trivial evaluation, hence we deduce that
\bea \sG = \frac{(-)^{l_2+l_3+l_5+l_6}}{(2l_4+1)} \delta_{l_1,l_4}, \eea
and hence that the ribbon graph evaluation is 
\begin{multline} \R[{\cal G}] = N^{-2}\sum_{l_1l_2l_3 l_5l_6}(-)^{l_2+l_3+l_5+l_6}(2l_1+1)(2l_2+1)(2l_3+1)(2l_5+1)(2l_6+1)\times \\ \times \wsjj{l_1}{l_2}{l_3}\wsjj{l_1}{l_2}{l_3}\wsjj{l_1}{l_5}{l_6}\wsjj{l_1}{l_5}{l_6}. \end{multline}
This expression can be evaluated to $N^0=1$ by using the identity \refb{eq:sixj2}.

\section{A three-dimensional interpretation of the Hermitian matrix model }\label{sec:HMMPR} 

We have expressed every ribbon graph generated by a Hermitian matrix model correlator as a sum over representations of $\lsu$ weighted by Wigner $3j$ and $6j$ symbols, and as sums over the representation labels $j_i$ weighted by $6j$s and dimension factors. We can find a three-dimensional interpretation of these sums by using the Ponzano-Regge model of quantum gravity, 
which was first introduced in \cite{pr} and also reviewed in \cite{jwbing, dowdall}.
In this section, we show that the partition functions of the Ponzano-Regge model correspond exactly to 
these ribbon graph sums when certain boundary conditions are applied.

\subsection{The Ponzano-Regge model}

The Ponzano-Regge model is defined by assigning a partition function to any triangulation of a 3-manifold, possibly with boundary, with a spin label $j_i$ assigned to each edge, and a sum performed over all possible values of the spin labels corresponding to internal edges. The sum is weighted by the function
\bea W = \prod_{\rm interior\ edges} (-)^{2j_i}(2j_i+1) \prod_{\rm tetrahedra}\wsj{j_1}{j_2}{j_3}{j_4}{j_5}{j_6}, \eea
hence the partition function 
\bea Z = \sum_{\rm interior\ edges} W \eea
is a function of the values of the spin labels on the boundary.

This partition function takes a very similar form to the sums associated to planar graphs in the Hermitian matrix model. By using a judicious choice of labelled cell complex (i.e. a labelled triangulation of a manifold), we can reproduce exactly the ribbon graph sums for any graph generated by a correlator, which gives us a way of interpreting the zero-dimensional combinatoric theory as a three-dimensional topological theory of gravity. We review the features of this model necessary to show this correspondence.

\begin{figure}[H]
\centering
\includegraphics[width=0.3\textwidth]{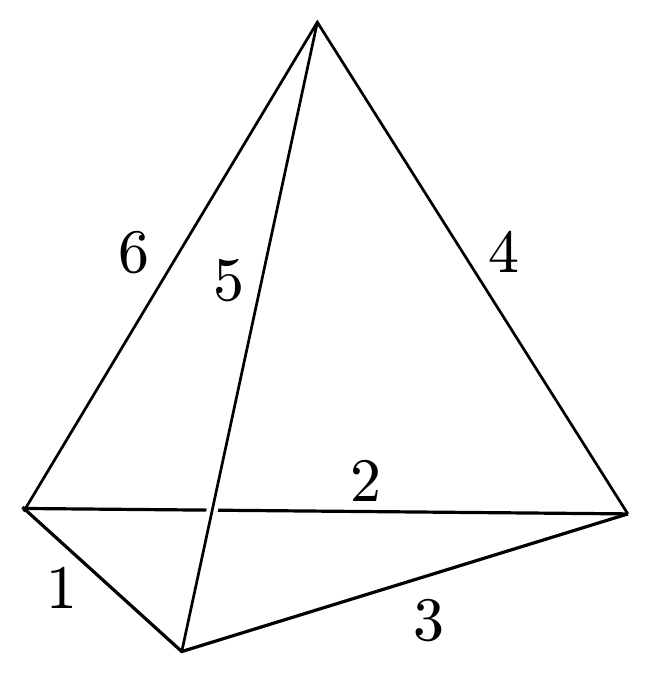}
\caption{Each $6j$ symbol in the Ponzano-Regge state sum corresponds to a labelled tetrahedron.}\label{fig:egtet}
\end{figure}

In the Hermitian matrix model, the range of summation of the spins is $0\leq l_i \leq 2j$, where $l_i\in \mathbb{Z}$, but in the Ponzano-Regge state sum exhibited above, the spin labels in general range over all possible half-integer values up to infinity. However, there are constraints on the ranges that these labels can take, which are imposed by the $6j$ symbols. Recall that the $6j$ symbol
\bea\label{eq:eg6j} \wsj{j_1}{j_2}{j_3}{j_4}{j_5}{j_6} \eea
is zero unless the triples $\{j_1, j_2, j_3\}$, $\{j_1, j_5, j_6\}$, $\{j_2, j_4, j_6\}$, $\{j_3, j_4, j_5\}$ all satisfy the triangle inequalities. These triples correspond to the four triangles in the cell complex that border each tetrahedron\footnote{Note that the tetrahedron associated with a $6j$ in the Ponzano-Regge model is different from the tetrahedral network associated to a $6j$ in the previous section. The labels associated to edges meeting at a vertex of a trivalent graph satisfy a triangle constraint, while the labels associated to a face of a tetrahedron satisfy a triangle constraint in the Ponzano-Regge model. These two tetrahedra are dual to each other.}. This means that if two out of three edges in a triangle are constrained to a finite range, then a triangle inequality constrains the third edge to a finite range. Thus, noting that the boundary labels are fixed, we see that there is an iterative way of deducing the set of edges in a complex whose labels span a finite range. In particular, if all the edges in the complex span a finite range, then the state sum must converge, but if any labels are unconstrained then the state sum will in general diverge. It can be shown that any triangulation with a vertex in the interior must possess spin labels that diverge. The converse statement, that any triangulation with no internal vertices must converge, is not true in general, but does hold for all the cases that we consider in this paper. 

We wish to construct labelled tetrahedral complexes whose state sums reproduce the ribbon graph sums in a systematic manner. While it is possible to do this without introducing internal vertices, it is clearer and more systematic to use triangulations with a single internal vertex, and to introduce a method of regularising these sums. In the next section we introduce the Turaev-Viro partition function, which is a state sum model similar to the Ponzano-Regge model, that naturally constrains all the ranges of summation to be finite. We can use this model to recover the Ponzano-Regge state sums, and hence the ribbon graph sums, in the `classical' limit.

\subsection{The Turaev-Viro model as a regulator for the Ponzano-Regge model}\label{sec:tv}

The Ponzano-Regge model assigns partition functions to complexes labelled by the irreducible representations of $\lsu$.
We can deform this model by replacing the representations of $\lsu$ in the complex with labelled representations of the quantum deformation of the Lie algebra $\qsu$, where $q$ is a deformation parameter. The classical algebra is recovered when $q$ is set to 1.
This deformed algebra has representations analagous to the irreducible representations of $\lsu$, labelled by half-integers $j$, and containing $(2j+1)$ states, which can be recoupled to generate quantum $3j$ and quantum $6j$ symbols \cite{ruegg}. Unlike $\lsu$, however, the number of representations of the quantum algebra is finite whenever $q$ is a root of unity not equal to 1.

Thus, if we demand that $q$ is a root of unity, and replace all the representation-dependent expressions in the Ponzano-Regge state sums with their quantum analogues, then the sums over representations have finitely many terms and are thus well-defined. This quantity will diverge as $q$ tends towards 1 (while still being a root of unity), but there is a natural way of regulating this divergence that gives a $q$-independent limit which coincides with the Ponzano-Regge model for the cases where both state sums are convergent. We can thus define the Ponzano-Regge model for divergent sums as being the classical limit of the quantum state sum model.

A more detailed treatment of the quantum state sum model is given in \cite{turaevviro}. We present here the details of how the relevant quantities, such as summation ranges, representation dimensions, and the $6j$ symbols, deform after being taken to their quantum analogues.

We take an integer $r\geq3$, and set $q:=e^{2\pi i/r}$, an $r^\text{th}$ root of unity. We define the `quantum integer'
\bea [n]  := \frac {q^{n/2}-q^{-n/2}}{q^{1/2} - q^{-1/2}}, \eea
that has the property that $[n]\to n$ as $r\to \infty$ and $q\to 1$, and define the quantum factorials 
\bea [n]! := [n][n-1]\ldots[2][1]. \eea
We say that a triple of spin labels $\{j_1, j_2, j_3\}$ satisfy the quantum triangle constraints if they satisfy
the classical triangle constraints with the extra conditions
\bea j_i\leq (r-2)/2, \qquad j_1+j_2+j_3 \leq r-2. \eea
By taking the explicit expression of a $6j$ symbol in terms of sums and products of factorials with triangle constraints
given in \cite{edmonds, rose}, we can replace the factorials in the definition of a $6j$ symbol with the quantum factorials, 
and upgrade the triangle constraints to quantum triangle constraints,
to generate the quantum $6j$ symbol
\bea \wsjq{j_1}{j_2}{j_3}{j_4}{j_5}{j_6}. \eea
This definition coincides with the definition of a quantum $6j$ given in terms of the recouplings of representations of the quantum algebra \cite{kirillovreshetikhin}.
The quantum $6j$ symbol converges to the classical $6j$ as $q\to 1$, but crucially is only non-zero for finitely many $j_i$ for each value of $r$. This means that if we replace all the $6j$ symbols in the Ponzano-Regge state sum with quantum $6j$s, we arrive at an always convergent state sum with weight
\bea W_q = \prod_{\rm interior\ edges} (-)^{2j_i}[2j_i+1]\prod_{\rm tetrahedra}\wsjq{j_1}{j_2}{j_3}{j_4}{j_5}{j_6}. \eea

Each term in this expression converges to the classical analogue as $q\to1$, hence this partition function reproduces the original Ponzano-Regge state sum in the $q\to1$ limit in the cases where the original Ponzano-Regge state sum converges. For the complexes with interior vertices, we use the quantum normalisation factor $w^2 = -2r/(q^{1/2}-q^{-1/2})^2$ and define the Turaev-Viro partition function
\bea Z_q = w^{-2v}\sum_{\mathrm{interior\ edges}} W_q, \eea
where $v$ is the number of internal vertices in the triangulation.

By its construction this partition function is finite for any root of unity $q$, and converges to the Ponzano-Regge partition function when the Ponzano-Regge partition function is finite. It is more difficult to prove that $Z_q$ tends to a finite value for more general complexes, but for the classes of manifolds we will be constructing, the $q\to1$ limit is well-defined, and hence we will take this as the definition of the regularised Ponzano-Regge partition function for complexes with interior vertices.

One of the most important properties of the Turaev-Viro and Ponzano-Regge models is triangulation independence. Any two triangulations of a 3-manifold that are equal on the boundary can be deformed from one to the other by a series of operations on the complex called Pachner moves \cite{pachner}. These moves are mergings and splittings of glued tetrahedra that will change the terms that appear in the $6j$ sums, but due to two identities relating sums of products of quantum $6j$ symbols, these operations will not change the overall value of the partition function.

\begin{figure}[h]
\centering
{\includegraphics[width=0.3\textwidth]{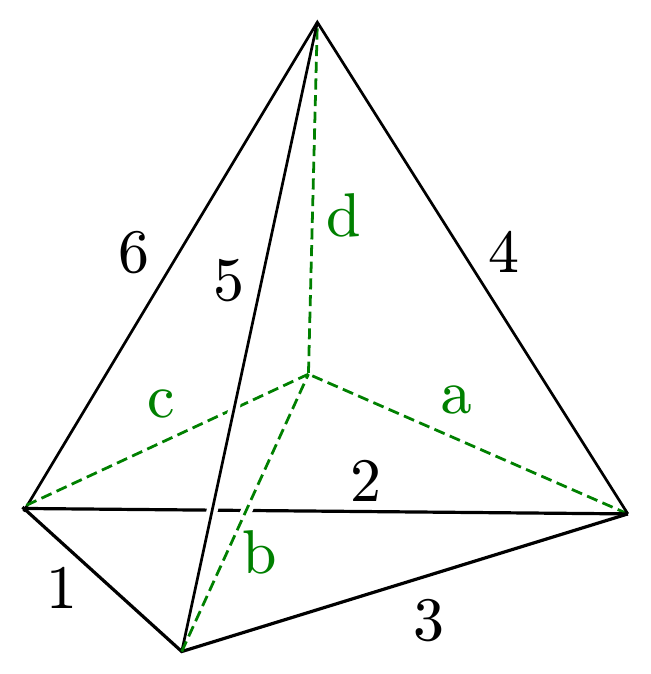}}\raisebox{45pt}{\includegraphics[width=0.15\textwidth]{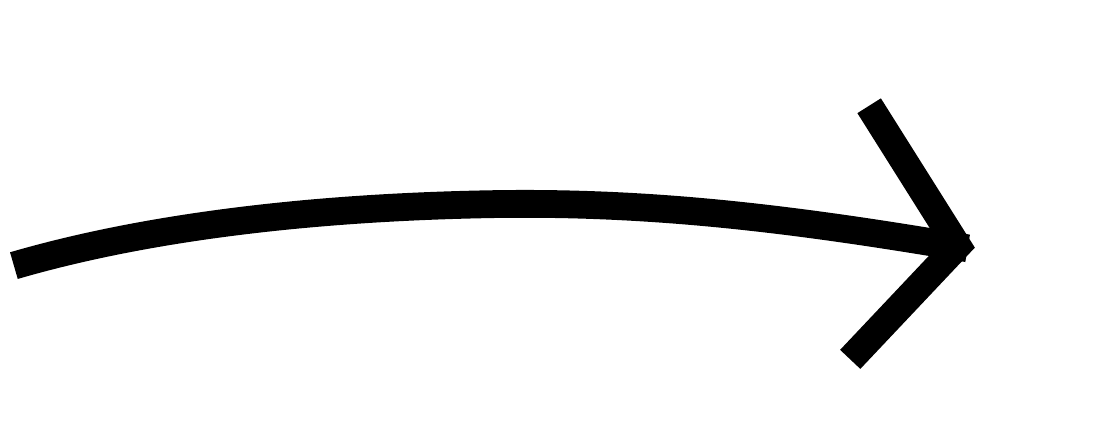}}
{\includegraphics[width=0.3\textwidth]{6jint3}}
\caption{The 4-1 Pachner Move.}
\label{fig:pach1}
\end{figure}

\begin{figure}[h]
\centering
{\includegraphics[width=0.36\textwidth]{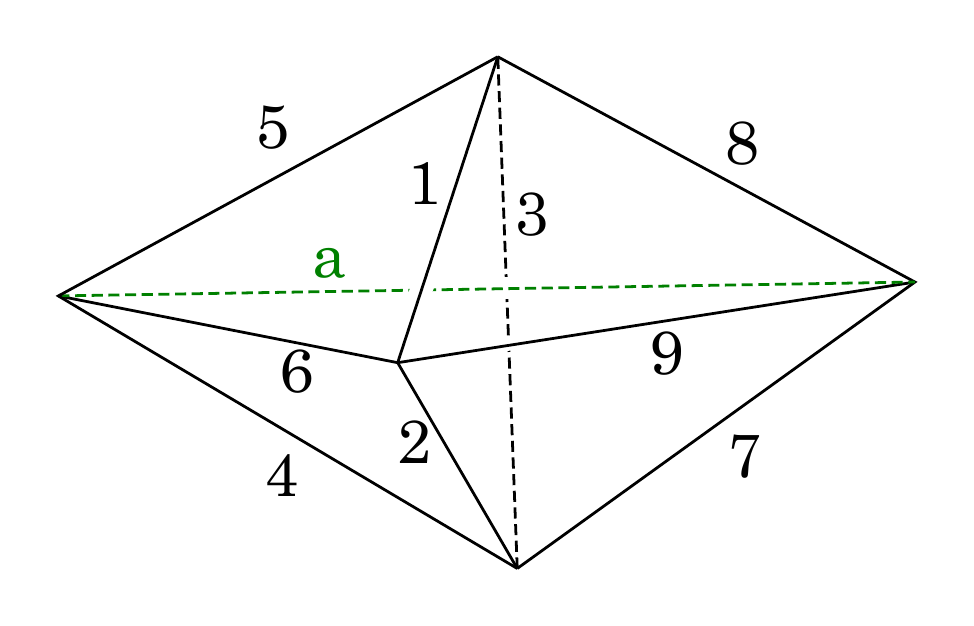}}\quad \raisebox{30pt}{\includegraphics[width=0.18\textwidth]{arrow2}}
{\includegraphics[width=0.36\textwidth]{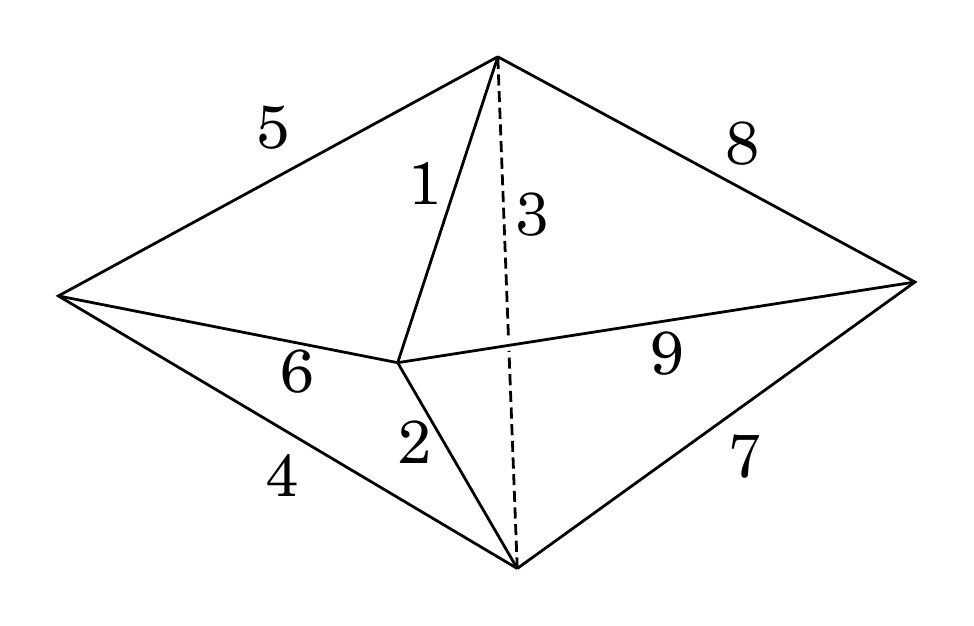}}
\caption{The 3-2 Pachner Move.}
\label{fig:pach2}
\end{figure}

The identity corresponding to the 4-1 move is 
\begin{multline}
w^{-2}\sum_{a,b,c,d}(-)^{2a+2b+2c+2d}[2a+1][2b+1][2c+1][2d+1]\times \\ \wsjq{j_1}{j_2}{j_3}{a}{b}{c}\wsjq{j_1}{j_5}{j_6}{d}{c}{b} \wsjq{j_2}{j_4}{j_6}{d}{c}{a}\wsjq{j_3}{j_4}{j_5}{d}{b}{a} = \wsjq{j_1}{j_2}{j_3}{j_4}{j_5}{j_6}, 
\end{multline}
and the identity corresponding to the 3-2 move is the Biedenharn-Elliot identity (which also holds in the $q\to1$ limit),
\bea \label{eq:bhid}\sum_a(-)^{2a}[2a+1]\wsjq{j_1}{j_5}{j_6}{a}{j_9}{j_8}\wsjq{j_2}{j_4}{j_6}{a}{j_9}{j_7}\wsjq{j_3}{j_4}{j_5}{a}{j_8}{j_7} = \wsjq{j_1}{j_2}{j_3}{j_4}{j_5}{j_6}\wsjq{j_1}{j_2}{j_3}{j_7}{j_8}{j_9}.\qquad \eea

\subsection{Constructing the manifolds associated to planar graphs}

We  present a prescription for constructing a 3-complex from a graph $\G$ in such a way that its Ponzano-Regge partition function corresponds to the ribbon graph sum \refb{eq:totalsum}. We denote this construction as a mapping $\mG$ from graphs to labelled triangulations of manifolds, and the partition function $Z[\cal M]$ of a labelled triangulation $\cal M$ as a function from labelled manifolds into $\mathbb{C}$.  We find that there is a very simple relation between the matrix integral evaluation of a 
graph $\rG$ and the composition $ Z \circ {\cal M} [ \G ] \equiv Z [ \mG ]  $. Letting $V$ be the number of vertices of the graph, we prove that
\boxeq{ \label{eq:prrelation}  N^{V} Z [ \mG ]  \equiv \rG  . }
In this section we present and discuss the construction of $\mG$ for a general planar graph.
We will see that the construction has the inner Belyi triangulation in its interior 
and the outer Belyi triangulation on its boundary, hence we call $\mG$ the {\bf Belyi 3-complex} of 
the graph $\G$. We  prove that it reproduces the ribbon graph sum $\rG$ in the next subsection, and discuss 
the construction of manifolds for non-planar ribbon graphs in Section \ref{sec:PRNP}.

We construct the labelled complex $\mG$ piecewise from the graph by defining 3-complexes associated to each graph vertex, and then gluing the complexes together using the data of the graph edges. First, assign an orientation to each edge of the graph. For each vertex, create a complex with two tetrahedra glued together on a single face, and then glue on a tetrahedron for each half-edge directed towards the vertex. 
Thus, for a vertex with three incoming half-edges, define
\bea\label{eq:complex1}  \m{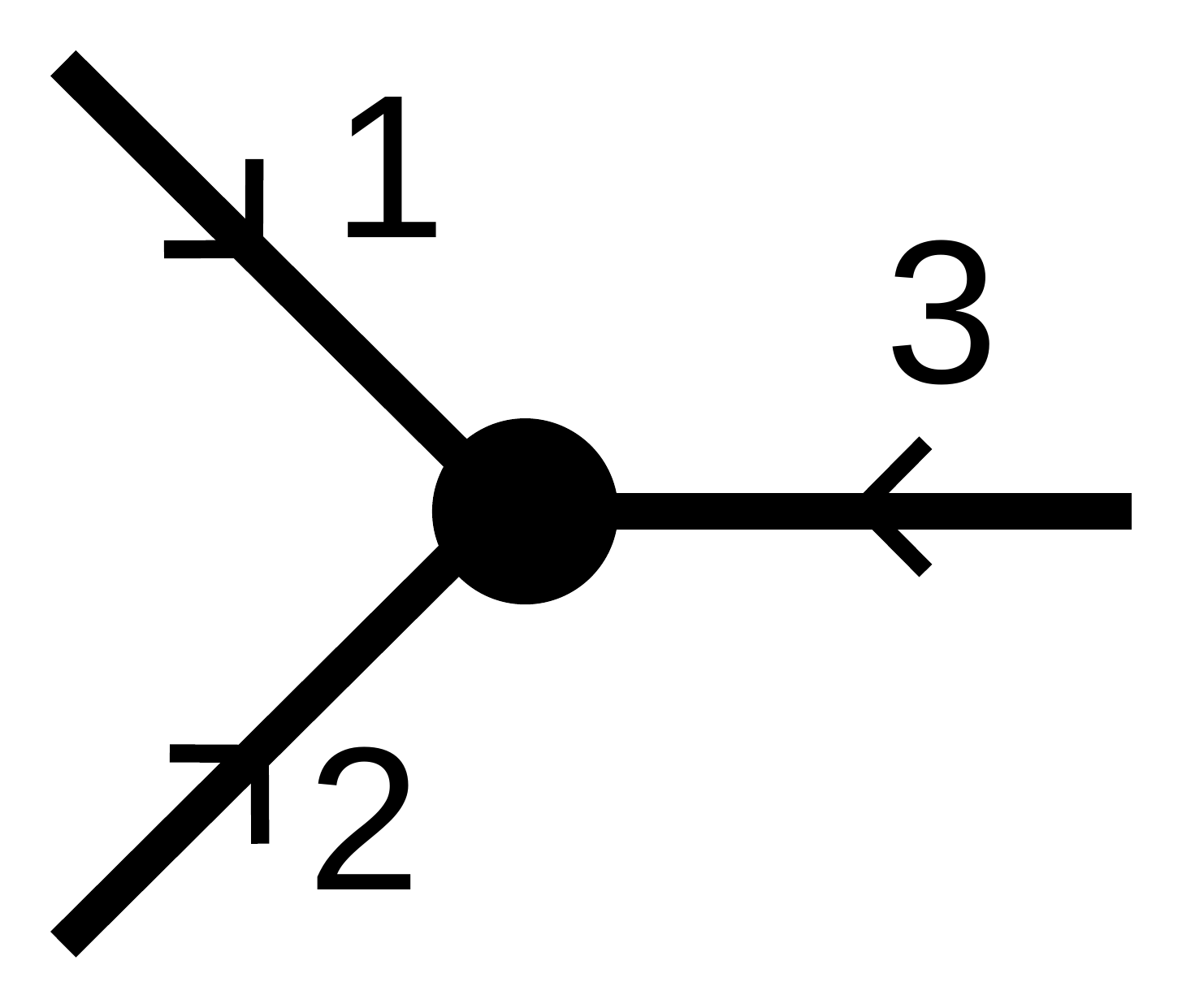} = \raisebox{-0.20\textwidth}{\includegraphics[width=0.4\textwidth]{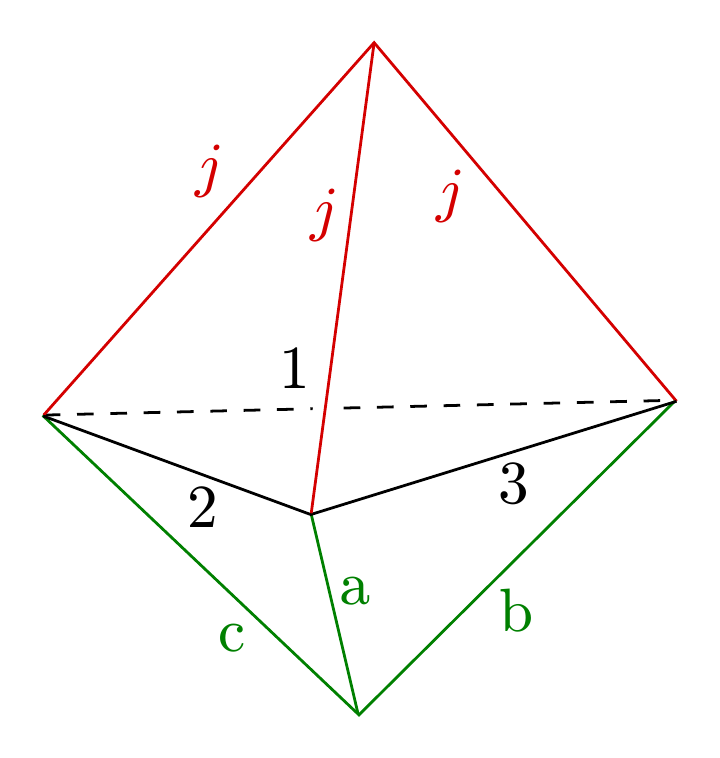}} \eea
and for a vertex with three outgoing half-edges, define
\bea\label{eq:complex2}   \m{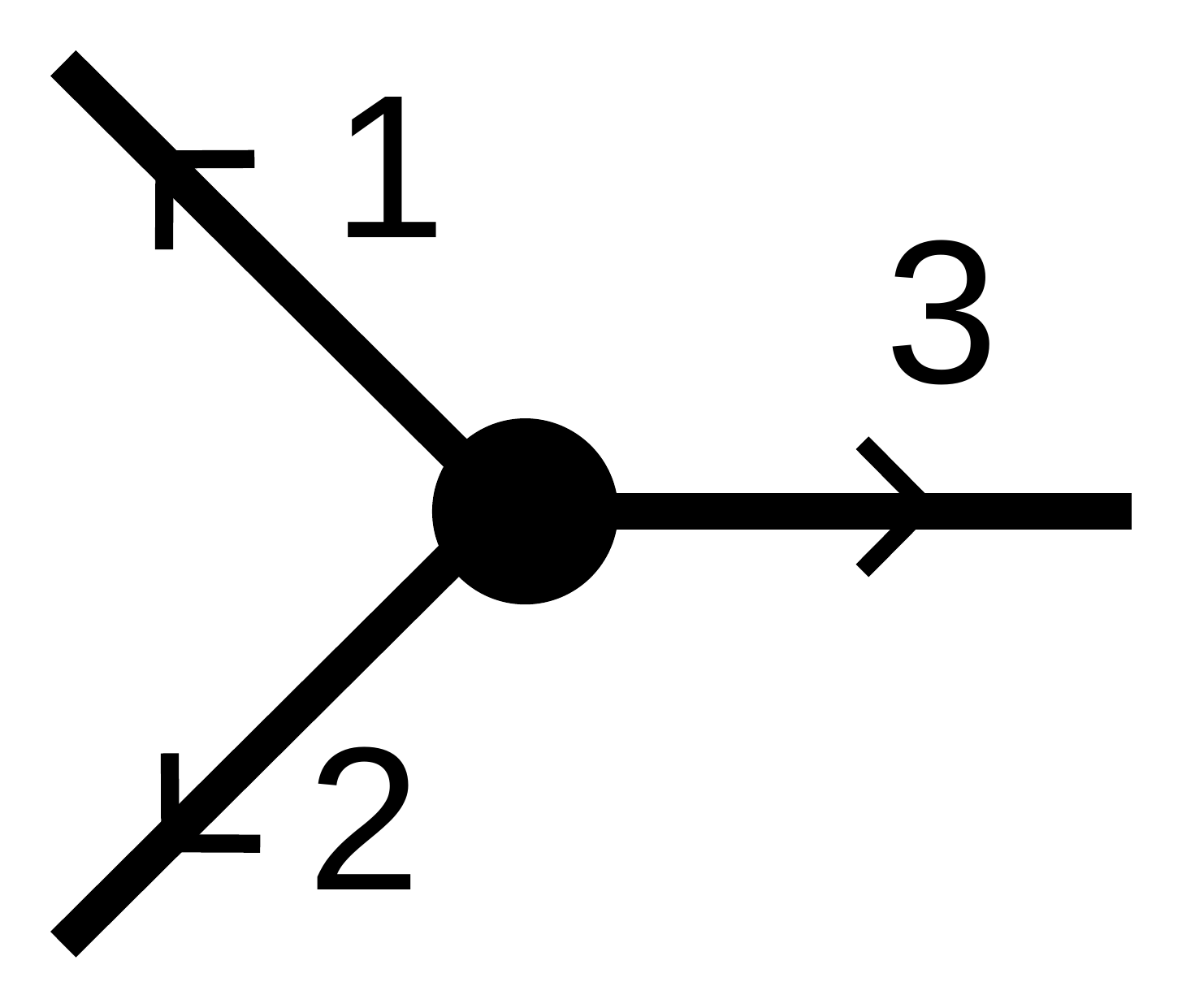} = \raisebox{-0.20\textwidth}{\includegraphics[width=0.4\textwidth]{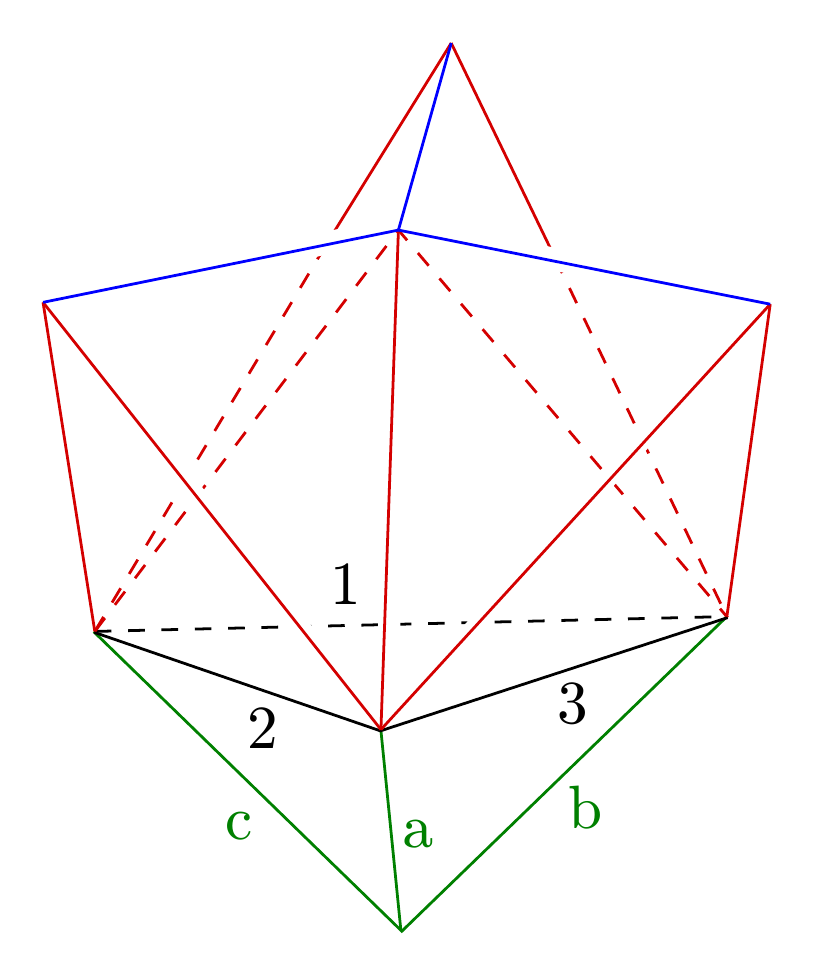}} \eea
and similarly for vertices with two or three incoming half-edges.

In these and all subsequent diagrams of 3-complexes, we use different colours to denote different types of spin labels. The red edges will always have the assigned label $j$, the blue edges will have the assigned label 0, and both will always be on the boundary of the complex. The black edges in the complex inherit the labels $l_i$ from the graph edges incident to the graph vertex. They will be in the interior of the glued manifold, and the presence of the tetrahedra with red $j$ labels will constrain these labels to run over the integers $0\leq l_i\leq 2j$. Finally, in this section, the green edges in the glued manifold will always be interior edges connected to an interior vertex, and their labels, denoted by Roman indices, will run over infinitely many values in general.

We glue together the complexes associated to a pair of graph vertices connected by an edge labelled $i$ by identifying the pair of triangles incident to the corresponding labelled edge $i$.

\bea  \m{tinyvc} \m{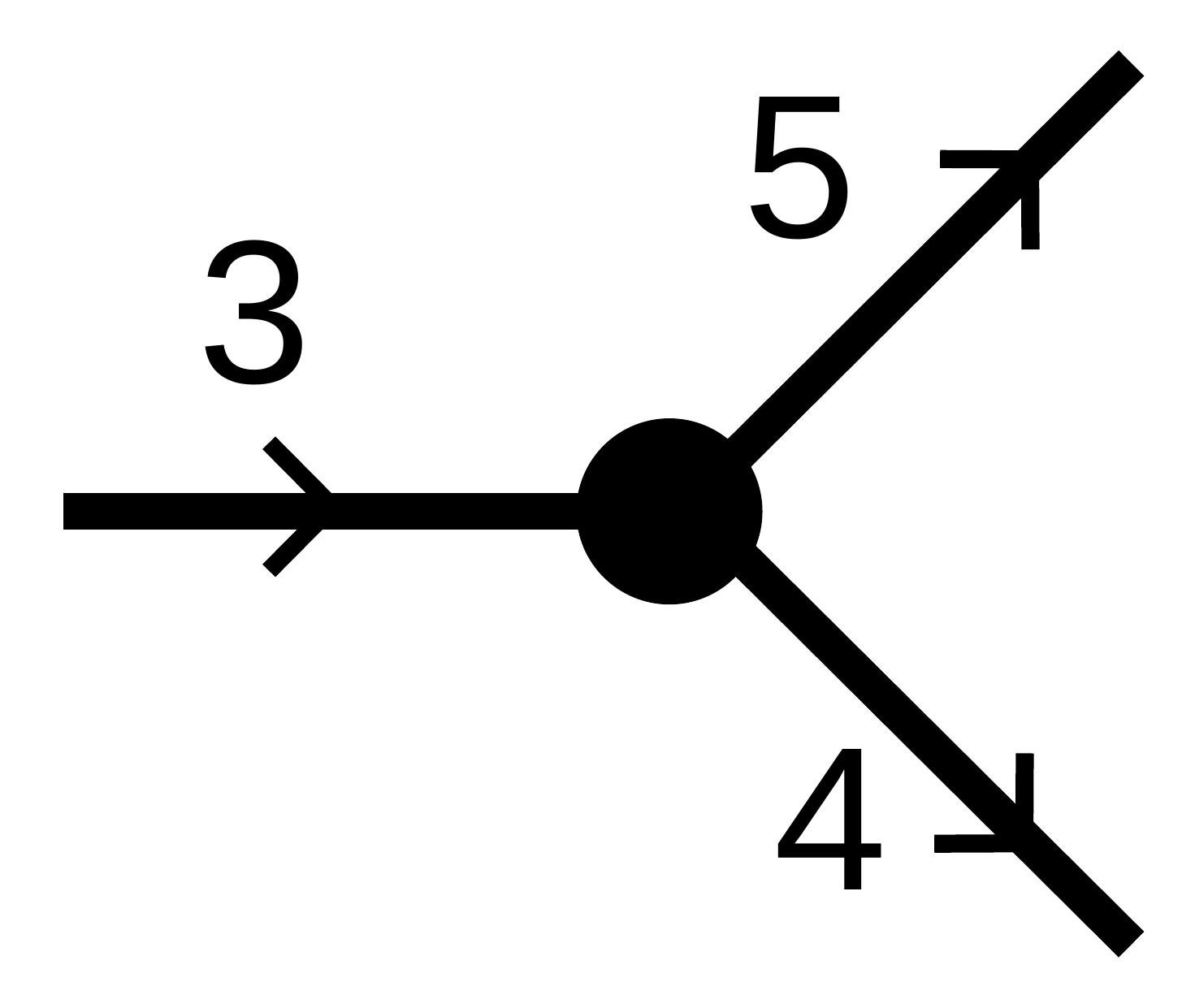} \quad = \raisebox{-.1\textwidth}{\includegraphics[width=0.6\textwidth]{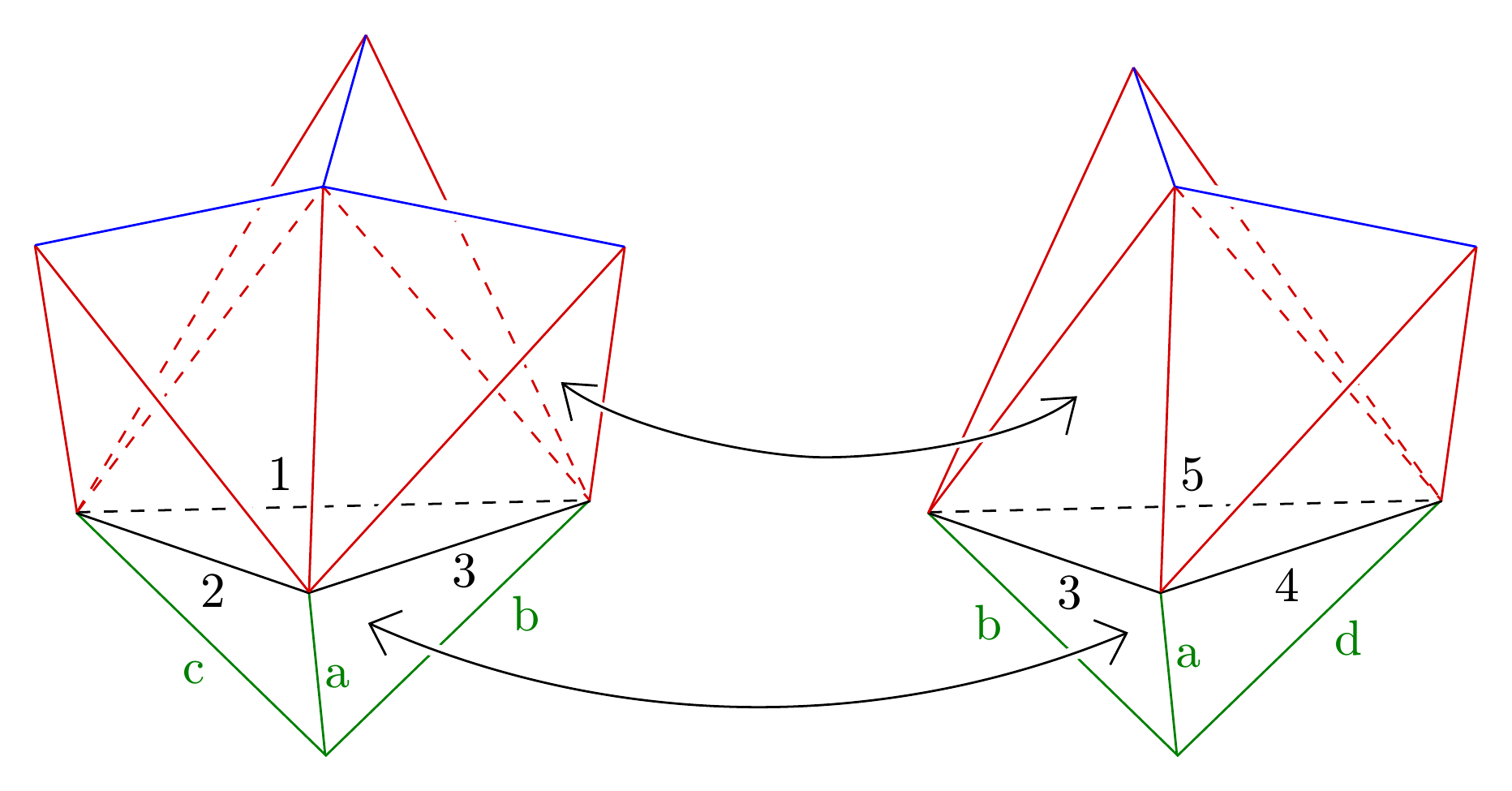}} \eea

\bea  \m{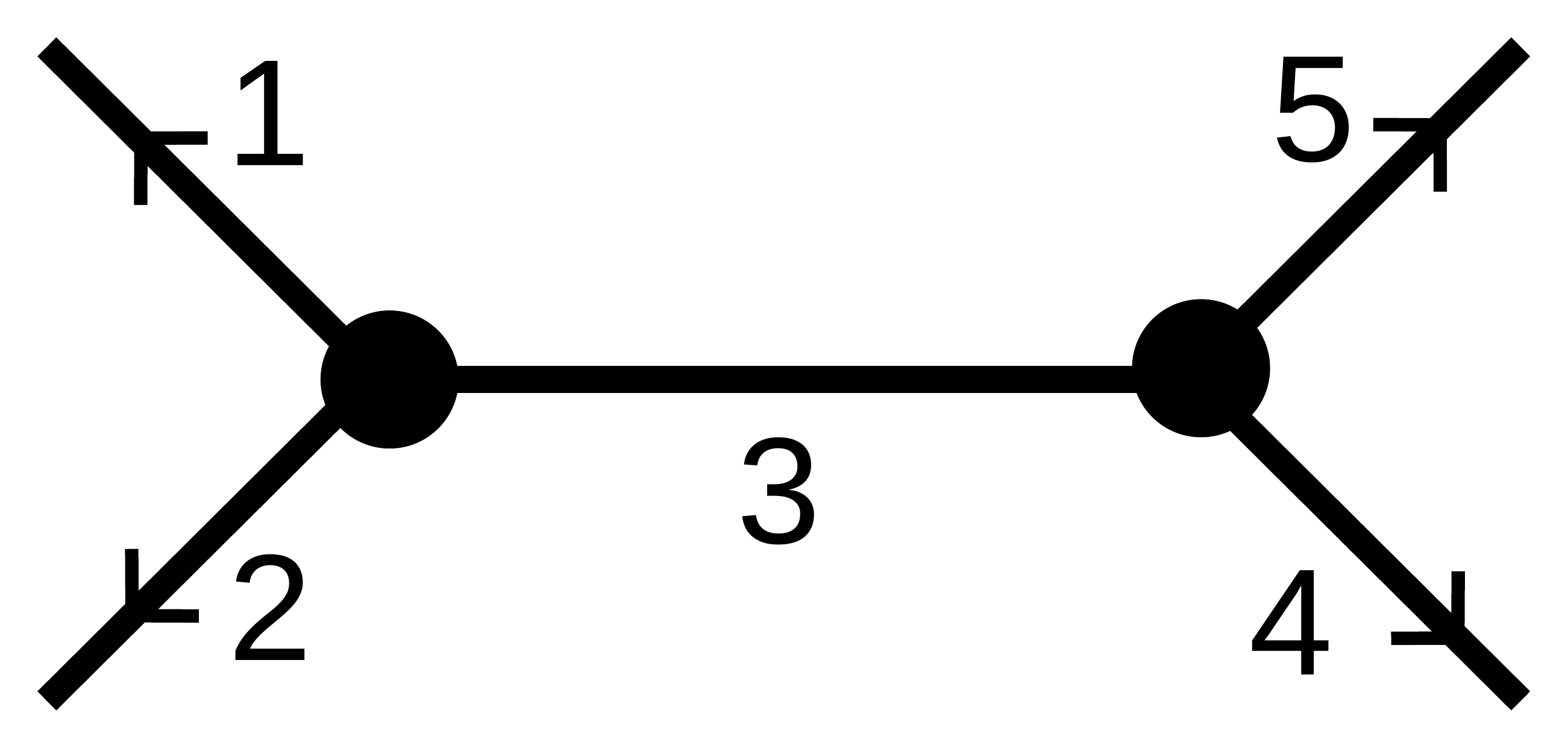} \quad = \raisebox{-.2\textwidth}{\includegraphics[width=0.45\textwidth]{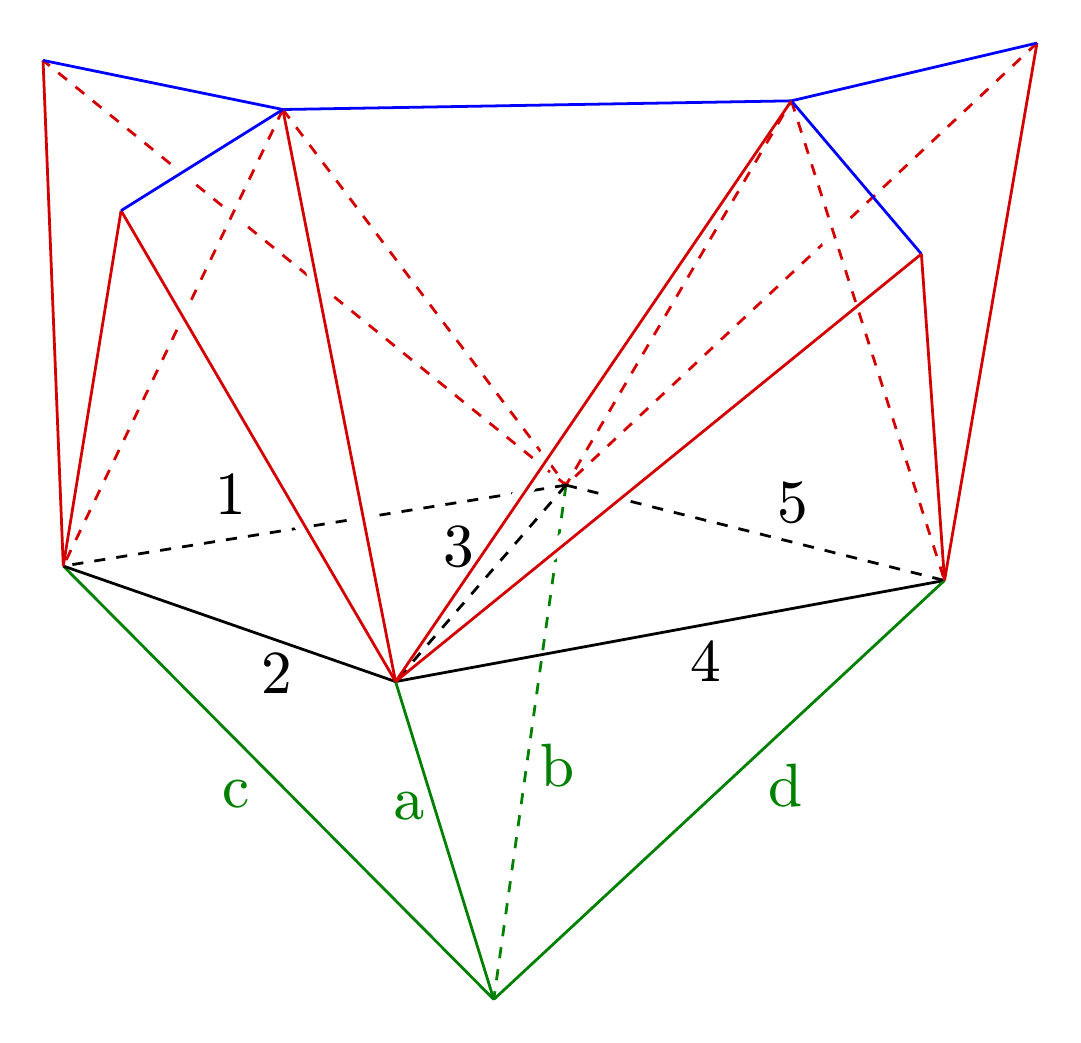}} \eea
Carrying out this gluing procedure for all connections between vertices for the planar graph ${\cal G}$ will produce a labelled triangulation of the 3-dimensional ball, which will be our definition of $\mG$. The black edges will correspond to a triangulation of the sphere dual to the graph $\G$ and will lie in the interior of the manifold. The blue edges trace out a copy of the graph $\G$ on the boundary of the ball.

We can see that this combination of red, blue, and black edges in this complex form the Belyi triangulations of a sphere discussed in Section \ref{sec:belyi}. Considering the ribbon graph $\G$ as a dessin d'enfant, we see that the blue and red edges of the 3-complex form the outer Belyi triangulation associated to the graph, embedded on the boundary of the complex. The black edges form the inner Belyi triangulation associated to the graph, which is embedded in the interior of the 3-complex. 
We can therefore view the construction of the labelled complex $\mG$ as `lifting' the Belyi triangulation of a sphere into three dimensions. For this reason, we call the labelled complex $\mG$ the {\bf Belyi 3-complex} associated to a ribbon graph $\G$. The edges
in a complex that form the outer Belyi triangulation will always have the same colour-dependent spin labels, $j$ for red edges and $0$ for blue edges. Hence, in the following text when we refer to an outer Belyi triangulation in a complex, we shall also implicitly include the colour-dependent spin labellings assigned to the edges.

\subsection{Evaluating the partition function of a Belyi 3-complex}\label{sec:prproof}

We evaluate the Ponzano-Regge partition function of a Belyi 3-complex by taking the $q\to1$ limit of its Turaev-Viro state sum.
The term-by-term limit of a sum can be taken if all the labels in the summation are constrained to a finite range by the classical triangle
constraints. Since we shall always take the large $r$ limit, we assume throughout that the noncommutativity parameter $j\ll r$.

The partition functions $Z_q$, and its classical limit $Z$, are multiplicative. For a pair of disjoint labelled complexes ${\cal M}_1$ and ${\cal M}_2$, we have 
\bea Z_q[{\cal M}_1\amalg{\cal M}_2] = Z_q[{\cal M}_1]Z_q[{\cal M}_2]. \eea
This multiplicative rule is modified slightly when the subcomplexes ${\cal M}_1$ and ${\cal M}_2$ are not disjoint, but share a boundary. If all the edges on the shared boundary remain on the boundary of the glued complex, then the above relation still holds. For a more general gluing, edges that are on the boundary of ${\cal M}_1$ may be in the interior of ${\cal M}_1\cup {\cal M}_2$, so new weight factors and sums need to be introduced. The general gluing procedure for $Z$ is
\bea Z_q[{\cal M}_1\cup {\cal M}_2] = w^{-2v}\sum_{j_i\in\cal B}\left(\prod_{j_i \in \cal B}(-)^{2j_i}[2j_i+1]\right)Z_q[{\cal M}_1]Z_q[{\cal M}_2], \eea
where ${\cal B}$ is the subset of spin labels assigned to lines in $\partial{\cal M}_1\cap\partial{\cal M}_2$ that are in the interior of ${\cal M}_1\cup {\cal M}_2$, and $v$ is the number of vertices that were on the boundary of ${\cal M}_1$ and ${\cal M}_2$ but in the interior of the glued manifold.

This property means that $Z[\mG]$ is a sum over the labels associated to the green and black edges in the Belyi 3-complex, weighted by $6j$s. The zero spin label assigned to the blue edge and the identity 
\bea\label{eq:trivialtet} \wsjq{0}{j}{j}{l_i}{j}{j} = \frac{(-)^{2j}}{[2j+1]} \eea
means that many tetrahedra in the complex have a trivial state sum evaluation, leaving two non-trivial tetrahedra associated to each vertex of the ribbon graph. One of these tetrahedra generates a $6j$ with three repeated $j$ labels, but the other tetrahedron, the `internal' tetrahedron, is not as straightforward to interpret. The complex composed of all these interior tetrahedra gives a triangulation of a ball bounded by the inner Belyi triangulation, and its associated partition function is actually equal to $\sG$, the spin network state sum associated to the labelled ribbon graph. This is proved in \cite{moussouris} and reviewed in \cite{hcruiz}, and is only valid for planar graphs. 

In this paper, however, we adopt another approach to showing the equivalence of the partition function of a manifold and its ribbon graph evaluation.
We extend the definition of the ribbon graph evaluation $\R$ to include fragments of graphs, and calculate the corresponding changes made to $Z\circ{\cal M}$ and $\R$ under the trivalent graph moves of Section \ref{sec:moves} on fragments of the graph. We recall the 2-2 move on a pair of vertices,
\bea \raisebox{-12pt}{\includegraphics[height=32pt]{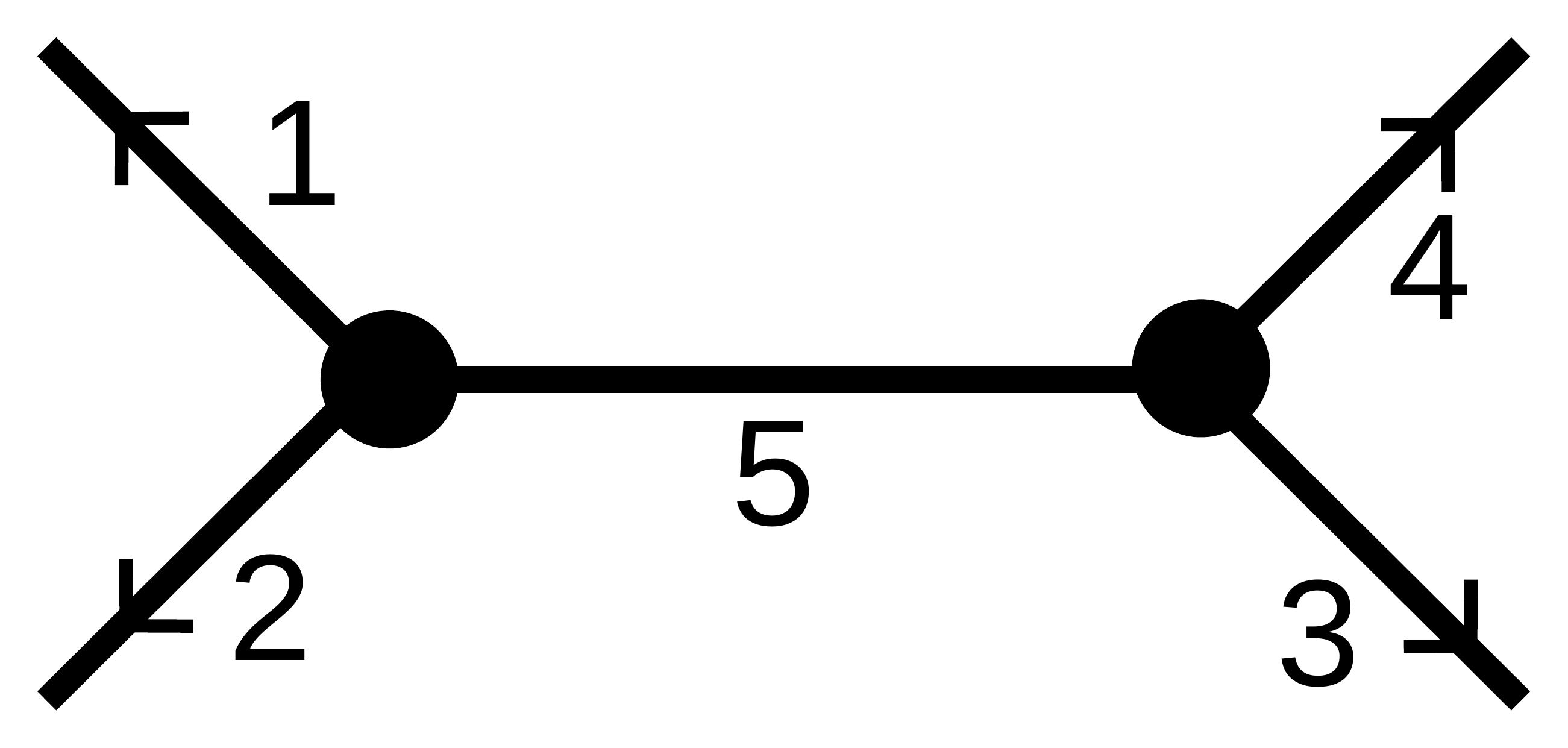}}  \quad \longrightarrow \quad \raisebox{-28pt}{\includegraphics[height=64pt]{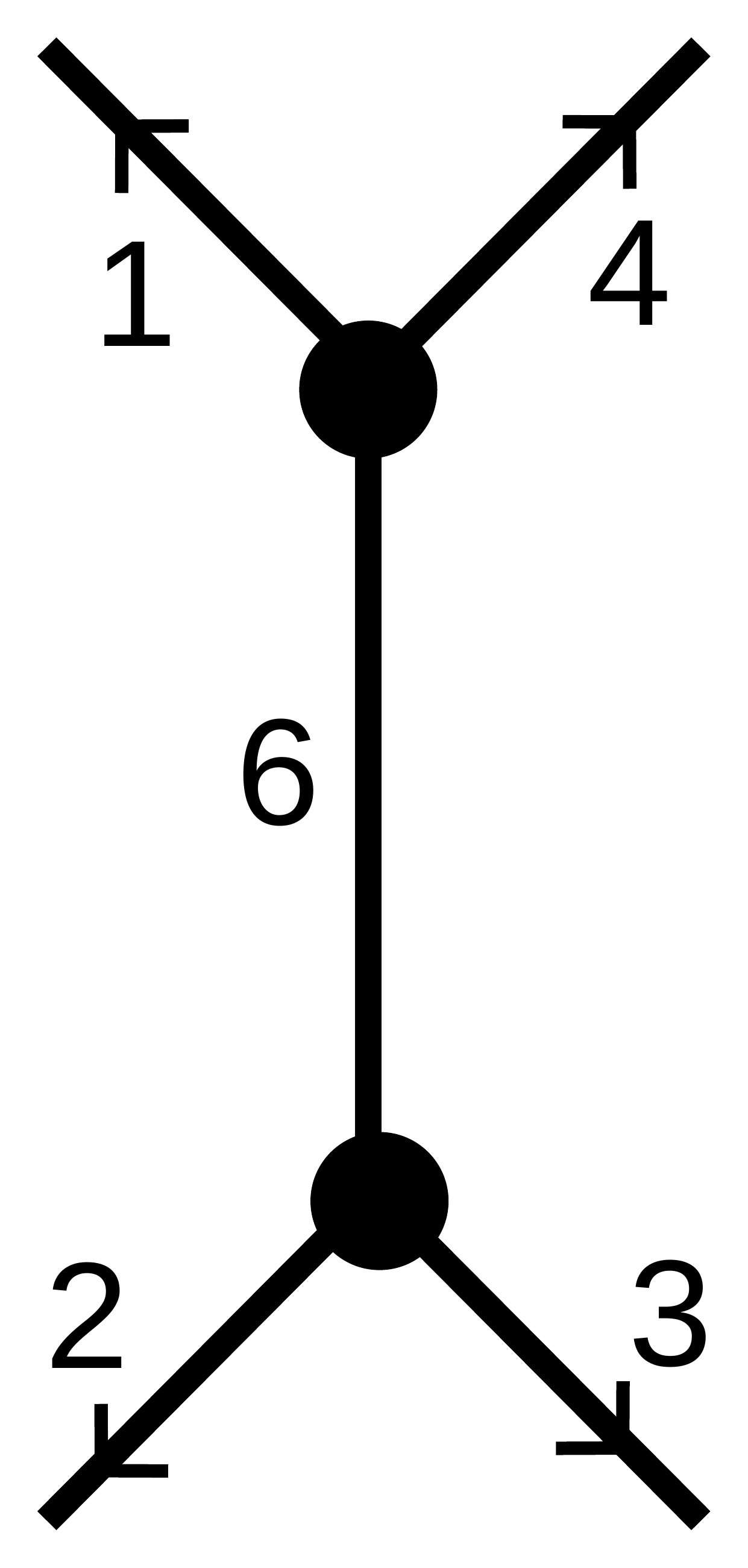}} \eea
and the 3-1 move
\bea \raisebox{-16pt}{\includegraphics[height=40pt]{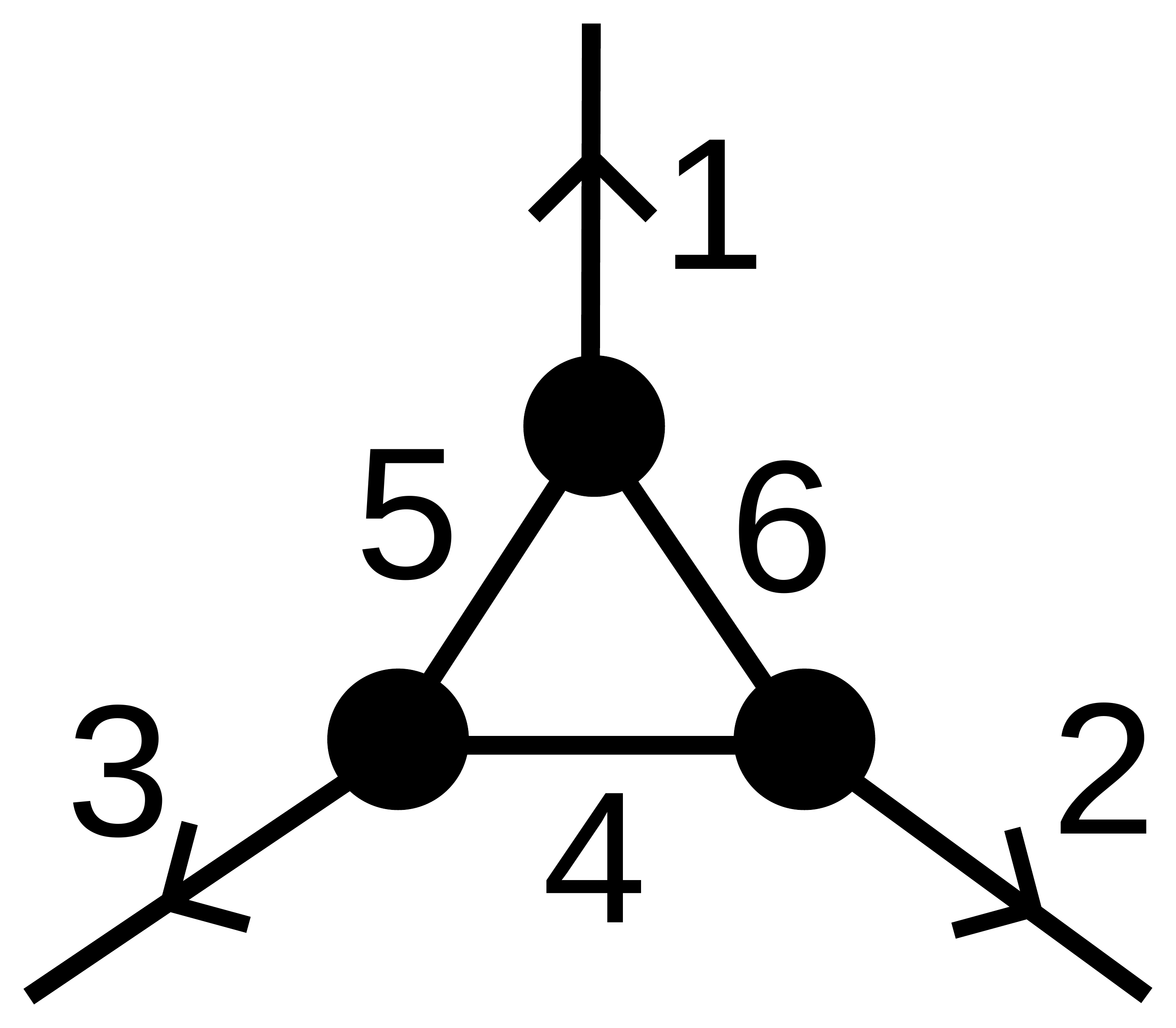}}  \quad \longrightarrow \quad \raisebox{-16pt}{\includegraphics[height=40pt]{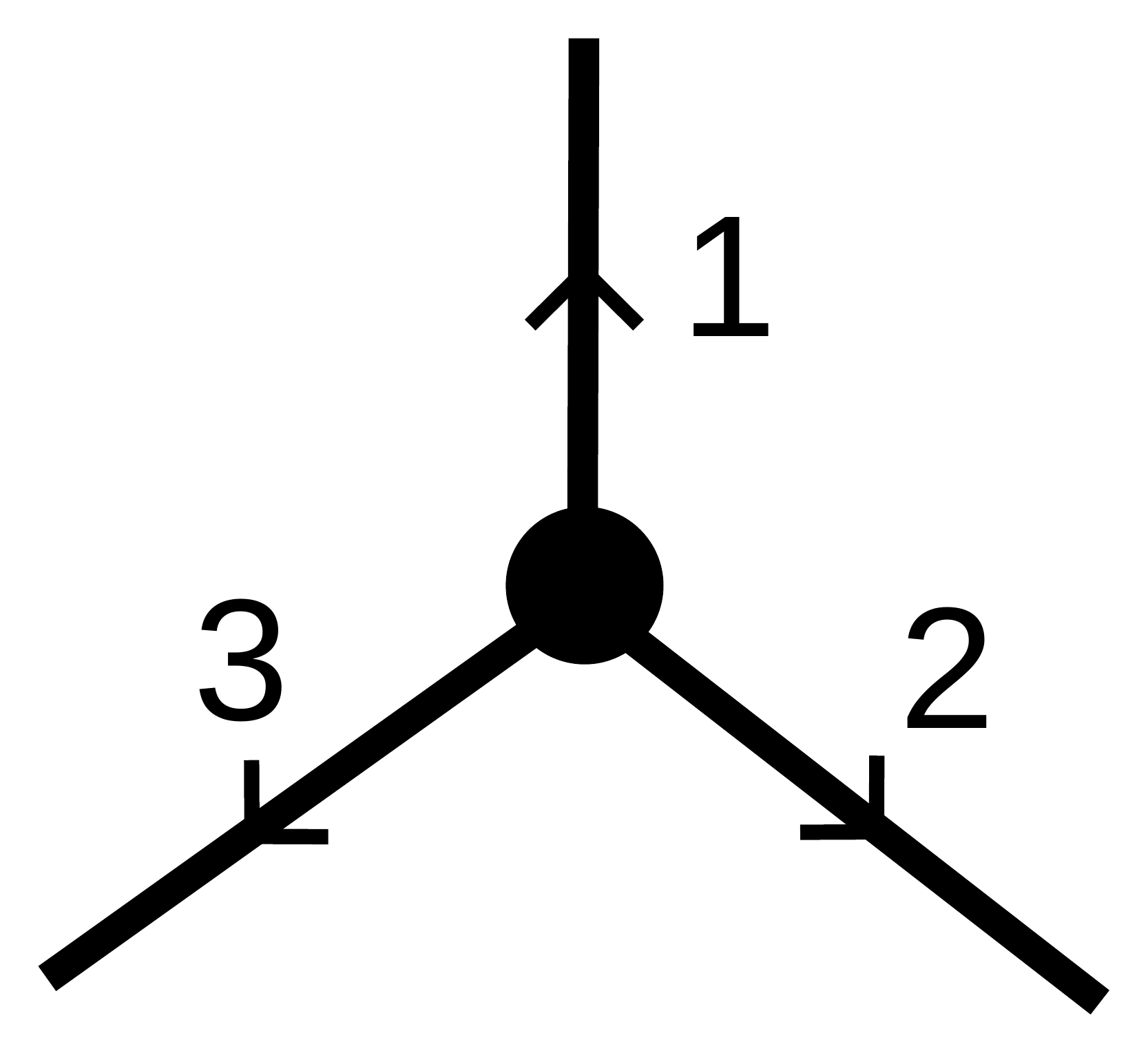}}. \eea
These moves will not change the genus of a graph, so we have
\bea \ra{tiny2-2a}  = \rb{tiny2-2b}, \eea  
and
\bea \ra{tiny3-1a}  =  \ra{tiny3-1b}. \eea  
Hence, we wish to show that the corresponding result holds for $N^V Z\circ{\cal M}$, which are
\bea N^2 Z\circ\m{tiny2-2a}  = N^2 Z\circ\mtwo{tiny2-2b} \eea  
and
\bea N^3 Z\circ\m{tiny3-1a}  =  N Z\circ\m{tiny3-1b}. \eea  
These results can be proved by evaluating the partition function $Z$ on fragments of the graphs before and after the moves are applied. This calculation is shown explicitly in Appendix \ref{sec:planarmove}.

Now we employ the algorithm for reducing down planar graphs that was described in Section \ref{sec:algsg}. For planar graphs with more than two vertices, only the 3-1 and 2-2 moves are required to reduce a trivalent graph down to the two-vertex theta graph, so we can perform these moves to arrive at
\bea N^V Z[\mG]= N^2 Z[{\cal M}[\Theta]]. \eea
\begin{figure}[H]
\centering
{\includegraphics[width=0.36\textwidth]{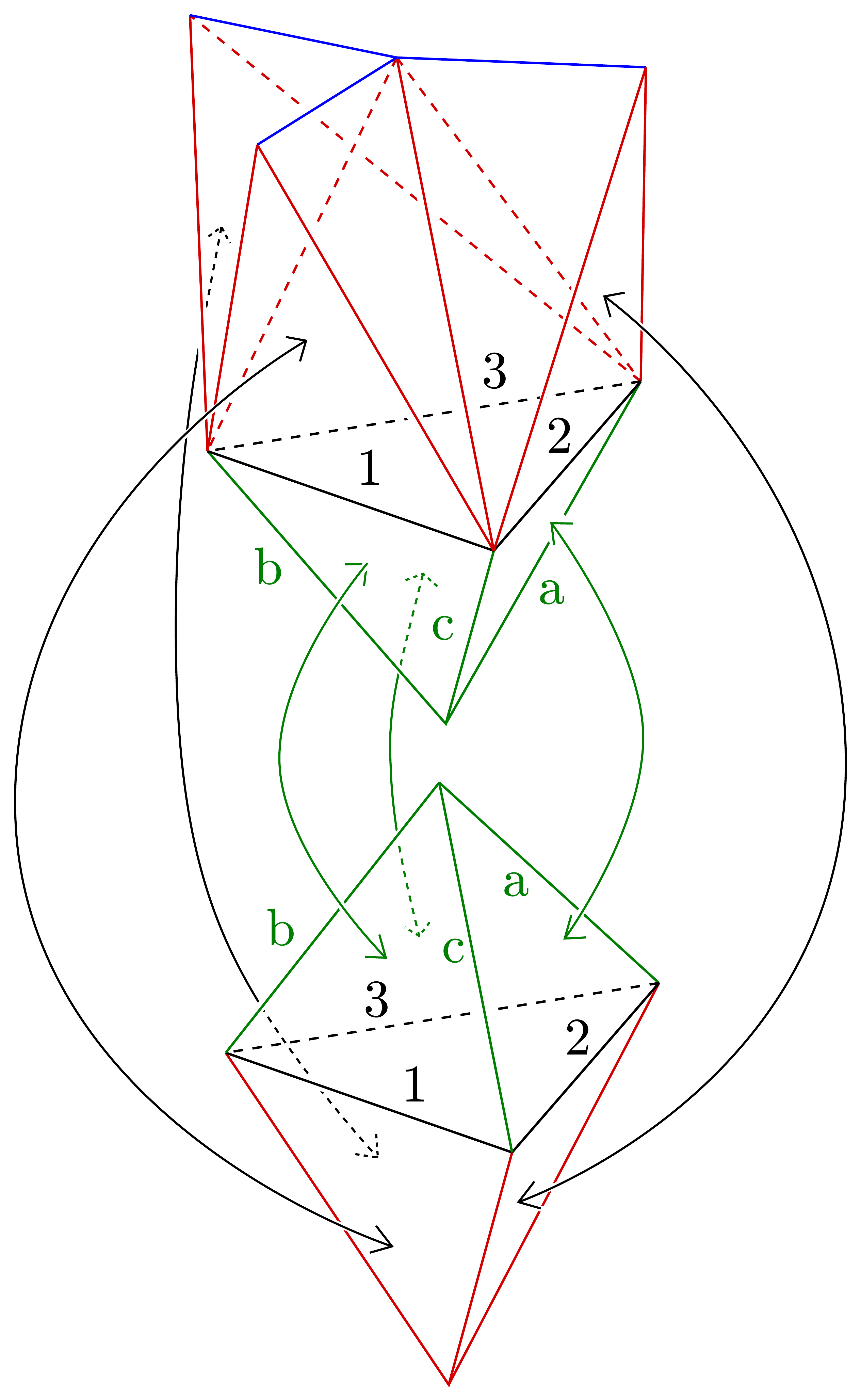}}\qquad \raisebox{100pt}{\includegraphics[width=0.18\textwidth]{arrow2}}
{\includegraphics[width=0.36\textwidth]{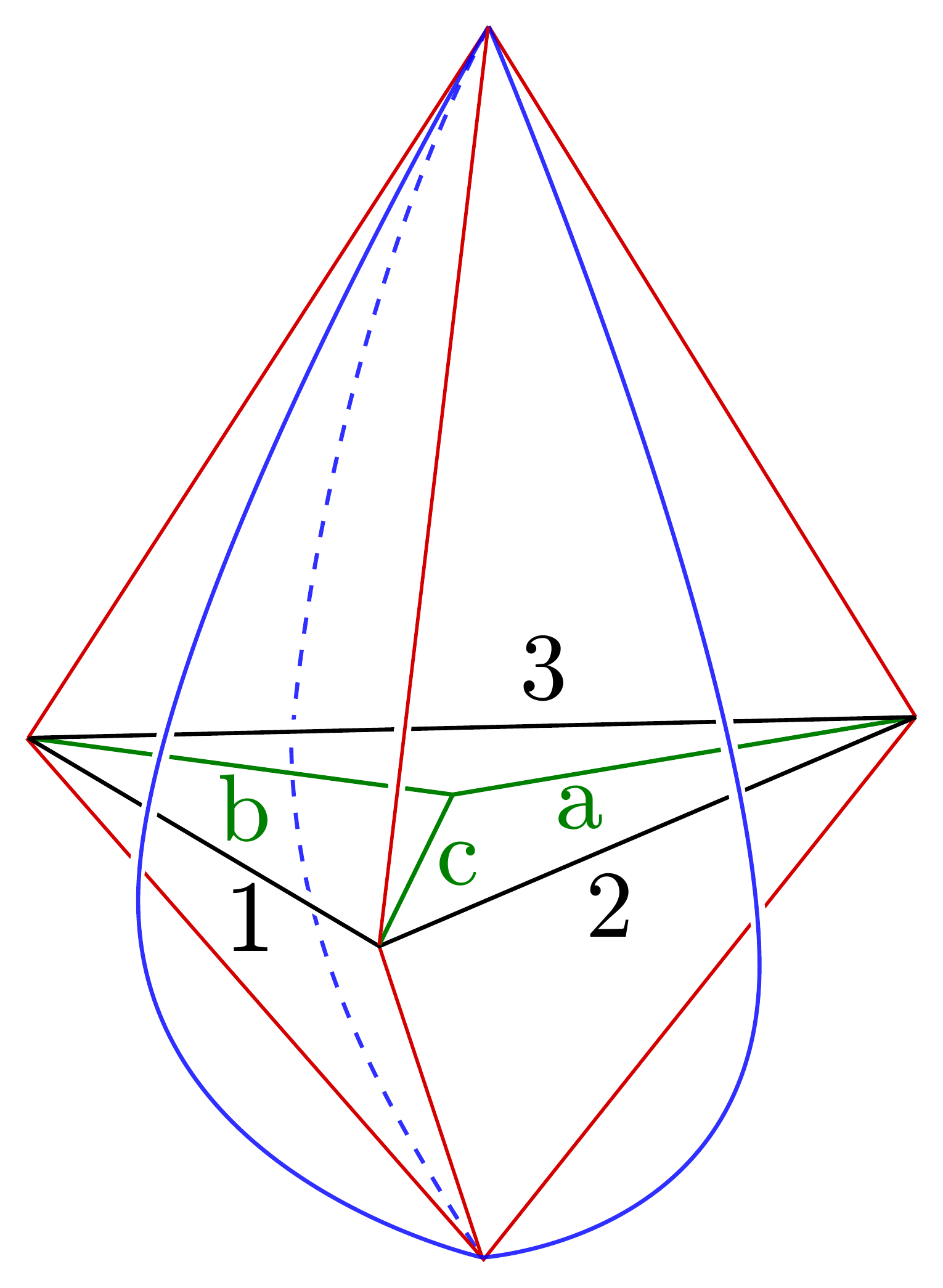}}
\caption{The gluing of the complex associated to the theta graph $\Theta$.}
\label{fig:thetacomplex}
\end{figure}

The 3-complex associated with the two-vertex theta graph is shown in Figure \refb{fig:thetacomplex}, and consists of seven tetrahedra, six interior edges, and an interior vertex. Its partition function is derived by the limit of the Turaev-Viro partition function, and is 

\begin{multline} N^2Z = \lim_{q\to 1}N^2 w^{-2}\sum_{\substack{l_1l_2l_3 \\ a b c}}(-)^{2a+2b+2c}[2a+1][2b+1][2c+1][2l_1+1][2l_2+1][2l_3+1] \times \\ 
\times \wsjq{0}{j}{j}{l_1}{j}{j} \wsjq{0}{j}{j}{l_2}{j}{j} \wsjq{0}{j}{j}{l_3}{j}{j} \wsjjq{l_1}{l_2}{l_3}\wsjjq{l_1}{l_2}{l_3}\wsjq{l_1}{l_2}{l_3}{a}{b}{c}\wsjq{l_1}{l_2}{l_3}{a}{b}{c} 
\end{multline}

We have omitted here the factors of $(-)^{2l_i}$ since the $6j$ triangle constraints force these phase factors to be equal to 1. The sums over the edges labelled $a$, $b$, and $c$ are unbounded as $q$ tends to 1, so we must perform these sums before taking the classical $6j$ limit. We first use the orthogonality relation
\bea \sum_c(-)^{2c}[2c+1]\wsjq{l_1}{l_2}{l_3}{a}{b}{c}\wsjq{l_1}{l_2}{l_3}{a}{b}{c} = \frac{1}{[2l_3+1]}\Delta(a,b,l_3)\Delta(l_1,l_2,l_3), \eea
where the function $\Delta$ is 1 if the spin labels satisfy a triangle constraint, and zero otherwise. We next use the relation 
\bea \sum_{a b}(-)^{2a+2b+2l_3}[2a+1][2b+1]\Delta(a,b,l_3) = w^2(-)^{2l_3}[2l_3+1], \eea
which generates the quantum factor $w^2$ of the Turaev-Viro model, to deduce that 
\begin{multline} N^2Z = \lim_{q\to 1}N^2 \sum_{l_1l_2l_3}[2l_1+1][2l_2+1][2l_3+1] \times \\
\times \wsjq{0}{j}{j}{l_1}{j}{j} \wsjq{0}{j}{j}{l_2}{j}{j} \wsjq{0}{j}{j}{l_3}{j}{j} \wsjjq{l_1}{l_2}{l_3}\wsjjq{l_1}{l_2}{l_3},
\end{multline}
where the previously generated $\Delta(l_1,l_2,l_3)$ has been absorbed into a $6j$.
All ranges of summation are now finite, so we can take the limit term-by-term of this expression, using Equation \refb{eq:trivialtet} to remove the three `trivial' $6j$s, to get
\bea N^2Z = \frac{(-)^{2j}}{(2j+1)}\sum_{l_1l_2l_3}(2l_1+1)(2l_2+1)(2l_3+1)\wsjj{l_1}{l_2}{l_3}\wsjj{l_1}{l_2}{l_3} \label{eq:same6j} \eea

This is exactly the same $6j$ sum given in Section \ref{sec:sixjex} for the ribbon graph evaluation of the theta graph. Hence, we can quote the above to state that the evaluation of this partition function is 
\bea N^2Z[{\cal M}[\Theta]] = (2j+1)^2 = N^2. \eea
Comparing this partition function with the known evaluation of a ribbon graph of genus zero, we can therefore deduce that for any planar graph, we have the relation
\bea N^V Z [\mG] = N^2. \eea
As we know that the ribbon graph evaluation of any planar ribbon graph is $N^2$, we have therefore proved that equation \refb{eq:prrelation} holds over all planar graphs.

\subsection{Example: The Tetrahedral Graph}
As an example of the equivalence of the Ponzano-Regge state sums and ribbon graph $6j$ sums, we apply the construction to the tetrahedral planar graph.

\bea {\cal M} \left[  \raisebox{-45pt}{\includegraphics[width=0.25\textwidth]{6jno}} \right] \qquad = \quad \raisebox{-0.22\textwidth}{\includegraphics[width=0.5\textwidth]{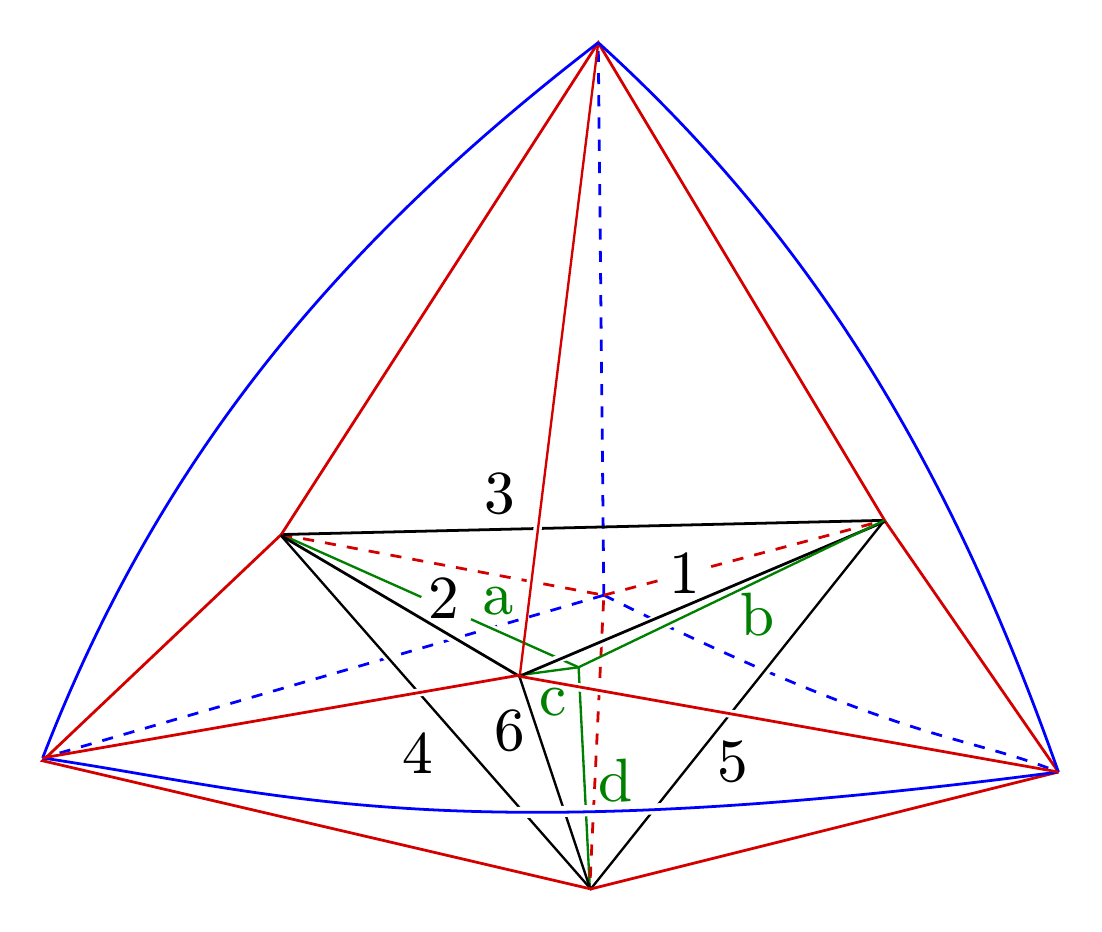}} \eea

This graph has partition function

\begin{multline}
N^4 Z = N^4 \lim_{q\to1} w^{-2}\sum_{\mathrm{internal\ labels}}[2l_1+1]\ldots[2l_6+1](-)^{2a+2b+2c+2d}[2a+1][2b+1][2c+1][2d+1] \times \\
\times \wsjq{0}{j}{j}{l_1}{j}{j} \wsjq{0}{j}{j}{l_2}{j}{j} \wsjq{0}{j}{j}{l_3}{j}{j}\wsjq{0}{j}{j}{l_4}{j}{j} \wsjq{0}{j}{j}{l_5}{j}{j} \wsjq{0}{j}{j}{l_6}{j}{j} \times \\ 
\times \wsjjq{l_1}{l_2}{l_3} \wsjjq{l_2}{l_4}{l_6} \wsjjq{l_3}{l_4}{l_5} \wsjjq{l_1}{l_5}{l_6} \times \\
\times \wsjq{l_1}{l_2}{l_3}{a}{b}{c} \wsjq{l_2}{l_4}{l_6}{d}{c}{a}\wsjq{l_3}{l_4}{l_5}{d}{b}{a} \wsjq{l_1}{l_5}{l_6}{d}{c}{b}.
\end{multline}
By using \refb{eq:trivialtet}, the $6j$s corresponding to the six tetrahedra with a blue line immediately evaluate to $[N]^{-6}$. The four tetrahedra surrounding the interior vertex correspond to the $6j$s containing the labels $a,b,c$, and $d$. These four tetrahedra can in fact be reduced to a single tetrahedron using the 4-1 Pachner move and its associated $6j$ identity from Section \ref{sec:tv}. This removes all the dependence on the quantum regularisation factor $w^{-2}$ and the spin labels $a,b,c$, and $d$ from the sum, and all the remaining labels are summed over a finite range, so we can take the term-by-term classical limit to state that
\begin{multline}\label{eq:same6j2} 
N^4 Z = N^{-2}\sum_{\substack{l_1l_2l_3 \\ l_4l_5l_6}}(2l_1+1)(2l_2+1)(2l_3+1)(2l_4+1)(2l_5+1)(2l_6+1) \times \\
\times \wsjj{l_1}{l_2}{l_3} \wsjj{l_2}{l_4}{l_6} \wsjj{l_3}{l_4}{l_5} \wsjj{l_1}{l_5}{l_6} \wsj{l_1}{l_2}{l_3}{l_4}{l_5}{l_6}.
\end{multline}
This is the same $6j$ sum that appears in Section \ref{sec:sixjex} for the associated ribbon graph.

\begin{figure}[H]
\centering
{\includegraphics[width=0.36\textwidth]{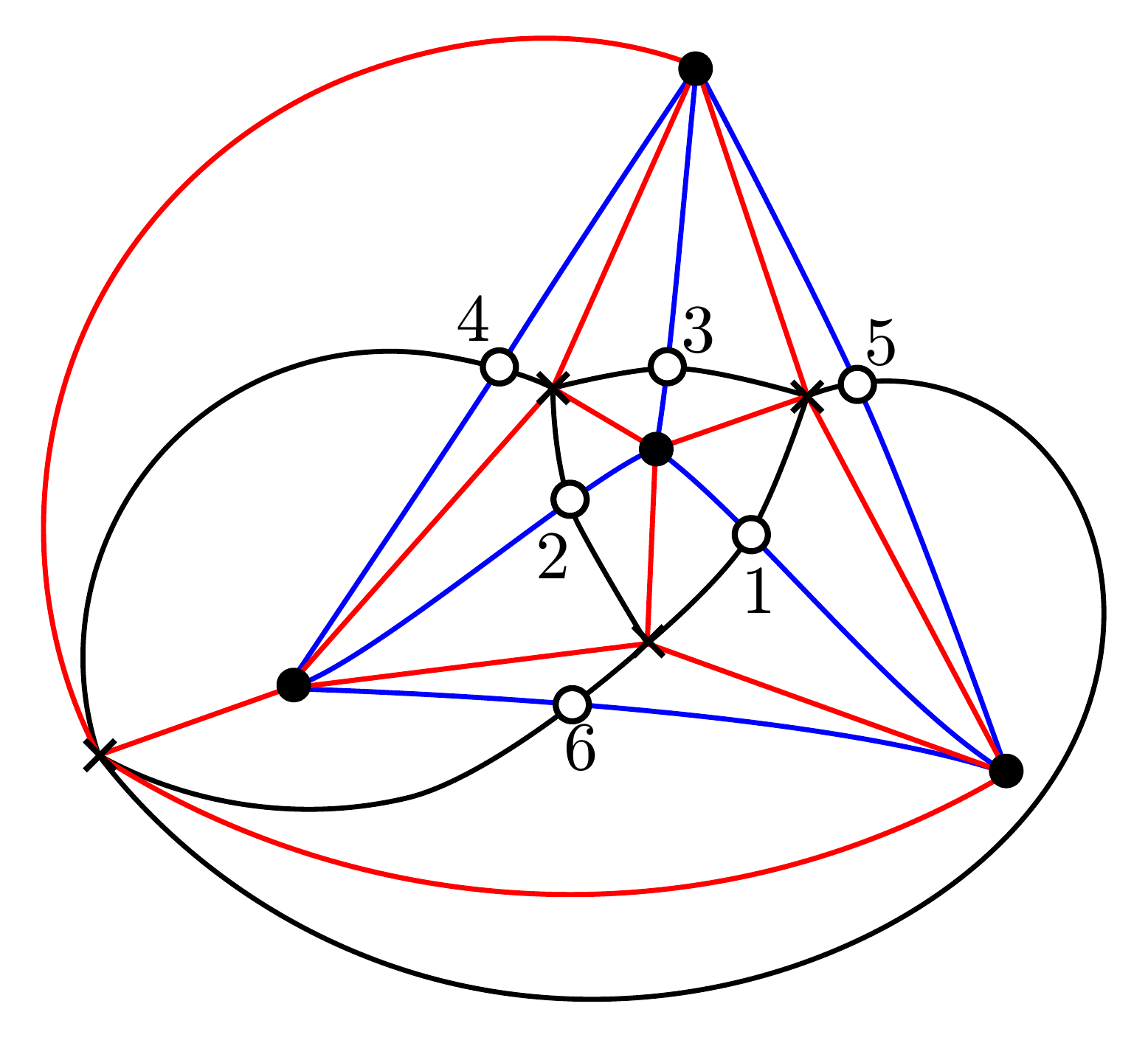}}
\caption{The manifold construction contains the Belyi triangulation.}
\label{fig:belyi2a}
\end{figure}

Finally, we note that the Belyi triangulation associated to the ribbon graph is present in this complex. The red and blue boundary edges of the complex form the outer Belyi triangulation of the sphere, and the black edges form an inner Belyi triangulation of the sphere embedded in the interior. If we project the black edges from the interior onto the boundary, we generate the full Belyi triangulation shown in Figure \ref{fig:belyi2a}.

\section{Higher genus ribbon graphs} \label{sec:PRNP}

In section \ref{sec:belyi}, we defined the inner and outer Belyi triangulations associated with a Belyi map. 
In the previous section, we described a method of generating a 3-complex 
from a trivalent planar ribbon graph, with an outer Belyi triangulation on its boundary and an inner
Belyi triangulation in its interior. We then showed that it has the same Ponzano-Regge state sum 
evaluation as its ribbon graph sum. In addition, we can see in examples \refb{eq:same6j} and \refb{eq:same6j2} 
that after summing out all the labels associated to the green interior lines, the Ponzano-Regge $6j$ sum takes the same
form as the ribbon graph $6j$ sums \refb{eq:thetaribbon} and \refb{eq:tetribbon}.
In general, the Ponzano-Regge $6j$ sum of a Belyi 3-complex will take the same form as the $6j$ sum associated to a
{\it planar} ribbon graph. This is because the Belyi 3-complex contains a sub-complex, homeomorphic to a ball, bounded by 
the $l_i$-labelled inner Belyi triangulation, whose Ponzano-Regge partition function is equal to $\sG$.

The aim of this section is to extend the correspondence between Hermitian matrix model ribbon graphs and Belyi 3-complexes to 
non-planar graphs. 
We show that a labelled 3-complex, homeomorphic to a handlebody, with an outer Belyi triangulation on its boundary, can be generated 
from a ribbon graph of arbitrary genus $g$, and that the Ponzano-Regge partition function of this complex and its associated $6j$ sum satisfy the relation
\bea \label{eq:npcorres} N^V Z\circ \mG = N^{2-2g} = \rG. \eea 

The main obstacle to this generalisation is that the vertex-by-vertex construction of $\mG$ given in Section 
\ref{sec:HMMPR} is no longer valid in the non-planar case. We must therefore present an alternative construction of the 
complex  $\mG$ for a graph of higher genus, but this construction will no longer have the spin network state sum $\sG$ appearing explicitly 
in its partition function. However, the non-planar complex $\mG$ will possess an outer Belyi triangulation on its 
boundary and an inner Belyi triangulation in their interior, as in the planar case, and we will prove that its normalised 
Ponzano-Regge partition function is equal to its ribbon graph sum. We thus consider our alternative construction as the appropriate 
generalisation of the planar Belyi 3-complex construction.

\subsection{Belyi 3-complexes of handlebodies}\label{sec:prnp1}

The construction of Section \ref{sec:HMMPR} associates a complex of tetrahedra to each vertex from a trivalent
ribbon graph, or equivalently to each black vertex in a dessin d'enfant. The complexes are then glued
together to construct a triangulation of a ball with an outer Belyi triangulation on its boundary, which we
call the Belyi 3-complex. 
We can then use moves on the tetrahedra to show that the Ponzano-Regge state sum evaluation 
of this complex is identical to the $6j$ sum.

\begin{figure}[h]
\centering
\subfloat[]{\label{fig:np1a}\includegraphics[width=0.4\textwidth]{np1}}\hspace{20mm}
\subfloat[]{\label{fig:np1b}\includegraphics[height=0.3\textwidth]{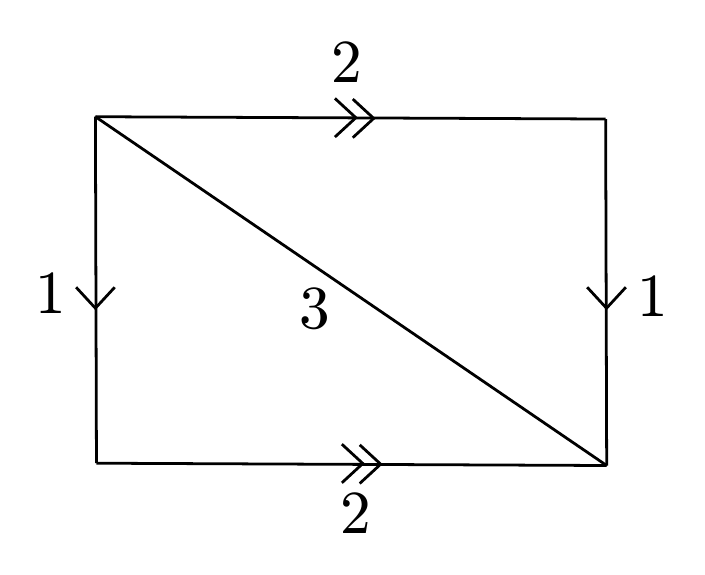}}
\caption{The simplest non-planar trivalent graph has genus 1, and its dual is a triangulation of the torus, where the labelled edges are identified.}\label{fig:np1}
\end{figure}

We run into several problems when we try to apply the same approach to graphs of higher genus.
Firstly, for even the simplest cases, the complex constructed from the gluing of the complexes 
given in Equations \refb{eq:complex1} and \refb{eq:complex2}  does not produce the anticipated 
genus-dependent result. For the non-planar graph shown in Figure \ref{fig:np1}, 
numerical computation of the Turaev-Viro partition function and consideration of $6j$ asymptotics 
both show that the partition function equals zero in the $q\to1$ limit. This cannot be the correct result,
as the ribbon graph evaluates to $N^{2-2g} = 1$. In addition, the generated 
3-complex is not a topological manifold. The interior vertex is a conical singularity,
which is not present in the generated triangulations of the ball.

Due to the triangulation-independence property of the Ponzano-Regge model, once a manifold
and its boundary triangulation are specified, any triangulation of its interior will yield the same partition function.
The previous construction generates a 3-complex with the correct boundary triangulation, but the wrong partition
function. We can therefore interpret the failure of the previous method for higher genus as being caused by the 
wrong interior manifold being generated by the previous construction.

A natural guess for the correct manifold to consider is a handlebody. A handlebody can be formed in three dimensions
by embedding a surface of genus $g$ into $\mathbb{R}^3$ and considering the volume enclosed within the surface.
This manifold is compact and has the genus $g$ surface as its boundary. 
We aim to construct from a general genus graph a triangulation of a handlebody where the edges of the triangulation on the boundary correspond to the outer Belyi triangulation. We call any such triangulation of a handlebody with $j$-labelled red edges and 0-labelled 
blue edges a Belyi 3-complex of the graph, and write $\mG$ to denote a Belyi 3-complex of a graph $\G$. 

To create a triangulation of a handlebody with the boundary data given by the ribbon graph, 
we need to use a different method of constructing the complex.
For a general higher genus graph, we can construct a Belyi 3-complex in three stages. 
First, find a triangulation of the handlebody with the same genus as the graph, and whose triangulation on 
the boundary is dual to the ribbon graph. This boundary triangulation is
the inner Belyi triangulation of the ribbon graph.
In keeping with the conventions established in the previous
section, we will colour the boundary inner Belyi triangulation edges 
in black, and the remaining interior edges of this complex in green.

\begin{figure}[h]
\centering
\includegraphics[width=0.3\textwidth]{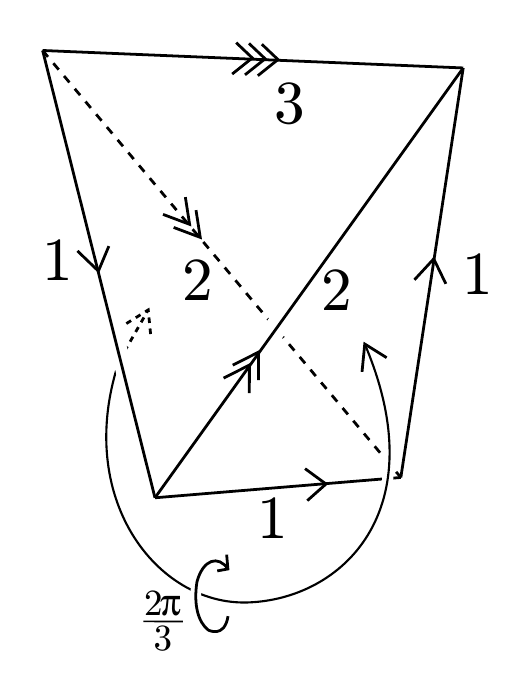}
\caption{The solid torus can be triangulated with a single tetrahedron by gluing two faces after a $2\pi/3$ rotation.}\label{fig:np2}
\end{figure}
Next, for each triangle on the boundary of the handlebody, glue on a tetrahedron with three red edges labelled $j$.
Finally, for each edge of the inner Belyi triangulation, add on a tetrahedron with four red $j$-labelled edges 
and one blue 0-labelled edge, by gluing the tetrahedron on to the faces adjacent to the edge of the inner Belyi
triangulation. This gluing will move the inner Belyi triangulation into the interior of the complex, 
and hence generate the required triangulated handlebody with an outer Belyi triangulation on its boundary. 

\begin{figure}[h]
\centering
\includegraphics[width=0.4\textwidth]{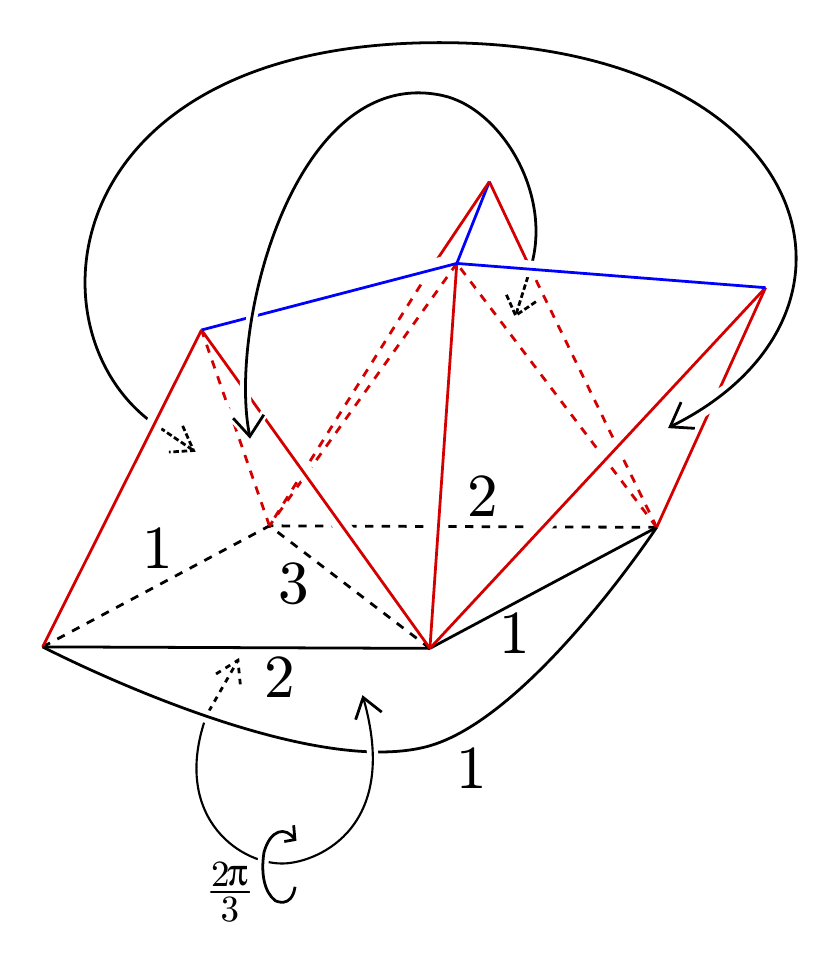}
\caption{The triangulation of the solid torus generated by the simplest non-planar graph. Three pairs of triangles in the diagram are identified.}\label{fig:np3}
\end{figure}

To exhibit this procedure, we consider the simplest example of a non-planar ribbon graph, given in Figure \ref{fig:np1}.
The graph is of genus one, so we seek a triangulation of the solid torus with two triangles on the boundary.
In fact, a particularly simple triangulation of the solid torus with two boundary triangles is known in the literature
on triangulations of 3-manifolds. The single tetrahedron triangulation of the solid torus given in \cite{jacorubinstein}, 
displayed here in Figure \ref{fig:np2}, is constructed by gluing two triangles together after a twist by $2\pi/3$.
In the diagram, the triangles labelled by the triples $(l_1, l_1, l_2)$ are identified, and the two distinct triangles
labelled $(l_1, l_2, l_3)$ form the boundary of the solid torus.
The boundary triangles of this tetrahedron form the inner Belyi triangulation of Figure \ref{fig:np1}. 
Hence we can add on two tetrahedra with black and red edges, and three tetrahedra with black, red and blue edges,
to generate the final Belyi 3-complex shown in Figure \ref{fig:np3}. 

We can check that this complex has the expected Ponzano-Regge evaluation.
It is known that a genus one ribbon graph evaluates to $N^{0}=1$, so we expect
that $N^V Z[\mG] = 1$ for any genus one ribbon graph with $V$ vertices.
With the appropriate normalisation $N^V$ added in, the Ponzano-Regge partition function of
the complex given in Figure \ref{fig:np3} is
\bea N^2Z= \frac{(-)^{2j}}{(2j+1)}\sum_{l_1l_2l_3}(2l_1+1)(2l_2+1)(2l_3+1)\wsj{l_1}{l_2}{l_3}{l_1}{l_2}{l_1}\wsjj{l_1}{l_2}{l_3}\wsjj{l_1}{l_2}{l_3}.\quad \eea
We can now apply the Biedenharn-Eliot identity \refb{eq:bhid2} on the sum over $l_3$ to get
\bea N^2Z= \frac{(-)^{2j}}{(2j+1)}\sum_{l_1l_2}(2l_1+1)(2l_2+1)\wsjj{l_1}{l_1}{l_2}\wsjj{l_1}{l_1}{l_2}, \eea
and the orthogonality relation of the $6j$s \refb{eq:6jorthog} reduces this sum to
\bea N^2 Z = \frac{1}{(2j+1)^2}\sum_{l_1}\Delta(l_1,j,j)(2l_1+1). \eea
The range of summation of $l_1$ is constrained by $\Delta$ to the integers $0\leq l_1 \leq 2j$, and hence the
final evaluation of this partition function is
\bea N^2 Z = 1, \eea
as required for consistency with \refb{eq:npcorres}.

In this example, the triangulation of the solid torus with the correct inner Belyi triangulation was simply stated.
In general, however, it is hard to find by inspection a triangulation of a handlebody whose boundary corresponds to a given 
inner Belyi triangulation, and it is necessary to employ an algorithmic procedure to construct such a complex. 

For a genus one graph, we can construct such a complex by starting from the one-tetrahedron triangulation 
of the solid torus given in Figure \ref{fig:np2} and repeatedly adding new tetrahedra onto its boundary.
This layering of tetrahedra will alter the triangulation on the boundary of the complex, but will not alter the topology
of the complex. In fact, the layering will alter the boundary triangulation by the Alexander moves introduced in Section \ref{sec:moves}.
The fact that it is possible to relate any two triangulations of a (2D) torus by a series of Alexander moves means that, given any 
genus one ribbon graph, there exists a sequence of layerings of tetrahedra onto the one-tetrahedron triangulation of the solid torus 
that will produce a solid torus triangulation with the inner Belyi triangulation associated to the graph on its boundary.

\begin{figure}[H]
\centering
\subfloat[]{{\label{fig:m1-3}\includegraphics[width=0.38\textwidth]{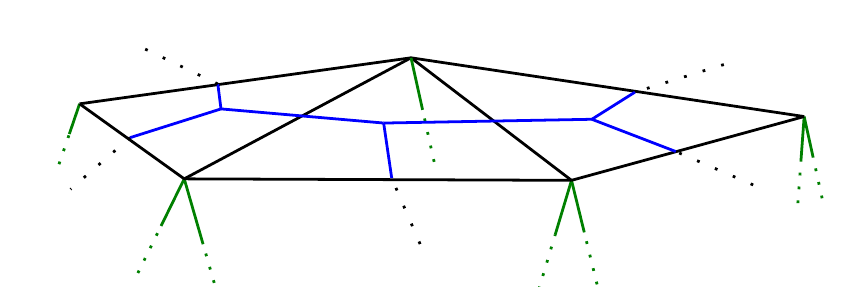}}\qquad \raisebox{25pt}{\includegraphics[width=0.18\textwidth]{arrow2}}
{\includegraphics[width=0.38\textwidth]{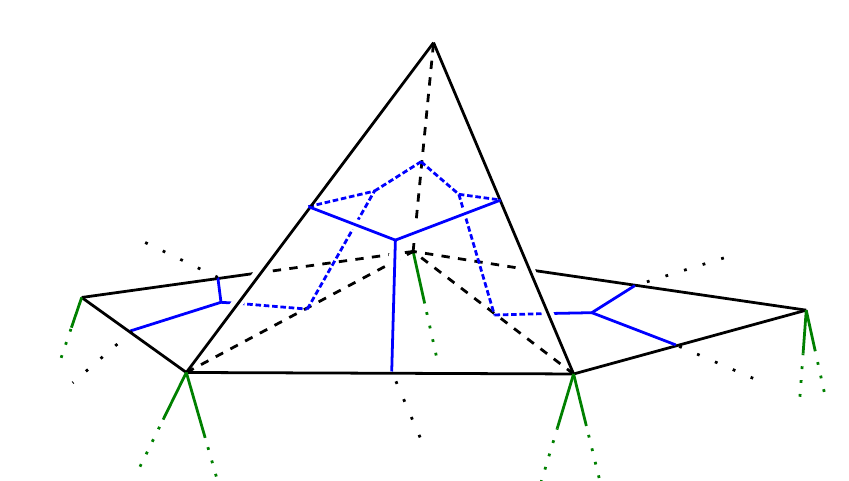}}}

\subfloat[]{{\label{fig:m2-2}\includegraphics[width=0.38\textwidth]{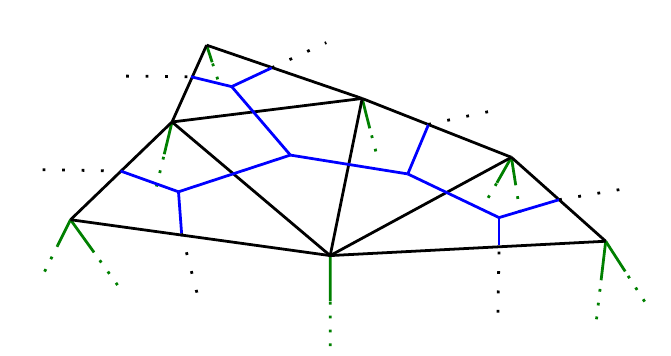}}\qquad \raisebox{25pt}{\includegraphics[width=0.18\textwidth]{arrow2}}
{\includegraphics[width=0.38\textwidth]{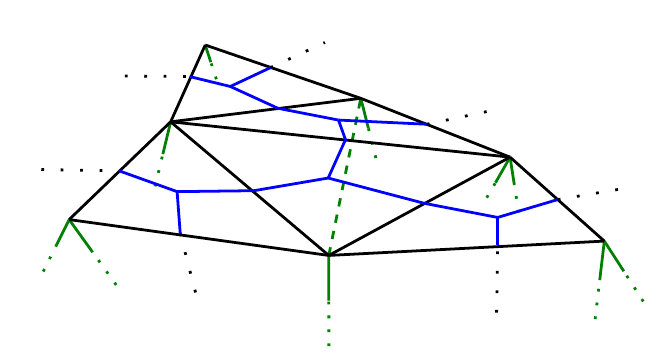}}}

\subfloat[]{{\label{fig:m3-1}\includegraphics[width=0.38\textwidth]{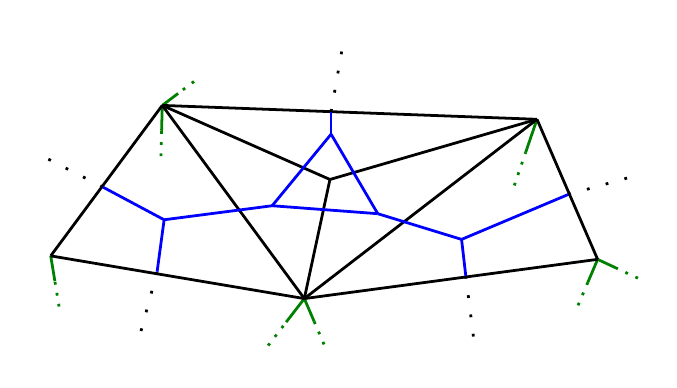}}\qquad \raisebox{25pt}{\includegraphics[width=0.18\textwidth]{arrow2}}
{\includegraphics[width=0.38\textwidth]{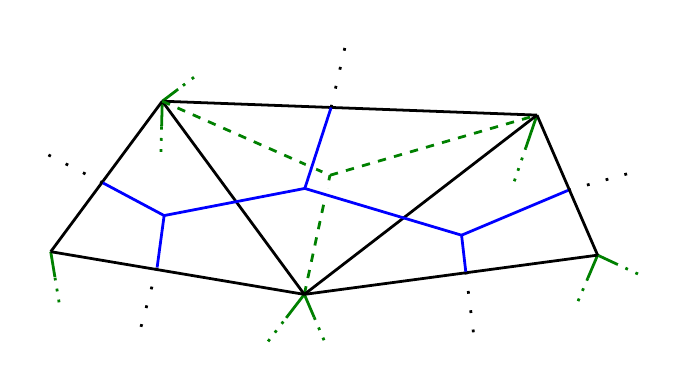}}}

\caption{The change to the boundary triangulation from gluing on a tetrahedron across one or two, or three faces is dual to a 1-3, 2-2, or 3-1 move. The boundary triangulation is shown in black, its dual graph is superimposed in blue, and the other interior edges are drawn in green.}
\label{fig:mlayers}
\end{figure}

In Figure \ref{fig:mlayers}, we have shown three ways of layering a tetrahedron on to a complex
that do not change its topology. The black edges denote the boundary of this complex, and the green edges lie in the interior
of the complex. In addition, we have superimposed the trivalent graph dual to the boundary triangulation in blue.
We see from the figure that layering a tetrahedron on either one, two, or three faces of the complex will alter the 
boundary triangulation of the complex by a 1-3, 2-2, or 3-1 Alexander move. 
The Alexander moves, introduced at the end of Section \ref{sec:moves}, are dual to the trivalent graph moves.
A summary of these two-dimensional moves, along with the three-dimensional Pachner moves of Section \ref{sec:tv}, 
is presented in a diagram in Appendix \ref{sec:allmoves}.

We can now state a procedure for generating a Belyi 3-complex associated to any graph of genus one. Given a genus one ribbon graph,
we can employ a series of trivalent graph moves that will reduce the graph down to the two-vertex graph in Figure \ref{fig:np1a}.
By considering the triangulations dual to this sequence of graphs, and reversing the ordering of the sequence, we generate a 
sequence of triangulations of the 2D torus, each related to its successor by an Alexander move. 
The initial term in this sequence is the triangulation of the torus given in Figure \ref{fig:np1b},
and the final term is the inner Belyi triangulation of the original ribbon graph. 
Each of the moves in this sequence can be associated with a layering of a tetrahedron on to a complex.
By starting from the one-tetrahedron triangulation of the solid torus, and sequentially carrying out the layering
associated to each Alexander move, we generate a series of triangulations of the solid torus.
The final complex in the sequence will be the required solid torus triangulation with the inner Belyi 
triangulation on its boundary.
Finally, we can add on the required red and blue labelled tetrahedra to generate a full Belyi 3-complex
with the topology of a solid torus.

We demonstrate this procedure by considering the following example of a genus one graph with six vertices.
This example will be very useful in the next subsection, as it provides a means of constructing triangulations of 
handlebodies of arbitrary genus, and it will be used in the evaluation of the partition function of a Belyi 3-complex 
of general genus.

\begin{figure}[h]
\centering
\subfloat[]{\label{fig:st1a}\includegraphics[height=0.3\textwidth]{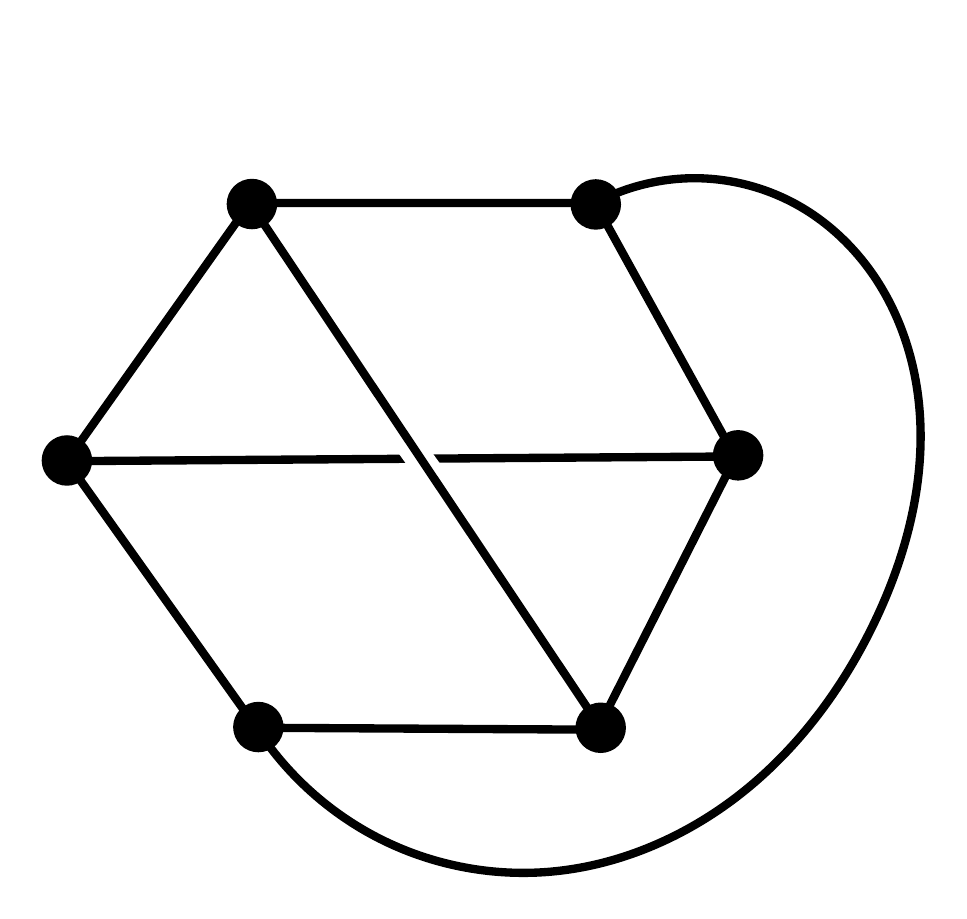}}\hspace{20mm}
\subfloat[]{\label{fig:st1b}\includegraphics[height=0.3\textwidth]{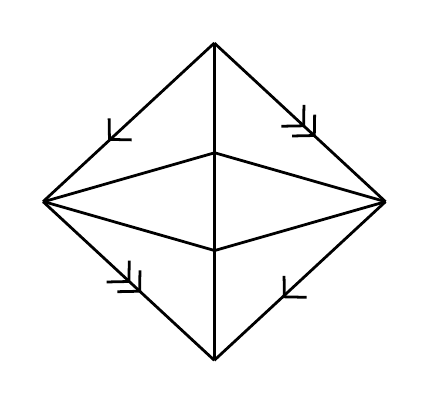}}
\caption{A six-vertex genus one graph and its dual triangulation of the torus.}\label{fig:st1}
\end{figure}

\begin{figure}[h]
\centering
\includegraphics[width=1\textwidth]{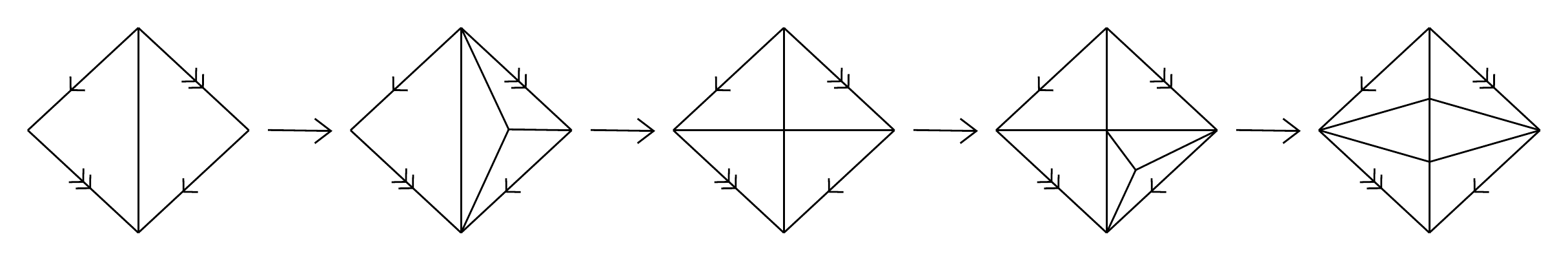}
\caption{The sequence of Alexander moves that generates the triangulation of the torus in Figure \ref{fig:st1b} from Figure \ref{fig:np1b}.}\label{fig:st2}
\end{figure}

\begin{figure}[H]
\centering
\subfloat[]{\label{fig:st3a}\includegraphics[width=.4\textwidth]{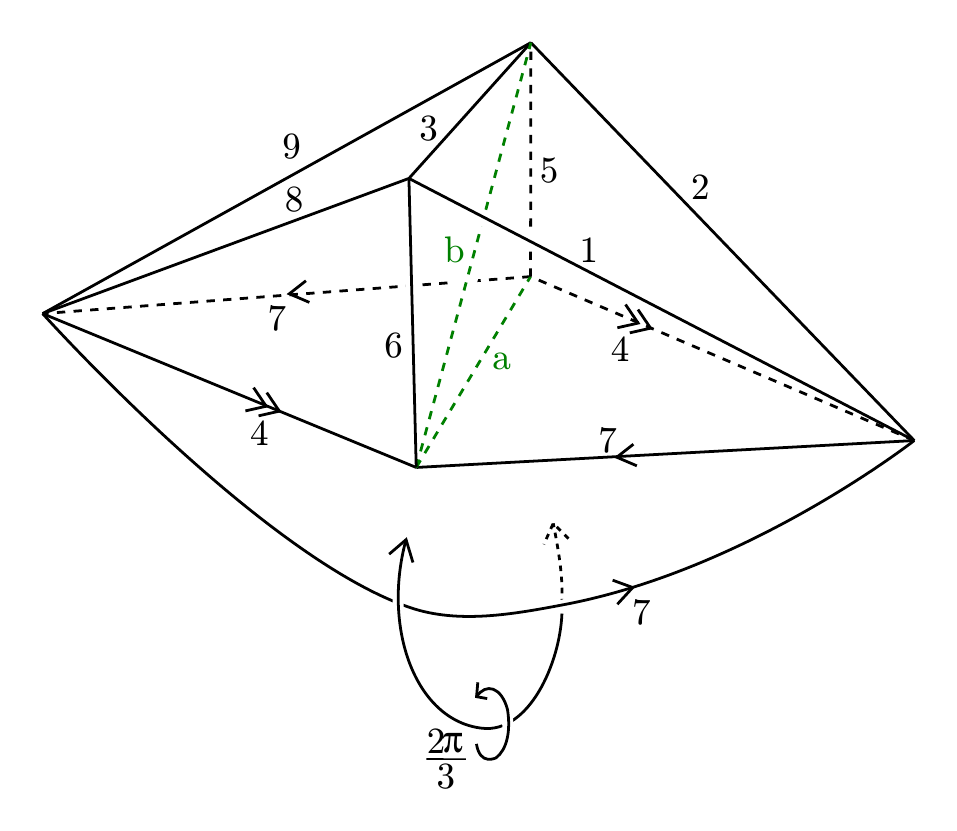}}\hspace{10mm}
\subfloat[]{\label{fig:st3b}\includegraphics[width=.4\textwidth]{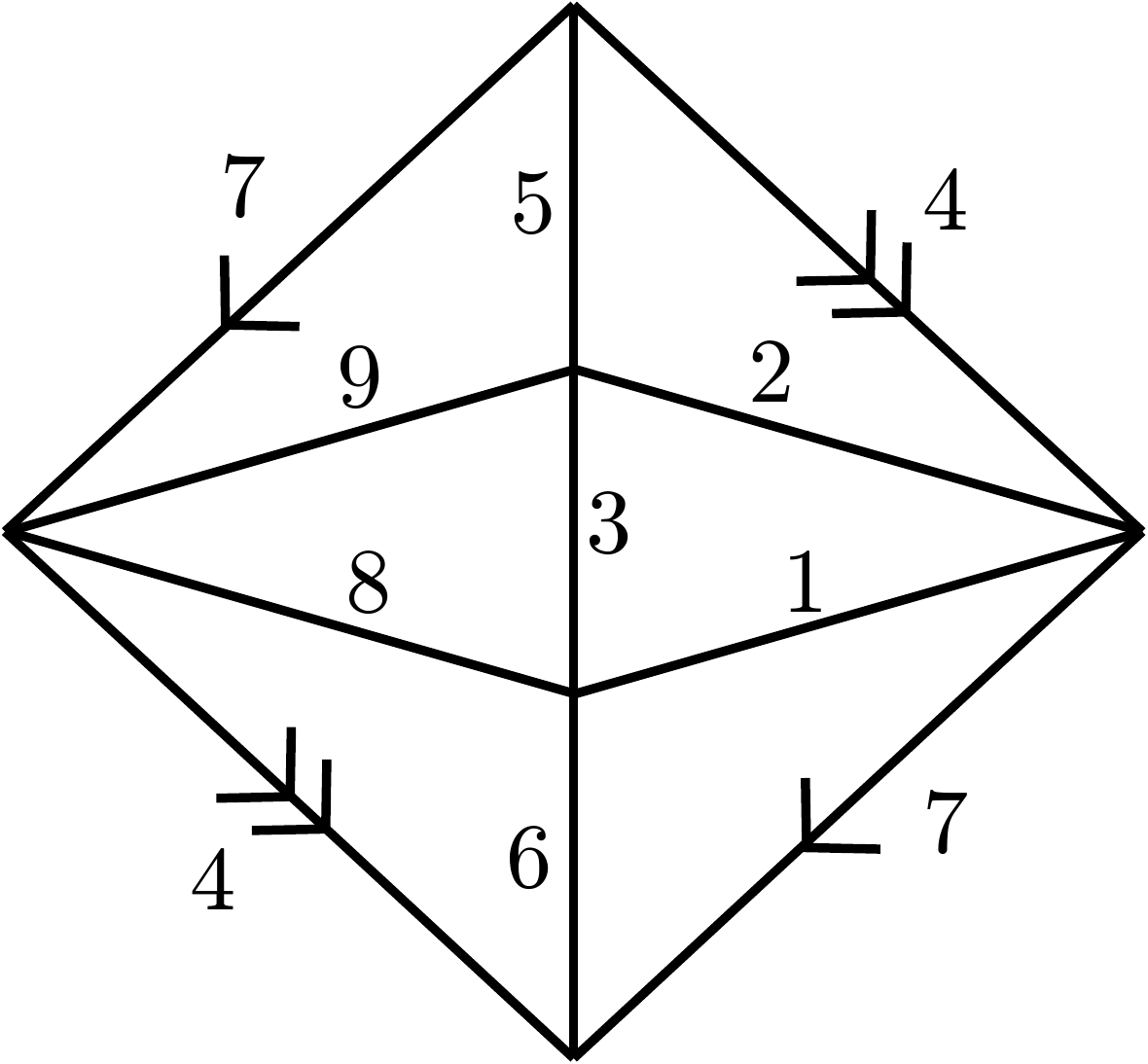}}
\caption{The labelled triangulation of the solid torus and its boundary, which is an inner Belyi triangulation, that matches Figure \ref{fig:st1b}. The interior edges are drawn in green.}\label{fig:st3}
\end{figure}

\begin{figure}[h]
\centering
\includegraphics[width=1\textwidth]{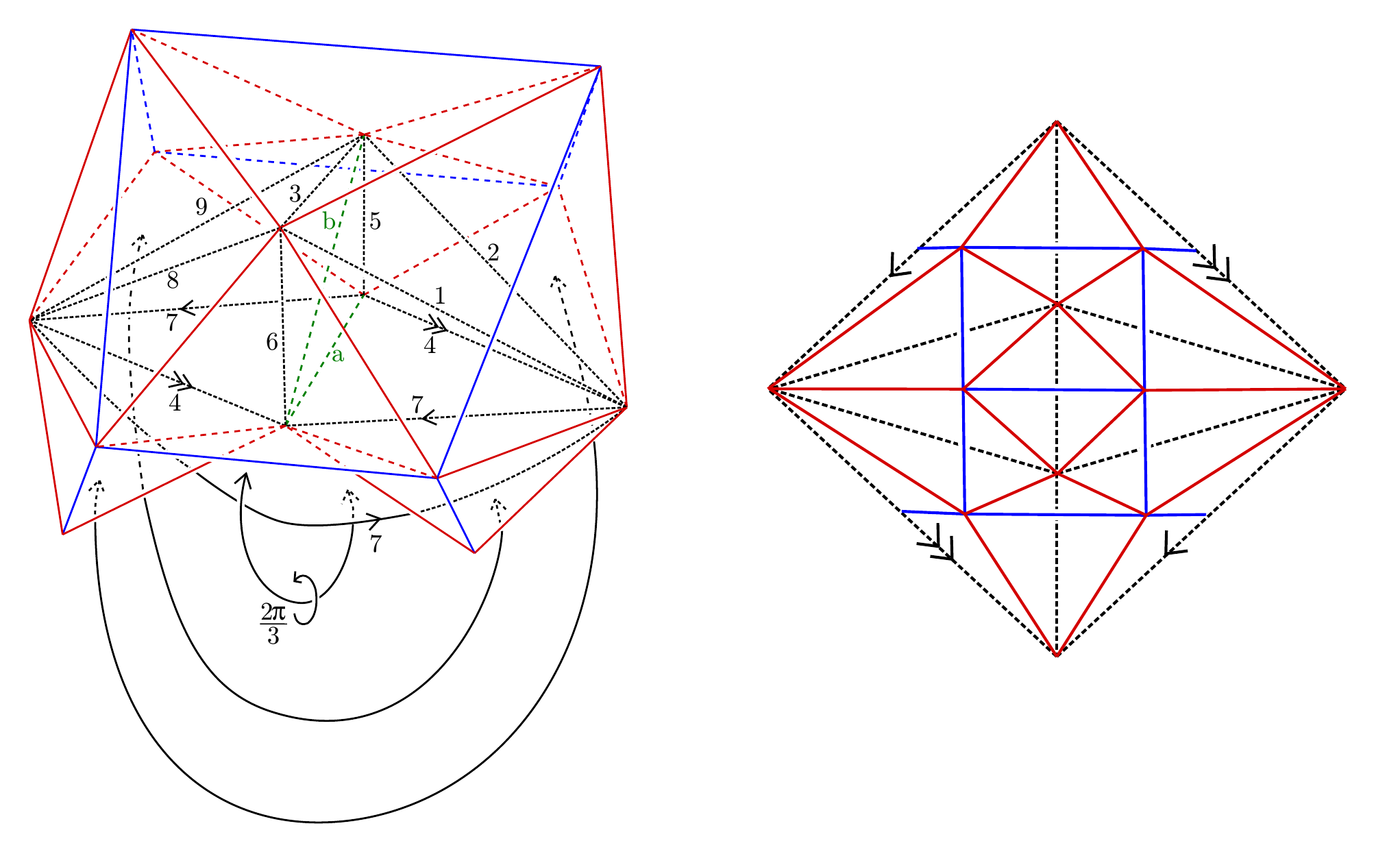}
\caption{The twenty-tetrahedron Belyi 3-complex associated to the ribbon graph in Figure \ref{fig:st1}, where three pairs of triangles are identified. Its boundary is an outer Belyi triangulation, and the black edges trace out an inner Belyi triangulation.}\label{fig:stmonster}
\end{figure}

Consider the graph shown in Figure \ref{fig:st1a}. Its dual triangulation, shown in Figure \ref{fig:st1b}, is related to the 
two-triangle triangulation of the torus by the sequence of Alexander moves shown in Figure \ref{fig:st2}. 
Hence, starting from the one-tetrahedron
triangulation of the torus, and layering on a tetrahedron corresponding to each of these moves, we can construct a 
triangulation of the solid torus whose boundary matches Figure \ref{fig:st1b}. The complex constructed from these layerings
is given in Figure \ref{fig:st3a}, and its labelled boundary is shown in Figure \ref{fig:st3b}.
We can create a Belyi 3-complex from this complex by adding on a tetrahedron with 
three red edges for each of the six faces, and a tetrahedron with a blue edge for each of the nine labelled black edges on the boundary. This generated Belyi 3-complex contains twenty tetrahedra in total, displayed in Figure \ref{fig:stmonster}, and has the partition function

\begin{multline}\label{eq:solidt}
N^6Z[\mG] = \frac{(-)^{2j}}{(2j+1)^3}\sum_{a,b,l_i} (-)^{2a+2b}(2a+1)(2b+1)\prod_{i=1,\ldots, 9}(2l_i+1)\times \\ \times \wsj{l_4}{l_7}{a}{l_4}{l_7}{l_7}  \wsj{l_4}{l_7}{a}{b}{l_5}{l_2}  \wsj{l_4}{l_7}{a}{l_5}{b}{l_9} \wsj{l_2}{l_7}{b}{l_6}{l_3}{l_1}  \wsj{l_4}{l_9}{b}{l_3}{l_6}{l_8} \times \\ \times 
\wsjj{l_2}{l_4}{l_5}\wsjj{l_5}{l_7}{l_9}\wsjj{l_3}{l_8}{l_9}\wsjj{l_4}{l_6}{l_8}\wsjj{l_1}{l_6}{l_7}\wsjj{l_1}{l_2}{l_3},
\end{multline}
where we have used the identity \refb{eq:trivialtet} on the nine tetrahedra containing a blue edge.

This sum can be calculated using the Biedenharn-Eliot and orthogonality relations on the $6j$s. For example,
considering the $6j$s containing the label $a$, we can use
\bea \sum_a(-)^{2a}(2a+1)\wsj{l_4}{l_7}{a}{l_4}{l_7}{l_7}  \wsj{l_4}{l_7}{a}{b}{l_5}{l_2}  \wsj{l_4}{l_7}{a}{l_5}{b}{l_9} = 
\wsj{l_2}{l_7}{l_9}{l_4}{b}{l_7}\wsj{l_2}{l_7}{l_9}{l_7}{l_5}{l_4} \eea
to remove the sum over $a$ and a $6j$. This also reduces the number of $6j$s containing $b$ by one, so we can now apply
the Biedenharn-Eliot identity \refb{eq:bhid2} on the sum over the label $b$, to obtain
\begin{multline}
N^6Z = \frac{(-)^{2j}}{(2j+1)^3}\sum_{l_i}\prod_{i=1,\ldots, 9}(2l_i+1)  
\wsj{l_2}{l_7}{l_9}{l_7}{l_5}{l_4} \wsj{l_1}{l_7}{l_8}{l_4}{l_6}{l_7} \wsj{l_1}{l_7}{l_8}{l_9}{l_3}{l_2} \times \\ \times 
\wsjj{l_1}{l_2}{l_3}\wsjj{l_2}{l_4}{l_5}\wsjj{l_5}{l_7}{l_9}\wsjj{l_3}{l_8}{l_9}\wsjj{l_4}{l_6}{l_8}\wsjj{l_1}{l_6}{l_7},
\end{multline}
Applying the same identity successively on the labels $l_6$, $l_4$, and $l_8$, we generate
\begin{multline}
N^6Z = \frac{(-)^{2j}}{(2j+1)^3}\sum_{\substack{l_1l_2l_3\\ l_5l_7l_9}}(2l_1+1)(2l_2+1)(2l_3+1)(2l_5+1)(2l_7+1)(2l_9+1) \times \\ \times 
\wsjj{l_1}{l_2}{l_3}\wsjj{l_1}{l_2}{l_3}\wsjj{l_2}{l_7}{l_9}\wsjj{l_2}{l_7}{l_9}\wsjj{l_5}{l_7}{l_9}\wsjj{l_5}{l_7}{l_9}.
\end{multline}
Using the orthogonality relation \refb{eq:6jorthog}  on $l_5$, then $l_7$, we obtain
\begin{multline}\label{eq:shortcut}
N^6Z = \frac{(-)^{2j}}{(2j+1)^5}\sum_{l_1l_2l_3l_9}(2l_1+1)(2l_2+1)(2l_3+1)(2l_9+1)\Delta(l_9,j,j)\wsjj{l_1}{l_2}{l_3}\wsjj{l_1}{l_2}{l_3} \\
= \frac{(-)^{2j}}{(2j+1)^3}\sum_{l_1l_2l_3}(2l_1+1)(2l_2+1)(2l_3+1)\wsjj{l_1}{l_2}{l_3}\wsjj{l_1}{l_2}{l_3}.
\end{multline}

This is now the same sum as in \refb{eq:thetaribbon}, so we can state that the final evaluation of 
the partition function of this complex is 
\bea N^6Z[\mG] = N^0 = 1. \eea
This is the desired result, as it is consistent with the relation \refb{eq:npcorres}.

\subsection{Evaluating the partition function of any Belyi 3-complex}\label{sec:npproof}

In the previous subsection, we gave a method of constructing a triangulation of a 
solid torus from the Belyi triangulation of a graph. We know that ribbon graphs of genus $g$ have the evaluation 
$\rG = N^{2-2g}$, and found that the normalised Ponzano-Regge  partition function $N^VZ$
corresponded with the ribbon graph evaluation in both the examples calculated in the previous section.
We made the hypothesis that the partition function of a general Belyi-triangulated handlebody of genus 
$g$ is given by $N^V Z = N^{2-2g}$. In this subsection, we first discuss how to construct Belyi 3-complexes
for a graph of genus greater than one, and then prove that the Ponzano-Regge partition function is equal 
to the ribbon graph evaluation for all graphs.

It is possible to construct a genus $g$ handlebody complex by gluing $g$ solid torus complexes together.
A genus $g$ handlebody can be created from a solid torus and a genus $(g-1)$ handlebody by 
identifying the two manifolds at contractible discs on their boundaries.
Hence it is possible to generate a genus $g$ handlebody by identifying $g$ solid tori at contractible discs on their boundaries.
A triangle on the boundary of a solid torus complex is contractible if it is homeomorphic to a disc, or equivalently if it has three distinct vertices. 
Therefore if we take $g$ copies of a triangulation of the solid torus with at least two contractible triangles on its boundary, then these
complexes can be glued together to produce a triangulation of a genus $g$ handlebody. The complex given in Figure \ref{fig:st3} has two
contractable triangles on its boundary, and hence can be used to construct a handlebody of arbitrary genus.

From this initial genus $g$ handlebody complex, we can construct a triangulation of the genus $g$ handlebody with 
any genus $g$ inner Belyi triangulation on its boundary. This complex can be constructed similarly to the genus one case
by using the layerings shown in Figure \ref{fig:mlayers}. Any two triangulations of a genus $g$ surface can be 
related by a series of Alexander moves; in particular, there exists a sequence of moves that relate the boundary triangulation
of the initial complex to any inner Belyi triangulation. By sequentially applying the tetrahedral layerings to the initial handlebody complex,
we will construct a series of complexes whose boundary triangulations are related by a series of Alexander moves, and thus can construct 
handlebody complexes with any required inner Belyi triangulation on its boundary.
Finally, from this complex we create the required Belyi 3-complex by gluing on the labelled red and blue tetrahedra.

Having discussed how to construct a Belyi 3-complex associated to any ribbon graph, we now prove that
the Ponzano-Regge partition function of the Belyi 3-complex is determined by its genus. 
We start by showing that the partition functions of any two Belyi 3-complexes for graphs of the same genus are equal. 
We can do this by employing a similar method as was used in the planar case, which is by 
showing that the partition function of $\mG$ does not change under the application of the 2-2
and 3-1 trivalent graph moves to $\G$. We do this by taking
two general Belyi 3-complexes that are almost identical except for at a few
tetrahedra, and whose boundaries are outer Belyi triangulations of two ribbon graphs
related by a trivalent graph move. We then show that their partition functions are equal; that is,
we show the equalities

\bea N^2Z \left[ \raisebox{-0.1\textheight}{\includegraphics[height=0.2\textheight]{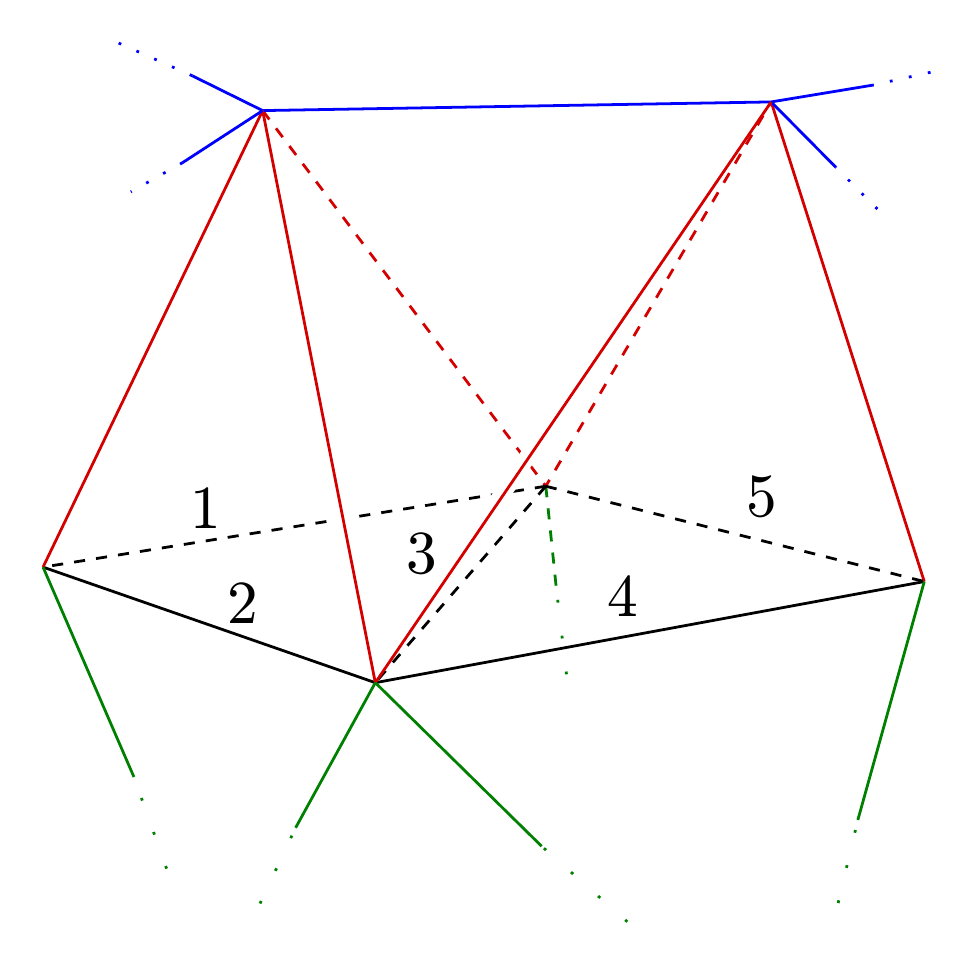}} \right] \quad &=& \quad N^2Z \left[ \raisebox{-0.1\textheight}{\includegraphics[height=0.2\textheight]{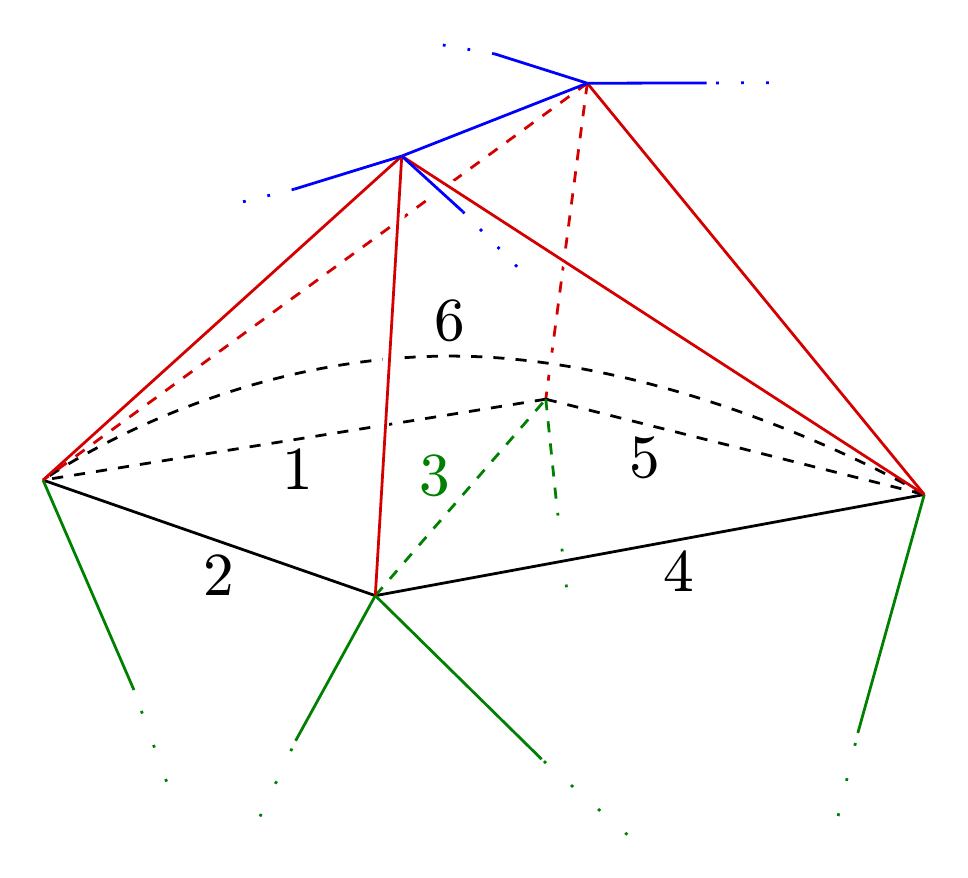}} \right], \qquad \\ \ret
N^3Z \left[ \raisebox{-0.1\textheight}{\includegraphics[height=0.2\textheight]{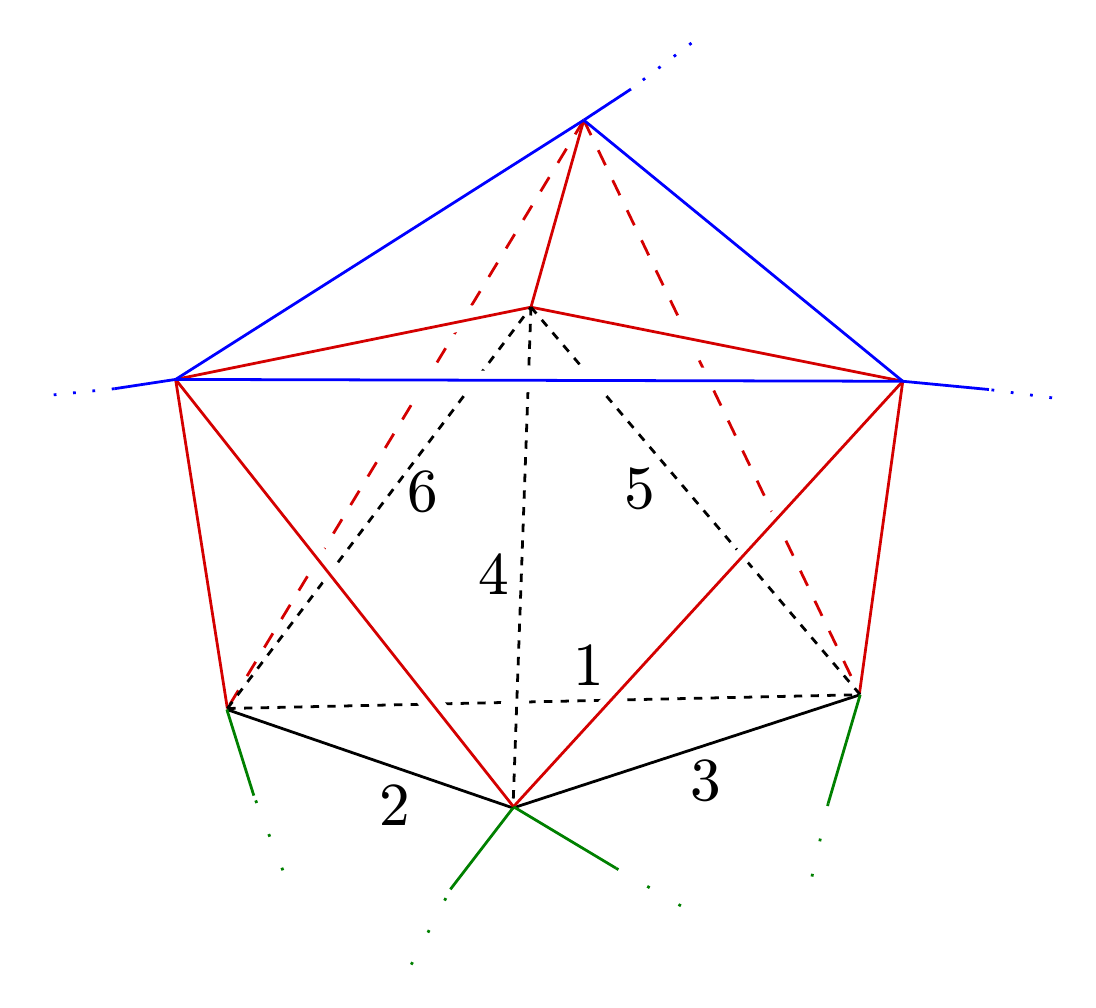}} \right] \quad &=& \quad NZ \left[ \raisebox{-0.1\textheight}{\includegraphics[height=0.20\textheight]{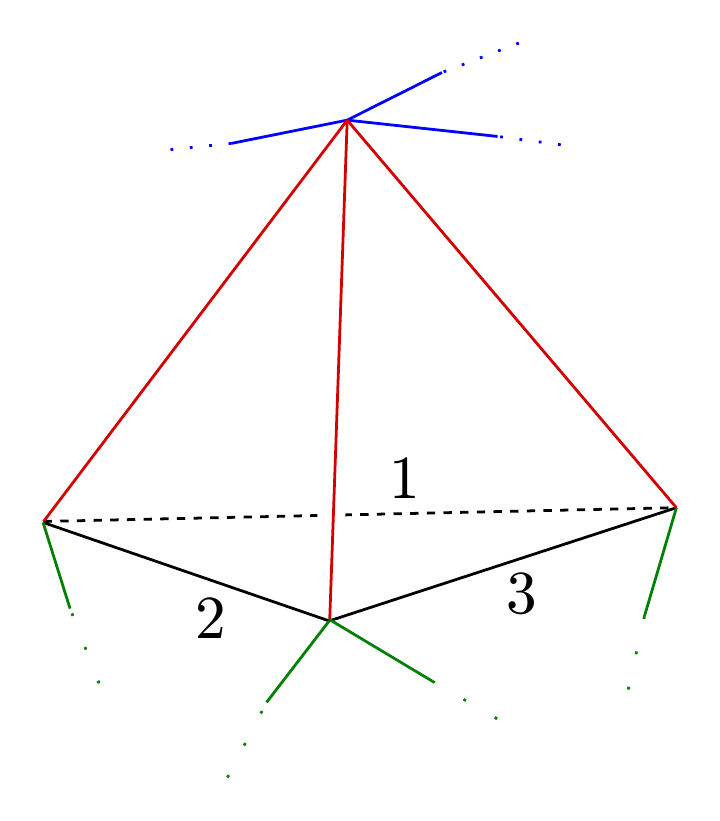}} \right]. \qquad\eea
These calculations are performed in Appendix \ref{sec:nplanarmove}.

As the genus-preserving trivalent graph moves are dual to the Alexander moves, it is possible to relate any two trivalent 
graphs of the same genus by a series of the 2-2, 3-1, and 1-3 moves. Therefore,
for any two graphs, there exists a sequence of Belyi-triangulated complexes ${\cal M}[\G_i]$ that differ from their
predecessor by a 3-1, 1-3, or a 2-2 move, and hence all have the same partition function. 
Finally, we note that the Ponzano-Regge partition function of a complex depends only on the topology of the manifold
and the labelled triangulation on its boundary, and hence that any two Belyi 3-complexes of the same ribbon graph have
the same partition function. We can therefore conclude that the partition
functions of any two Belyi 3-complexes generated from graphs of the same genus are equal.

As a consequence of this, we now need only calculate the partition function of a single Belyi 3-complex of each genus
to show that any Belyi 3-complex of genus $g$ has the evaluation $N^{2-2g}$.
We can do this by induction on $g$. 
First, we take a handlebody Belyi 3-complex of a ribbon graph $\G$ of genus $g$ graph 
with $V$ vertices, and demand that one of the triangles of the inner Belyi triangulation is contractible
(or equivalently that one of the vertices of the ribbon graph lies at the intersection of three distinct faces
of the graph). As an inductive hypothesis, we assume that its partition function evaluates to $N^{2-2g}$.

\begin{figure}[h]
\centering
\subfloat[]{\label{fig:cuta}\includegraphics[width=0.39\textwidth]{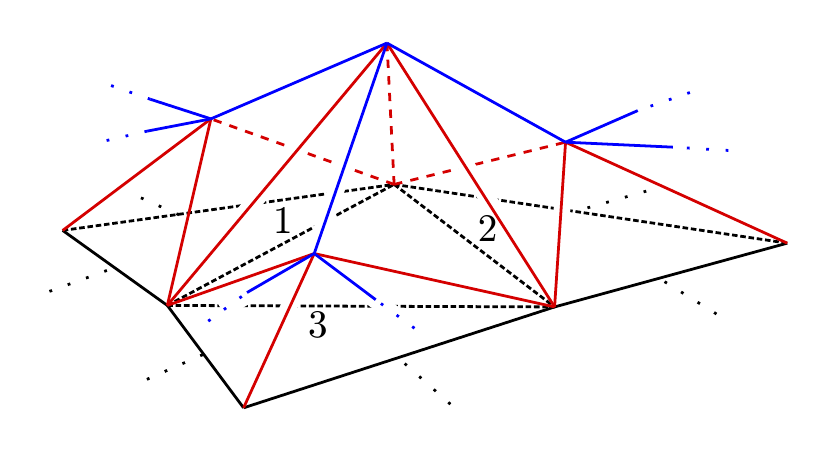}}\raisebox{0.05\textheight}{\quad \includegraphics[width=0.18\textwidth]{arrow2}}
\subfloat[]{\label{fig:cutb}\includegraphics[width=0.39\textwidth]{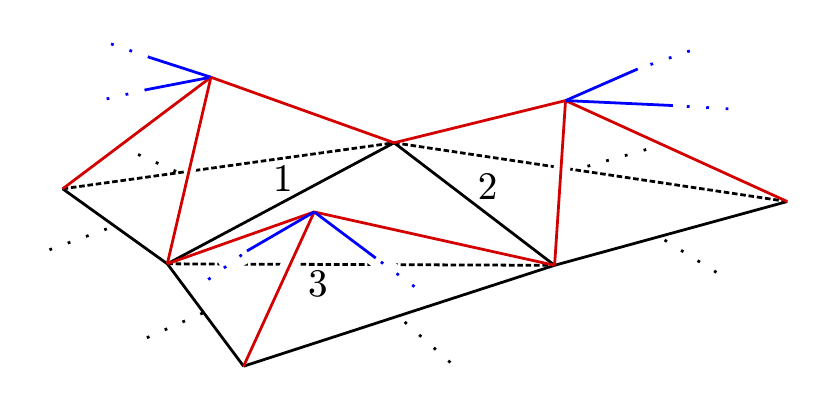}}
\caption{The complex near the contractible triangle, and the complex after the removal of the four tetrahedra.}\label{fig:cut}
\end{figure}

Consider the part of the complex near the contractible triangle, which we label $(l_1, l_2, l_3)$.
The complex must consist of a tetrahedron containing the triangle $(l_1, l_2, l_3)$ and three red edges, and 
also three tetrahedra with blue and red edges containing $l_1$, $l_2$, and $l_3$ respectively. These tetrahedra are shown in 
Figure \ref{fig:cuta}. We can separate the partition function of the complex
$N^V Z$ into the partition function of these four tetrahedra and the partition function of the rest of the complex by denoting
the partition function of the rest of the complex as $\bar{Z}(l_1,l_2,l_3)$, so we can write its partition function as
\bea \label{eq:ih} N^V Z  = \frac{(-)^{2j}}{(2j+1)^3}\sum_{l_1l_2l_3}(2l_1+1)(2l_2+1)(2l_3+1) \wsjj{l_1}{l_2}{l_3}\bar{Z}(l_1,l_2,l_3) = N^{2-2g}. \eea
Note that $\bar{Z}(l_1,l_2,l_3)$ is the partition function of a complex formed from the original complex by the removal of the four tetrahedra, and has $l_1$, $l_2$, and $l_3$ on its boundary. This reduced complex is shown in Figure \ref{fig:cutb}.

We can glue the solid torus shown in Figure \ref{fig:st3} on to the black triangle of Figure \ref{fig:cutb} by identifying the two 
triangles labelled $(l_1,l_2,l_3)$. Since both triangles are contractible, the new complex is topologically the connected sum 
of a genus $g$ handlebody and a solid torus, and hence is a labelled triangulation of a genus $(g+1)$ handlebody.
At this stage, its boundary is neither an outer nor inner Belyi triangulation, as its boundary contains five triangles with black edges in addition to triangles with blue and red edges.
However, we can make the complex a Belyi 3-complex by gluing on a collection of red and blue-labelled tetrahedra on to these black 
edges. As before, to each of the black triangles on the boundary we glue on a tetrahedron
with three red edges (labelled $j$) and associated $6j$
\bea \wsj{l_i}{l_j}{l_k}{j}{j}{j}, \eea
and to each of the black edges on the boundary we glue on a tetrahedra with a blue edge (labelled zero), which
has the associated $6j$
\bea \wsj{0}{j}{j}{l_i}{j}{j} = \frac{(-)^{2j}}{(2j+1)}. \eea

In total, we will glue on five tetrahedra that each possess three red edges, 
and nine tetrahedra that each possess a blue edge. The boundary triangulation of the glued complex
will be the outer Belyi triangulation of a genus $(g+1)$ ribbon graph $\G'$ with four more vertices than the previous
ribbon graph $\G$. As before, we weight the partition function of a Belyi complex associated to a ribbon graph with
$V'=V+4$ vertices by a factor of $N^{V'}$. We can therefore see that its partition function is similar to the 
expression given in \refb{eq:solidt} for the triangulation of a solid torus, but containing a factor of $\bar{Z}$, one fewer $6j$, 
and a different power of the prefactor $N$. The new partition function is
\begin{multline} \label{eq:handlebody} N^{V'} Z' := N^{V+4} Z' = \frac{(-)^{2j}}{(2j+1)^5}\sum_{a,b,l_i} (-)^{2a+2b}(2a+1)(2b+1)\prod_{i=1,\ldots, 9}(2l_i+1)\times \\ 
\times \wsj{l_4}{l_7}{a}{l_4}{l_7}{l_7}  \wsj{l_4}{l_7}{a}{b}{l_5}{l_2}  \wsj{l_4}{l_7}{a}{l_5}{b}{l_9} \wsj{l_2}{l_7}{b}{l_6}{l_3}{l_1}  \wsj{l_4}{l_9}{b}{l_3}{l_6}{l_8} \times \\ \times 
\wsjj{l_2}{l_4}{l_5}\wsjj{l_5}{l_7}{l_9}\wsjj{l_3}{l_8}{l_9}\wsjj{l_4}{l_6}{l_8}\wsjj{l_1}{l_6}{l_7} \bar{Z}(l_1,l_2,l_3).
\end{multline}

Since the partition function $\bar{Z}(l_1,l_2,l_3)$ has no dependence on the labels $l_4, l_5, \ldots, l_9, a, b$, the calculation proceeds identically to in the previous example. Hence, we can follow the previous steps \refb{eq:solidt} - \refb{eq:shortcut} to state that
\bea
N^{V'} Z' = \frac{(-)^{2j}}{(2j+1)^5}\sum_{l_1l_2l_3}(2l_1+1)(2l_2+1)(2l_3+1)\wsjj{l_1}{l_2}{l_3}\bar{Z}(l_1,l_2,l_3).
\eea
Finally, using the inductive hypothesis \refb{eq:ih}, we can state that
\bea N^{V'}Z' = \frac{1}{N^2}N^{2-2g} = N^{2-2(g+1)}. \eea

This is sufficient to prove the conjecture \refb{eq:npcorres} by induction. Since we have shown in Section \ref{sec:prnp1} that 
a Belyi 3-complex of genus one with a contractible triangle in its inner Belyi triangulation
has the partition function evaluation $N^0=1$, we can therefore deduce that any Belyi 3-complex of
any ribbon graph of genus $g$ with $V$ vertices has the partition function
\bea N^V Z[\mG] = N^{2-2g} = \rG. \eea

\section{Discussion} \label{sec:discussions} 

We now discuss some further aspects of the link between Hermitian matrix models and
state sum models that we have uncovered, with a view to future research. 
We start by considering the known realisation of the Ponzano-Regge model in terms of Chern-Simons theory
with $ISO(3)$ gauge group, in the limit of large $k$, to express our result as a connection 
between the Hermitian matrix model and Chern-Simons.
A related line of inquiry would start by using the quantum group generalisation of the $\lsu$ algebra
to generalise the $3j$ coupling coefficients that appear in our formulation of the Hermitian matrix 
model. Finally, we consider the non-Hermitian complex matrix model, whose correlator contributions 
are described by bipartite graphs. The same methods as described above can be applied
to these graphs to generate topological state sums, so this suggests a  3D interpretation 
of the correlators of the half-BPS operators of ${\cal N}=4$ super Yang-Mills in 4D.

\subsection{Euclidean gravity and $ISO(3)$ Chern-Simons theory} 

The known links between Chern-Simons theory and the Turaev-Viro model
can be used to connect  our Belyi 3-complex construction to Chern-Simons theory. 
In this section we briefly review these links between Chern-Simons theories, 
quantum gravity in three dimensions and the Turaev-Viro model. More detailed reviews can be found in \cite{freidelkrasnov, carlipreview}.

 Quantum gravity in three dimensions was reformulated as a  Chern-Simons quantum field theory in 
 the  eighties \cite{witten88jones, witten88soluble}. In the frame formalism,
the $SU(2) $ one-forms $e^a$ and $\omega^a$ give a Euclidean metric and spin connection on a
manifold ${\cal M}$, and the Einstein-Hilbert action with cosmological constant $\Lambda$ takes the form
\bea \label{eq:eh} S_{EH} = 2\int_{\cal M} \left[ e^a \wedge( d\omega_a + \frac{1}{2}\epsilon_{abc}\omega^b\wedge\omega^c) + \frac{\Lambda}{6}\epsilon_{abc}e^a\wedge e^b\wedge e^c \right]. \eea
It was shown that this action is equivalent to a Chern-Simons theory with gauge group $G$, 
given by the action
\bea S_{CS}[A] = \frac{k}{4\pi}\int_{\cal M} \tr\left[ A\wedge dA + \frac{2}{3}A\wedge A\wedge A \right], \eea
where the gauge group $G$ is $SO(4) \sim SU(2)\times SU(2)$ for positive cosmological constant $\Lambda$, or $ISO(3)$ when 
$\Lambda$ is zero.
It was also shown that the action \refb{eq:eh} can be rewritten as the difference of two Chern-Simons 
actions with gauge group $SU(2) $,
\bea S_{EH} = S_{CS}[A_+] - S_{CS}[A_-], \eea
where
\bea  A^a_{\pm} = \omega^a \pm \sqrt{\Lambda}e^a, \qquad \Lambda = \left(\frac{4\pi}{k}\right)^2. \eea

We can therefore see that 3D gravity with $\Lambda>0$ corresponds to Chern-Simons theory with gauge group $SO(4)$ 
and finite level $k$, and that 3D gravity with $\Lambda=0$ corresponds to Chern-Simons with gauge group $ISO(3)$ and $k\to\infty$.
After path integration, the partition function $Z_{EH}$ associated with the Einstein-Hilbert action with positive cosmological constant
and the partition functions of $ISO(3)$ and $SU(2)$ Chern-Simons theory satisfy the relation
\bea Z_{EH} = Z_{CS}^{SO(4)} = |Z_{CS}^{SU(2)} |^2. \eea
and in the large $k$ limit, this becomes 
\bea Z_{EH}^{\Lambda=0} = Z_{CS}^{ISO(3)} = \lim_{k\to\infty}|Z_{CS}^{SU(2)} |^2 . \eea

Later, the Reshetikhin-Turaev topological invariant $Z_{RT}^G $ for a gauge group $G$ was introduced in \cite{reshturaev}. 
This is an invariant of a closed 3-manifold, and is also dependent on a quantum deformation parameter $k$.  It is  equivalent to the path integral
of Chern-Simons theory, which can be expressed as  $Z_{RT}^{G  } = Z_{CS}^{ G } $.

In \cite{turaev95}, the Turaev-Viro partition function was shown to be equal to 
the absolute square of the Reshetikhin-Turaev invariant. Specializing to $G=SU(2)$, 
\bea \label{eq:tvrt} 
Z_{TV}^{ SU(2) }  = |Z_{RT}^{ SU(2)} |^2. \eea
This can be interpreted as meaning that the Turaev-Viro partition function defines 
quantised gravity in three dimensions given by the action \refb{eq:eh}.
The equivalence of the Turaev-Viro model with 3D quantum gravity was demonstrated 
explicitly in \cite{ooguri91} by showing that the Hilbert space of
states of the Turaev-Viro model in the $k\to\infty$ limit, which is the Ponzano-Regge model, is isomorphic to the Hilbert space of
states of the theory given by the action \refb{eq:eh} with $\Lambda$ set to zero. The space of states in the Ponzano-Regge
model is spanned by the labellings of a triangulation of a closed 2D surface $\Sigma$, and the Ponzano-Regge partition 
function for manifolds with two $\Sigma$ boundary components extends to an inner product on this space. 
The phase space of the $ISO(3)$ Chern-Simons theory is the moduli space of flat $ISO(3)$ connections on $\Sigma$.
This is equal to the cotangent bundle of the moduli space of flat $SU(2)$ connections on $\Sigma$.
This leads to a Hilbert space of the $ISO(3)$  Chern-Simons theory  spanned by the $SU(2)$  labellings of  trivalent graphs of the surface $\Sigma$. An  inner product on this space is given by inserting the labelled trivalent graphs in a 3-manifold 
as Wilson operators in the path integral.  

We have thus arrived at an interpretation of the Ponzano-Regge partition 
functions of our Belyi 3-complexes as the partition functions for $ISO(3)$ Chern Simons theory 
with $SU(2)$-labelled Wilson lines on the boundary.  This provides the link between the Gaussian matrix model 
and $ISO(3)$ Chern-Simons. 

There is also an interpretation of the Gaussian matrix model in terms of $SU(2)$ Chern Simons theory. 
The correspondence between the Reshetikhin-Turaev and Turaev-Viro partition functions of closed manifolds can 
be extended to manifolds with boundary.
In \cite{freidelkrasnov}, a generalisation of \refb{eq:tvrt} is given for a handlebody $H$, whose labelled boundary triangulation
is dual to a labelled trivalent graph $\Gamma$. It is stated that 

\bea \label{eq:tvbdy} 
Z^{ SU(2)} _{TV}[H, \Gamma] = Z_{RT}^{SU(2)} [H\cup -H, \Gamma],
\eea
i.e. the Turaev-Viro partition function of the handlebody is equal to the Reshetikhin-Turaev invariant of the closed manifold $H\cup-H$,
which is two copies of the handlebody trivially glued on their boundary, and with the trivalent graph embedded in the interior. 
From the point of view of the $SU(2)$ Chern-Simons theory, this labelled trivalent graph is a Wilson operator embedded in the manifold
$H\cup-H$. Hence, we can write 
\bea 
Z^{ISO(3)}_{CS} [ H, \Gamma]  =  Z_{CS }^{SU(2)} [H\cup -H, \Gamma].
\eea 

In our 3D interpretation of the Hermitian matrix model, 
 Belyi 3-complexes were introduced as triangulations of  handlebodies
with  boundary triangulations generated from ribbon graphs.  
From the point of view of the Chern-Simons theory with $SU(2)$ gauge group, this boundary 
triangulation is a network of Wilson lines, each carrying an $\lsu$ representation
label  $j$ or 0, and embedded in a closed 3-manifold.  We therefore also  have a link between the Hermitian matrix model correlators 
 and  $SU(2)$  Chern Simons theory on closed 3-manifolds 
 with  embedded Wilson lines.   
 Chern-Simons terms are known to arise in the worldvolumes of M-theory membranes \cite{ABJM}.
 Our  new connections between the Hermitian matrix model and Chern Simons theories
 will hopefully shed light on the embedding of the  holographic hierachy of matrix model, Belyi maps, and Ponzano-Regge 
 into  M-theory.

There are other  known links between theories in zero, two and three dimensions that may 
be usefully compared with  our results. For example, Chern-Simons theory with gauge group $U(N)$ 
is related via a matrix model to 2D Yang-Mills theory with $U(N)$ gauge group in \cite{szabotierz}.
This differs from our correspondence, as we have related  $U(N)$-invariant observables in the matrix model 
to $ISO(3)$  (or $SU(2)$)  Wilson operators in 3D. Chern-Simons theory has also been related to a matrix
model in \cite{marino}. 
 Another link  between two and three dimensions was observed 
 in connection with the perturbative expansion of the Hermitian matrix model   in \cite{RRW}. The refined counting of  ribbon graphs
  has a three-dimensional interpretation as a topological field theory with permutation groups as 
  gauge groups.  Here, our approach using the fuzzy sphere has focused on individual ribbon 
  graphs and has lead to other 3D topological field theories with  continuous  groups ($ISO(3)$ and $SU(2)$)  as 
  gauge groups. Going beyond individual graphs, constructing the generating function of the corresponding 
  boundary triangulations using a local boundary term in  the Chern Simons action  is an interesting problem. 
   A more complete picture  of these diverse  0-2-3 dimensional correspondences would also be highly desirable.

\subsection{A $q$-deformation of the Hermitian matrix model correlators }

In our correspondence between Belyi triangulations and the Ponzano-Regge model, we used the Turaev-Viro model
to regulate the divergent Ponzano-Regge state sums. When considering triangulations of
manifolds with interior vertices, we deformed the $6j$ symbols representing the coupling of representations of $\lsu$
to the quantum $6j$ symbols representing the coupling of representations of $\qsu$, where the quantum parameter $q$ is a root of unity. 
After summing out the spin labels that were not
bounded from above by the parameter $2j$, we took the classical limit $q\to1$ to obtain a finite sum over classical $6j$s, and hence
recovered the Ponzano-Regge state sum. The Turaev-Viro model has thus been a  mathematical tool to regulate 
the divergent Ponzano-Regge state sum models. 

It is natural to ask if the matrix model correspondence can be extended beyond 
Ponzano-Regge to Turaev-Viro beyond $q=1$.  
For the Belyi complexes of genus $g$ constructed in Sections  \ref{sec:HMMPR} and \ref{sec:PRNP}, 
 we have  checked
that the weighted Turaev-Viro partition function evaluates to $[N]^{2-2g}$,
 where the noncommutativity parameter $N=2j+1$ has been replaced by a quantum integer.
A matrix model analogue would require an appropriate  $q$-deformation of the 
Gaussian matrix model, which would be defined so as  to  produce  sums over  quantum $6j$s of the type arising in Turaev-Viro. 
 One possible prescription would be to use the $q$-deformed fuzzy
spherical harmonics described in \cite{qfuzzy}, but it is not clear that this will lead to an evaluation for ribbon graphs of genus $g$
of  $[N]^{2-2g}$. A successful outcome would lead to a matrix model equivalent of finite $k$ Chern-Simons theory with 
$SO(4)\sim SU(2)\times SU(2)$ gauge group.

\subsection{Half-BPS operators in the complex matrix model } 

In this paper we have developed a correspondence between the Hermitian matrix model and the Ponzano-Regge state sum model.
We could have instead considered the more general complex matrix model with correlator functions 
\bea \cor{ \tr (\sigma Z^{\otimes d}) \tr (\tau Z^{\dagger \otimes d})} = \int DZ\ DZ^\dagger\  e^{- N tr Z Z^\dagger}  \tr (\sigma Z^{\otimes d}) \tr (\tau Z^{\dagger \otimes d}), \eea
where the integration is now performed over all $N\times N$ complex matrices, and the permutations $\sigma, \tau \in S_d$ act on the indices of the matrices. This generalisation doubles the integrated degrees of freedom from $N^2$ real variables to 2$N^2$ real variables. This model is relevant in ${\cal N}=4$ super Yang-Mills in 4D, as the matrix model correlators generate the half-BPS operators of a single complex scalar field transforming in the adjoint of U(N) \cite{gg, halfbps}. 

The expansion of a general complex matrix in terms of fuzzy spherical harmonics proceeds analogously to the Hermitian matrix model, but with a doubling of the degrees of freedom of the matrices. The variables $a_{lm}$ and $a_{lm}^*$ are independent in the complex matrix model, and their correlators are
\bea \cor{ a_{lm}\  a_{l'm'}^*} = \frac{1}{N^2}\delta_{l,l'}\delta_{m,m'},  \qquad \cor{ a_{lm}\  a_{l'm'}} = \cor{ a_{lm}^*\  a_{l'm'}^*} =0 \eea 
These relations, together with the separation of traces of $Z$ and $Z^\dagger$ in the correlator, will cause all the graphs generated from the correlators in the complex matrix model to be bipartite.
As we have not altered the fuzzy spherical harmonics algebra in switching from Hermitian to complex matrices, the $3j$ and $6j$ sums assigned to each graph remain unchanged. 
Thus, the generalisation to complex matrices results in a more constrained theory where only bipartite graphs appear. The 2D-3D correspondence from Belyi triangulations to the Ponzano-Regge model still holds for this subfamily of the graphs, so
this gives a 3D $ISO(3)$  Chern-Simons interpretation of half-BPS correlators of 4D SYM theory. It would be interesting to 
see  the path integal of the Chern-Simons theory arising more directly from the half-BPS sector.

\section{ Summary and outlook }\label{sec:summary} 

The Gaussian Hermitian matrix model has been a useful toy model for gauge-string duality
since the nineties \cite{BK,DS,GM}.
The earliest interpretations recognised it as a model of $c=-2$ matter 
coupled to Liouville theory, which suggested that the strings exist, in some formal sense, 
in a dual space-time of $-2$ dimensions. The link between correlators of the Gaussian model and Belyi maps was highlighted 
recently, and used to propose that there is a string interpretation with $S^2$ as target space \cite{dMR2010}. 
As a way to develop the idea of $S^2$ as a stringy target space, we considered the 
fuzzy sphere construction of matrix algebras, where the matrix becomes a quantum field on the fuzzy sphere
and the matrix integral becomes a path integral for a scalar field on the fuzzy sphere. 
The computation of correlators in the fuzzy sphere approach leads to a labelling of the 
ribbon graphs of the Hermitian matrix model with $\lsu$ spins, and a sum over weights 
determined by these spins. 
 
We focused on trivalent graphs which arise in the computation of correlators of $(\tr X^3)^V$. 
More general correlators can be expressed in terms of these, but the  case of cubic graphs 
allows the simplest statement of the key ideas here. The weights of the spin-labelled ribbon graphs 
arising from the multiplication of fuzzy spherical harmonics are given in terms of $3j$ and 
$6j$ symbols. We showed that the sums over the $3j$s can be performed to leave 
us with sums weighted by $6j$ symbols. We drew on ideas from spin networks,
which are closely related to the spin-labelled ribbon graphs.

To describe Belyi maps, it is natural to consider some triangulations of the surface supporting 
the ribbon graph that are generated from the graph. In addition to the standard
Belyi triangulation, we distinguished two other triangulations generated from trivalent ribbon graphs,
which we called the inner and outer Belyi triangulation.  

In Section \ref{sec:HMMPR}, we considered the Belyi triangulations of spheres associated to planar ribbon graphs,
and extended these to generate triangulations of the three-dimensional ball, with a view to describing a 
precise connection with the Ponzano-Regge model of 3D gravity.
The Ponzano-Regge model is a state sum model that assigns to a spin-labelled 3-complex a partition function,
which is a sum over $\lsu$ spins weighted by $6j$ symbols \cite{pr}. The triangulations of the ball are three-dimensional complexes of 
tetrahedra, which we called Belyi 3-complexes, whose edges are labelled by representations of $\lsu$.
Our construction guarantees that the boundary of the ball is equipped with the outer Belyi triangulation
of the sphere, while the inner Belyi triangulation also appears in the interior of the Belyi 3-complex. 
Computing the partition function of this Belyi 3-complex using identities and sums involving $6j$s
yields precisely the evaluation of the ribbon graph in the Hermitian matrix model. 
In addition, the Belyi 3-complex possesses a sub-complex, also a triangulation of the ball, 
whose boundary is the inner Belyi triangulation.
The partition function of this sub-complex corresponds to the evaluation 
of the $3j$ sum as a product of $6j$s. Hence the same sum over $6j$s is performed in
both the Ponzano-Regge and Hermitian matrix models.

The correspondence between the Ponzano-Regge and Hermitian matrix models continues to hold 
for higher genus ribbon graphs, where the triangulation of a higher genus surface is extended to a 
triangulation of a handlebody. 
The details of this construction are somewhat different, as there is no longer a vertex-by-vertex 
construction of the complex associated to the ribbon graph, as there is in the planar case.
In addition, the product of $6j$s corresponding to the evaluation of the $3j$ sum does not appear 
explicitly in the partition function of a handlebody. However, we exhibited the construction of a 
Belyi 3-complex associated to any ribbon graph, which contains the labelled outer Belyi triangulation
on its boundary and the inner Belyi triangulation, dual to the ribbon graph, in its interior. 
Finally, we proved that the Hermitian matrix model evaluation of any ribbon graph matches up
to the partition function of the handlebody Belyi 3-complex generated from the graph.

Several possible extensions of this work have been discussed in Section  \ref{sec:discussions}.
Given  the existing conjectures on  hierarchies of holographic dualities on M-theory,  
an embedding of the holographic hierarchy of the Gaussian Matrix model into M-theory 
would be particularly fascinating.

\vskip.1in 
{ \Large 
{ \centerline { \bf Acknowledgements }  } } 

\vskip.1in 
We are grateful  to Robert de Mello Koch for very helpful discussions and  collaboration in the early stages of this work. 
We thank David Berman, Chong-Sun Chu, Reza Doobary, Marcos Mari\~{n}o and Jurgis Pasukonis for discussions. 
We are grateful to the Institute for Advanced Study, Princeton for hospitality while part of this work was done. 
SR  is supported by STFC Grant ST/J000469/1, String theory, gauge theory, and duality. 

\vskip.5in 

\appendix
\section{Appendix}
\subsection{Properties of the Wigner $3j$ and $6j$ symbols}\label{sec:awig}

In this appendix we give a summary of the representation theory of $\lsu$ used in this paper,
taking definitions, expressions and identities from \cite{varshalovich, edmonds, rose}.
The finite dimensional irreducible representations of the Lie algebra $\lsu$ are labelled by a half-integer $j\in \{0,\frac{1}{2}, 1, \frac{3}{2}, \ldots\}$, where the dimension of each representation is $(2j+1)$. A basis of each representation is $\ket{jm}$, where $m$ is a half-integer in $\{-j,-j+1,\ldots,j-1,j\}$.
The tensor product of two representations of $\lsu$ can be decomposed into a direct sum of irreducible representations of $\lsu$. This coupling of representations gives rise to the Clebsch-Gordan coefficients to describe how states in the coupled representation can be expressed as sums in the uncoupled representation, and vice versa. The Wigner $3j$ symbols are more symmetric versions of these coefficients, and the Wigner $6j$ symbols are a generalisation to describe the coupling of three representations. 

\subsubsection{Clebsch-Gordan coefficients}
For two representations labelled by integer or half-integer $j_1$, $j_2$, a state in the product representation $\ket{JM}$ can be written
\bea
\ket{JM} = \sum_{m_1m_2}C_{j_1m_1\ j_2m_2}^{JM}\ket{j_1m_1}\otimes\ket{j_2m_2},
\eea
where the coefficients $C_{j_1m_1\ j_2m_2}^{JM}$ are the Clebsch-Gordan coefficients of this representation. 
From constraints such as unitarity and the orthonormality of both representations, and demanding the Clebsch-Gordan coefficients to be real, the inverse equation is
\bea
\ket{j_1m_1}\otimes\ket{j_2m_2} = \sum_{JM}C_{j_1m_1\ j_2m_2}^{JM}\ket{JM}.
\eea
The values of $J$ and $M$ in the coupled representation must satisfy 
\bea
M=m_1+m_2, \qquad J \in \{|j_1-j_2|, |j_1-j_2|+1, \ldots, j_1+j_2\},
\eea
so the Clebsch-Gordan coefficients are defined to be zero if these conditions are not upheld. Note that the conditions
\bea \label{eq:triangle}
j_3 \in \{|j_1-j_2|, |j_1-j_2|+1, \ldots, j_1+j_2\}, \qquad j_1+j_2+j_3 \mathrm{\ is\ an\ integer} 
\eea
are invariant under any permutation of the labels $\{j_1,j_2,j_3\}$. As the symmetry group on three objects is the triangle symmetry group $S_3$, we call these the triangle constraints on the labels $\{j_1, j_2, j_3\}$. We introduce the function $\Delta(j_1,j_2,j_3)$ that is defined to be equal to 1 if the labels $\{j_1,j_2,j_3\}$ satisfy the triangle constraints, and zero otherwise.
This means that $\Delta(j_1,j_2,j_3)=1$ if the Clebsch-Gordan coefficients $C_{j_1m_1\ j_2m_2}^{j_3m_3}$ are non-vanishing.

The coefficients also satisfy the two orthogonality conditions
\bea\label{eq:orthoga} \sum_{m_1m_2} C_{j_1m_1\ j_2m_2}^{JM}C_{j_1m_1\ j_2m_2}^{J'M'} = \delta_{J,J'}\delta_{M,M'}
\eea
and 
\bea\label{eq:orthogb} \sum_{J, M} C_{j_1m_1\ j_2m_2}^{JM}C_{j_1m_1'\ j_2m_2'}^{JM} = \delta_{m_1,m_1'}\delta_{m_2,m_2'}. \eea
An explicit expression for the Clebsch-Gordan coefficients appears in \cite{rose}, with which one can derive some relations of the coefficients under switching and reversing of their labels. For example, most relations can be derived from the three relations
\begin{eqnarray} \label{eq:rose}
C_{j_1m_1\ j_2m_2}^{j_3m_3} &=& (-)^{j_1+j_2-j_3} C_{j_1-m_1\ j_2-m_2}^{j_3-m_3} \ret
&=& (-)^{j_1+j_2-j_3} C_{j_2m_2\ j_1m_1}^{j_3m_3} \ret 
&=& (-)^{j_1-m_1} \left(\frac{2j_3+1}{2j_2+1}\right)^\frac{1}{2} C_{j_1m_1\  j_3-m_3}^{j_2-m_2}.  
\end{eqnarray}

\subsubsection{Wigner $3j$ symbols}
The Wigner $3j$ symbols are a more symmetric version of the Clebsch-Gordan coupling coefficients. We define
\bea \wigthreej{j_1}{j_2}{j_3}{m_1}{m_2}{m_3} = (-)^{j_1-j_2-m_3}(2j_3+1)^{-\frac{1}{2}}C_{j_1m_1\ j_2m_2}^{j_3-m_3}. \eea
This symbol must still obey the constraint \refb{eq:triangle} to take non-zero values, but now also satisfies $m_1+m_2+m_3=0$, and has a simplified symmetry under permutation of labels:
\bea \wigthreej{j_1}{j_2}{j_3}{m_1}{m_2}{m_3} =  \wigthreej{j_2}{j_3}{j_1}{m_2}{m_3}{m_1} =  \wigthreej{j_3}{j_1}{j_2}{m_3}{m_1}{m_2} \nn \eea
\bea = (-)^{j_1+j_2+j_3}\wigthreej{j_1}{j_3}{j_2}{m_1}{m_3}{m_2} = (-)^{j_1+j_2+j_3}\wigthreej{j_2}{j_1}{j_3}{m_2}{m_1}{m_3} = (-)^{j_1+j_2+j_3}\wigthreej{j_3}{j_2}{j_1}{m_3}{m_2}{m_1} \nn \eea
\bea\label{eq:wtjsym}
= (-)^{j_1+j_2+j_3}\wigthreej{j_1}{j_2}{j_3}{-m_1}{-m_2}{-m_3}.
\eea 
They now satisfy the orthogonality relations
\bea\label{eq:orthogc} \sum_{m_1m_2}(2j_3+1)\wigthreej{j_1}{j_2}{j_3}{m_1}{m_2}{m_3}\wigthreej{j_1}{j_2}{j_3'}{m_1}{m_2}{m_3'} = \delta_{j_3,j_3'}\delta_{m_3,m_3'}\Delta(j_1,j_2,j_3)  \eea
and
\bea\label{eq:orthogd} \sum_{j_3m_3}(2j_3+1)\wigthreej{j_1}{j_2}{j_3}{m_1}{m_2}{m_3}\wigthreej{j_1}{j_2}{j_3}{m_1'}{m_2'}{m_3} = \delta_{m_1,m_1'}\delta_{m_2,m_2'}\Delta(j_1,j_2,j_3) . \eea

\subsubsection{Wigner $6j$ symbols}
The Wigner $3j$s are associated with the direct sum of two representations with labels $j_1$ and $j_2$. We now consider the direct sum of three representations. We could first take the sum of the representations labelled $j_1$ and $j_2$ to make a representation labelled by $j_{12}$, and then couple on the $j_3$ to make a triply coupled representation $J$. Alternatively, we could start by adding the $j_2$ and $j_3$ to make a $j_{23}$ representation, and then couple on the $j_1$.
\begin{figure}[H]
\centering
\includegraphics[width=0.7\textwidth]{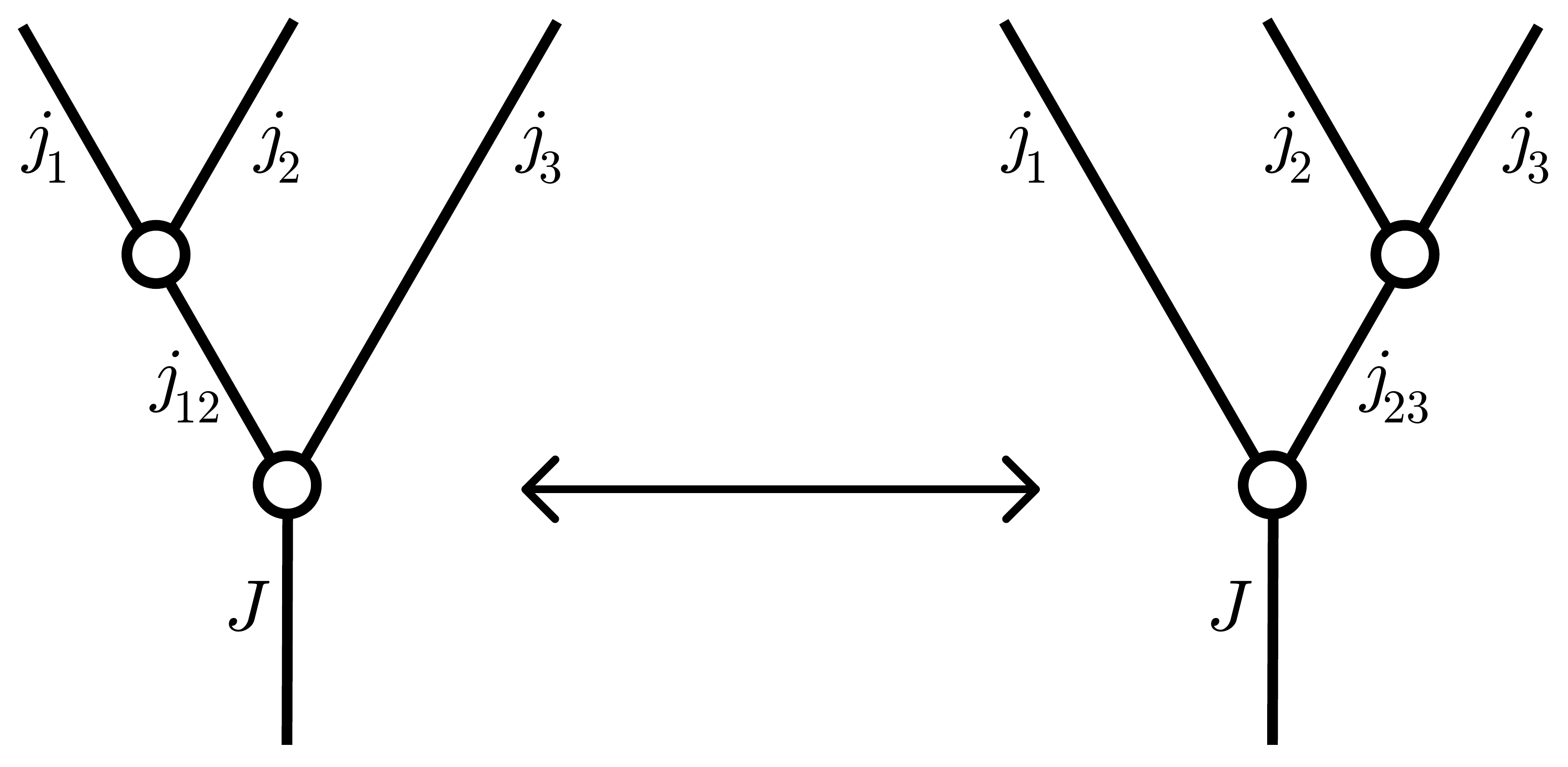}
\caption{Two methods of adding the $j_1$, $j_2$, and $j_3$ representations}\label{fig:coupling}
\end{figure}
In both cases, we arrive at a triply coupled representation labelled by $J$. Consider a state in the representation $\ket{JM}_{12}$ arising from the coupling of $j_{12}$ to $j_3$. This can be written
\bea \ket{JM}_{12} &=& \sum_{m_{12}m_3}C_{j_{12}m_{12}\ j_3m_3}^{JM}\ket{j_{12}m_{12}}\otimes\ket{j_3m_3} \ret
&=& \sum_{m_1m_2m_3m_{12}}C_{j_{12}m_{12}\ j_3m_3}^{JM}C_{j_1m_1\ j_2m_2}^{j_{12}m_{12}}\ket{j_1m_1}\otimes\ket{j_2m_2}\otimes\ket{j_3m_3}
\eea
The state corresponding to the coupling of $j_{23}$ to $j_1$ is similarly written
\bea \ket{JM}_{23} = \sum_{m_1m_2m_3m_{23}}C_{j_1m_1\ j_{23}m_{23}}^{JM}C_{j_2m_2\ j_3m_3}^{j_{23}m_{23}}\ket{j_1m_1}\otimes\ket{j_2m_2}\otimes\ket{j_3m_3}. \eea
We can consider the difference between the two methods of coupling by considering their inner product,
\bea \langle JM|_{12}JM\rangle_{23} = \sum_{\substack{m_1m_2m_3 \\ m_{12} m_{23}}} C_{j_{12}m_{12}\ j_3m_3}^{JM}C_{j_1m_1\ j_2m_2}^{j_{12}m_{12}}
C_{j_1m_1\ j_{23}m_{23}}^{JM}C_{j_2m_2\ j_3m_3}^{j_{23}m_{23}} \eea
The state label $M$ has not been summed over in this inner product, but it is shown in \cite{edmonds} that this expression is actually independent of $M$. Therefore, this inner product is a function purely of the six $j$ labels. We can
now define the Wigner $6j$ symbol as a symmetrised version of this inner product,
\bea \wsj{j_1}{j_2}{j_{12}}{j_3}{J}{j_{23}} = [(2j_{12}+1)(2j_{23}+1)]^{-\frac{1}{2}}(-)^{j_{12}+j_{23}} \langle JM|_{12}JM\rangle_{23} \nn \eea
\bea =  [(2j_{12}+1)(2j_{23}+1)]^{-\frac{1}{2}}(-)^{j_{12}+j_{23}} \sum_{\substack{m_1m_2m_3 \\ m_{12} m_{23}}} C_{j_1m_1\ j_2m_2}^{j_{12}m_{12}}C_{j_{12}m_{12}\ j_3m_3}^{JM}
C_{j_2m_2\ j_3m_3}^{j_{23}m_{23}}C_{j_1m_1\ j_{23}m_{23}}^{JM} \nn \eea
\begin{multline} = \sum_{\substack{m_1,\ m_2,\ m_3 \\ m_{12},\ m_{23},\ M}}(-)^{m_1+m_2+m_{12}+m_3+M+m_{23}}
\wigthreej{j_1}{j_2}{j_{12}}{m_1}{m_2}{m_{12}}\wigthreej{j_1}{J}{j_{23}}{-m_1}{M}{-m_{23}} \times \\ \times \wigthreej{j_{12}}{j_3}{J}{-m_{12}}{m_3}{-M}\wigthreej{j_2}{j_{23}}{j_3}{-m_2}{m_{23}}{-m_3} .\quad \end{multline}

The $6j$ symbol is either real or pure imaginary, depending on whether $j_{12}+j_{23}$ is integer or half-integer.
They are invariant under any permutation of their columns, 
\bea \label{eq:sym1} \wsj{j_1}{j_2}{j_3}{j_4}{j_5}{j_6} = \wsj{j_3}{j_1}{j_2}{j_6}{j_4}{j_5} = \wsj{j_2}{j_3}{j_1}{j_5}{j_6}{j_4}  \ret \ret
= \wsj{j_1}{j_3}{j_2}{j_4}{j_6}{j_5} = \wsj{j_3}{j_2}{j_1}{j_6}{j_5}{j_4} = \wsj{j_2}{j_1}{j_3}{j_5}{j_4}{j_6},
\eea
and also under the interchange of the upper and lower arguments in any two of their columns, 
\bea \label{eq:sym2} \wsj{j_1}{j_2}{j_3}{j_4}{j_5}{j_6} = \wsj{j_4}{j_5}{j_3}{j_1}{j_2}{j_6} = \wsj{j_4}{j_2}{j_6}{j_1}{j_5}{j_3} = \wsj{j_1}{j_5}{j_6}{j_4}{j_2}{j_3}.
\eea
As they are composed of Clebsch-Gordan coefficients which obey the triangular constraints, the above $6j$ is only non-zero when the sets of variables 
\bea \label{eq:triangles} \{j_1, j_2, j_3\}, \quad \{j_1, j_5, j_6\}, \quad \{j_2, j_4, j_6\}, \quad \{j_3, j_4, j_5\} \eea
all obey the constraint \refb{eq:triangle}. These triples are mapped into each other by the action of the symmetries \refb{eq:sym1} and \refb{eq:sym2}.

\begin{figure}[H]
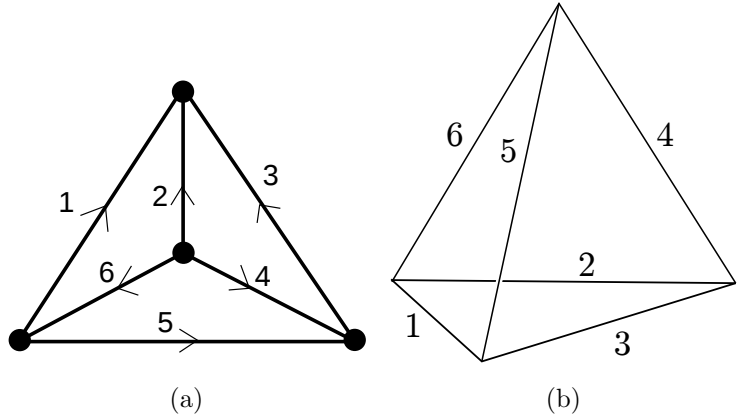

\centering
\subfloat[]{\label{fig:tet6ja}\includegraphics[width=0.3\textwidth]{6jno}}
\subfloat[]{\label{fig:tet6jb}\includegraphics[height=0.3\textwidth]{6jint3}}
\caption{The $6j$ symbol as a planar spin network and as a tetrahedron.}
\label{fig:tet6j}
\end{figure}

A $6j$ symbol can be viewed as a weight associated with a planar spin network (graph embedded on a plane, with edges marked by spins)
or as a tetrahedron.  The 24 symmetries of the $6j$ generated by \refb{eq:sym1} and \refb{eq:sym2} correspond to the symmetry group of the tetrahedron $S_4$. The spin network can be obtained by a 2D duality on the surface of  the tetrahedron. In Figure \ref{fig:tet6ja}, the
trivalent vertices represent the couplings of the triples of representations, and hence the triangle constraints are satisfied by the
labels meeting at a vertex. This interpretation of the spin network as a $6j$ is used extensively in Section \ref{sec:HMMandFS}
in describing the $3j$ sums from ribbon graphs. Alternatively, as in Figure \ref{fig:tet6jb}, the couplings of representations
can be interpreted as the triangles of the tetrahedron. This interpretation of the  $6j$ as a tetrahedron  is used in the Ponzano-Regge
model in Section \ref{sec:HMMPR}. 

There are many relations and identities regarding sums of $3j$s and $6j$s, many of which are listed in \cite{varshalovich}. Perhaps the most useful $6j$ relations in this paper are the orthogonality relation
\bea \label{eq:6jorthog} \sum_{j_3}(-)^{2j_3+2j_6}(2j_3+1)(2j_6+1) \wsj{j_1}{j_2}{j_3}{j_4}{j_5}{j_6}\wsj{j_1}{j_2}{j_3}{j_4}{j_5}{j_7} = \Delta(l_1,l_5,l_6) \Delta(l_2,l_4,l_6) \delta_{j_6,j_7},\qquad \eea
and the Biedenharn-Eliot identity, 
\bea \label{eq:bhid2} \sum_{j_a}(-)^{2j_a}(2j_a+1)\wsj{j_1}{j_5}{j_6}{j_a}{j_9}{j_8} \wsj{j_2}{j_4}{j_6}{j_a}{j_9}{j_7} \wsj{j_3}{j_4}{j_5}{j_a}{j_8}{j_7} = \wsj{j_1}{j_2}{j_3}{j_4}{j_5}{j_6} \wsj{j_1}{j_2}{j_3}{j_7}{j_8}{j_9}.\qquad  \eea
The Biedenharn-Eliot identity has an interpretation as the equivalence of the Ponzano-Regge state sum model under the 
3-2 Pachner Move on tetrahedra given in Figure \ref{fig:pach2}.

Also, these relations are useful in explicitly performing calculations of contributions to correlators on the fuzzy sphere.
\bea\label{eq:sixj0}
\sum_{l }(2l+1)\Delta(l,j,j) = (2j+1)^2,
\eea
\bea\label{eq:sixj1}
\sum_{l_1 }(2l_1+1)\wsjj{l_1}{l_2}{l_3}\wsjj{l_1}{l_2}{l_3} = \frac{(-)^{2j}}{(2j+1)}\Delta(l_2,j,j)\Delta(l_3,j,j),
\eea
\bea\label{eq:sixj2}
\sum_{l_1 l_2}(2l_1+1)(2l_2+1)\wsjj{l_1}{l_2}{l_3}\wsjj{l_1}{l_2}{l_3} = (-)^{2j}(2j+1)\Delta(l_3,j,j),
\eea
\bea\label{eq:sixj4}
\sum_{l_1 l_2}(-)^{l_1+l_2}(2l_1+1)(2l_2+1)\wsjj{l_1}{l_2}{l_3}\wsjj{l_1}{l_2}{l_3} = (2j+1)\delta_{l_3,0},
\eea
\bea\label{eq:sixj3}
\sum_{l_1}(-)^{3j+l_1}(2l_1+1)^\frac{3}{2}\wsjj{l_1}{l_1}{0} = (2j+1)^\frac{3}{2}. 
\eea

\subsection{The planar 2-2 and 3-1 moves in the Ponzano-Regge model}\label{sec:planarmove}

In this section we demonstrate that the planar moves on ribbon graphs do not change the value of the normalised Ponzano-Regge partition function on their associated constructed manifolds. We show this locally by considering the complexes associated with the relevant fragments of the graph. As in the main text, we freely interchange $N\equiv2j+1$ throughout. For the 2-2 move, we aim to show that

\bea N^2 Z\circ\m{tiny2-2a}  = N^2 Z\circ\mtwo{tiny2-2b}. \eea  

Noting that
\bea \m{tiny2-2a} \qquad = \raisebox{-0.25\textwidth}{\includegraphics[width=0.5\textwidth]{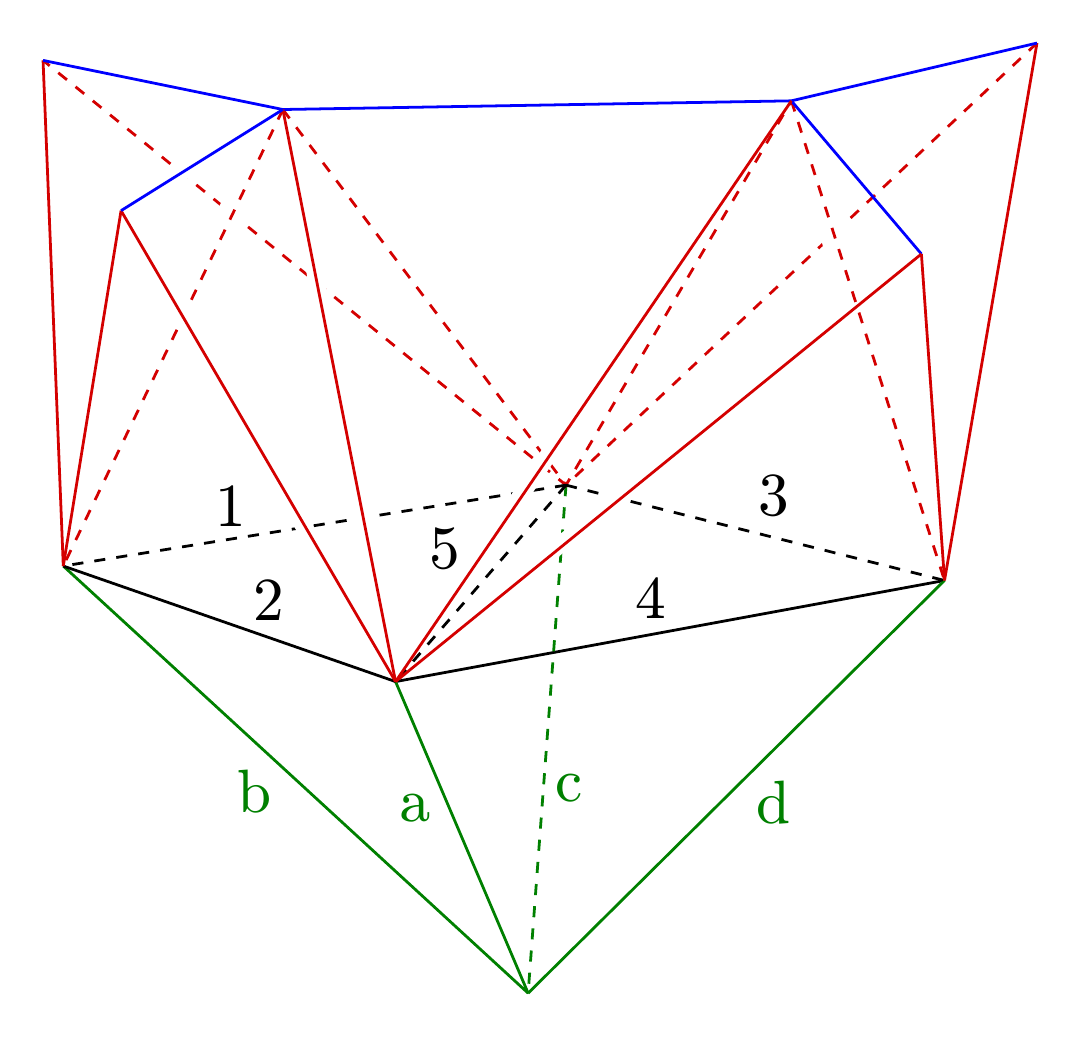}}, \eea
we calculate the partition function $Z$ to be
\begin{multline} N^2 Z\circ\m{tiny2-2a} = \frac{(-)^{2j}}{(2j+1)^3}\sum_{l_5}(2l_5+1)\wsj{l_1}{l_2}{l_5}{a}{c}{b}\wsj{l_3}{l_4}{l_5}{c}{a}{d}\times \\
\times \wsj{l_1}{l_2}{l_5}{j}{j}{j}\wsj{l_3}{l_4}{l_5}{j}{j}{j}, \quad \end{multline}
and employing the 2-3 move on the two tetrahedra with green lines, we get
\bea\hspace{-5mm} = \frac{(-)^{2j}}{(2j+1)^3}\sum_{l_5l_6}(2l_5+1)(-)^{2l_6}(2l_6+1)\wsj{l_1}{l_2}{l_5}{l_3}{l_4}{l_6}\wsj{l_2}{l_3}{l_6}{d}{b}{a}\wsj{l_1}{l_4}{l_6}{d}{b}{c} \wsj{l_1}{l_2}{l_5}{j}{j}{j}\wsj{l_3}{l_4}{l_5}{j}{j}{j}\nn \eea
\bea = N^2 Z \left[ \raisebox{-0.2\textwidth}{\includegraphics[width=0.45\textwidth]{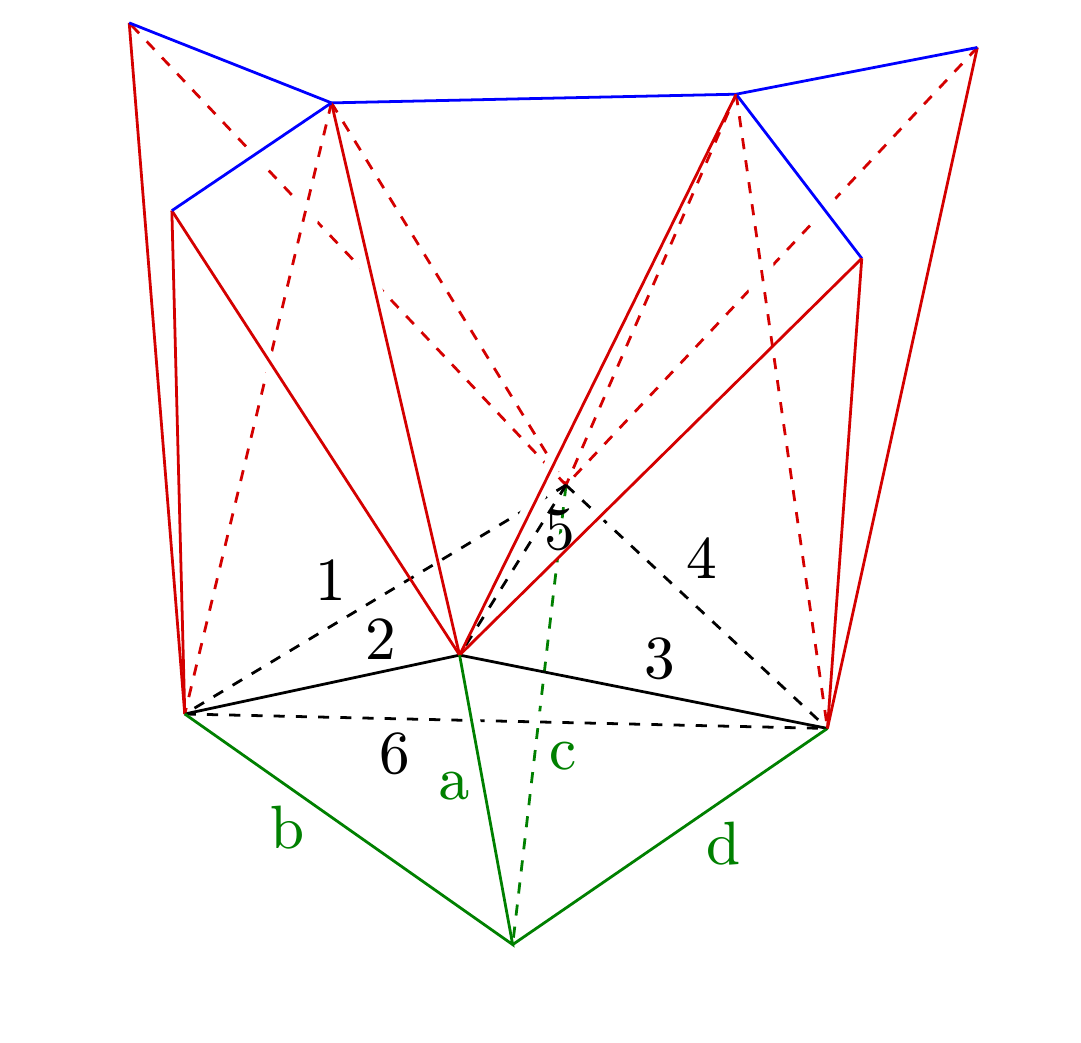}} \right]  \eea
Using the gluing rules of the partition function, we can split this complex into two parts and consider the sums separately.
\bea \hspace{-10mm}[\ldots ]= N^2\sum_{l_6}(-)^{2l_6}(2l_6+1) Z \left[ \raisebox{-0.13\textwidth}{\includegraphics[width=0.32\textwidth]{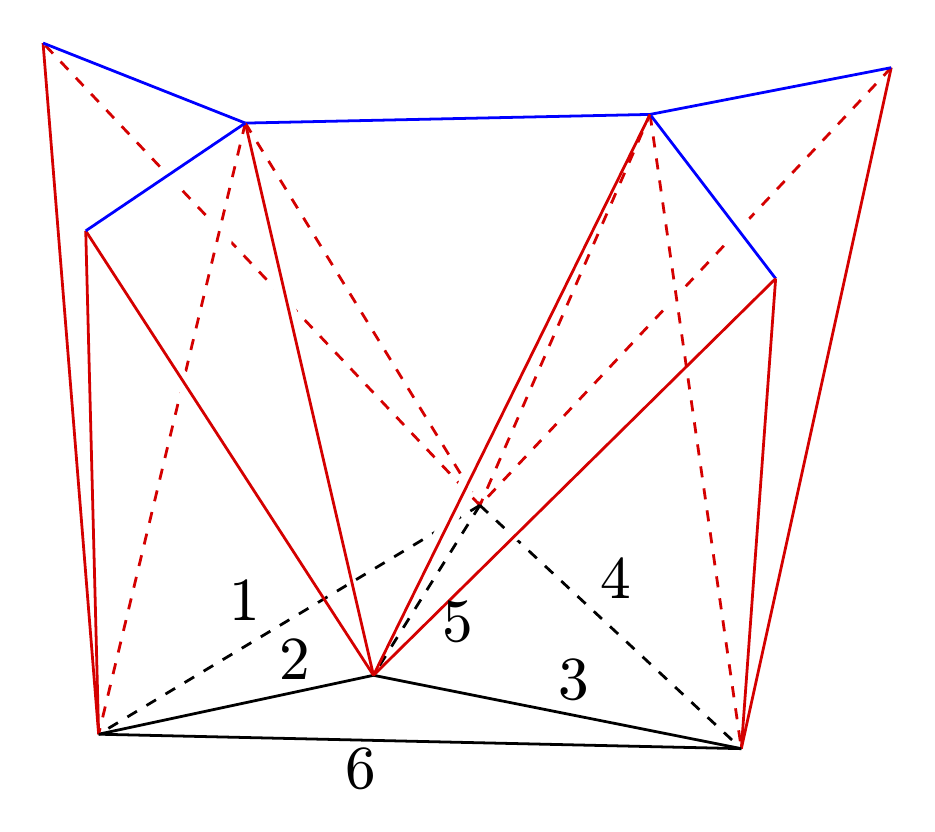}} \right]  Z \left[ \raisebox{-0.13\textwidth}{\includegraphics[width=0.32\textwidth]{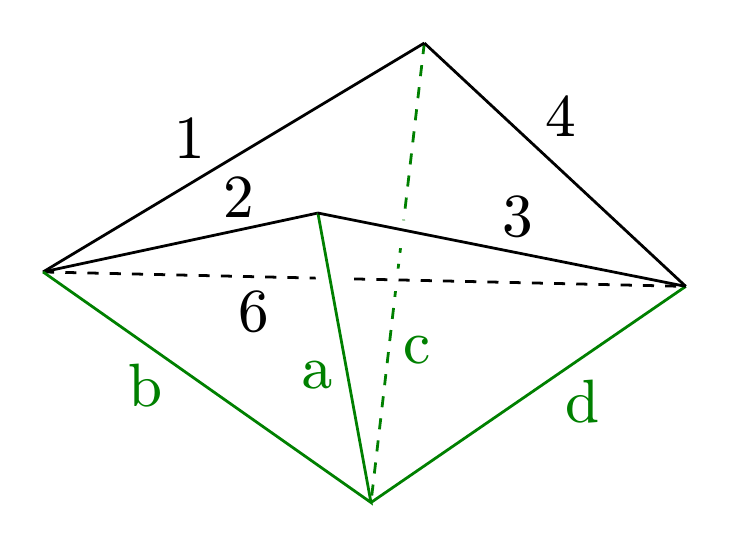}} \right] \qquad\eea
The first of these two factors can be manipulated using the 3-2 move on its constituent $6j$s,
\bea \hspace{-5mm} Z \left[ \raisebox{-0.15\textwidth}{\includegraphics[width=0.35\textwidth]{amfrag2a}} \right] = \frac{(-)^{2j}}{(2j+1)^5}\sum_{l_5}(2l_5+1)\wsj{l_1}{l_2}{l_5}{l_3}{l_4}{l_6} \wsj{l_1}{l_2}{l_5}{j}{j}{j}\wsj{l_3}{l_4}{l_5}{j}{j}{j}\nn \eea
\begin{multline} = \frac{(-)^{2j}}{(2j+1)^5} \quad Z\left[ \raisebox{-0.12\textwidth}{\includegraphics[width=0.25\textwidth]{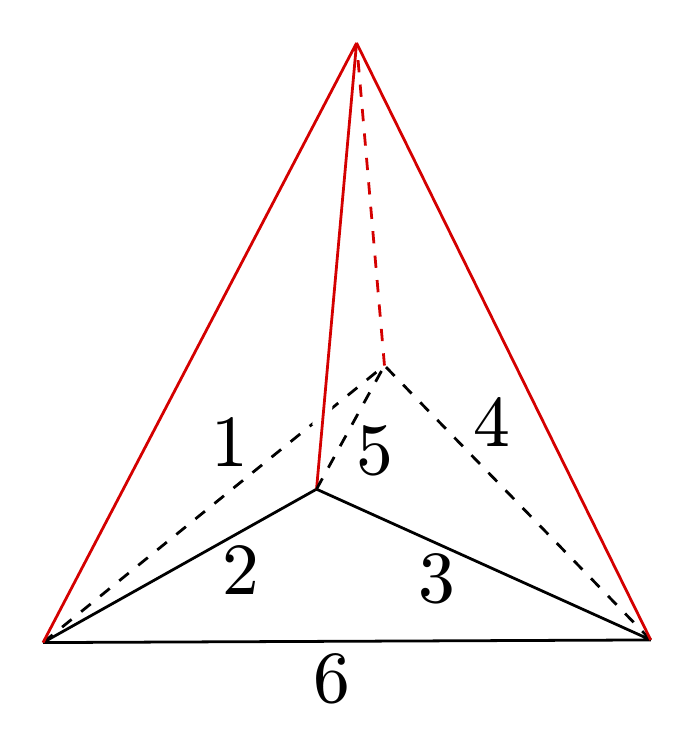}} \right]  = \frac{(-)^{2j}}{(2j+1)^5}\quad Z\left[ \raisebox{-0.10\textwidth}{\includegraphics[width=0.25\textwidth]{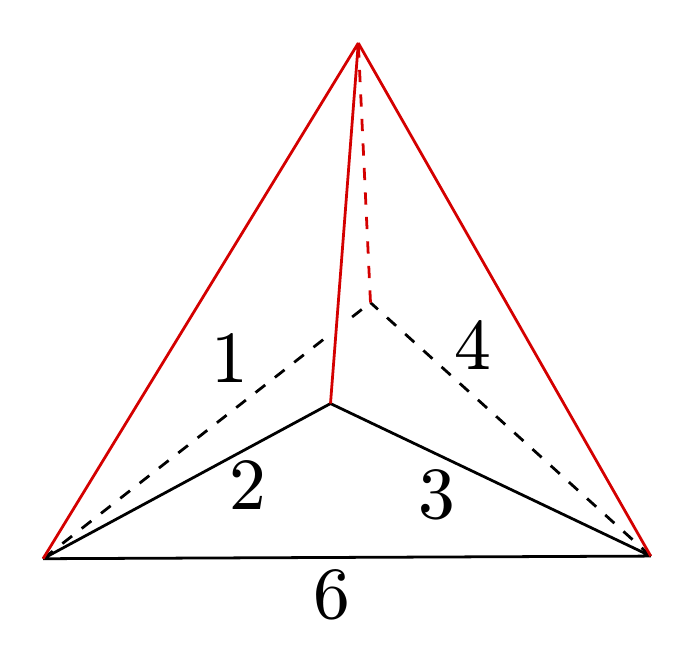}} \right] \\  \\
= \frac{(-)^{2j}}{(2j+1)^5} \wsj{l_2}{l_3}{l_6}{j}{j}{j}\wsj{l_1}{l_4}{l_6}{j}{j}{j}. \end{multline}
We thus deduce that
\begin{multline} \hspace{-5mm}N^2 Z\circ\m{tiny2-2a} =  \frac{(-)^{2j}}{(2j+1)^3} \sum_{l_6}(2l_6+1)\wsj{l_2}{l_3}{l_6}{j}{j}{j}\wsj{l_1}{l_4}{l_6}{j}{j}{j}\wsj{l_2}{l_3}{l_6}{d}{b}{a}\wsj{l_1}{l_4}{l_6}{d}{b}{c} \\ \\ = N^2 Z \left[ \raisebox{-0.12\textwidth}{\includegraphics[width=0.25\textwidth]{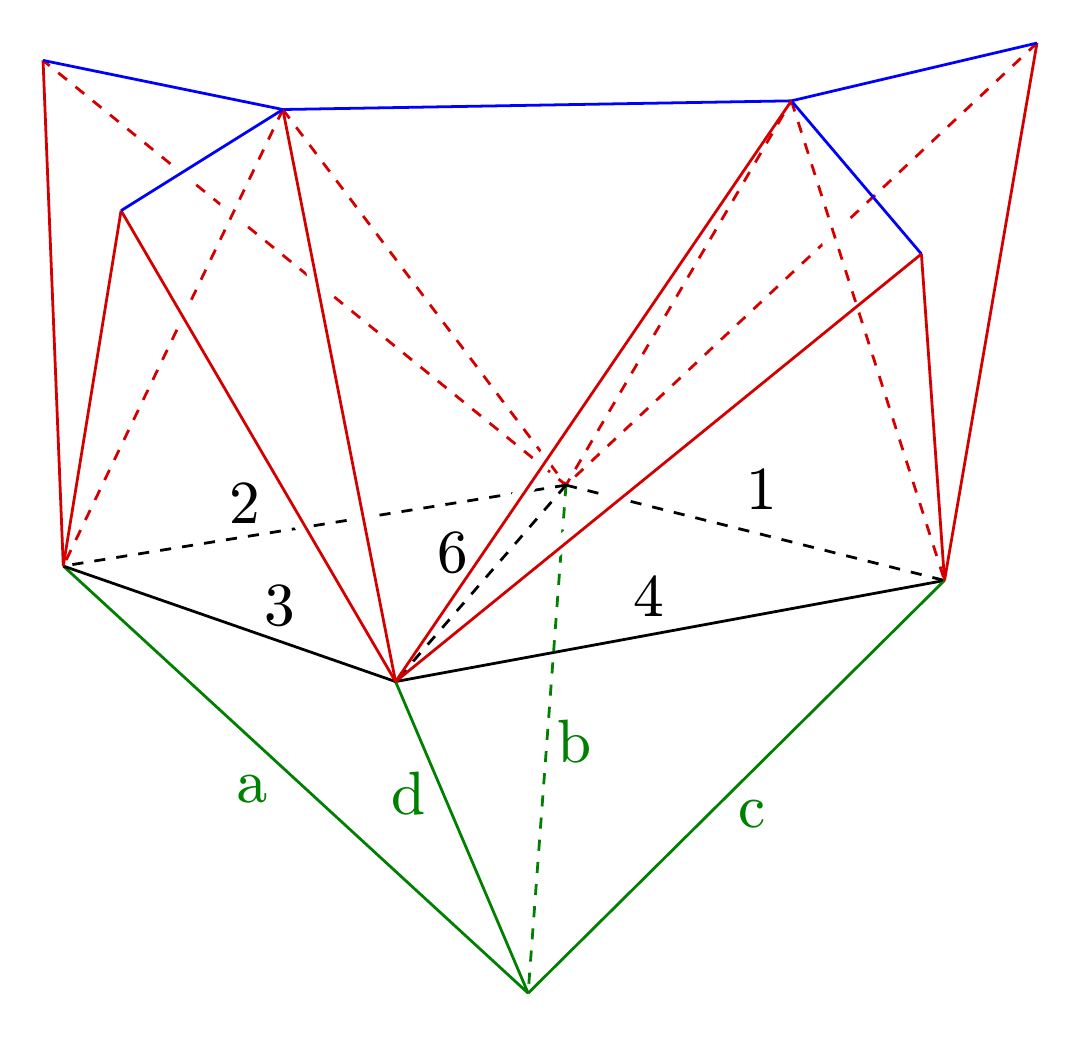}} \right]  = N^2 Z \circ\mtwo{tiny2-2b}, \end{multline}
as required.

We next consider the 3-1 move
\bea \ra{tiny3-1a}  = \ra{tiny3-1b}, \eea  
and attempt to deduce the corresponding move in terms of $Z\circ{\cal M}$. Starting from 
\bea \m{tiny3-1a} \qquad = \raisebox{-0.30\textwidth}{\includegraphics[width=0.6\textwidth]{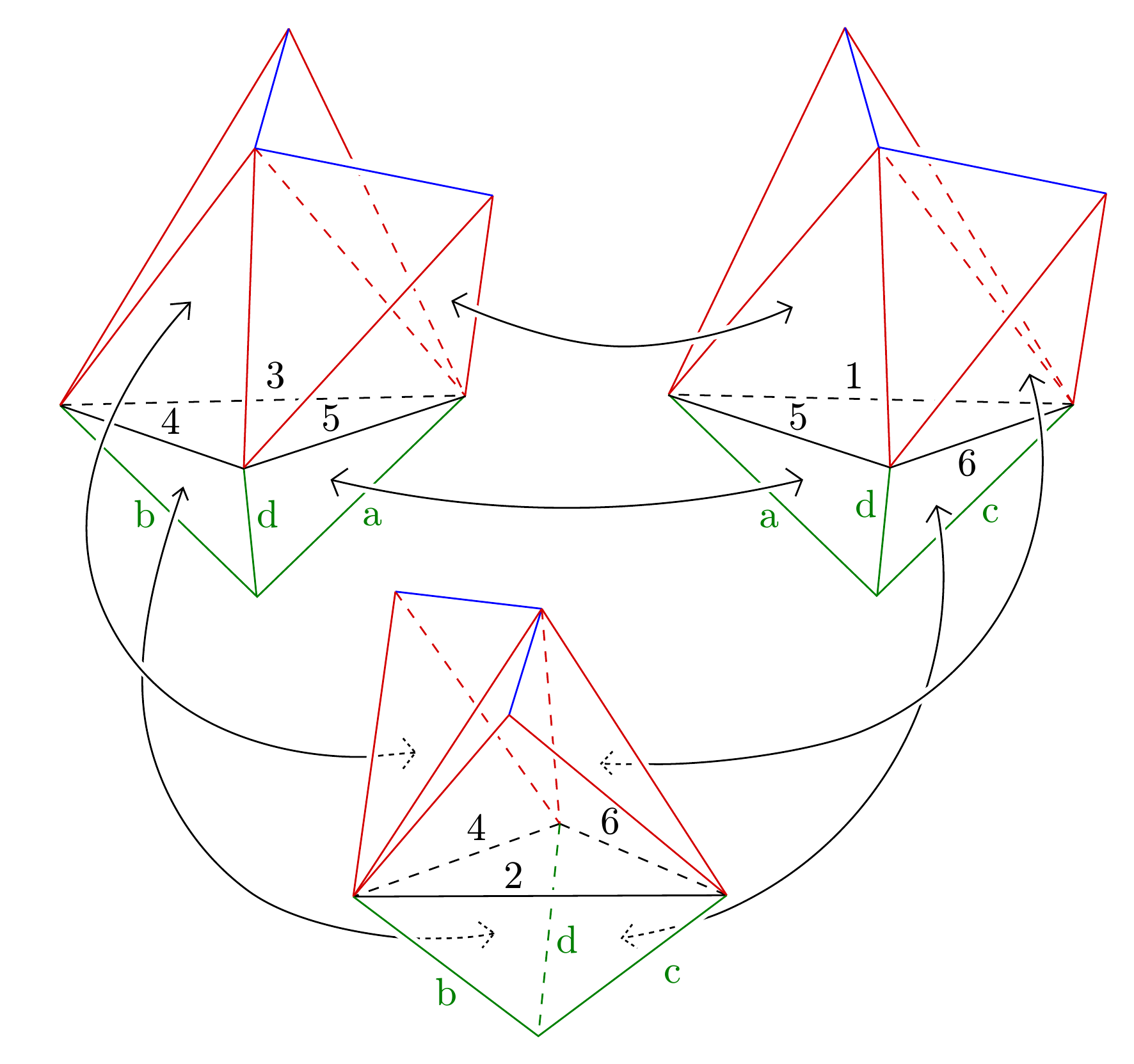}}\nn \eea
\bea = \raisebox{-0.20\textwidth}{\includegraphics[width=0.4\textwidth]{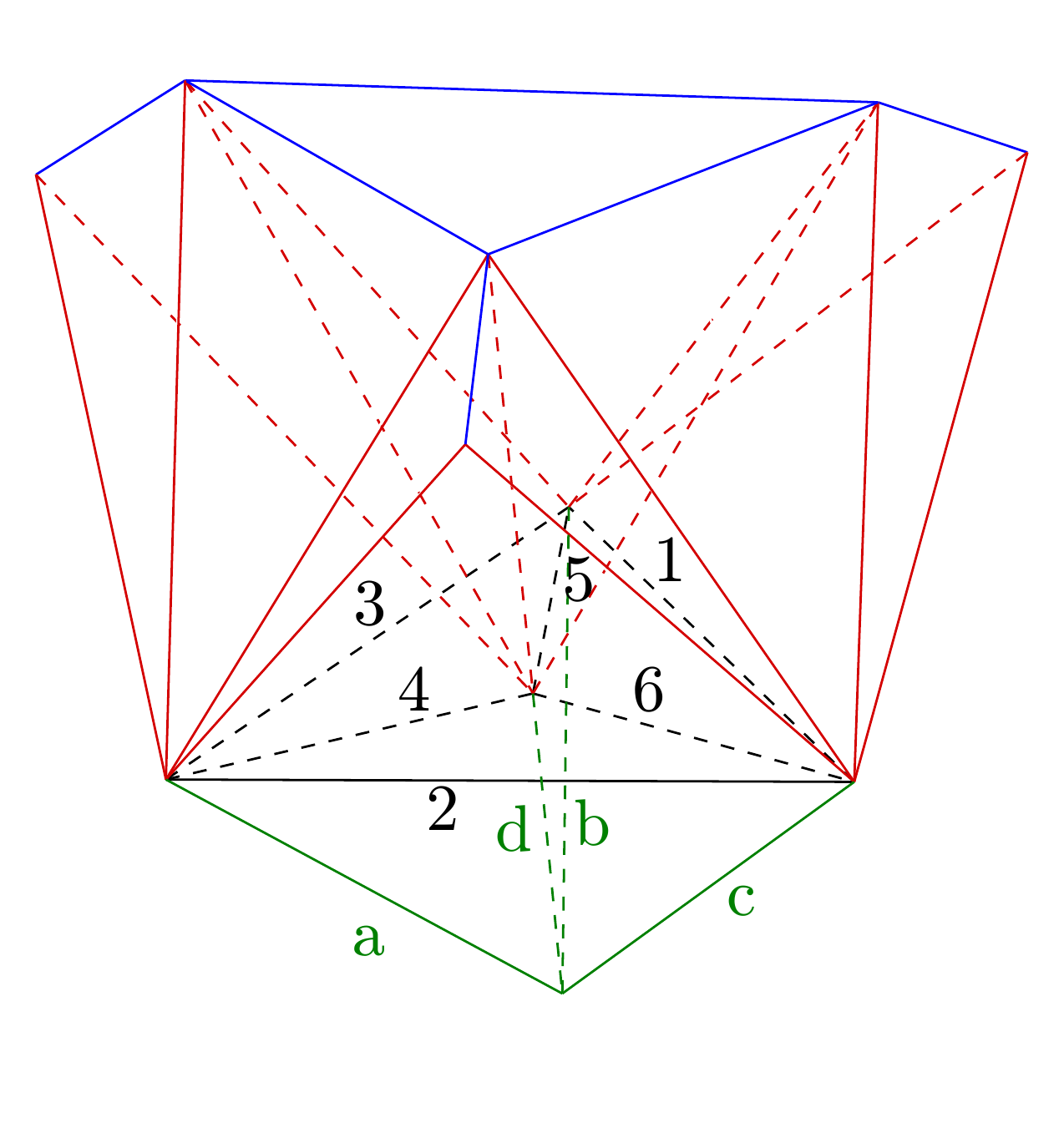}} \eea
and using the 3-2 move on the three tetrahedra connected to the green line labelled $d$, 
\begin{multline}
N^3 Z\circ \m{tiny3-1a} = \left(\frac{1}{2j+1}\right)^3 \sum_{l_4,l_5,l_6}(2l_4+1)(2l_5+1)(2l_6+1) \times \\
\times \wsj{l_1}{l_5}{l_6}{j}{j}{j} \wsj{l_3}{l_4}{l_5}{j}{j}{j} \wsj{l_2}{l_4}{l_6}{j}{j}{j} \wsj{l_1}{l_2}{l_3}{l_4}{l_5}{l_6} \wsj{l_1}{l_2}{l_3}{a}{b}{c}
\end{multline}
We perform the sum over the label $l_6$ using the Biedenharn-Eliot identity,
\bea \sum_{l_6} (2l_6+1)\wsj{l_1}{l_5}{l_6}{j}{j}{j} \wsj{l_2}{l_4}{l_6}{j}{j}{j} \wsj{l_1}{l_2}{l_3}{l_4}{l_5}{l_6} = \wsj{l_1}{l_2}{l_3}{j}{j}{j} \wsj{l_3}{l_4}{l_5}{j}{j}{j} \eea
and use the orthogonality of the $6j$s,
\bea  \sum_{l_4l_5}(2l_4+1)(2l_5+1) \wsj{l_3}{l_4}{l_5}{j}{j}{j} \wsj{l_3}{l_4}{l_5}{j}{j}{j} = (-)^{2j}(2j+1) \Delta(l_3,j,j)  \eea
to deduce that 
\bea N^3 Z\circ \m{tiny3-1a} = \frac{(-)^{2j}}{(2j+1)^2} \wsj{l_1}{l_2}{l_3}{j}{j}{j}  \wsj{l_1}{l_2}{l_3}{a}{b}{c}\nn  \eea
\bea = (2j+1)\quad Z \left[ \raisebox{-0.15\textwidth}{\includegraphics[width=0.25\textwidth]{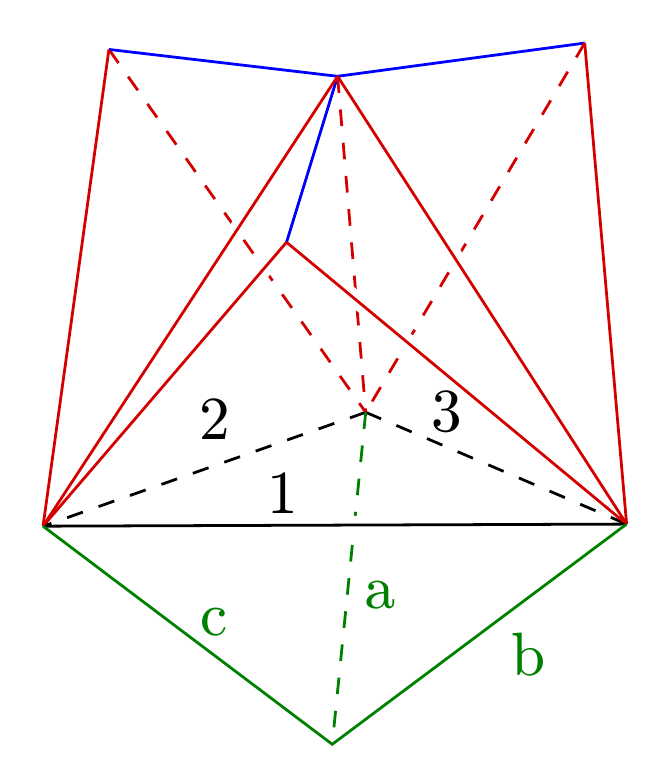}} \right] \eea
Hence
\bea N^3 Z\circ \m{tiny3-1a} =  N  Z\circ \m{tiny3-1b}. \eea 

\subsection{The non-planar 2-2 and 3-1 moves}\label{sec:nplanarmove}

In this section, we show that the partition functions of handlebodies generated from ribbon graphs related by a trivalent
graph move are equal. 

For the 2-2 move, we take two Belyi 3-complexes that are identical except at several tetrahedra on and near their boundary.
They have the partition functions

\bea N^2Z \left[ \raisebox{-0.1\textheight}{\includegraphics[height=0.2\textheight]{npmove1}} \right] \quad 
= (-)^{2j}(2j+1)\sum_{\substack{l_1l_2l_3\\ l_4l_5l_6}} (2l_1+1)(2l_2+1)(2l_3+1)\times \ret \times (2l_4+1)(2l_5+1) \wsjj{l_1}{l_2}{l_3}\wsjj{l_3}{l_4}{l_5} {\bar Z}(l_1,l_2,l_3,l_4,l_5), \qquad \eea
\bea \label{eq:npap2} N^2Z \left[ \raisebox{-0.1\textheight}{\includegraphics[height=0.2\textheight]{npmove2}} \right] \quad 
= (-)^{2j}(2j+1) \sum_{\substack{l_1l_2l_3\\ l_4l_5}} (2l_1+1)(2l_2+1)(2l_3+1)\times \ret \times (2l_4+1)(2l_5+1)(2l_6+1) \wsjj{l_2}{l_4}{l_6}\wsjj{l_1}{l_5}{l_6}\wsj{l_1}{l_2}{l_3}{l_4}{l_5}{l_6}{\bar Z}(l_1,l_2,l_3,l_4,l_5),  \qquad \eea
where ${\bar Z}(l_1,l_2,l_3,l_4,l_5)$ is the partition function of the remainder of the complex not shown in the diagram. 
Since both complete complexes are Belyi 3-complexes, they contain the red and blue edges of an outer Belyi triangulation 
on their boundary. By noting that the blue edges trace out two trivalent graphs, we see that
the two Belyi triangulated 3-complexes are associated to ribbon graphs related by a 2-2 trivalent graph move.

We can perform the sum over $l_6$ in the second partition function using the Biedenharn-Eliot identity
\bea\label{eq:ashortcut} \sum_{l_6}(2l_6+1) \wsjj{l_2}{l_4}{l_6}\wsjj{l_1}{l_5}{l_6}\wsj{l_1}{l_2}{l_3}{l_4}{l_5}{l_6} =  \wsjj{l_1}{l_2}{l_3}\wsjj{l_3}{l_4}{l_5} \eea
and after plugging this identity into \ref{eq:npap2}, we can deduce that the partition functions of the two complexes are equal.

Similarly, for the 3-1 move, we can consider two Belyi 3-complexes that differ only on and near their boundary, with
partition functions

\begin{multline}\label{eq:athree} N^3Z \left[ \raisebox{-0.1\textheight}{\includegraphics[height=0.2\textheight]{npmove3}} \right] \quad  
= (-)^{2j} \sum_{\substack{l_1l_2l_3\\ l_4l_5l_6}} (2l_1+1)(2l_2+1)(2l_3+1)(2l_4+1)\times \\ \times (2l_5+1)(2l_6+1)
\wsjj{l_1}{l_5}{l_6} \wsjj{l_2}{l_4}{l_6} \wsjj{l_3}{l_4}{l_5} \wsj{l_1}{l_2}{l_3}{l_4}{l_5}{l_6}{\bar Z}(l_1,l_2,l_3), \qquad \end{multline}
\begin{multline} N Z \left[ \raisebox{-0.1\textheight}{\includegraphics[height=0.2\textheight]{npmove4}} \right] \quad  
= (2j+1) \sum_{l_1l_2l_3} (2l_1+1)(2l_2+1)(2l_3+1) \wsjj{l_1}{l_2}{l_3}{\bar Z}(l_1,l_2,l_3). \qquad \end{multline}
We can again apply the identity \ref{eq:ashortcut} to sum out the label $l_6$ from \refb{eq:athree}, resulting in 
\begin{multline}\label{eq:athree2} N^3Z \left[ \raisebox{-0.1\textheight}{\includegraphics[height=0.2\textheight]{npmove3}} \right] \quad  
= (-)^{2j} \sum_{\substack{l_1l_2l_3\\ l_4l_5}} (2l_1+1)(2l_2+1)(2l_3+1)\wsjj{l_1}{l_2}{l_3} \times \\ \times (2l_4+1)(2l_5+1)
 \wsjj{l_3}{l_4}{l_5}  \wsjj{l_3}{l_4}{l_5} {\bar Z}(l_1,l_2,l_3). \qquad \end{multline}
 Finally, we apply the orthogonality relation \refb{eq:sixj2} on the sum over the labels $l_4$ and $l_5$ to deduce that the partition functions are equal.

\subsection{Summary of the trivalent graph and Pachner moves}\label{sec:allmoves}

\begin{figure}[H]
\centering
\includegraphics[width=1\textwidth]{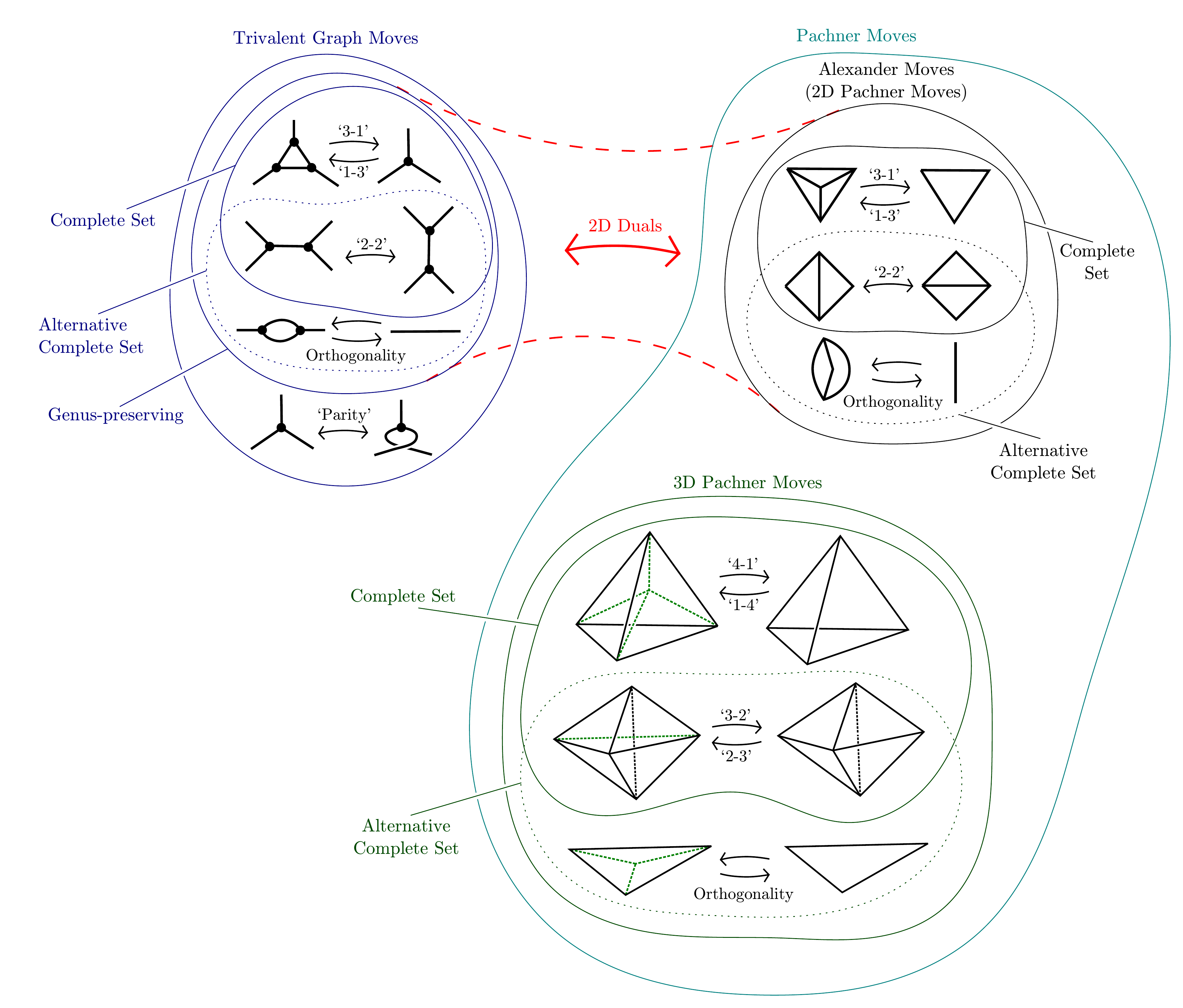}
\end{figure}
{\bf Trivalent graph moves} are operations that relate trivalent ribbon graphs on 2D surfaces. {\bf Genus-preserving trivalent graph moves} do not change the genus of the ribbon graph. A {\bf complete set} of genus-preserving trivalent graph moves are sufficient to relate any two trivalent graphs of the same genus by a series of moves. An {\bf alternative complete set} of the genus-preserving moves is also given. 

{\bf Pachner moves} are operations that relate triangulations of compact manifolds. Any two triangulations of
a compact manifold can be related to each other by a series of Pachner moves.

{\bf 2D Pachner moves} are sometimes called {\bf Alexander moves}, and can relate any two triangulations
of a compact surface. A {\bf complete set} of Alexander moves is sufficient to relate any two triangulations 
of a surface of the same genus (oriented surfaces of the same genus are homeomorphic).
An {\bf alternative complete set} is also given. The genus-preserving moves on ribbon graphs are {\bf dual} to the
Alexander moves.

The {\bf 3D Pachner moves} can relate any two triangulations of a compact 3-manifold with boundary,
provided that the boundary triangulations are identical. A {\bf complete set} of 3D Pachner moves is 
given that is sufficient to relate any two triangulations 
of the same 3-manifold.  An {\bf alternative complete set} is also given. (The orthogonality 3D Pachner move reduces a pair of tetrahedra, glued together on three faces, to a single triangle.)

\end{document}